\documentclass[12pt]{report}
\usepackage{psfig,cite}

\def\be{\begin{equation}}
\def\ee{\end{equation}}
\def\bea{\begin{eqnarray}}
\def\eea{\end{eqnarray}}
\def\ap{\alpha_{I\!\!P}}
\def\pom{I\!\!P}
\def\pro{\mbox{\scriptsize p}}
\def\max{\mbox{\scriptsize max}}
\def\sca{\mbox{\scriptsize scalar}}
\def\loc{\mbox{\scriptsize loc}}
\def\cl{\mbox{\scriptsize cl}}
\def\aem{\alpha_{\mbox{\scriptsize em}}}
\def\xbj{x_{\mbox{\scriptsize Bj}}}
\def\ds{\displaystyle}
\def\x{x_\perp}
\def\y{y_\perp}
\def\sc{\scriptsize}
\def\qqg{$q {\bar q}g$}
\def\qq{$q {\bar q}$}
\def\pp{$p_{\perp}$}
\def\nn{\nonumber \\}
\def\lsim{\mbox{\raisebox{-.5mm}{$\,\stackrel{<}{\scriptstyle\sim}\,$}}}
\newcommand{\mst}[2]{\mbox{\raisebox{-1mm}{$\,\stackrel{#1}{\scriptstyle 
#2}\,$}}}
\def\sb{\sqrt{{}\,\,\,\mbox{\raisebox{0cm}[.3cm][0cm]{}}}\!\!\!\!\beta}
\newcommand{\ks}{k\!\!\!/}
\newcommand{\As}{A\!\!\!/}
\newcommand{\ls}{l\!\!/}
\newcommand{\as}{a\!\!\!/}
\newcommand{\qs}{q\!\!\!/}
\newcommand{\ppl}{p\!\!/}
\newcommand{\epsilons}{\epsilon\!\!/}
\newcommand{\ri}{\mbox{\footnotesize I}}

\newcommand{\gnufig}[3]{
\begin{figure}[t]
\begin{center}
{#3}
\end{center}
\vspace*{.2cm}
\refstepcounter{figure}
\label{fig:#1}
{\bf Figure \ref{fig:#1}:}{#2}
\end{figure}}

\makeatletter
\newcommand\mychapter{\vspace*{1cm}\@startsection{chapter}{1}{\z@}%
                                   {-3.5ex \@plus -1ex \@minus -.2ex}%
                                   {4.3ex \@plus.2ex}%
                                   {\reset@font\huge\bfseries}}
\makeatother

\begin{document}

\vspace*{.5cm}
{\large HD-THEP-99-12\hfill May 1999}
\vspace*{2.8cm}

\begin{center}
{\Large\bf Diffraction in Deep Inelastic Scattering}\\[2.5cm]
{\large A. Hebecker}\\[.5cm]
{\it Institut f\"ur Theoretische Physik der Universit\"at Heidelberg\\
Philosophenweg 16, 69120 Heidelberg, Germany}\\[2.1cm]

{\bf Abstract}\end{center}
\noindent
Different theoretical methods used for the description of diffractive 
processes in \linebreak small-$x$ deep inelastic scattering are reviewed. 
The semiclassical approach, where a partonic fluctuation of the incoming 
virtual photon scatters off a superposition of target colour fields, is 
used to explain the basic physical effects. In this approach, diffraction 
occurs if the emerging partonic state is in a colour singlet, thus 
fragmenting independently of the target. Other approaches, such as the idea 
of the pomeron structure function and two gluon exchange calculations, are 
also discussed in some detail. Particular attention is paid to the close 
relation between the semiclassical approach and the method of diffractive 
parton distributions, which is linked to the relation between the target 
rest frame and the Breit frame point of view. While the main focus is on 
diffractive structure functions, basic issues in the diffractive production 
of mesons and of other less inclusive final states are also discussed. 
Models of the proton colour field, which can be converted into predictions 
for diffractive cross sections using the semiclassical approach, are 
presented. The concluding overview of recent experimental results is very 
brief and mainly serves to illustrate implications of the theoretical 
methods presented. 
\vspace*{2cm}

\setcounter{page}{1}
\thispagestyle{empty}
\newpage

${\mbox{}}$
\thispagestyle{empty}
\newpage

\tableofcontents
\newpage

\mychapter{Introduction}
\section{Preface}
The term diffraction is derived from optics, where it describes the 
deflection of a beam of light and its decomposition into components with 
different frequencies. In high energy physics it was originally used for 
small-angle elastic scattering of hadrons. If one of the hadrons, say the 
projectile, is transformed into a set of two or more final state particles, 
the process is called diffractive dissociation or inelastic diffraction. 
Good and Walker have pointed out that a particularly intuitive physical 
picture of such processes emerges if the projectile is described as a 
superposition of different components which scatter elastically off the 
target~\cite{gw}. Since the corresponding elastic amplitude is different 
for each component, the outgoing beam will contain a new superposition of 
these components and therefore, in general, new physical states. These are 
the dissociation products of the projectile. 

Even at very high energy, the above processes are generically soft, i.e., 
the momentum transfer is small and the dissociation products have small 
$p_\perp$. Therefore, no immediate relation to perturbative QCD is apparent. 

By contrast, diffractive jet production, observed at the CERN S$p\bar{p}$S 
collider in proton-antiproton collisions~\cite{ua8}, involves a hard scale. 
Although one of the hadrons escapes essentially unscathed, a high-$p_\perp$ 
jet pair, which is necessarily associated with a high virtuality in the 
intermediate states, is produced in the central rapidity range. The cross 
section of the process is parametrically unsuppressed relative to 
non-diffractive jet production. This seems to contradict a na\"\i ve partonic 
picture since the colour neutrality of the projectile is destroyed if one 
parton is removed to participate in the hard scattering. The interplay of 
soft and hard physics necessary to explain the effect provides one of the 
main motivations for the study of these `hard diffractive' processes. 

The present review is focussed on diffraction in deep inelastic scattering 
(DIS), which is another example of a hard diffractive process. This process 
became experimentally viable with the advent of the electron-proton 
collider HERA, where DIS at very small values of the Bjorken variable $x$ 
can be studied. In the small-$x$ or high-energy region, a significant 
fraction of the observed DIS events have a large rapidity gap between the 
photon and the proton fragmentation region~\cite{rg1,rg2}. In contrast to 
the standard DIS process $\gamma^*p\to X$, the relevant reaction reads 
$\gamma^*p\to XY$, where $X$ is a high-mass hadronic state and $Y$ is the 
elastically scattered proton or a low-mass excitation of it. Again, these 
events are incompatible with the na\"\i ve picture of a partonic target 
and corresponding simple ideas about the colour flow. Na\"\i vely, the 
parton struck by the virtual photon destroys the colour neutrality of the 
proton, a colour string forms between struck quark and proton remnant, and 
hadronic activity is expected throughout the detector. Nevertheless, the 
observed diffractive cross section is not power suppressed at high 
virtualities $Q^2$ with respect to standard DIS. 

The main theoretical interest is centered around the interplay of soft and 
hard physics represented by the elastic or almost elastic scattering of the 
proton and the scattering of the highly virtual photon respectively. 
Diffractive DIS is much simpler than hard diffraction in hadronic reactions 
since only one non-perturbative object is involved. In a large fraction of 
the events, the momentum transfer to the proton is very small. Therefore, 
one can hope to gain a better understanding of the bound state dynamics of 
the proton by studying diffractive DIS. Inclusive reactions of the virtual 
photon with hadronic targets are well-studied theoretically and constrained 
by a large amount of DIS data. It is a challenge to utilize this knowledge 
for the investigation of the non-perturbative dynamics of diffraction and, 
thereby, of the proton structure. 

The paper is organized as follows. In the remainder of the Introduction, 
different approaches to diffractive DIS are put into historical and 
physical perspective. The semiclassical model, which is particularly close 
to the interests of the author, is emphasized. Chapter~\ref{sect:basic} is 
concerned with the fundamental observations, the basic concepts required 
for their understanding, and the necessary kinematic considerations. The 
semiclassical approach, which is used as the starting point for the 
discussion of other models, is introduced in Chapter~\ref{sect:sc}. In 
Chapter~\ref{sect:sp}, the concepts of soft pomeron, triple pomeron vertex 
and pomeron structure function are discussed. Diffractive parton 
distributions, which are the more fundamental objects from the point of 
view of perturbative QCD, are introduced. The semiclassical approach is put 
in relation to the method of diffractive parton distributions, and explicit 
formulae for these quantities are derived. In Chapter~\ref{sect:tge}, two 
gluon exchange calculations are discussed. The emphasis is on their 
rigorous validity in specific kinematic situations and on their partial 
correspondence to the semiclassical approach. Chapter~\ref{sect:mod} 
introduces three models for the colour field of the proton, which can be 
converted into predictions for diffractive cross sections using the methods 
of Chapters~\ref{sect:sc} and~\ref{sect:sp}. A discussion of recent 
experimental results and their description by theoretical models is given 
in Chapter~\ref{sect:exp}, followed by the Conclusions.

\section{Models for diffraction}
The claim that diffraction in DIS should be a leading twist effect, and the 
understanding of the fundamental mechanism underlying such processes can be 
traced back to the famous paper of Bjorken and Kogut~\cite{bk}. Their 
argument is based on a qualitative picture of DIS in the target rest frame, 
where the incoming virtual photon can be considered as a superposition of 
partonic states. The large virtuality $Q^2$ sets the scale, so that states 
with low-$p_\perp$ partons, i.e., aligned configurations, are suppressed in 
the photon wave function. However, in contrast to high-$p_\perp$ 
configurations, these aligned states have a large hadronic interaction 
cross section with the proton. Therefore, their contribution to the DIS 
cross section is expected to be of leading twist. Naturally, part of this 
leading twist contribution is diffractive since the above low-$p_\perp$ 
configurations represent transversely extended, hadron-like objects, which 
have a large elastic cross section with the proton.

Note that the basic technical methods date back even further, namely, to the 
calculation of $\mu^+\mu^-$ pair electroproduction off an external 
electromagnetic field by Bjorken, Kogut and Soper~\cite{bks}. They derive 
the transition amplitude of the incoming virtual photon into a $\mu^+\mu^-$ 
pair, which corresponds to the $q\bar{q}$ wave function of the virtual 
photon employed by Nikolaev and Zakharov in their seminal work on 
diffraction at HERA~\cite{wf}.

The above physical picture, commonly known as the aligned jet model of 
diffractive and non-diffractive DIS, can naturally be extended to account 
for a soft energy growth of both processes. (Note that we are dealing with 
the underlying energy dependence of the soft scattering process, not with 
the additional $x$-dependence induced by Altarelli-Parisi evolution, that is 
added on top of it.) Since a large sample of hadronic cross sections can be 
consistently parametrized using the concept of the Donnachie-Landshoff or 
soft pomeron~\cite{dl}, it is only natural to assume that this concept can 
also be used to describe the energy dependence of the interaction of the 
aligned jet component of the small-$x$ virtual photon and the target. This 
gives rise to the desired non-trivial energy dependence although, as will 
be discussed in more detail later on, this energy dependence is not 
sufficiently steep in the diffractive case. Note, however, that the above 
soft energy dependence may also be logarithmic, which has the advantage of 
explicit consistency with unitarity at arbitrarily high energies. 

A more direct way of applying the concept of the soft pomeron to the 
phenomenon of hard diffraction was suggested by Ingelman and Schlein in 
the context of diffractive jet production in hadronic collisions~\cite{is}. 
Their idea of a partonic structure of the pomeron, which can be tested in 
hard processes, applies to the case of diffractive DIS as well~\cite{dl1}. 
Essentially, one assumes that the pomeron can, like a real hadron, be 
characterized by a parton distribution. This distribution factorizes from 
the pomeron trajectory and the pomeron-proton vertex, which are both 
obtained from the analysis of purely soft hadronic reactions. The 
above non-trivial assumptions are often referred to as `Regge hypothesis' or 
`Regge factorization'. 

The Ingelman-Schlein approach described above is based on the intuitive 
picture of a pomeron flux associated with the proton beam and on the 
conventional partonic description of the pomeron photon collision. In the 
limit where not only the total proton-photon center-of-mass energy but also 
the energy of the pomeron-photon collision becomes large, the concept of 
the triple pomeron vertex can be applied~\cite{dl2}. At HERA, this limit 
corresponds to rapidity gap events with very large diffractive masses. 

The concept of fracture functions of Veneziano and Trentadue~\cite{vt} or, 
more specifically, the diffractive parton distributions of Berera and Soper 
\cite{bs} provide a framework for the study of diffractive DIS that is 
firmly rooted in perturbative QCD. Loosely speaking, diffractive parton 
distributions describe the probability of finding, in a fast moving proton, 
a parton with a certain momentum fraction $x$, under the additional 
requirement that the proton remains intact losing only a certain fraction of 
its momentum. This idea is closely related to the concept of a partonic 
pomeron described above, but it gives up the Regge hypothesis, thus being 
less predictive. This weaker factorization assumption is sometimes referred 
to as `diffractive factorization', as opposed to the stronger Regge 
factorization assumption. 

An essential feature of diffractive DIS is the colour singlet exchange in 
the $t$ channel, necessary to preserve the colour neutrality of the target. 
None of the above approaches address this requirement explicitly 
within the framework of QCD. Instead, the colour singlet exchange is 
postulated by assuming elastic scattering in the aligned jet model, by 
using the concept of the pomeron, or by defining diffractive parton 
distributions. If one insists on using the fundamental degrees of freedom 
of QCD, the simplest possibility of realizing colour singlet exchange at 
high energy is the exchange of two gluons~\cite{low}. One could say that two 
gluons form the simplest model for the pomeron. 

The fundamental problem with this approach is the applicability of 
perturbative QCD to the description of diffractive DIS. As will be 
discussed in more detail below, the hard scale $Q^2$ of the photon does 
not necessarily justify a perturbative description of the $t$ channel 
exchange with the target. 

Diffractive processes where the $t$ channel colour singlet exchange is 
governed by a hard scale include the electroproduction of heavy vector 
mesons~\cite{rys}, electroproduction of light vector mesons in the case of 
longitudinal polarization~\cite{bro} or at large $t$~\cite{fr}, and virtual 
Compton scattering~\cite{mea,ji,rad}. In the leading logarithmic 
approximation, the relevant two-gluon form factor of the proton can be 
related to the inclusive gluon distribution~\cite{rys}. Accordingly, a very 
steep energy dependence of the cross section, which is now proportional to 
the square of the gluon distribution, is expected from the known steep 
behaviour of small-$x$ structure functions. The origin of this steep rise 
itself may be attributed to a combination of the $x$-dependence of the 
input distributions and their Altarelli-Parisi evolution, or to the BFKL 
resummation of large logarithms of $x$. 

To go beyond leading logarithmic accuracy, the non-zero momentum transferred 
to the proton has to be taken into account. This requires the use of 
`non-forward' or `off-diagonal' parton distributions (see~\cite{mea,dit} and 
refs. therein), which were discussed in~\cite{ji,rad} within the 
present context. Although their scale dependence is predicted by well-known 
evolution equations, only limited information about the relevant input 
distributions is available. In particular, the simple proportionality to 
the square of the conventional gluon distribution is lost. 

The perturbative calculations of meson electroproduction discussed above 
were put on a firmer theoretical basis by the factorization proof of 
\cite{cfs1}, where the conditions required for the applicability of 
perturbation theory are discussed in detail. As shown explicitly in the 
simple model calculation of~\cite{hl}, QCD gauge invariance ensures that 
the non-perturbative meson formation process takes place {\it after} the 
scattering off the hadronic target. 

The situation becomes even more complicated if two gluon exchange 
calculations are applied to more general diffractive final states. Examples 
are the exclusive electroproduction of heavy quark pairs~\cite{ccnz,ccle,
cclo,ccd} or high-$p_\perp$ jets~\cite{nz,di,blw,pt1}. A straightforward 
perturbative analysis shows that the virtual photon side of the process is 
dominated by small transverse distances, establishing two gluon exchange as 
the dominant mechanism. However, it is not obvious whether the very 
definition `exclusive', meaning exclusive on a partonic level, can be 
extended to all orders in perturbation theory. As will become clear later 
on, additional soft partons in the final state may destroy the hardness 
provided by the large~$p_\perp$ or the heavy quark mass. 

From the point of view of the full diffractive cross section, perturbative 
two gluon exchange corresponds to a higher twist effect. However, it can be 
argued that it dominates the kinematic region where the mass $M$ of the 
diffractively produced final state $X$ is relatively small, $M\ll 
Q$~\cite{ht}. More generally, higher twist effects in diffraction, 
their calculability and possibilities for their experimental observation 
are discussed in~\cite{ht,bert,bekw}. 

Eventually it is possible to attempt the description of the full cross 
section of diffractive DIS within the framework of two gluon exchange. 
In such a general setting, perturbation theory can not be rigorously 
justified and two gluons are basically used as a model for the soft colour 
singlet exchange in the $t$ channel. Nevertheless, it is interesting to 
see to what extent concepts and formulae emerging from perturbation theory 
can describe a wider range of phenomena. In particular, much attention has 
been devoted to the energy dependence of diffractive cross sections both in 
the framework of conventional BFKL summation of gluon 
ladders~\cite{bwtr,blwtr} and within the colour dipole approach to 
small-$x$ resummation~\cite{nz,bptp}. 

An apparently quite different approach emerged with the idea that soft 
colour interactions might be responsible for the large diffractive cross 
section at HERA. The starting point is Buchm\"uller's observation of the 
striking similarity between $x$ and $Q^2$ dependence of diffractive and 
inclusive DIS at small $x$~\cite{b}. It is then natural to assume that the 
same hard partonic processes underlie both cross sections and that 
differences in the final states are the result of non-perturbative soft 
colour exchange~\cite{bh1,eir}. 

In Ref.~\cite{bh1}, boson-gluon fusion was proposed as the dominant partonic 
process, and diffraction was claimed to be the result of soft colour 
neutralization of the produced $q\bar{q}$ pair. A simple statistical 
assumption about the occurrence of this colour neutralization lead to a 
surprisingly good fit to the data. Note that similar ideas concerning the 
rotation of quarks in colour space had previously been discussed by 
Nachtmann and Reiter~\cite{nr} in connection with QCD vacuum effects on 
hadron-hadron scattering. 

A closely related approach was introduced in~\cite{eir}, where the 
assumption of soft colour neutralization was implemented in a Monte Carlo 
event generator based on perturbation theory. Normally, the partonic 
cascades underlying the Monte Carlo determine the colour of all partons 
produced and, with a certain model dependence, the hadronic final state. 
The introduction of an ad-hoc probability for partons within the cascade to 
exchange colour in a non-perturbative way leads to a significant increase 
of final states with rapidity gaps, resulting in a good description of the 
observed diffractive events. 

Both of the above models share a fundamental theoretical problem: the soft 
colour exchange is introduced in an ad-hoc manner, independently of 
kinematic configuration or space-time distances of the partons involved. 
If all relevant distances are large, our ignorance of the true mechanism of 
non-perturbative interactions justifies the na\"\i ve assumption of random 
colour exchange. However, in many other situations, e.g., if a small colour 
dipole is involved, colour exchange is suppressed in a perturbatively 
calculable way. This is known as colour transparency. The above simple soft 
colour models do not properly account for this effect. One possibility of 
incorporating the knowledge of the perturbative aspects of QCD while keeping 
the essential idea of soft colour exchange as the source of leading twist 
diffraction is the semiclassical treatment.

\section{The semiclassical method and its relation to other approaches} 
The semiclassical approach to diffractive electroproduction evolved from the 
attempt to justify the phenomenologically successful idea of soft colour 
exchange~\cite{bh1,eir} within the framework of QCD. The starting point is 
the proton rest frame picture of DIS advocated long ago in~\cite{bk} and 
developed since by many authors (see in particular~\cite{wf}). To describe 
diffraction within this framework, the scattering of energetic partons in 
the photon wave function off the hadronic target needs to be understood. 
While the scattering of small transverse size configurations is calculable 
perturbatively, i.e., via two gluon exchange, different models may be 
employed for larger size configurations. A rather general framework is 
provided by the concept of the dipole cross section $\sigma(\rho)$ 
utilized in the analysis of~\cite{wf}. 

It is helpful to begin by recalling the discussion of high-energy 
electroproduction of $\mu^+\mu^-$ pairs off atomic targets given by 
Bjorken, Kogut and Soper in~\cite{bks}. There, the closely related QED 
problem was solved by treating the high-energy scattering of $\mu^+\mu^-$ 
pairs off a given electromagnetic field in the eikonal approximation. An 
analogous QCD process was considered by Collins, Soper and Sterman in 
\cite{css}, where the production of heavy quark pairs in an external colour 
field was calculated. In Ref.~\cite{nac}, Nachtmann developed the idea of 
eikonalized interactions with soft colour field configurations as a method 
for the treatment of elastic high-energy scattering of hadrons. 

The semiclassical approach to diffraction, introduced in~\cite{bh2}, 
combines the concepts and methods outlined above. The proton is modelled by 
a soft colour field, and the interactions of the fast partons in the photon 
wave function with this field are treated in the eikonal approximation. As a 
result, the non-perturbative proton structure is encoded in a combination 
of non-Abelian phase factors associated with the partons. In the simplest 
case of a $q\bar{q}$ fluctuation, two such phase factors are combined in a 
Wegner-Wilson loop, which carries all the information about the target. 
Since the outgoing $q\bar{q}$ pair can be either in a colour singlet or 
in a colour octet state, both diffractive and non-diffractive events are 
naturally expected in this approach. Thus, both types of events are 
described within the same framework. The colour state of the produced pair 
is the result of the fundamentally non-perturbative interaction with the 
proton colour field, very much in the spirit of the soft colour 
neutralization of~\cite{bh1}. 

The essential difference between the semiclassical calculation and the soft 
colour proposal is the recognition that the possibility of soft colour 
exchange is intimately related to parton level kinematics. The 
semiclassical approach allows for a consistent treatment of the scattering 
of both small and large transverse size configurations. The requirement of 
colour neutrality in the final state of diffractive events is explicitly 
shown to suppress the former ones. This confirms the familiar qualitative 
arguments on which the aligned jet model is based~\cite{bk}. 

A further essential step is the inclusion of higher Fock states in the 
photon wave function. In the framework of two gluon exchange, corresponding 
calculations for the $q\bar{q}g$ state were performed in~\cite{nz}. The 
semiclassical analysis of the $q\bar{q}g$ state~\cite{bhm} demonstrates 
that, to obtain leading twist diffraction, at least one of the three partons 
has to have small $p_\perp$ and has to carry a small fraction of the 
photon's longitudinal momentum. This is a natural generalization of the 
aligned jet model, which was previously discussed on a qualitative level in 
\cite{afs}. 

A fundamental prediction derived from the semiclassical treatment of the 
$q\bar{q}g$ state is the leading twist nature of diffractive 
electroproduction of heavy quark pairs~\cite{bhmcc} and of high-$p_\perp$ 
jets~\cite{bhmpt}. These large cross sections arise from $q\bar{q}g$ 
configurations where the gluon is relatively soft and has small $p_\perp$, 
so that the $t$ channel colour singlet exchange remains soft in spite of the 
additional hard scale provided by the $q\bar{q}$ pair. 

As mentioned previously, considerable work has been done attempting to 
employ the additional hardness provided by final states with jets or heavy 
quarks in order to probe mechanisms of perturbative colour singlet exchange 
\cite{ccnz,ccle,cclo,ccd,nz,di,blw,pt1}. In fact, the semiclassical approach 
is also well suited to follow this line of thinking. As demonstrated in 
\cite{bhmpt}, the features of diffractive jet production from $q\bar{q}$ 
and $q\bar{q}g$ fluctuations of the photon, both treated consistently in 
the semiclassical framework, are very different. In the $q\bar{q}$ case, a 
small-size colour dipole tests the small distance structure of the proton 
colour field. It can be shown within the semiclassical framework that the 
relevant operator is the same as in inclusive DIS, thus reproducing the 
well-known relation with the square of the gluon distribution~\cite{rys}. 
In the $q\bar{q}g$ case, the additional soft gluon allows for a 
non-perturbative colour neutralization mechanism even though the final 
state contains two high-$p_\perp$ jets. 

The semiclassical treatment of leading twist diffraction is equivalent to a 
treatment based on diffractive parton distributions~\cite{h}. To be more 
precise, the results of~\cite{bhm} lead to the conclusion that a leading 
twist contribution to diffraction can arise only from virtual photon 
fluctuations that include at least one soft parton. Starting from this 
premise, the diffractive production of final states with one soft parton 
and a number of high-$p_\perp$ partons was considered in~\cite{h}. It was 
shown that the cross section can be written as a convolution of a hard 
partonic cross section and a diffractive parton distribution. The latter is 
given explicitly in terms of an average over the target colour field 
configurations underlying the semiclassical calculation. 

From this point of view, the semiclassical prediction of the dominance of 
$q\bar{q}g$ final states and soft colour neutralization in high-$p_\perp$ 
jet and heavy quark production appears natural. It corresponds simply to 
the dominant partonic process, boson-gluon fusion, with the gluon taken from 
the diffractive gluon distribution. The additional gluon in the final state 
is, from the point of view of parton distributions, merely a consequence of 
the preserved colour neutrality of the target. From the point of view of the 
semiclassical calculation, it is the scattering product of the original 
gluon from the photon wave function that is necessary for the softness 
of the colour singlet exchange. 

Even without an explicit model for the proton colour field, the 
semiclassical approach provides an intuitive overall picture of inclusive 
and diffractive small-$x$ DIS and predicts a number of qualitative features 
of the final state. However, the ultimate goal is to develop the 
understanding of non-perturbative colour dynamics that is needed to 
calculate the required field averages. 

A very successful description of a number of hadronic reactions was 
obtained in the framework of the stochastic vacuum. Different effects of 
the non-trivial structure of the QCD vacuum in high-energy hadron-hadron 
scattering were considered in~\cite{nr}. Field strength correlators in a 
theory with vacuum condensate were employed in~\cite{ln,ad} to describe 
the pomeron. The model of the stochastic vacuum was introduced by Dosch 
and Simonov \cite{ds} in the context of gluon field correlators in the 
Euclidean theory. Based on ideas of \cite{ln,ad} and, in particular, on 
Nachtmann's suggestion to describe high-energy hadron-hadron scattering 
in an eikonal approach~\cite{nac}, the stochastic vacuum model was adapted 
to Minkowski space and applied to high-energy processes by Dosch and 
Kr\"amer~\cite{dk}. In the present context of diffractive and 
non-diffractive electroproduction, the work on inclusive DIS~\cite{dr}, 
$C$- and $P$-odd contributions to diffraction~\cite{cp}, and exclusive 
vector meson production~\cite{dgkp,dkp,rue} is particularly relevant. 
Results on diffractive structure functions have recently been 
reported~\cite{ram}. 

In the special case of a very large hadronic target, the colour field 
averages required in the semiclassical approach can be calculated without 
specifying the details of the non-perturbative colour dynamics involved. 
McLerran and Venugopalan observed that the large target size, realized, 
e.g., in an extremely heavy nucleus, introduces a new hard scale into the 
process of DIS~\cite{mv}. From the target rest frame point of view, this 
means that the typical transverse size of the partonic fluctuations of the 
virtual photon remains perturbative~\cite{hw}. Thus, the perturbative 
treatment of the photon wave function in the semiclassical calculation is 
justified. Note that the small size of the partonic fluctuations of the 
photon does not imply a complete reduction to perturbation theory. The long 
distance which the partonic fluctuation travels in the target compensates 
for its small transverse size, thus requiring the eikonalization of gluon 
exchange. 

For a large target it is natural to introduce the additional assumption 
that the gluonic fields encountered by the partonic probe in distant regions 
of the target are not correlated. In this situation, inclusive and 
diffractive DIS cross sections become completely calculable. A corresponding 
analysis of inclusive and diffractive parton distributions was performed in 
\cite{bgh}. Starting from the above large target model, expressions for the 
inclusive quark distribution, which had previously been discussed in a 
similar framework in~\cite{bhq}, and for the inclusive gluon distribution 
were obtained. Diffractive quark and gluon distributions were calculated, 
within the same model for the colour field averaging, on the basis of 
formulae from~\cite{h}. The resulting structure functions, obtained from 
the above input distributions by Altarelli-Parisi evolution, provide a 
satisfactory description of the experimental data~\cite{bgh}. 

A number of further approaches to diffractive DIS were proposed by several 
authors. For example, it was argued in~\cite{whi} that a super-critical 
pomeron with colour-charge parity $C_c=-1$ plays an essential role. As 
a result, the main features of the na\"\i ve boson-gluon fusion model 
of~\cite{bh1} are reproduced while its problems, which were outlined above, 
are avoided. In a different approach, a geometric picture of diffraction, 
based on the idea of colourless gluon clusters, was advertised (see, 
e.g.,~\cite{ber}). There can be no doubt that far more material relevant to 
the present subject exists which, because of the author's bias or 
unawareness, is not appropriately reflected in this introduction or the 
following chapters. 

\newpage

\mychapter{Basic Concepts and Phenomena}\label{sect:basic}
In this chapter, the kinematics of diffractive electroproduction at small 
$x$ is explained in some detail, and the notation conventionally used for 
the description of this phenomenon is introduced. The main experimental 
observations are discussed, and the concept of the diffractive structure 
function, which is widely used in analyses of inclusive diffraction, is 
explained. This standard material may be skipped by readers familiar with 
the main HERA results on diffractive DIS.

\section{Kinematics}
To begin, recall the conventional variables for the description of DIS. 
An electron with momentum $k$ collides with a proton with momentum $P$. 
In neutral current processes, a photon with momentum $q$ and virtuality 
$q^2=-Q^2$ is exchanged, and the outgoing electron has momentum $k'=k-q$. 
In inclusive DIS, no questions are asked about the hadronic final state 
$X_W$, which is only known to have an invariant mass square $W^2=(P+q)^2$. 
The Bjorken variable $x=Q^2/(Q^2+W^2)$ characterizes, in the na\"\i ve 
parton model, the momentum fraction of the incoming proton carried by the 
quark that is struck by the virtual photon. If $x\ll 1$, which is the 
relevant region in the present context, $Q$ is much smaller than the photon 
energy in the target rest frame. In this sense, the photon is almost real 
even though $Q^2\gg\Lambda^2$ (where $\Lambda$ is some soft hadronic scale). 
It is then convenient to think in terms of a high-energy $\gamma^*p$ 
collision with centre-of-mass energy $W$. 

The small-$x$ limit of DIS became experimentally viable only a few years 
ago, with the advent of the HERA accelerator (Hadron-Elektron-Ring-Anlage)
at the DESY accelerator centre in Hamburg. In this machine, 27 GeV electrons 
(or positrons) are collided with 920 GeV protons at interaction points 
inside the H1 and ZEUS detectors. The large centre-of-mass energy of the 
$ep$ collision, $\sqrt{s}\simeq 300$ GeV, allows for a very high hadronic 
energy $W$ and thus for the observation of events with both very small $x$ 
and high $Q^2$. For values of $Q^2$ well in the perturbative domain, a 
statistically usable number of events with $x$ as small as $10^{-4}$ is 
observed. 

\begin{figure}[t]
\begin{center}
\vspace*{.0cm}
\parbox[b]{7.5cm}{\psfig{width=7.5cm,file=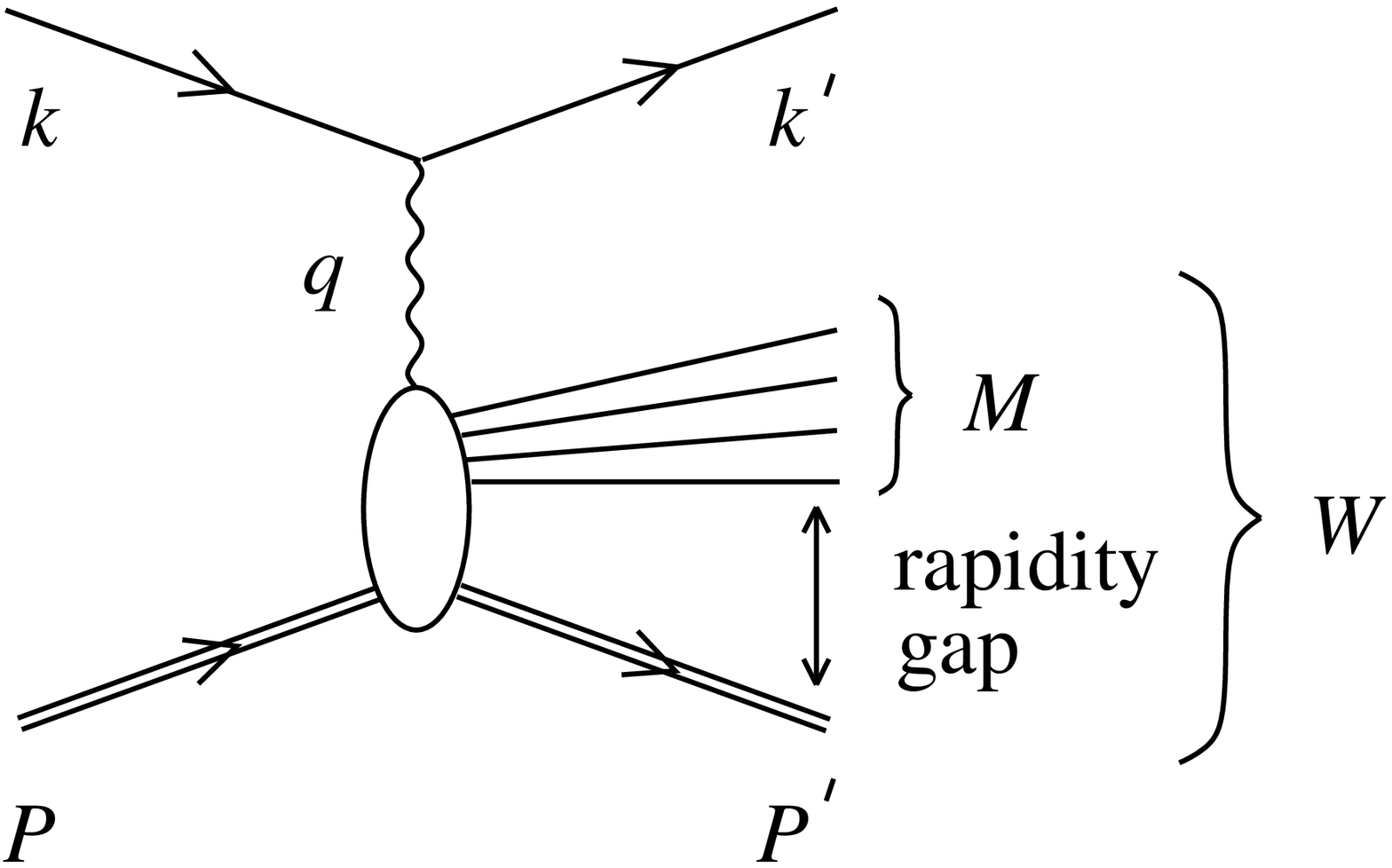}}\\
\end{center}
\refstepcounter{figure}
\label{fig:dep}
{\bf Figure \ref{fig:dep}:} Diffractive electroproduction. The full hadronic 
final state with invariant mass $W$ contains the elastically scattered 
proton and the diffractive state with invariant mass $M$. 
\end{figure}

Loosely speaking, diffraction is the subset of DIS characterized by a 
quasi-elastic interaction between virtual photon and proton. A particularly 
simple definition of diffraction is obtained by demanding that, in the 
$\gamma^*p$ collision, the proton is scattered elastically. Thus, in 
diffractive events, the final state contains the scattered proton with 
momentum $P'$ and a diffractive hadronic state $X_M$ with mass $M$ (see 
Fig.~\ref{fig:dep}). Since diffractive events form a subset of DIS events, 
the total invariant mass of the outgoing proton and the diffractive state 
$X_M$ is given by the standard DIS variable $W$. 

The following parallel description of inclusive and diffractive DIS suggests 
itself. 

In the former, virtual photon and proton collide to form a hadronic 
state $X_W$ with mass $W$. The process can be characterized by the 
virtuality $Q^2$ and the scaling variable $x=Q^2/(Q^2+W^2)$, the momentum 
fraction of the struck quark in the na\"\i ve parton model.

In the latter, a colour neutral cluster of partons is stripped off the 
proton. The virtual photon forms, together with this cluster, a hadronic 
state $X_M$ with mass $M$. The process can be characterized by $Q^2$, as 
above, and by a new scaling variable $\beta=Q^2/(Q^2+M^2)$, the momentum 
fraction of this cluster carried by the struck quark. 

Since diffraction is a subprocess of inclusive DIS, the struck quark from 
the colour neutral cluster also carries a fraction $x$ of the proton 
momentum. Therefore, the ratio $\xi=x/\beta$ characterizes the momentum 
fraction that the proton loses to the colour neutral exchange typical of an 
elastic reaction. This exchanged colour neutral cluster loses a momentum 
fraction $\beta$ to the struck quark that absorbs the virtual photon. As 
expected, the product $x=\beta\xi$ is the fraction of the original proton's 
momentum carried by this struck quark. Since the name pomeron is frequently 
applied to whichever exchange with vacuum quantum numbers dominates the 
high-energy limit, many authors use the notation $x_{\pom}=\xi$, thus 
implying that the proton loses a momentum fraction $\xi$ to the exchanged 
pomeron. 

Therefore, $x$, $Q^2$ and $\beta$ or, alternatively, $x$, $Q^2$ and $\xi$ 
are the main kinematic variables characterizing diffractive DIS. A further 
variable, $t=(P-P')^2$, is necessary if the transverse momenta of the 
outgoing proton and the state $X_M$ relative to the $\gamma^*P$ axis are 
measured. Since the proton is a soft hadronic state, the value of $|t|$ is 
small in most events. The small momentum transferred by the proton also 
implies that $M\ll W$. 

To see this in more detail, introduce light-cone co-ordinates $q_\pm=q_0\pm 
q_3$ and $q_\perp=(q_1,q_2)$. It is convenient to work in a frame where the 
transverse momenta of the incoming particles vanish, $q_\perp=P_\perp=0$. 
Let $\Delta$ be the momentum transferred by the proton, $\Delta=P-P'$, and 
$m_{\pro}^2=P^2=P'^2$ the proton mass squared. For forward scattering, 
$P_\perp'=0$, the relation 
\be
t=\Delta^2=\Delta_+ \Delta_-=-\xi^2m_{\pro}^2
\ee
holds. Since $\xi=(Q^2+M^2)/(Q^2+W^2)$, this means that small $M$ implies 
small $|t|$ and vice versa. Note, however, that the value of $|t|$ is larger 
for non-forward processes, where $t=\Delta_+ \Delta_--\Delta_\perp^2$. 

So far, diffractive events have been characterized as those DIS events 
which contain an elastically scattered proton in their hadronic final 
state. An even more striking feature is the large gap of hadronic activity 
seen in the detector between the scattered proton and the diffractive state 
$X_M$. It will now be demonstrated that this feature, responsible for the 
alternative name `rapidity gap events', is a direct consequence of the 
relevant kinematics. 

Recall the definition of the rapidity $y$ of a particle with momentum 
$k$,
\be
y=\frac{1}{2}\ln\frac{k_+}{k_-}=\frac{1}{2}\ln\frac{k_0+k_3}{k_0-k_3}\,.
\ee
This is a convenient quantity for the description of high-energy collisions 
along the $z$-axis. Massless particles moving along this axis have rapidity 
$-\infty$ or $+\infty$, while all other particles are characterized by some 
finite intermediate value of $y$ (for a detailed discussion of the role 
of rapidity in the description of diffractive kinematics see, e.g., 
\cite{cr}.) 

In the centre-of-mass frame of the $\gamma^*p$ collision, with the $z$-axis 
pointing in the proton beam direction, the rapidity of the incoming 
proton is given by $y_{\pro}=\ln(P_+/m_{\pro})$. At small $\xi$, the 
rapidity of the scattered proton is approximately the same. This is to be 
compared with the highest rapidity $y_{\max}$ of any of the particles in the 
diffractive state $X_M$. Since the total plus component of the 4-momentum
of $X_M$ is given by $(\xi-x)P_+$, and the pion, with mass $m_\pi$, is the 
lightest hadron, none of the particles in $X_M$ can have a rapidity above 
$y_{\max}=\ln((\xi-x)P_+/m_\pi)$. Thus, a rapidity gap of size $\Delta y=\ln 
(m_\pi/(\xi-x)m_{\pro})$ exists between the outgoing proton and the state 
$X_M$. For typical values of $\xi \sim 10^{-3}$ (cf. the more detailed 
discussion of the experimental setting below) the size of this gap can be 
considerable. 

Note, however, that the term `rapidity gap events' was coined to describe the 
appearance of diffractive events in the HERA frame, i.e., a frame defined by 
the electron-proton collision axis. The rapidity in this frame is, in 
general, different from the photon-proton frame rapidity discussed above. 
Nevertheless, the existence of a gap surrounding the outgoing proton in the 
$\gamma^*p$ frame clearly implies the existence of a similar gap in the 
$ep$ frame. The exact size of the $ep$-frame rapidity gap follows from the 
specific event kinematics. The main conclusion so far is the kinematic 
separation of outgoing proton and diffractive state $X_M$ in diffractive 
events with small $\xi$. 

\begin{figure}[ht]
\begin{center}
\parbox[b]{15cm}{\psfig{width=15cm,file=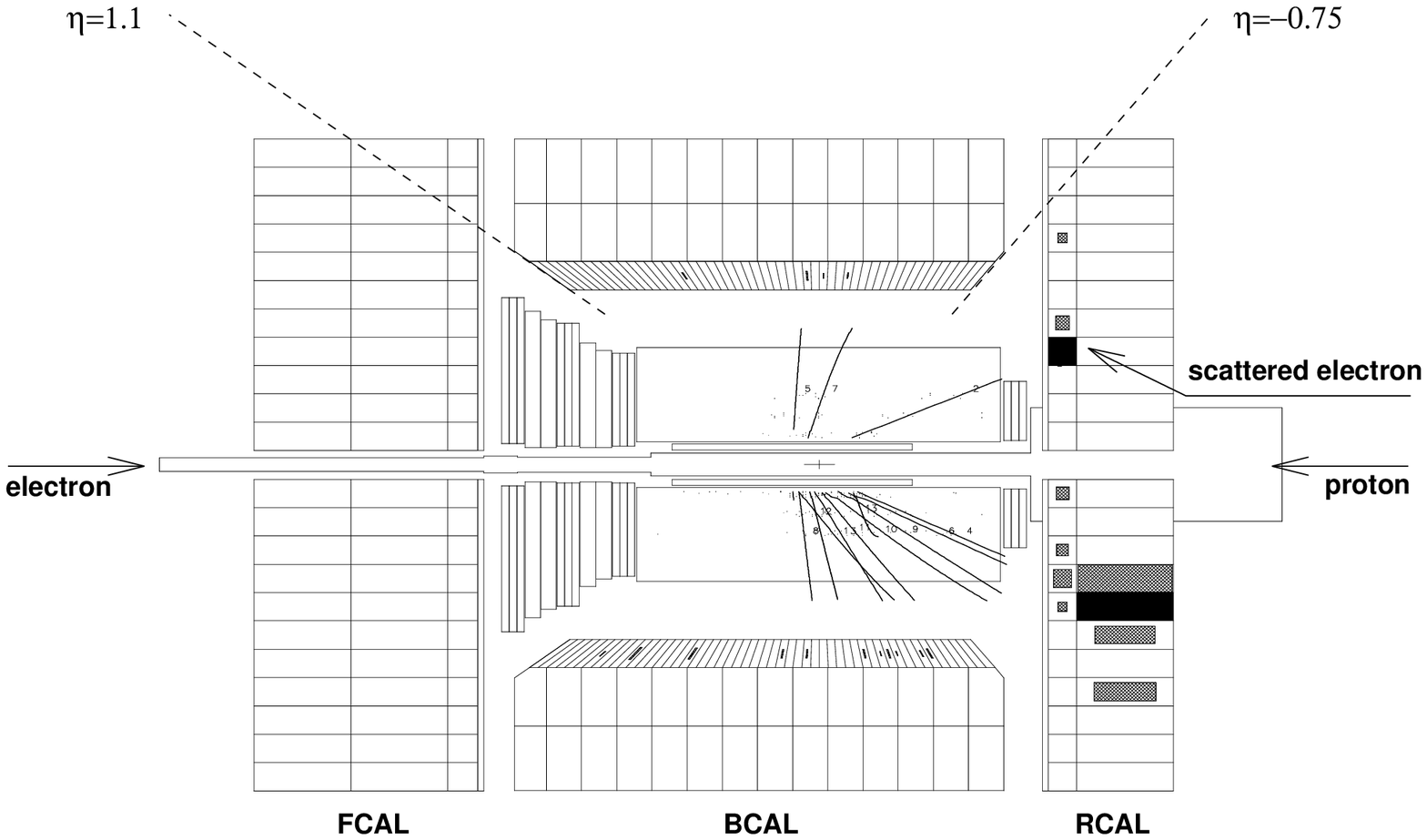}}\\
\vspace*{-.3cm}
\end{center}
\refstepcounter{figure}
\label{fig:hf}
{\bf Figure \ref{fig:hf}:} Diffractive event in the ZEUS detector
(figure from~\cite{rg1}).
\end{figure}

The appearance of a typical event in the ZEUS detector is shown in 
Fig.~\ref{fig:hf}, where FCAL, BCAL, and RCAL are the forward, central 
(barrel) and rear calorimeters. The absence of a significant energy deposit 
in the forward region is the most striking feature of this DIS event. In the 
na\"\i ve parton model of DIS, a large forward energy deposit is expected 
due to the fragmentation of the proton remnant, which is left after a quark 
has been knocked out by the virtual photon. In addition, the whole rapidity 
range between proton remnant and current jet is expected to fill with the 
hadronization products of the colour string that develops because of the 
colour charge carried by the struck quark. Thus, the event in 
Fig.~\ref{fig:hf} shows a clear deviation from typical DIS, even though the 
outgoing proton left through the beam pipe and remains undetected. Hence, 
the name rapidity gap events is also common for what was called diffractive 
electroproduction in the previous paragraphs. 

The definition of diffraction used above is narrower than necessary. 
Without losing any of the qualitative results, the requirement of a final 
state proton $P'$ can be replaced by the requirement of a low-mass hadronic 
state $Y$, well separated from the diffractive state $X_M$. In this case, 
the argument connecting elastically scattered proton and rapidity gap has 
to be reversed: the existence of a gap between $X_M$ and $Y$ becomes the 
distinctive feature of diffraction and, under certain kinematic 
conditions, the interpretation of $Y$ as an excitation of the incoming 
proton, which is now {\it almost} elastically scattered, follows. 

However, this wider definition of diffraction has the disadvantage of 
introducing a further degree of freedom, namely, the mass of the proton 
excitation. If one insists on using only the previous three parameters, 
$x$, $Q^2$ and $\xi$, the definition of a diffractive event becomes 
ambiguous. More details are found in the discussion of the experimental 
results below and in the relevant experimental papers.

\section{Fundamental observations}
Rapidity gaps are expected even if in all DIS events a quark is knocked out 
of the proton leaving a coloured remnant. The reason for this is the 
statistical distribution of the produced hadrons, which results in a small 
yet finite probability for final states with little activity in any 
specified detector region. However, the observations described below are 
clearly inconsistent with this explanation of rapidity gap events. 

The first analysis of rapidity gap events at HERA was performed by the ZEUS 
collaboration~\cite{rg1}. More than 5\% of DIS events were found to possess 
a rapidity gap. The large excess of the event numbers compared to na\"\i ve 
parton model expectations was soon confirmed by an H1 measurement 
\cite{rg2}. The analyses are based on the pseudo-rapidity $\eta=-\ln\tan 
(\theta/2)$, where $\theta$ is the angle of an outgoing particle relative to 
the beam axis. Pseudo-rapidity and rapidity are identical for massless 
particles; the difference between these two quantities is immaterial for 
the qualitative discussion below. 

\begin{figure}[ht]
\begin{center}
\vspace*{.2cm}
\parbox[b]{8.5cm}{\psfig{width=8.5cm,file=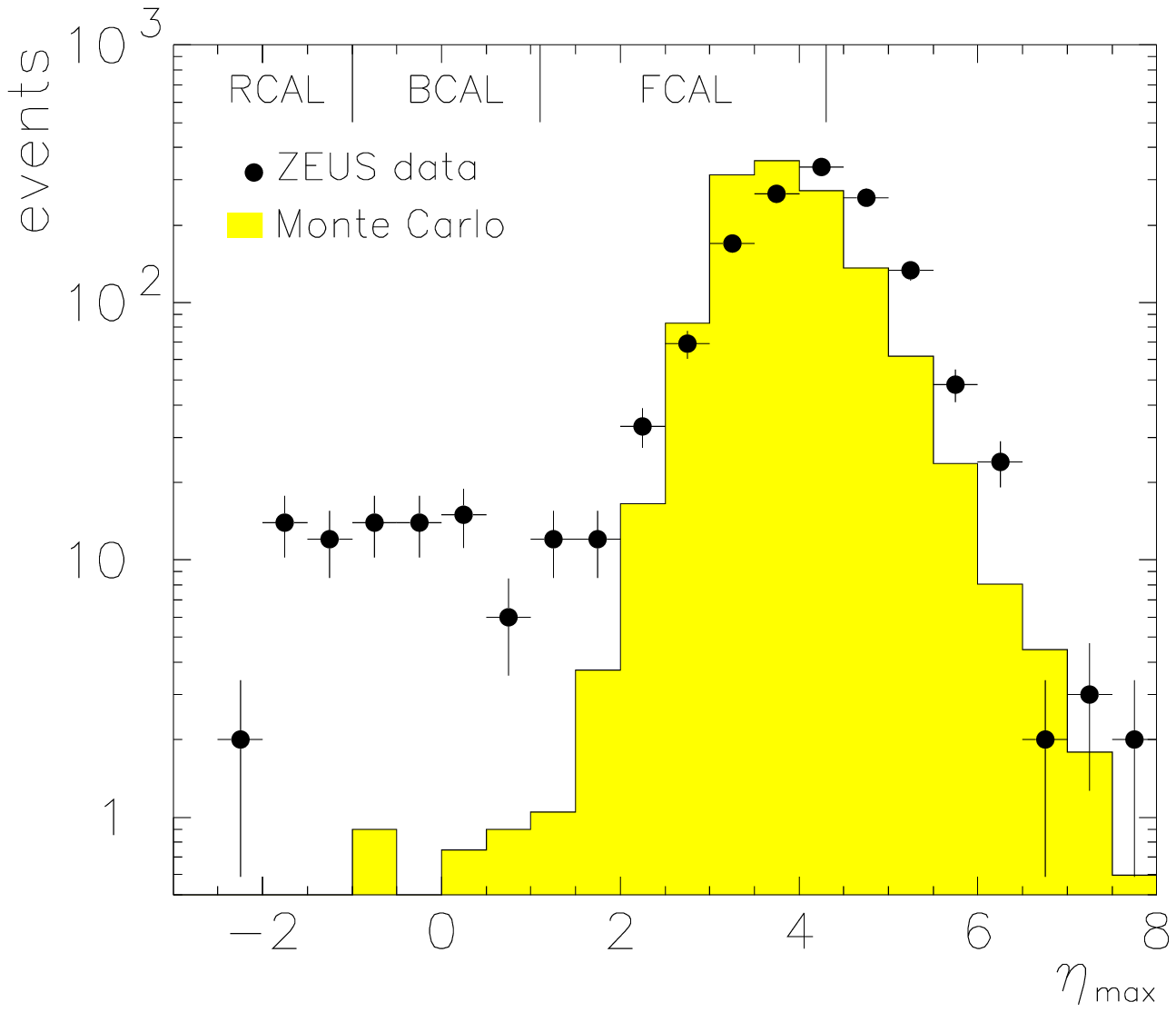}}\\
\end{center}
\refstepcounter{figure}
\label{fig:zeus}
{\bf Figure \ref{fig:zeus}:} Distribution of $\eta_{\max}$, the maximum 
rapidity of a calorimeter cluster in an event, measured at HERA (figure 
from~\cite{rg1}). 
\end{figure}

In the ZEUS analysis, a rapidity $\eta_{\max}$ was defined as the maximum 
rapidity of a calorimeter cluster in an event. A cluster was defined as an 
isolated set of adjacent cells with summed energy higher than 400 MeV. The 
measured $\eta_{\max}$ distribution is shown in Fig.~\ref{fig:zeus}. (Note 
that the smallest detector angle corresponds to $\eta_{\max}=4.3$; larger 
values are an artifact of the clustering algorithm.)

To appreciate the striking qualitative signal of diffraction at HERA, the 
measured $\eta_{\max}$ distribution has to be compared with na\"\i ve 
expectations based on a purely partonic picture of the proton. This is best 
done using a parton-model-based Monte Carlo event generator. The 
corresponding $\eta_{\max}$ distribution, which is also shown in 
Fig.~\ref{fig:zeus}, is strongly suppressed at small $\eta_{\max}$. This 
qualitative behaviour is expected since the Monte Carlo (for more details 
see~\cite{rg1} and refs. therein) starts from a partonic proton, 
calculates the hard process and the perturbative evolution of the QCD 
cascade, and finally models the hadronization using the Lund string model 
(see, e.g.,~\cite{agis}). According to the Lund model, the colour string, 
which connects all final state partons and the coloured proton remnant, 
breaks up via $q\bar{q}$ pair creation, thus producing the observed mesons. 
The rapidities of these particles follow a Poisson distribution, resulting 
in an exponential suppression of large gaps. 

It should be clear from the above discussion that this result is rather 
general and does not depend on the details of the Monte Carlo. QCD radiation 
tends to fill the rapidity range between the initially struck quark and the 
coloured proton remnant with partons. A colour string connecting these 
partons is formed, and it is highly unlikely that a large gap emerges in 
the final state after the break-up of this string.

However, the data shows a very different behaviour. The expected exponential 
decrease of the event number with $\eta_{\max}$ is observed only above 
$\eta_{\max}\simeq 1.5$; below this value a large plateau is seen. Thus, 
the na\"\i ve partonic description of DIS misses an essential qualitative 
feature of the data, namely, the existence of non-suppressed large 
rapidity gap events. 

To give a more specific discussion of the diffractive event sample, it is 
necessary to define which events are to be called diffractive or rapidity 
gap events. It is clear from Fig.~\ref{fig:zeus} that, on a qualitative 
level, this can be achieved by an $\eta_{\max}$ cut separating the events 
of the plateau. The resulting qualitative features, observed both by the 
ZEUS~\cite{rg1} and H1 collaborations~\cite{rg2}, are the following. 

There exists a large rapidity interval where the $\eta_{\max}$ distribution 
is flat. For high $\gamma^*p$ energies $W$, the ratio of diffractive 
events to all DIS events is approximately independent of $W$. The $Q^2$ 
dependence of this ratio is also weak, suggesting a leading-twist 
contribution of diffraction to DIS. Furthermore, the diffractive mass 
spectrum is consistent with a $1/M^2$ distribution. 

A number of additional remarks are in order. Note first that the 
observation of a flat $\eta_{\max}$ distribution and of a $1/M^2$ spectrum 
are interdependent as long as masses and transverse momenta of final state 
particles are much smaller than $M^2$. To see this, observe that the plus 
component of the most forward particle momentum and the minus component of 
the most backward particle momentum are largely responsible for the total 
invariant mass of the diffractive final state. This gives rise to the 
relation $dM^2/M^2=d\ln M^2 \sim d\eta_{\max}$, which is equivalent to the 
desired result. 

Furthermore, it has already been noted in~\cite{rg2} that a significant 
contribution from exclusive vector meson production, e.g., the process 
$\gamma^*p\to\rho\,p$, is present in the rapidity gap event sample. A more 
detailed discussion of corresponding cross sections, which have by now been 
measured, and of relevant theoretical considerations is given in 
Chapters~\ref{sect:tge} and \ref{sect:exp}. The discussion of other, more 
specific features of the diffractive final state, such as the presence of 
charmed mesons or high-$p_\perp$ jets, is also postponed.

\section{Diffractive structure function}\label{sect:dsf}
The diffractive structure function, introduced in~\cite{ip} and first 
measured by the H1 collaboration~\cite{h1f2d}, is a powerful concept for the 
analysis of data on diffractive DIS, which is now widely used by 
experimentalists and theoreticians. 

Recall the relevant formulae for inclusive DIS (see, e.g.,~\cite{esw}). The 
cross section for the process $ep\to eX$ can be calculated if the hadronic 
tensor, 
\be
W_{\mu\nu}(P,q)=\frac{1}{4\pi}\sum_X<P|j_\nu^\dagger(0)|X><X|j_\mu(0)|P>
(2\pi)^4\delta^4(q+P-p_X)\,,
\ee
is known. Here $j$ is the electromagnetic current, and the sum is over all 
hadronic final states $X$. Because of current conservation, $q\cdot W=W\cdot 
q=0$, the tensor can be decomposed according to 
\be
W_{\mu\nu}(P,q)=\left(g_{\mu\nu}-\frac{q_\mu q_\nu}{q^2}\right)W_1(x,Q^2)+
\left(P_\mu+\frac{1}{2x}q_\mu\right)\left(P_\nu+\frac{1}{2x}q_\nu\right)
W_2(x,Q^2)\,.\label{dec}
\ee
The data is conveniently analysed in terms of the two structure functions 
\bea
F_2(x,Q^2)&=&(P\cdot q)\,W_2(x,Q^2)\\
F_L(x,Q^2)&=&(P\cdot q)\,W_2(x,Q^2)-2xW_1(x,Q^2)\,.
\eea
Introducing the ratio $R=F_L/(F_2-F_L)$, the electron-proton cross section 
can be written as 
\be
\frac{d^2\sigma_{ep\to eX}}{dx\,dQ^2}=\frac{4\pi\aem^2}{xQ^4}\left\{1-y+
\frac{y^2}{2[1+R(x,Q^2)]}\right\}\,F_2(x,Q^2)\,,
\ee
where $y=Q^2/sx$, and $s$  is the electron-proton centre-of-mass energy 
squared. In the na\"\i ve parton model or at leading order in $\alpha_s$ in 
QCD, the longitudinal structure function $F_L(x,Q^2)$ vanishes, and $R=0$. 
Since $R$ corresponds to the ratio of longitudinal and transverse virtual 
photon cross sections, $\sigma_L/\sigma_T$, it is always positive, and the 
corrections associated with a non-zero $R$ are small at low values of $y$. 

In the simplest definition of diffraction, the inclusive final state $X$ is 
replaced by the state $X_M\,P'$, which consists of a diffractively produced 
hadronic state with mass $M$ and the scattered proton. This introduces 
the two additional kinematic variables $\xi$ and $t$. However, no 
additional independent 4-vector is introduced as long as the measurement is 
inclusive with respect to the azimuthal angle of the scattered proton. 
Therefore, the decomposition in Eq.~(\ref{dec}) remains valid, and the two 
diffractive structure functions $F_{2,L}^{D(4)}(x,Q^2,\xi,t)$ can be 
defined. The diffractive cross section reads 
\be
\frac{d^2\sigma_{ep\to epX_M}}{dx\,dQ^2\,d\xi\,dt}=\frac{4\pi\aem^2}{xQ^4}
\left\{1-y+\frac{y^2}{2[1+R^{D(4)}(x,Q^2,\xi,t)]}\right\}\,F_2^{D(4)}
(x,Q^2,\xi,t)\,,\label{d4s}
\ee
where $R^D=F_L^D/(F_2^D-F_L^D)$. In view of the limited precision of the 
data, the dominance of the small-$y$ region, and the theoretical expectation 
of the smallness of $F_L^D$, the corrections associated with a non-zero 
value of $R^D$ are neglected in the following. 

\begin{figure}[t]
\begin{center}
\vspace*{-.8cm}
\parbox[b]{14cm}{\psfig{width=14cm,file=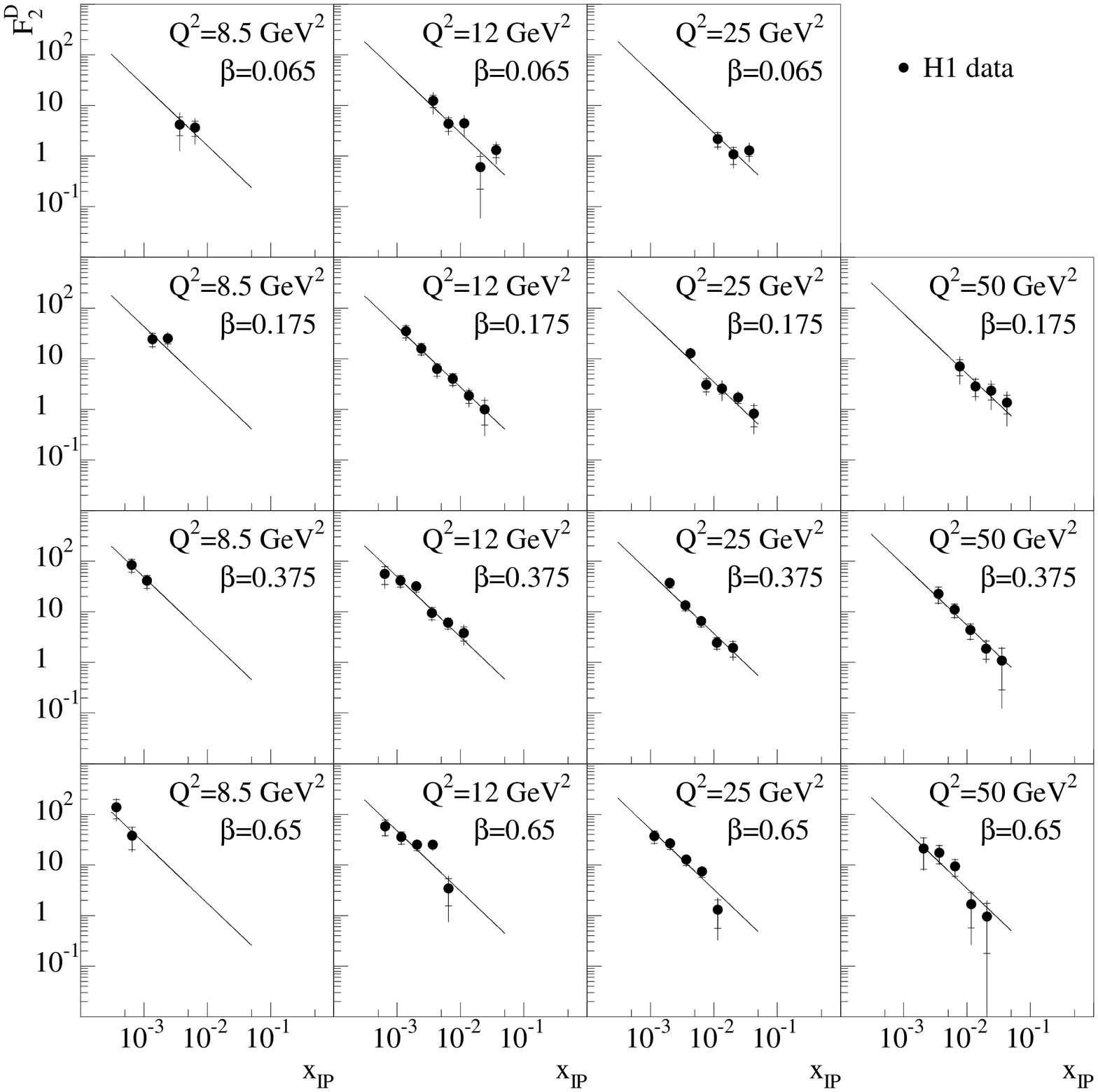}}\\
\end{center}
\refstepcounter{figure}
\label{fig:h1}
{\bf Figure \ref{fig:h1}:} First measurement of the diffractive structure 
function $F_2^{D(3)}(x,Q^2,\xi)$. The fit is based on a factorizable $\xi$ 
dependence of the form $\xi^{-1.19}$ (figure from~\cite{h1f2d}).
\end{figure}

A more inclusive and experimentally more easily accessible quantity can be 
defined by performing the $t$ integration, 
\be
F_2^{D(3)}(x,Q^2,\xi)=\int dt\, F_2^{D(4)}(x,Q^2,\xi,t)\,.
\ee
The results of the first measurement of this structure function, performed 
by the H1 collaboration, are shown in Fig.~\ref{fig:h1}. (The underlying 
cross section includes events with small-mass excitations of the proton in 
the final state.) Far more precise measurements, since performed by both 
the H1 and ZEUS collaborations, are discussed in Chapter~\ref{sect:exp} 
together with different theoretical predictions. 

The main qualitative features of diffractive electroproduction, already 
discussed in the previous section, become particularly apparent if 
the functional form of $F_2^{D(3)}$ is considered. The $\beta$ and $Q^2$ 
dependence of $F_2^{D(3)}$ is relatively flat. This corresponds to the 
observations discussed earlier that diffraction is a leading twist effect 
and that the mass distribution is consistent with a $1/M_X^2$ spectrum. 
The success of a $1/\xi^n$ fit, with $n$ a number close to 1, reflects the 
approximate energy independence of the diffractive cross section. More 
specifically, however, and in view of more precise recent measurements, it 
can be stated that the fitted exponent is above 1, so that a slight energy 
growth of the diffractive cross section is observed. This will be discussed 
in more detail later on. 

Note finally that the formal definition of $F_2^{D(3)}$ as an integral of 
$F_2^{D(4)}$~\cite{h1f2d} is not easy to implement since the outgoing proton 
or proton excitation is usually not tagged. Therefore, most measurements 
rely on different kinematic cuts, in particular an $\eta_{\max}$ cut, and 
on models of the non-diffractive DIS background. A somewhat different 
definition of $F_2^{D(3)}$, based on the subtraction of `conventional DIS' 
in the $M_X^2$ distribution, was introduced in the ZEUS analysis 
of~\cite{subt}. From a theoretical perspective, the direct measurement of 
$F_2^{D(4)}$ by tagging the outgoing proton appears most desirable. 
Recently, such a measurement has been presented by the ZEUS 
collaboration~\cite{f2d4} although the statistics are, at present, far 
worse than in the best available direct analyses of 
$F_2^{D(3)}$~\cite{nh1,nzeus}. 

\newpage

\mychapter{Semiclassical Approach}\label{sect:sc}
In this chapter, the main physical idea and the technical methods of the 
semiclassical approach to diffraction are introduced. Although historically 
the semiclassical approach is not the first relevant model, it is, in the 
author's opinion, well suited as a starting point for the present review. On 
the one hand, it is sufficiently simple to be explained on a technical level 
within the limited space available. On the other hand, it allows for a clear 
demonstration of the interplay between the relevant kinematics and the 
fundamental QCD degrees of freedom. 

The underlying idea is very simple. From the proton rest frame point of 
view, the very energetic virtual photon develops a partonic fluctuation 
long before the target. The interaction with the target is modelled as the 
scattering off a superposition of soft target colour fields, which, in the 
high-energy limit, can be calculated in the eikonal approximation. 
Diffraction occurs if this partonic system is quasi-elastically scattered 
off the proton. This means, in particular, that both the target and the 
partonic fluctuation remain in a colour singlet state.

\section{Eikonal formulae for high-energy scattering}\label{sect:ef}
The amplitude for an energetic parton to scatter off a given colour field 
configuration is a fundamental building block in the semiclassical 
approach. This amplitude is the subject of the present section. The two 
other basic ingredients for the diffractive cross section, i.e., the 
amplitudes for the photon to fluctuate into different partonic states and 
the integration procedure over all colour field configurations of the 
target proton are discussed in the remainder of this chapter. 

The essential assumptions are the softness and localization of the colour 
field and the very large energy of the scattered parton. Localization means 
that, in a suitable gauge, the colour field potential $A_\mu(x)$ vanishes 
outside a region of size $\sim 1/\Lambda$, where $\Lambda$ is a typical 
hadronic scale. Later on, it will be assumed that typical colour field 
configurations of the proton fulfil this condition. Softness means that 
the Fourier decomposition of $A_\mu(x)$ is dominated by frequencies much 
smaller than the energy of the scattered parton. The assumption that this 
holds for all fields contributing to the proton state is a non-trivial one. 
It will be discussed in more detail at the end of this chapter. 

\begin{figure}[ht]
\begin{center}
\vspace*{.2cm}
\parbox[b]{6cm}{\psfig{width=6cm,file=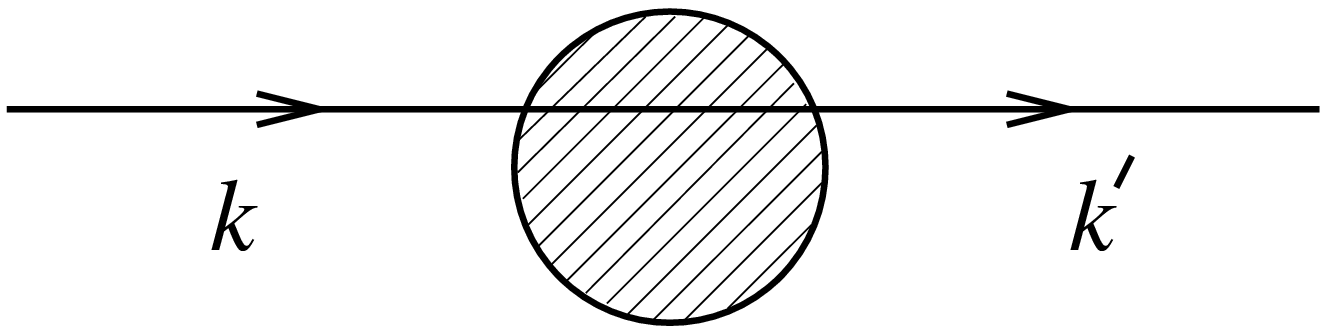}}\\
\end{center}
\refstepcounter{figure}
\label{fig:qs}
{\bf Figure \ref{fig:qs}:} Scattering of a quark off the target colour 
field. 
\end{figure}

The relevant physical situation is depicted in Fig.~\ref{fig:qs}, where the 
blob symbolizes the target colour field configuration. Consider first the 
case of a scalar quark that is minimally coupled to the gauge field via the 
Lagrangian 
\be
{\cal L}_{\sca}=\left(D_\mu \Phi\right)^*\left(D^\mu\Phi\right)
\ee
with the covariant derivative 
\be
D_\mu=\partial_\mu+igA_\mu\,.
\ee
In the high-energy limit, where the plus component of the quark momentum 
becomes large, the amplitude of Fig.~\ref{fig:qs} then reads 
\be
i2\pi\delta(k_0'-k_0)T=2\pi\delta(k_0'-k_0)2k_0\left[\tilde{U}
(k_\perp'-k_\perp)-(2\pi)^2\delta^2(k_\perp'-k_\perp)\right]\,.\label{qs} 
\ee
It is normalized as is conventional for scattering processes off a fixed 
target (see, e.g.,~\cite{nb}). The expression in square brackets is the 
Fourier transform of the impact parameter space amplitude, $U(x_\perp)-1$, 
where 
\be
U(x_\perp)=P\exp\left(-\frac{ig}{2}\int_{-\infty}^{\infty}A_-(x_+,x_\perp)
dx_+\right)\label{um}
\ee
is the non-Abelian eikonal factor. The unit matrix $1\!\in SU(N_c)$, with 
$N_c$ the number of colours, subtracts the field independent part, and the 
path ordering operator $P$ sets the field at smallest $x_+$ to the 
rightmost position. 

This formula or, more precisely, its analogue in the more realistic case of 
a spinor quark was derived by many authors. In the Abelian case, the 
high-energy amplitude was calculated in~\cite{bks} in the framework of 
light-cone quantization. This result was taken over to QCD in~\cite{css}. A 
derivation in covariant gauge, based on the solution of the equation of 
motion for a particle in the colour background field, was given in 
\cite{nac}. In~\cite{bhm}, the amplitude for the scattering of a fast gluon 
off a soft colour field was derived by similar methods. 

For completeness, a derivation of the amplitude in Eq.~(\ref{qs}), based on 
the summation of diagrams of the type shown in Fig.~\ref{fig:gsu}, is given 
in Appendix~\ref{sect:eik} of the present review. For the purpose of this 
section, it is sufficient to explain the main elements of Eq.~(\ref{qs}) in 
a physical way, without giving the technical details of the derivation. 

\begin{figure}[ht]
\begin{center}
\vspace*{.2cm}
\parbox[b]{7cm}{\psfig{width=7cm,file=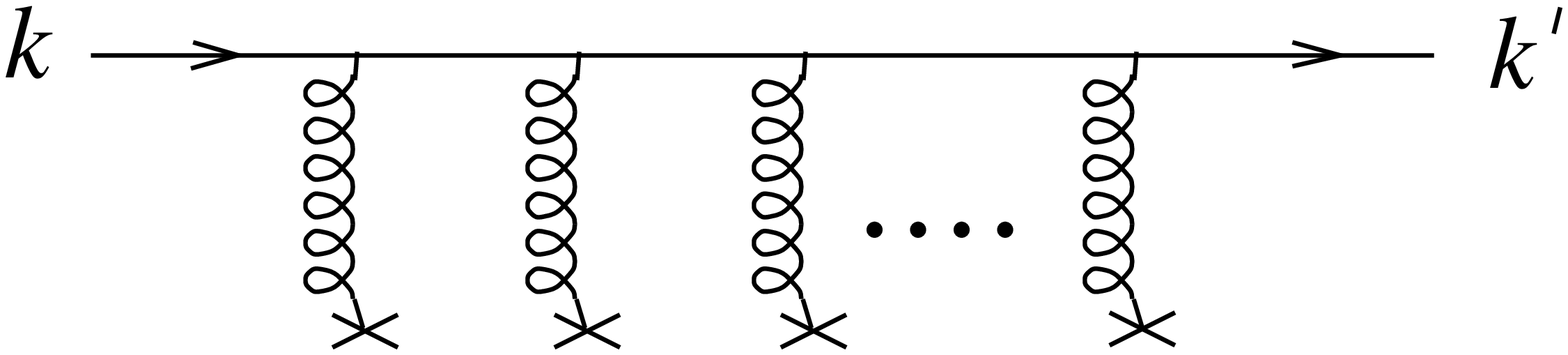}}\\
\end{center}
\refstepcounter{figure}
\label{fig:gsu}
{\bf Figure \ref{fig:gsu}:} Typical diagrammatic contribution to the eikonal 
amplitude, Eq.~(\ref{qs}). Attachments of gluon lines with crosses 
correspond to vertices at which the classical external field appears. 
\end{figure}

To begin with, it is intuitively clear that the eikonal factor $U(x_\perp)$ 
appears since the fast parton, travelling in $x_+$ direction and passing 
through the target at transverse position $x_\perp$, is rotated in colour 
space by the field $A_\mu(x)$ that it encounters on its way. 

Furthermore, the above amplitude is given for the situation in which the 
parton is localized at $x_-\simeq 0$. This co-ordinate of the fast parton 
does not change during the scattering process. Thus, $A_\mu$ is always 
evaluated at $x_-\simeq 0$, and the $x_-$ dependence is not shown 
explicitly. 

The energy $\delta$-function in Eq.~(\ref{qs}) is an approximate one. It 
appears because the energy of the parton can not be significantly changed 
by the soft colour field. 

Finally, due to the explicit factor $k_0$, the amplitude grows linearly 
with the parton energy, as is expected for a high-energy process with 
$t$-channel exchange of vector particles, in this case, of gluons. 

The amplitude of Eq.~(\ref{qs}) is easily generalized to the case of a 
spinor quark, where the new spin degrees of freedom are characterized by 
the indices $s$ and $s'$ (conventions of Appendix~\ref{sect:me}). In the 
eikonal approximation, which is valid in the high-energy limit, helicity 
flip contributions are suppressed by a power of the quark energy (see 
Appendix~\ref{sect:eik} for more details). Thus, the generalization of 
Eq.~(\ref{qs}) reads 
\be
i2\pi\delta(k_0'-k_0)T_{ss'}=2\pi\delta(k_0'-k_0)2k_0\left[\tilde{U}
(k_\perp'-k_\perp)-(2\pi)^2\delta^2(k_\perp'-k_\perp)\right]\delta_{ss'}\,.
\label{qss}
\ee

\begin{figure}[ht]
\begin{center}
\vspace*{.2cm}
\parbox[b]{6cm}{\psfig{width=6cm,file=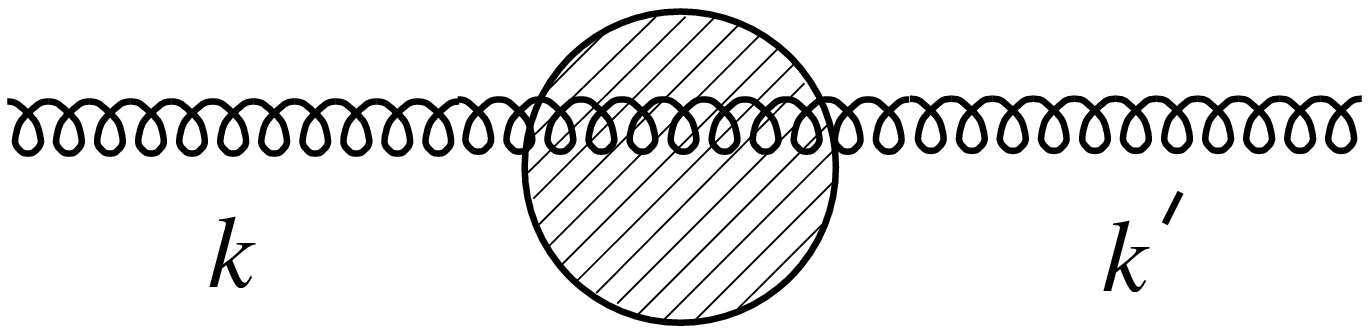}}\\
\end{center}
\refstepcounter{figure}
\label{fig:gs}
{\bf Figure \ref{fig:gs}:} Scattering of a gluon off the target colour 
field. 
\end{figure}

Similarly, the amplitude for the scattering of a very energetic gluon off 
a soft colour field is readily obtained from the basic formula, 
Eq.~(\ref{qs}). Note that, although the fast gluon and the gluons of the 
target colour field are the same fundamental degrees of freedom of QCD, the 
semiclassical approximation is still meaningful since an energy cut can be 
used to define the two different types of fields. The polarization of the 
fast gluon is conserved in the scattering process. The main difference to 
the quark case arises from the adjoint representation of the gluon, which 
determines the representation of the eikonal factor. Thus, the amplitude 
corresponding to Fig.~\ref{fig:gs} reads 
\be
i2\pi\delta(k_0'-k_0)T_{\lambda\lambda'}=2\pi\delta(k_0'-k_0)2k_0
\left[\tilde{U}^{\cal A}(k_\perp'-k_\perp)-(2\pi)^2\delta^2(k_\perp'-k_\perp)
\right]\delta_{\lambda\lambda'}\,,\label{gss}
\ee
where $\lambda,\,\lambda'$ are the polarization indices and $\tilde{U}^{\cal 
A}$ is the Fourier transform of $U^{\cal A}(x_\perp)$, the adjoint 
representation of the matrix $U(x_\perp)$ defined in Eq.~(\ref{um}). 

Clearly, the above eikonal amplitudes have no physical meaning on their own 
since free incoming quarks or gluons cannot be realized. However, they 
serve as the basic building blocks for the high-energy scattering of 
colour neutral objects discussed in the next section.

\section{Production of $q\bar{q}$ pairs}\label{sect:qq}
In this section, the eikonal approximation is used for the calculation of 
the amplitude for $q\bar{q}$ pair production off a given target colour 
field~\cite{bhm}. Both diffractive and inclusive cross sections are 
obtained from the same calculation, diffraction being defined by the 
requirement of colour neutrality of the produced pair. The qualitative 
results of this section are unaffected by the procedure of integrating over 
all proton colour field configurations, which is discussed in 
Sect.~\ref{sect:av}. 

\begin{figure}[ht]
\begin{center}
\vspace*{.2cm}
\parbox[b]{6.3cm}{\psfig{width=6.3cm,file=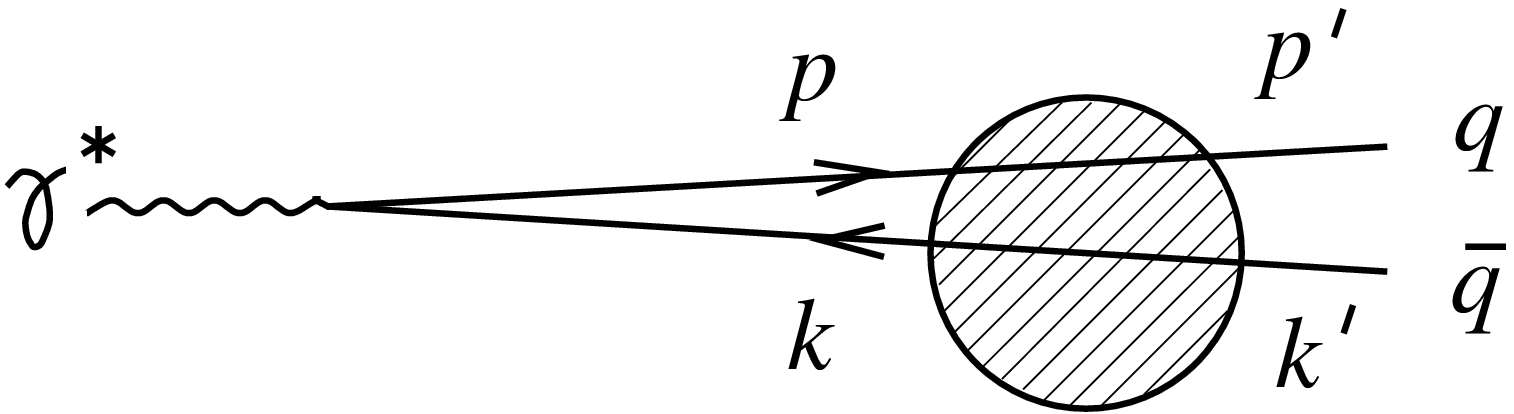}}\\
\end{center}
\refstepcounter{figure}
\label{fig:qq}
{\bf Figure \ref{fig:qq}:} Electroproduction of a $q\bar{q}$ pair off the 
target colour field.
\end{figure}

The process is illustrated in Fig.~\ref{fig:qq}. The corresponding $T$ 
matrix element has three contributions, 
\be
T=T_{q\bar{q}}+T_q+T_{\bar{q}}\,,\label{t1}
\ee
where $T_{q\bar{q}}$ corresponds to both the quark and antiquark interacting 
with the field, while $T_q$ and $T_{\bar{q}}$ correspond to only one of the 
partons interacting with the field. 

Let $V_q(p',p)$ and $V_{\bar{q}}(k',k)$ be the effective vertices for an 
energetic quark and antiquark interacting with a soft gluonic field. For 
quarks with charge $e$, the first contribution to the amplitude of 
Eq.~(\ref{t1}) reads 
\be
i2\pi\delta(k_0'+k_0-q_0)T_{q\bar{q}}=ie\int\frac{d^4k}{(2\pi)^4}
\bar{u}_{s'}(p')V_q(p',p)\frac{i}{\ppl-m}\epsilons(q)\frac{i}{-\ks-m}
V_{\bar{q}}(k,k')v_{r'}(k')\,,\label{tqq}
\ee
where $q=p+k$ by momentum conservation, $\epsilon(q)$ is the polarization 
vector of the incoming photon, and $r',s'$ label the spins of the outgoing 
quarks. 

At small $x$, the quark and antiquark have large momenta in the proton rest 
frame. Hence, the propagators in Eq.~(\ref{tqq}) can be treated in a
high-energy approximation. In a co-ordinate system where the photon momentum 
is directed along the $z$-axis, the large components are $p_+$ and $k_+$. 
It is convenient to introduce, for each vector $k$, a vector $\bar{k}$ 
whose minus component satisfies the mass shell condition,
\be
\bar{k}_-={k_{\perp}^2+m^2\over k_+}\,,
\ee
while the other components are identical to those of $k$. The propagators 
in Eq.~(\ref{tqq}) can be rewritten according to the identities
\bea
{1\over \ks +m} &=& \frac{\sum_r v_r(\bar{k})\bar{v}_r(\bar{k})}
{k^2-m^2} + \frac{\gamma_+}{2 k_+}\ ,\label{prop1}\\
{1\over \ppl - m} &=& \frac{\sum_s u_s(\bar{p}) \bar{u}_s(\bar{p})}
{p^2-m^2} + \frac{\gamma_+}{2 p_+}\ .\label{prop2}
\eea
To obtain the first term in a high-energy expansion of the scattering 
amplitude $T_{q\bar{q}}$, the terms proportional to $\gamma_+$ in 
Eqs.~(\ref{prop1}) and (\ref{prop2}) can be dropped. 

After inserting Eqs.~(\ref{prop1}) and (\ref{prop2}) into Eq.~(\ref{tqq}), 
the relations 
\bea
\hspace*{-1cm}\bar{u}_{s'}(p')V_q(p',p)u_s(p)&\!\!=\!\!&2\pi\delta(p_0'-
p_0)2p_0\left[\tilde{U}(p_\perp'-p_\perp)-(2\pi)^2\delta^2(p_\perp'-
p_\perp)\right]\delta_{ss'}\,,\label{v1}
\\
\hspace*{-1cm}\bar{v}_r(k)V_{\bar{q}}(k,k')v_{r'}(k')&\!\!=\!\!&2\pi
\delta(k_0'-k_0)2k_0\left[\tilde{U}^\dagger(k_\perp\!-\!k_\perp')-(2\pi)^2
\delta^2(k_\perp'-k_\perp)\right]\delta_{rr'}\,,\label{v2}
\eea
which correspond to Eq.~(\ref{qss}) and its antiquark analogue, can be 
applied. Writing the loop integration as $d^4k=(1/2)dk_+dk_-d^2k_\perp$ and 
using the approximation $\delta(l_0)\simeq 2\delta(l_+)$ for the energy 
$\delta$-functions, the $k_+$ integration becomes trivial. The $k_-$ 
integral is done by closing the integration contour in the upper or lower 
half of the complex $k_-$ plane. The result reads 
\bea
T_{q\bar{q}} &=&  -{ie\over 4\pi^2}\ q_+
  \int d^2k_{\perp}\ {\alpha(1-\alpha)\over N^2 + k_{\perp}^2}
  \ \bar{u}_{s'}(\bar{p})\epsilons(q)v_{r'}(\bar{k})\label{tqbq}\\
&&\hspace*{.2cm}\times \left[\tilde{U}(p_\perp'-p_\perp)-(2\pi)^2
\delta^2(p_\perp'-p_\perp)\right]\,\,\left[\tilde{U}^\dagger(k_\perp-
k_\perp')-(2\pi)^2\delta^2(k_\perp'-k_\perp)\right]\nonumber
\eea
where
\be
p'_+ = (1-\alpha)\ q_+\ ,\quad k'_+ = \alpha\ q_+\ ,\quad 
N^2 = \alpha (1 - \alpha) Q^2 + m^2\ .
\ee
Thus, $\alpha$ and $1\!-\!\alpha$ characterize the fractions of the photon 
momentum carried by the two quarks, while $(N^2+k_\perp^2)$ measures the 
off-shellness of the partonic fluctuation before it hits the target. In 
the following, the quark mass is set to $m=0$. 

The above expression for $T_{q\bar{q}}$ contains terms proportional to 
$UU^\dagger$, $U$, and $U^\dagger$, as well as a constant term. The 
amplitudes $T_q$ and $T_{\bar{q}}$, which contain terms proportional to 
$U$ and $U^\dagger$ and a constant term, are derived by the same methods. 
Calculating the full amplitude according to Eq.~(\ref{t1}), the terms 
proportional to $U$ and $U^\dagger$ cancel. Thus, the colour field 
dependence of $T$ is given by the expression 
\be
\left[\tilde{U}(p_\perp'-p_\perp)\tilde{U}^\dagger(k_\perp-k_\perp')-(2
\pi)^4\delta^2(p_\perp'-p_\perp)\delta^2(k_\perp'-k_\perp)\right]\,.
\label{uudd}
\ee
Introducing the fundamental function 
\be
W_{x_\perp}(y_\perp)=U(x_\perp)U^\dagger(x_\perp+y_\perp)-1\,,\label{wdef} 
\ee
which encodes all the information about the external field, the complete 
amplitude can eventually be given in the form 
\be
T = -{ie\over 4\pi^2}\ q_+
  \int d^2k_{\perp}\ {\alpha(1-\alpha)\over N^2 + k_{\perp}^2}
  \ \bar{u}_{s'}(\bar{p})\epsilons(q)v_{r'}(\bar{k})
\int_{x_\perp}e^{-i\Delta_\perp x_\perp}
\tilde{W}_{x_\perp}(k_\perp'-k_\perp)\,,\label{tfi}
\ee
where $\tilde{W}_{x_\perp}$ is the Fourier transform of $W_{x_\perp}
(y_\perp)$ with respect to $y_\perp$, and $\Delta_\perp=k_\perp'+p_\perp'$ 
is the total transverse momentum of the final $q\bar{q}$ state. 

{}From the above amplitude, the transverse and longitudinal virtual photon 
cross sections are calculated in a straightforward manner using the 
explicit formulae for $\bar{u}_{s'}(\bar{p})\epsilons(q)v_{r'}(\bar{k})$ 
given in Appendix~\ref{sect:me}. Summing over all $q\bar{q}$ colour 
combinations, as appropriate for the inclusive DIS cross section, the 
following result is obtained, 
\bea
\frac{d\sigma_L}{d\alpha\,dk_\perp'^2}&=&
{2e^2 Q^2\over (2\pi)^6}(\alpha(1-\alpha))^2 \int_{x_{\perp}} 
\left|\int d^2 k_{\perp} \frac{\tilde{W}_{x_{\perp}}(k_{\perp}'-k_{\perp})}
{N^2 + k_{\perp}^2}\right|^2\,,\label{dsl}
\\ \nonumber\\
\frac{d\sigma_T}{d\alpha\,dk_\perp'^2}&=&{e^2\over 2(2\pi)^6 }(\alpha^2 + 
(1-\alpha)^2)\int_{x_{\perp}} \left|\int d^2 k_\perp \frac{k_\perp 
\tilde{W}_{x_\perp}(k_\perp'-k_\perp)}{N^2 + k_\perp^2}\right|^2\!\!.
\label{dst}
\eea
Note that only a single integration over transverse coordinates appears. 
This is a consequence of the $\delta$-function induced by the phase space 
integration over $\Delta_{\perp}$, applied to $\exp[-i\Delta_\perp 
x_\perp]$ from Eq.~(\ref{tfi}) and to the corresponding exponential from 
the complex conjugate amplitude. The contraction of the colour indices of 
the two $W$ matrices is implicit. 

Consider the longitudinal cross section in more detail. The integrand can 
be expanded around $k_\perp=k_\perp'$. Shifting the integration variable 
$k_\perp$ to $l_\perp=k_\perp-k'_\perp$, the Taylor expansion of the 
denominator in powers of $l_\perp$ yields
\be
{1\over N^2 + (k'_\perp+l_\perp)^2} = {1\over N^2 + {k'}_\perp^2}
   - {2l_\perp k'_\perp \over (N^2 + {k'}_\perp^2)^2}\ +\ \ldots\ .
\ee
{}From the definition of the colour matrix $W_{x_\perp}(y_\perp)$ in 
Eq.~(\ref{wdef}), it is clear that
\bea
\int d^2l_\perp\ \tilde{W}_{x_\perp}(-l_\perp)&=&(2\pi)^2\,
W_{x_\perp}(0)\,\,=\,\,0\ ,\label{ex1}\\ 
\int d^2l_\perp\ l_\perp \tilde{W}_{x_\perp}(-l_\perp) &=& i (2\pi)^2\ 
       \partial_\perp W_{x_\perp}(0)\ .\label{ex2}
\eea
Using rotational invariance, i.e., $k'_i k'_j \rightarrow\frac{1}{2}
\delta_{ij}{k'}^2_\perp$, the result 
\be
\frac{d\sigma_L}{d\alpha\,dk_\perp'^2}=\frac{4e^2\,\alpha(1-\alpha)\,N^2
k_\perp'^2}{(2\pi)^2(N^2 + k_\perp'^2)^4}\int_{x_{\perp}} 
|\partial_\perp W_{x_\perp}(0)|^2 \label{sl}
\ee
is obtained. It evaluates to the total longitudinal cross section 
\be
\sigma_L={e^2\over 6\pi^2 Q^2}\ \int_{x_{\perp}} 
          \left|\partial_{\perp}W_{x_{\perp}}(0)\right|^2\,.\label{slt}
\ee

The transverse contribution can be evaluated in a similar way. In the 
perturbative region, where $\alpha(1-\alpha)\gg\Lambda^2/Q^2$ and 
$k_\perp'^2\gg\Lambda^2$, the integrand can again be expanded around 
$k_\perp=k_\perp'$. The analogue of Eq.~(\ref{sl}) reads 
\be
\frac{d\sigma_T}{d\alpha\,dk_\perp'^2}=
\frac{e^2[\alpha^2+(1-\alpha)^2]\,[N^4+k_\perp'^4]}
{2(2\pi)^2(N^2 + k_\perp'^2)^4}\int_{x_{\perp}} \left|\partial_{\perp}
W_{x_{\perp}}(0)\right|^2\ .\label{st}
\ee
While, in the longitudinal case, the $\alpha$ and $k_\perp'$ integrations 
were readily performed, a divergence is encountered in the transverse case, 
Eq.~(\ref{st}). The $k_\perp'$ integration gives rise to a factor 
$1/\alpha(1\!-\!\alpha)$, so that the $\alpha$ integral diverges 
logarithmically at $\alpha\to 0$ and at $\alpha\to 1$.
Thus, the region of small $\alpha(1\!-\!\alpha)$, where the expansion around 
$k_\perp=k_\perp'$ does not work, is important for the total cross section
$\sigma_T$, which therefore can not be obtained by integrating 
Eq.~(\ref{st}). Instead, the endpoint region, where the large distance 
structure of $W_{x_{\perp}}(y_\perp)$ is important, can be separated 
by a cutoff $\mu^2$ ($\Lambda^2\ll\mu^2\ll Q^2$). The complete leading 
twist result for the transverse cross section, where contributions 
suppressed by powers of $\Lambda^2/\mu^2$ or $\mu^2/Q^2$ have been 
dropped, reads 
\bea
\sigma_T&=&{e^2\over 6\pi^2 Q^2}\left(\ln\frac{Q^2}{\mu^2}-1\right)\ 
\int_{x_{\perp}}\left|\partial_{\perp}W_{x_{\perp}}(0)\right|^2\label{stt}
\\ \nonumber \\
&&+{e^2\over (2\pi)^6 }\int_0^{\mu^2/Q^2}d\alpha\int
d k_\perp'^2 \int_{x_{\perp}} \left|\int d^2 k_\perp \frac{k_\perp 
\tilde{W}_{x_\perp}(k_\perp'-k_\perp)}{N^2 + k_\perp^2}\right|^2\!\!.
\nonumber
\eea

The resulting physical picture can be summarized as follows. For 
longitudinal photon polarization, the produced $q\bar{q}$ pair has small 
transverse size and shares the photon momentum approximately equally. Only 
the small distance structure of the target colour field, characterized by 
the quantity $|\partial_{\perp}W_{x_{\perp}}(0)|^2$, is tested. For 
transverse photon polarization, an additional leading twist contribution 
comes from the region where $\alpha$ or $1\!-\!\alpha$ is small and 
$k_\perp'^2\sim\Lambda^2$. In this region, the $q\bar{q}$ pair penetrating 
the target has large transverse size, and the large distance 
structure of the target colour field, characterized by the function 
$W_{x_\perp}(y_\perp)$ at large $y_\perp$, is tested. This physical 
picture, known as the aligned jet model, was introduced in~\cite{bk} on a 
qualitative level and was used more recently for a quantitative discussion 
of small-$x$ DIS in~\cite{wf}. 

It is now straightforward to derive the cross sections for the production of 
colour singlet $q\bar{q}$ pairs corresponding, within the present approach, 
to diffractive processes. Note that the cross sections in Eqs.~(\ref{dsl}) 
and (\ref{dst}) can be interpreted as linear functionals of tr$(W_{x_\perp}
(y_\perp)W_{x_\perp}^\dagger(y_\perp'))$, where the trace appears because of 
the summation over all colours of the produced $q\bar{q}$ pair in the final 
state. Introducing a colour singlet projector into the underlying amplitude
corresponds to the substitution 
\be
\mbox{tr}\left(W_{\x}(\y)W_{\x}^{\dagger}(\y')\right)\rightarrow\frac{1}
{N_c}\mbox{tr}W_{\x}(\y)\mbox{tr}W_{\x}^{\dagger}(\y')\label{wsubs}
\ee
in Eqs.~(\ref{dsl}) and (\ref{dst}). This change of the colour structure 
has crucial consequences for the subsequent calculations. 

Firstly, the longitudinal cross section, given by Eqs.~(\ref{sl}) and 
(\ref{slt}), vanishes at leading twist since the derivative 
$\partial_{\perp}W_{x_{\perp}}(0)$ is in the Lie-algebra of $SU(N_c)$, and 
therefore tr$\,\partial_{\perp}W_{x_{\perp}}(0)=0$. 

Secondly, for the same reason the ln$Q^2$ term in the transverse cross 
section, given by Eq.~(\ref{stt}), disappears. The whole cross section is 
dominated by the endpoints of the $\alpha$ integration, i.e., the aligned 
jet region, and therefore determined by the large distance structure of the 
target colour field. At leading order in $1/Q^2$, the diffractive cross 
sections read
\bea
\sigma^D_L&=&0\\ \nopagebreak
\sigma^D_T&=&{e^2\over (2\pi)^6 N_c}\int_0^\infty d\alpha\int
d k_\perp'^2 \int_{x_{\perp}} \left|\int d^2 k_\perp \frac{k_\perp 
\mbox{tr}\tilde{W}_{x_\perp}(k_\perp'-k_\perp)}{N^2 + k_\perp^2}
\right|^2\!\!.\label{slst}
\eea
The cutoff of the $\alpha$ integration, $\mu^2/Q^2$, has been dropped 
since, due to the colour singlet projection, the integration is 
automatically dominated by the soft endpoint. 

In summary, the leading-twist cross section for small-$x$ DIS receives 
contributions from both small- and large-size $q\bar{q}$ pairs, the latter 
corresponding to aligned jet configurations. The requirement of colour 
neutrality in the final state suppresses the small-size contributions. 
Thus, leading twist diffraction is dominated by the production of pairs 
with large transverse size testing the non-perturbative large-distance 
structure of the target colour field.

\section{Higher Fock states}\label{sect:hfs}
\begin{figure}[ht]
\begin{center}
\vspace*{.2cm}
\parbox[b]{8cm}{\psfig{width=8cm,file=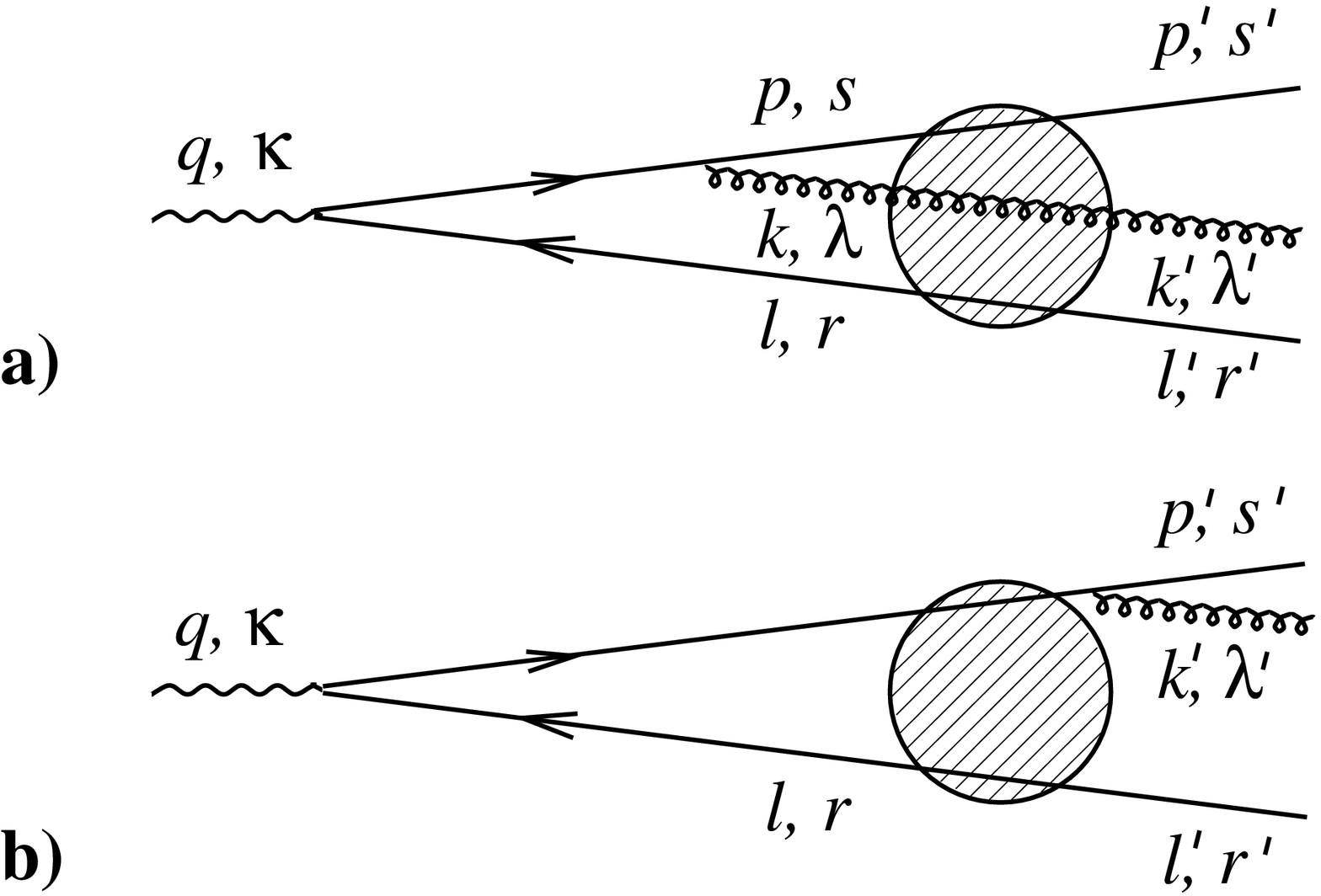}}\\
\end{center}
\refstepcounter{figure}
\label{fig:qqg}
{\bf Figure \ref{fig:qqg}:} Diagrams for the process $\gamma^*\to 
q\bar{q}g$. Two similar diagrams with the gluon radiated from the produced 
antiquark have to be added. 
\end{figure}

Given the leading order results for $q\bar{q}$ pair production of the last 
section, it is natural to ask about the importance of radiative 
corrections. A systematic procedure for calculating to all orders in 
perturbation theory does not yet exist in the semiclassical framework. 
However, as will be seen in the following chapters, the summation of 
leading logarithms in $Q^2$ is understood. Here, the particularly important 
case of the diffractive production of a quark-antiquark-gluon system (see 
Fig.~\ref{fig:qqg}) is discussed in some detail \cite{bhm}. The purpose of 
this discussion is to establish, as one of the essential features of 
diffractive DIS, the necessary presence of a wee parton in the wave 
function of the incoming virtual photon. 

Consider the sum of diagram Fig.~\ref{fig:qqg}a and its analogue with the 
gluon radiated from the antiquark. If the quark propagators with momenta 
$p$ and $l$ are rewritten according to Eqs.~(\ref{prop1}) and (\ref{prop2}) 
and the $\gamma_+$ terms are dropped, the corresponding amplitude can be 
given in the form
\be
i2\pi\delta(p_0'+k_0'+l_0'-q_0)T_{(\mbox{\sc a})}=\int\frac{d^4k}{(2\pi)^4}
\int\frac{d^4l}{(2\pi)^4}\sum_{sr\lambda}T^{(1)}_{s's,\,r'r,\,\lambda'
\lambda}\,T^{(2)}_{sr\lambda}\,.\label{ama}
\ee
Here $T^{(1)}$ is the amplitude for the scattering of the $q\bar{q}g$ 
system off the target colour field, and $T^{(2)}$ is the remainder of the 
diagram, describing the fluctuation of the virtual photon into the 
partonic state. In the following calculation, the $k_-$ integration will 
be performed in such a way that the gluon propagator goes on shell. 
Anticipating this procedure, the sum over the intermediate gluon 
polarizations $\lambda$ is restricted to the two physical polarizations, 
defined with respect to the on-shell vector $\bar{k}$. According to 
Eqs.~(\ref{qss}) and (\ref{gss}), the contribution to $T^{(1)}$ where 
quark, antiquark and gluon interact with the field reads 
\bea
\!\!\!\!\!\!\!\!\!\!\!\left(\,T^{(1)}_{qqg}\,\right)_{\,s's,\,r'r,\,
\lambda'\lambda}&\!=\!&
\quad\! i2\pi\delta(p_0'-p_0)2p_0\left[\tilde{U}
(p_\perp'-p_\perp)-(2\pi)^2\delta^2(p_\perp'-p_\perp)\right]
\delta_{s's}\\
\!\!\!&&\times i2\pi\delta(l_0'-l_0)2l_0\left[\,\tilde{U}^\dagger
(l_\perp-l_\perp')\,-\,(2\pi)^2\delta^2(l_\perp'-l_\perp)\right]
\delta_{r'r}\\
\!\!\!&&\times i2\pi\delta(k_0'-k_0)2k_0\left[\tilde{U}^{\cal A}
(k_\perp'-k_\perp)
-(2\pi)^2\delta^2(k_\perp'-k_\perp)\right]\delta_{\lambda'\lambda}\,.
\eea
As explained at the beginning of Sect.~\ref{sect:qq}, contributions where 
not all of the partons interact with the field have to be added. 

The photon-$q\bar{q}g$ transition amplitude $T^{(2)}$, with all colour 
indices suppressed, reads 
\be
T^{(2)}_{sr\lambda}=\frac{ieg_s}{p^2k^2l^2}\,\bar{u}_s(\bar{p})\left[
\epsilons_\lambda(\bar{k})\frac{i}{\ppl+\ks-m}\epsilons(q)+\epsilons(q)
\frac{i}{-\ls-\ks-m}\epsilons_\lambda(\bar{k})\right]v_r(\bar{l})\,.
\label{t2}
\ee

The colour structure of $T_{(\mbox{\sc a})}$ is given by the combination of 
the $U$ matrices in $T^{(1)}$, the indices of which are partially 
contracted according to the perturbative amplitude $T^{(2)}$. This colour 
structure is characterized by the colour tensor 
\be
(\tilde{W}^{(3)})^a_{\alpha\beta}=\int_{x_\perp,y_\perp,z_\perp} 
e^{i[x_\perp(p_\perp-p'_\perp)+y_\perp(k_\perp-k'_\perp)+z_\perp(l_\perp-
l'_\perp)]}(W^{(3)}(x_\perp,y_\perp,z_\perp))^a_{\alpha\beta}\, ,
\ee
where 
\be
(W^{(3)}(x_\perp,y_\perp,z_\perp))^a_{\alpha\beta}=
(U^{\cal A}(y_\perp))^{ab}(U(x_\perp)T^bU^\dagger(z_\perp))_{\alpha\beta}-
T^a_{\alpha\beta}\,,\label{wxyz}
\ee
and $T^a_{\alpha\beta}$ are the conventional $SU(N_c)$ generators with 
adjoint $(a=1\cdots N_c^2\!-\!1)$ and fundamental $(\alpha,\beta=1\cdots 
N_c)$ indices. The notation $W^{(3)}$ is chosen to stress the similarity 
with the $q\bar{q}$ case, where the external field is tested by the function 
$W\equiv W^{(2)}$, defined in Eq.~(\ref{wdef}). In analogy to this equation, 
the last term in Eq.~(\ref{wxyz}) subtracts the unphysical contribution 
where none of the partons is scattered by the external field. One can think 
of $x_\perp,y_\perp$ and $z_\perp$ as the transverse positions at which 
quark, gluon and antiquark penetrate the proton field, picking up 
corresponding non-Abelian eikonal factors. The indices $\alpha,\beta$ and 
$a$ correspond to the colours of the produced quark, antiquark and gluon. 

To obtain the complete result, the amplitude $T_{(\mbox{\sc b})}$, which is 
the sum of the diagram in Fig.~\ref{fig:qqg}b and its analogue with the 
gluon radiated from the antiquark, has to be added. The calculation is 
similar to the case of the amplitude $T_{(\mbox{\sc a})}$. Since the gluon is 
radiated after the quark pair passes the target field, the colour structure 
is determined by the same function $W$ that appeared in the $q\bar{q}$ 
production amplitude of the last section. 

The diagrams in Fig.~\ref{fig:qqg}, with summation over all possible 
colours of the $q\bar{q}g$ final state, represent an $\alpha_s$ correction 
to inclusive $q\bar{q}$ pair production according to Fig.~\ref{fig:qq}. 
The factor $\alpha_s$ is accompanied by a factor $\ln Q^2$. Higher Fock 
states lead to additional factors $\alpha_s\ln Q^2$, and an all-orders 
summation can be performed using standard renormalization group 
techniques, i.e., Altarelli-Parisi evolution. The implementation of this 
summation in the semiclassical framework is explained in more detail at the 
end of Sect.~\ref{sect:trf} and in Appendix~\ref{sect:isf}. Hence, the 
inclusive case is not further discussed in this section. 

\begin{figure}[ht]
\begin{center}
\vspace*{.2cm}
\parbox[b]{7.5cm}{\psfig{width=7.5cm,file=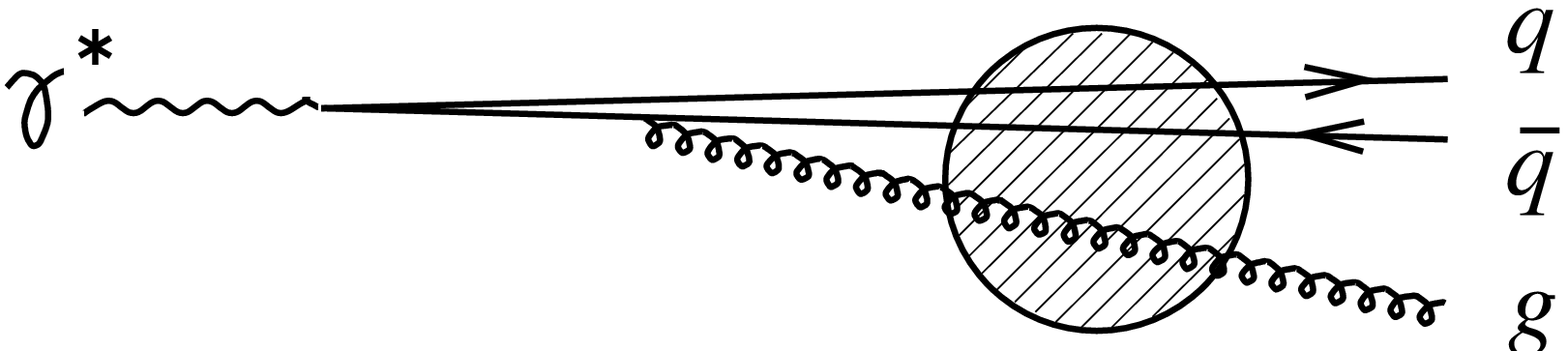}}\\
\end{center}
\refstepcounter{figure}
\label{fig:qqgg}
{\bf Figure \ref{fig:qqgg}:} Space-time picture in the case of fast, 
high-$p_\perp$ quark and antiquark, passing the proton at small transverse 
separation, with a relatively soft gluon further away. 
\end{figure}

\begin{figure}[ht]
\begin{center}
\vspace*{.2cm}
\parbox[b]{7.5cm}{\psfig{width=7.5cm,file=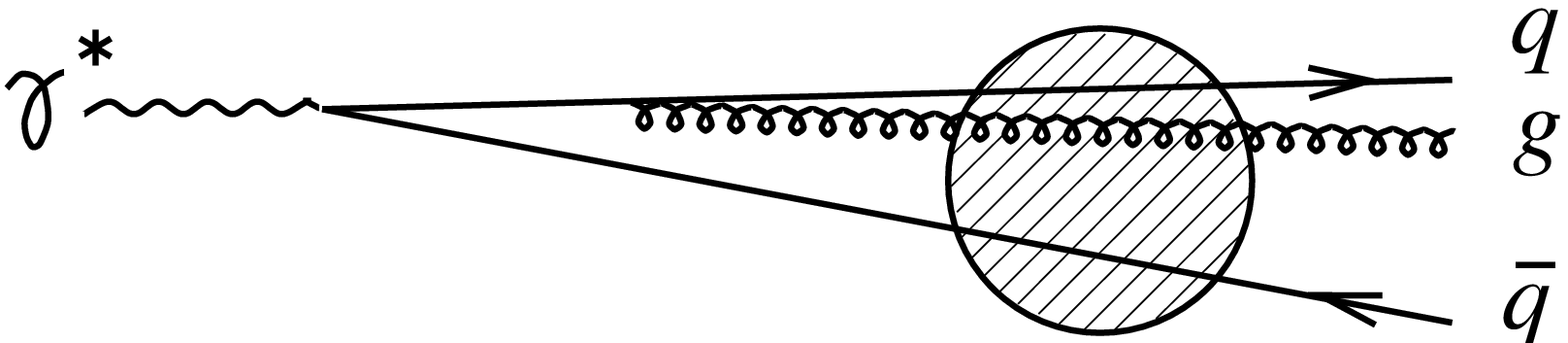}}\\
\end{center}
\refstepcounter{figure}
\label{fig:qqgq}
{\bf Figure \ref{fig:qqgq}:} Space-time picture in the case of fast, 
high-$p_\perp$ quark and gluon, passing the proton at small transverse 
separation, with a relatively soft antiquark further away. 
\end{figure}

For the diffractive case, the following situation can be anticipated on the 
basis of the experience of the last section. There are three 
qualitatively different kinematic configurations. First, if all three 
transverse momenta, $p_\perp',\,k_\perp'$ and $l_\perp'$ are soft, the 
large distance structure of the target field is tested, and a leading twist 
contribution to the DIS cross section results. Second, if all three momenta 
are large, $p_\perp'^2,\,k_\perp'^2,\,l_\perp'^2\gg\Lambda^2$, all three 
positions $x_\perp,\,y_\perp$ and $z_\perp$ in Eq.~(\ref{wxyz}) are close 
together, and an expansion in powers of the target colour field can be 
performed. The leading term corresponds to colour octet exchange, so that 
no leading twist contribution to diffractive DIS results. Third, if two 
of the produced partons have high transverse momenta and the remaining 
parton is soft, a leading twist diffractive contribution arises. The two 
possible situations, depicted in Figs.~\ref{fig:qqgg} and \ref{fig:qqgq}, 
are discussed quantitatively below. The results of this calculation justify 
the physical picture outlined above. 

Consider first the case of high-$p_\perp$ quark and antiquark, i.e., 
$p_\perp'^2,\, l_\perp'^2\gg\Lambda^2$, and relatively soft gluon, i.e., 
$k_\perp'^2\sim \Lambda^2$ and $\alpha'\ll 1$. Here $\alpha'=k_0'/q_0$ is 
the analogue of the variable $\alpha$ of the last section. 

The assumption of a smooth external field implies small transverse momentum 
transfer from the proton, i.e., $|l_\perp''| \sim \Lambda$, where $l_\perp'' 
= l_\perp'-l_\perp$.  The $l_\perp$ integration in Eq.~(\ref{ama}) can be 
replaced by an $l_\perp''$ integration, substituting at the same time 
\be
l_\perp=l_\perp'-l_\perp''\qquad\mbox{and}\qquad p_\perp=-l_\perp'+l_\perp''
-k_\perp\, .\label{pl}
\ee
Neglecting $l_\perp''$ in $T^{(2)}$, which is justified since 
$|l_\perp''^2|\ll|p_\perp'^2|,\, |l_\perp'^2|$, the only remaining 
$l_\perp''$ dependence is located in the colour factor $\tilde{W}^{(3)}$. 
This simplifies the $l_\perp''$ integration to
\be
\int\frac{d^2l_\perp''}{(2\pi)^2}\ (\tilde{W}^{(3)})^a_{\alpha\beta}\, .
\ee
Defining $\Delta\equiv p'+k'+l'-p-k-l$ to be the total momentum
transferred from the proton, $\tilde{W}^{(3)}$ can be given in the form 
\be
(\tilde{W}^{(3)})^a_{\alpha\beta} =\int_{x_\perp} e^{-ix_\perp\Delta_\perp}
\int_{y_\perp,z_\perp}e^{i[y_\perp(k_\perp-k'_\perp)+z_\perp(l_\perp-
l'_\perp)]}W^{(3)}(x_\perp,x_\perp+y_\perp,x_\perp+z_\perp)^a_{\alpha\beta}
\, ,
\ee
where $l_\perp$ is given by Eq.~(\ref{pl}). The $l_\perp''$ integration 
gives a $\delta$-function of the variable $z_\perp$, thus resulting in the 
final formula 
\be
\int\frac{d^2l_\perp''}{(2\pi)^2}\ (\tilde{W}^{(3)})^a_{\alpha\beta} = 
\int_{x_\perp}\ e^{-ix_\perp\Delta_\perp}\int_{y_\perp}\ e^{iy_\perp(
k_\perp-k_\perp')} W^{(3)}(x_\perp,x_\perp+y_\perp,x_\perp)^a_{\alpha\beta}
\, .\label{w3f}
\ee
This expression shows that in the kinematic situation with two 
high-$p_\perp$ quark jets and a relatively soft gluon the leading twist 
contribution is not affected by the transverse separation of the quarks. It 
is the transverse separation between quark-pair and gluon which tests large 
distances in the proton field and which can lead to non-perturbative effects.

The colour singlet projection of the colour tensor $W^{(3)}$ reads 
\be
S(W^{(3)})=\sqrt{\frac{2}{N_c^2-1}}\,\,(W^{(3)})^a_{\alpha\beta}
T^a_{\beta\alpha}\,.
\ee
Using the identity 
\be
(U^{\cal A})^{ab}=2\ \mbox{tr}[U^{-1}T^aUT^b]\,,
\ee
where $U\in SU(N_c)$, the contribution relevant for diffraction, i.e., the 
production of a colour singlet $q\bar{q}g$-system, takes the form 
\be
\int\frac{d^2l_\perp''}{(2\pi)^2}\ S(W^{(3)})=\int_{x_\perp} e^{-ix_\perp
\Delta_\perp}\,\frac{1}{\sqrt{2(N_c^2-1)}}\,\,\mbox{tr}[
\tilde{W}^{\cal A}_{x_\perp}(k_\perp-k_\perp')]\, ,\label{cf1}
\ee
where
\be
W^{\cal A}_{x_\perp}(y_\perp)=U^{\cal A}(x_\perp)
U^{{\cal A}\dagger}(x_\perp+y_\perp))-1\, .
\ee
This is analogous to the quark pair production of the previous section (cf. 
Eq.~(\ref{wdef})). However, now the two lines probing the field at 
positions $x_\perp$ and $x_\perp+y_\perp$ correspond to matrices in the 
adjoint representation. An intuitive explanation of this result is that 
the two high-$p_\perp$ quarks are close together and are rotated in colour 
space like a vector in the octet representation (cf. Fig.~\ref{fig:qqgg}). 

To make this last statement more precise, recall that an upper bound for the 
Ioffe-time of the fluctuation with two high-$p_\perp$ quarks is given by 
$q_0/p_\perp^2$. This means that the distance between the point 
where the virtual photon splits into the $q\bar{q}$-pair and the proton can 
not be larger than $q_0/p_\perp^2$. As long as the pair shares the 
longitudinal momentum of the photon approximately equally, i.e., 
$\alpha (1\!-\!\alpha) = {\cal O}(1)$, the opening angle is 
$\sim p_\perp/q_0$. Hence, when quark and antiquark hit the proton, their 
transverse distance is $\sim 1/|p_\perp|\ll 1/\Lambda$. 

The diagram in Fig.~\ref{fig:qqg}b and its analogue with the gluon radiated 
from the antiquark do not contribute to the diffractive cross section for 
high-$p_\perp$ $q\bar{q}$ pair production. This is obvious since the colour 
field is probed only by the $q\bar{q}$ system, the small transverse size of 
which prevents colour singlet exchange. Thus, the full cross section follows 
from Eq.~(\ref{ama}), the simplified colour tensor of Eq.~(\ref{cf1}), and 
the formulae of Appendix~\ref{sect:me}, which are used for the evaluation of 
the $q\bar{q}\gamma$ and $q\bar{q}g$ vertices in Eq.~(\ref{t2}). 

The results for longitudinal and transverse photon polarization read
\begin{eqnarray}
\frac{d\sigma_L}{d\alpha dp_\perp'^2d\alpha'dk_\perp'^2}&
=&\frac{\aem\alpha_s}{2\pi^2(N_c^2-1)}\,\frac{\alpha'Q^2
p_\perp'^2}{[\alpha(1\!-\!\alpha)]^2\hat{Q}^4}f_1(\alpha'\hat{Q}^2,k_\perp')
\,,\label{sle}
\\ \nonumber\\
\frac{d\sigma_T}{d\alpha dp_\perp'^2d\alpha'dk_\perp'^2}&
=&\frac{\aem\alpha_s}{16\pi^2(N_c^2-1)}\,\frac{\alpha'
[\alpha^2+(1\!-\!\alpha)^2]\,[p_\perp'^4+N^4]}
{[\alpha(1\!-\!\alpha)]^4\hat{Q}^4}f_1(\alpha'\hat{Q}^2,k_\perp')\,,
\label{ste}
\end{eqnarray}
with
\be
f_1(\alpha'\hat{Q}^2,k_\perp')=\int_{x_\perp}\left|\int\frac{d^2k_\perp}
{(2\pi)^2}\left(\delta^{ij}+\frac{2k_\perp^ik_\perp^j}{\alpha'\hat{Q}^2}
\right)\frac{\mbox{tr}\tilde{W}^{\cal A}_{x_\perp}(k_\perp'\!-\!k_\perp)}
{\alpha'\hat{Q}^2+k_\perp^2}\right|^2\,,
\label{eq:f}
\ee
and
\be
\hat{Q}^2=Q^2+\frac{p_\perp'^2}{\alpha(1\!-\!\alpha)}\quad,\quad N^2=
\alpha(1-\alpha)Q^2\,.\label{qhat}
\ee
At this point, the fundamental assumption of the softness of the gluon can 
be quantitatively justified. To achieve this, fix $\alpha$ and $p_\perp'$ 
and perform the integration over $\alpha'$ and $k_\perp'$. As in 
Eq.~(\ref{slst}), these two integrations are dominated by the region 
$\alpha'\ll 1$ and $k_\perp'^2\sim\Lambda^2$. Thus, the diffractive 
production of a $q\bar{q}g$ system with high-$p_\perp$ quark and antiquark 
occurs predominantly in a generalized aligned jet configuration, where the 
gluon is the wee parton in the photon wave function. It is this soft gluon 
that is responsible for the resulting leading twist cross section with 
colour singlet exchange. 

As already pointed out above, another leading twist contribution to 
diffraction arises from the kinematic region where the quark or the 
antiquark is the soft particle and, accordingly, the outgoing 
antiquark-gluon or quark-gluon system has large transverse momentum. 
The calculation proceeds along the same lines as in the soft gluon case. 
The main qualitative difference is in the colour structure. In analogy to 
the discussion leading to Eq.~(\ref{w3f}), the large $p_\perp$ of quark and 
gluon results in a small transverse separation (cf. Fig.~\ref{fig:qqgq}), 
so that effectively the high-$p_\perp$ quark-gluon system is colour rotated 
like a single quark. Accordingly, the field is tested by the same function 
$W$, built from $U$ matrices in the fundamental representation, that 
appeared in the $q\bar{q}$ production cross section. Notice also that, in 
contrast to the soft gluon case, the diagram in Fig.~\ref{fig:qqg}b has to 
be taken into account. 

The final results for the transverse and longitudinal cross sections read 
\begin{eqnarray}
\hspace*{-1cm}\frac{d\sigma_L}{d\alpha dp_\perp'^2d\alpha'dk_\perp'^2}&\!\!
=\!\!&\frac{\aem\alpha_s(N_c^2\!-\!1)}{2\pi^2N_c^2}\,\frac{Q^2}
{\alpha(1\!-\!\alpha)\hat{Q}^4}f_2(\alpha'\hat{Q}^2,k_\perp')
\,,
\\ \nonumber\\
\hspace*{-1cm}\frac{d\sigma_T}{d\alpha dp_\perp'^2d\alpha'dk_\perp'^2}&\!\!
=\!\!&\frac{\aem\alpha_s(N_c^2\!-\!1)}{8\pi^2N_c^2\,p_\perp^2
\hat{Q}^4}\left[\hat{Q}^4-2Q^2(\hat{Q}^2\!+\!Q^2)+\frac{\hat{Q}^4\!+\!Q^4}
{\alpha(1\!-\!\alpha)}\right]f_2(\alpha'\hat{Q}^2,k_\perp')\,,
\end{eqnarray}
with
\be
f_2(\alpha'\hat{Q}^2,k_\perp')=\int_{x_\perp}\left|\int\frac{d^2k_\perp}
{(2\pi)^2}\frac{k_\perp\mbox{tr}\tilde{W}_{x_\perp}(k_\perp'\!-
\!k_\perp)}{\alpha'\hat{Q}^2+k_\perp^2}\right|^2\,.
\ee
These cross sections include both the soft quark and the soft antiquark 
regions. The kinematic variables do not correspond to Fig.~\ref{fig:qqg}. 
They are generic in the sense that $p_\perp'$ and $-p_\perp'$ are the 
transverse momenta of the two hard jets, $\alpha$ and $1\!-\!\alpha$ are 
the corresponding momentum fractions, and the soft parton, in this case the 
quark or the antiquark, is characterized by the transverse momentum 
$k_\perp'$ and the longitudinal momentum fraction $\alpha'$. These 
conventions are chosen to emphasize the similarity with the soft gluon 
result of Eqs.~(\ref{sle})--(\ref{qhat}). Again, the $\alpha'$ and 
$k_\perp'$ integrations are dominated by the region $\alpha'\ll 1$ and 
$k_\perp'^2\sim\Lambda^2$.

\section{Field averaging}\label{sect:av}
So far, electroproduction off a fixed `soft' colour field has been 
considered. As a consequence, electroproduction cross sections approach 
constant values as $x\rightarrow 0$. However, a proper treatment of the 
target requires the integration over all relevant colour field 
configurations. 

Following the discussion of~\cite{bhm}, consider first the elastic 
scattering of a quark off a proton. Although this process is unphysical 
since quarks are confined, it can serve to illustrate the method of 
calculation. Therefore in the following, confinement is ignored, and 
quarks are treated as asymptotic states. The generalization to the 
physical case of electroproduction is straightforward and will be discussed 
subsequently. 

A point-like quark with initial momentum $q$ scatters off the proton, which 
is a relativistic bound state with initial momentum $p$. Let $m_{\pro}$ be 
the proton mass, and $s$ and $t$ the usual Mandelstam variables for a $2\to 
2$ process. In the high-energy limit, \hspace{.2cm}$s\gg t,\, m_{\pro}^2$, 
\hspace{.2cm} the contribution from the annihilation of the incoming quark 
with an antiquark of the proton is negligible. The amplitude is dominated 
by diagrams with a fermion line going directly from the initial to the 
final quark state. Therefore, the proton can be described by a 
Schr\"odinger wave functional $\Phi_P[A]$ (cf.~\cite{lus}) depending on the 
gluon field only. Quarks are integrated out, yielding a modification of the 
gluonic action. 

For a scattering process, the amplitude can be written in the proton rest 
frame as
\be \label{qamp}
<\!q'P'|qP\!>=\lim_{T\to\infty}\int DA_T DA_{-T} \Phi_{P'}^*[A_T] 
\Phi_P[A_{-T}]\int_{A_{-T}}^{A_T}DA\, e^{iS[A]}<\!q'|q\!>_A\, . 
\ee 
Here the fields $A_{-T}$ and $A_T$ are defined on three-dimensional surfaces 
at constant times $-T$ and $T$, and $A$ is defined in the four-dimensional 
region bounded by these surfaces. The field $A$ has to coincide with 
$A_{-T}$ and $A_T$ at the boundaries, and the action $S$ is defined by an 
integration over the domain of $A$. The amplitude $<\!q'|q\!>_A$ 
describes the scattering of a quark by the given external field $A$.

The initial state proton, having well defined momentum $\vec{P}$, 
is not well localized in space. However, the dominant field configurations
in the proton wave functional are localized on a scale $\Lambda \sim 
\Lambda_{QCD}$. The field configurations $A(\vec{x},t)$, which interpolate 
between initial and final proton state, are also localized in space at each 
time $t$. Assume that the incoming quark wave packet is localized such that 
it passes the origin $\vec{x}=0$ at time $t=0$. At this instant the field 
configuration $A(\vec{x},t)$ is centered at 
\be 
\vec{x}[A]\equiv\int d^3\vec{x} E_A(\vec{x})\cdot\vec{x}\Bigg/ 
\int d^3\vec{x} E_A(\vec{x})\, , 
\ee 
where $E_A(\vec{x})$ is the energy density of the field $A(\vec{x},t)$ at 
$t=0$. The amplitude (\ref{qamp}) can now be written as
\be 
<\!q'P'|qP\!>=\lim_{T\to\infty}\int d^3\vec{x}\int DA_T DA_{-T} 
\Phi_{P'}^*\Phi_P\int_{A_{-T}}^{A_T}DA\, e^{iS}
\delta^3(\vec{x}[A]-\vec{x})<\!q'|q\!>_A\, . 
\ee 
Using the transformation properties under translations,
\be
<\!q'|q\!>_{L_{\vec{x}}A}=e^{i(\vec{q}-\vec{q}\,')\vec{x}}<\!q'|q\!>_A
\quad,\quad\Phi_{P'}^*[L_{\vec{x}}A]\Phi_P[L_{\vec{x}}A]=
e^{i(\vec{P}-\vec{P}\,')\vec{x}}\Phi_{P'}^*[A]\Phi_P[A]\,\, ,
\ee
where 
\be
L_{\vec{x}}A(\vec{y})\equiv A(\vec{y}-\vec{x})\, ,
\ee
one obtains, 
\be
<\!q'P'|qP\!>=2m_{\pro}(2\pi)^3\delta^3(\vec{P}\,'+\vec{q}\,'-\vec{P}-\vec{q}
)\, \int_{\{A\}} <\!q'|q\!>_A\, .
\ee
Here $\int_{\{A\}}$ denotes the operation of averaging over all field 
configurations contributing to the proton state which are localized at 
$\vec{x}=0$ at time $t=0$. It is defined by 
\be
\int_{\{A\}} \Psi[A]\equiv\frac{1}{2m_{\pro}}\lim_{T\to\infty}
\int DA_T DA_{-T} \Phi_{P'}^*\Phi_P\int_{A_{-T}}^{A_T}DA\, e^{iS}
\delta^3(\vec{x}[A])\Psi[A]\label{idef}
\ee
for any functional $\Psi$. The normalization $\int_{\{A\}}\ 1=1$ follows 
from 
\be
<\!P'|P\!>=2P_0(2\pi)^3\delta^3(\vec{P}\,'-\vec{P}\,)\ .
\ee

More complicated processes can be treated in complete analogy as long as the
proton scatters elastically. In particular, the above arguments apply to 
the creation of colour singlet quark-antiquark pairs, 
\be
<\!q\bar{q}P'|\gamma^*P\!>=2m_{\pro}(2\pi)^3\delta^3(\vec{k}_f-\vec{k}_i)\ 
\int_{\{A\}} <\!q\bar{q}|\gamma^*\!>_A\, ,\label{qqa}
\ee
where $\vec{k}_i$ and $\vec{k}_f$ are the sums of the momenta in the initial 
and final states respectively. The generalization of this simplest 
diffractive process to a process with an additional fast final state gluon, 
$\gamma^*\to q\bar{q}g$ (cf.~Sect.~\ref{sect:hfs}), is straightforward. In 
contrast to the quark-proton scattering discussed above, here a colour 
neutral state is scattered off the proton. Therefore no immediate 
contradiction with colour confinement arises. However, it has to be assumed 
that the hadronization of the produced partonic state takes place after the 
interaction with the proton, which is described in terms of fast moving 
partons. 

The amplitude for the scattering off a soft external field contains an 
approximate energy $\delta$-function, giving rise to the definition of a 
functional $F$,
\be
<\,q\bar{q}\,|\,\gamma^*\,>_A=2\pi\delta(k_q^0+k_{\bar{q}}^0-q^0)\,F[A]
\,, 
\ee
where $q$, $k_q$, $k_{\bar{q}}$ are the momenta of the incoming photon and 
the outgoing quark and antiquark respectively. Since the energy transferred 
to the proton is small, Eq.~(\ref{qqa}) can now be written as 
\be
<\!q\bar{q}P'|\gamma^*P\!>=2m_{\pro}(2\pi)^4\delta^4(k_f-k_i)\ \int_{\{A\}} 
F[A]\, .\label{qqa1}
\ee

When calculating the cross section from Eq.~(\ref{qqa1}), the square of the 
4-momentum-conserving $\delta$-function translates into one 
4-momentum-conserving $\delta$-function using Fermi's trick. The spatial 
part of this $\delta$-function disappears after the momentum integration 
for the final state proton. The squared amplitude, integrated over the 
phase space $\Phi^{(1)}$ of the outgoing proton and normalized to the total 
space-time volume, reads 
\be
\frac{1}{VT}\,\int d\Phi^{(1)} |<\,q\bar{q}P'\,|\,\gamma^*P\,>|^2=4\pi 
m_{\pro}\delta(k_q^0+k_{\bar{q}}^0-q^0)\,\left|\,\int_{\{A\}}F[A]\,\right|^2
\,,\label{ts0}
\ee
as it should for the scattering off a superposition of external fields. 

{}From the above discussion, a simple recipe for the calculation of 
diffractive processes at high energy follows: 

The partonic process is calculated in a given external colour field, 
localized at $\vec{x}=0$ at time $t=0$. The weighted average over all 
colour fields contributing to the proton state is taken on the amplitude 
level. It is assumed that the typical contributing field is smooth on a 
scale $\Lambda$ and is localized in space on the same scale $\Lambda$. 
Finally, the cross section is calculated using standard formulae for 
the scattering off an external field. 

The above discussion was limited to the case of diffraction and did not 
address the important question of how inclusive and diffractive 
electroproduction processes are interrelated. To arrive at a combined 
treatment of both processes, additional assumptions concerning the 
treatment of the target colour field have to be introduced. A corresponding 
discussion, following the analysis of~\cite{bgh}, is outlined below. 

As explained previously, in the semiclassical framework it is natural to 
expect diffraction to occur whenever the produced $q\bar{q}$ pair emerges in 
a colour singlet state. Non-diffractive events are expected in the colour 
octet case. A combined treatment clearly forbids the use of the proton 
wave functional for the description of the final state. Thus, instead of 
the amplitude in Eq.~(\ref{qqa}), one has to consider the corresponding 
amplitude in a `mixed' representation, $<\,q\bar{q}\,A\,|\,\gamma^*P\,>$, 
where the final state consists of the outgoing $q\bar{q}$ pair and a colour 
field configuration $A$. This formalism allows for both the creation of 
colour singlet and non-singlet partonic states. 

For simplicity, time evolution of the field between the actual scattering 
process and the moment at which the final state field configuration $A$ is 
defined is neglected. The squared amplitude, summed over all fields $A$ and 
normalized to the total space-time volume, reads 
\be
\frac{1}{VT}\int DA\,|<\,q\bar{q}\,A\,|\,\gamma^*P\,>|^2=4\pi 
m_{\pro}\delta(k_q^0+k_{\bar{q}}^0-q^0)\int DA_{\loc}\,\Big|\,\Phi_p[
A_{\loc}]\,F[A_{\loc}]\,\Big|^2\,.\label{ts}
\ee
Here the integral over $A$ on the l.h. side replaces the phase space 
integral for the outgoing proton in Eq.~(\ref{ts0}). The index `loc' 
symbolizes that, on the r.h. side of Eq.~(\ref{ts}), the integration is 
restricted to fields localized at, say, $\vec{x}=0$. This can be achieved 
using translation covariance of the proton wave functional and of the 
matrix element $<q\bar{q}\,|\,\gamma^*>_A$, in an argument similar to 
the one leading to Eqs.~(\ref{idef}) and (\ref{ts0}). 

When writing $W_{\x}(\y)$, it has so far always been assumed that the 
functional dependence on the classical colour field configuration $A_{\cl}$ 
is implicit, so that one should really read $W_{\x}(\y)[A_{\cl}]$. 
As can be seen from Eq.~(\ref{ts}), the full inclusive cross sections are 
obtained from the previous formulae by the substitution 
\be
\mbox{tr}\!\left(W_{\x}(\y)[A_{\cl}]\,W^\dagger_{\x}(\y)
[A_{\cl}]\right)\rightarrow\int\!DA_{\loc}\Big|\,\Phi_P[A_{\loc}]\,\Big|^2
\,\,\mbox{tr}\!\left(W_{\x}(\y)[A_{\loc}]\,W^\dagger_{\x}(\y)
[A_{\loc}]\right).
\ee
The same applies to the diffractive cross sections, which are obtained by 
introducing a colour singlet projector on the amplitude level 
(cf.~Eq.~(\ref{wsubs})). 

Decomposing the field $A_{\loc}$ in Eq.~(\ref{ts}) into its Fourier modes 
$\tilde{A}_{\loc}(\vec{k})$, the path integral can be written as
\be
\int_{A_{\loc}} = \prod_{|\vec{k}|\ll|\vec{q}|}\int d\tilde{A}_{\loc}
(\vec{k})\,,
\ee
where the cutoff $|\vec{q}|$ is required to ensure that the basic 
precondition for the semiclassical treatment, the softness of the target 
colour field with respect to the momenta of the fast particles, is 
respected. This cutoff induces a non-trivial energy dependence of the 
squared amplitude in Eq.~(\ref{ts}) and therefore of both the inclusive 
and diffractive cross sections. 

At present, no complete derivation of the explicit form of that energy 
dependence from first principles exists. However, a number of interesting 
related developments, approaching the high-energy limit of QCD from the 
perspective of colour fields and eikonalized interactions, can be found, 
e.g., in~\cite{hes,jkmw}. 

On the basis of the above qualitative picture, a soft, non-perturbative 
energy growth was ascribed to the input parton distributions used in the 
phenomenological analysis of~\cite{bgh} (see Sects.~\ref{sect:lh} and 
\ref{sect:edsf} for more details). 

The discussion of the present section is closely related to Nachtmann's 
original proposal to treat high-energy hadron-hadron scattering in the 
eikonal approximation~\cite{nac}. The above viewpoint corresponds to the rest 
frame of the proton, where the colour field encountered by the fast partons 
of the projectile is naturally considered to be part of the proton state. 
The viewpoint of~\cite{nac} corresponds to the centre-of-mass frame of 
high-energy processes. In hadron-hadron scattering, both incoming particles 
are characterized by their parton content. These partons then interact via 
gluon fields which belong to neither of the two hadrons. So far, this is 
completely general. However, once the eikonal approximation is used to 
treat the parton propagators in the colour field background, one is forced 
to rely on modelling the light-cone wave function of the proton. In this 
case, the method becomes more predictive but less general than the 
wave functional treatment of the proton target discussed above. 

\newpage

\mychapter{From Soft Pomeron to Diffractive Parton Distributions}
\label{sect:sp}

In the previous chapter, small-$x$ DIS was described as the eikonalized 
interaction of a partonic fluctuation of the virtual photon with a 
superposition of colour field configurations of the proton. Diffraction 
occurs whenever the partonic fluctuation preserves its overall colour 
neutrality. An essential feature of this approach is its formulation 
exclusively in terms of the fundamental degrees of freedom of QCD. 

It is now appropriate to step back and take a look at diffraction from the 
historical perspective of soft hadronic physics (see~\cite{pred} for a 
recent review of hadronic diffraction). It will soon become clear how 
partonic ideas arise in this framework and in which way they are related to 
the results of the last chapter.

\section{Soft pomeron}

Before introducing the pomeron, a concept from soft hadronic physics, it 
is appropriate to give a simple argument why, even at very high photon 
virtualities, diffractive DIS is largely a soft process. Such an argument 
was presented by Bjorken and Kogut in the framework of their aligned jet 
model~\cite{bk} (see also~\cite{bjn}). 

\begin{figure}[ht]
\begin{center}
\vspace*{.2cm}
\parbox[b]{13cm}{\psfig{width=13cm,file=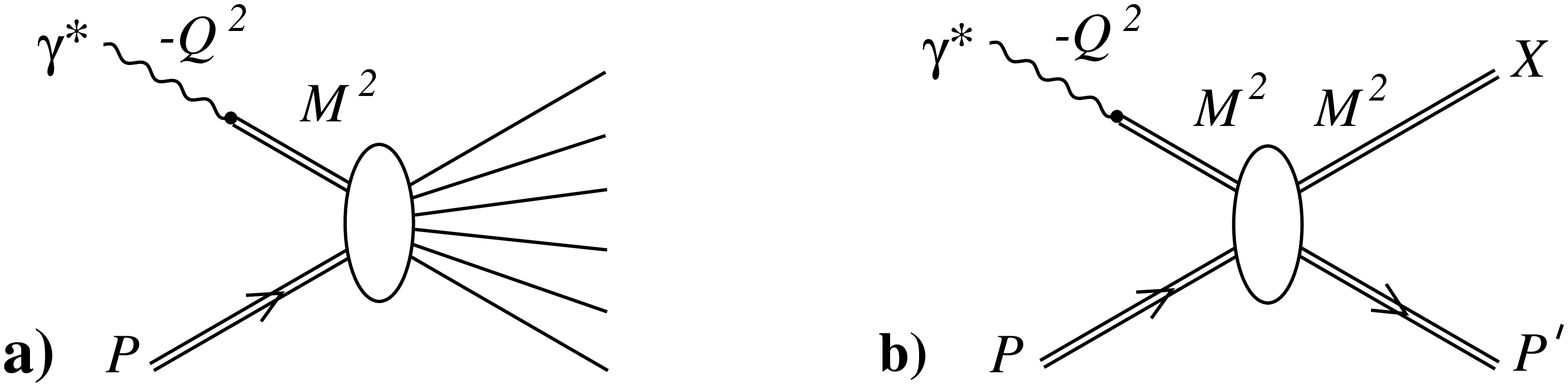}}\\
\end{center}
\refstepcounter{figure}
\label{fig:ajm}
{\bf Figure \ref{fig:ajm}:} Vector meson dominance inspired picture of 
inclusive (a) and diffractive (b) electroproduction.
\end{figure}

The underlying physical picture is based on vector meson dominance ideas. 
At high energy or small $x$, the incoming photon with virtuality $q^2=-Q^2$ 
fluctuates into a hadronic state with mass $M$, which then collides with 
the target (see Fig.~\ref{fig:ajm}a). The corresponding cross section for 
transverse photon polarization is estimated by 
\be
\frac{d\sigma_T}{dM^2}\sim \frac{dP(M^2)}{dM^2}\cdot\sigma(M^2)\,,
\ee
where the probability for the photon to develop a fluctuation with mass $M$ 
is given by 
\be
dP(M^2)\sim\frac{M^2dM^2}{(M^2+Q^2)^2}\,,
\ee
and $\sigma(M^2)$ is the cross section for this fluctuation to scatter off 
the target. The above expression for $dP(M^2)$ is most easily motivated in 
the framework of old-fashioned perturbation theory, where the energy 
denominator of the amplitude is proportional to the off-shellness of the 
hadronic fluctuation, $Q^2+M^2$. If this is the only source for a $Q^2$ 
dependence, the numerator factor $M^2$ is necessary to obtain a 
dimensionless expression. (See the original derivation of Gribov~\cite{gri} 
for more details.) 

Bjorken and Kogut assume that, for large $M^2$, the intermediate hadronic 
state typically contains two jets and that $\sigma(M^2)$ is suppressed for 
configurations with high $p_\perp$ (the latter effect being now known under 
the name of colour transparency). Consider hadronic fluctuations with a 
certain $M^2$, which, in their respective rest frames, are realized by two 
back-to-back jets. Under the assumption that the probability distribution 
of the direction of the jet axis is isotropic, simple geometry implies that 
aligned configurations, defined by $p_\perp^2<\Lambda^2$ (where $\Lambda^2$ 
is a soft hadronic scale), are suppressed by $\Lambda^2/M^2$. If only such 
configurations are absorbed with a large, hadronic cross section, the 
relations $\sigma(M^2) \sim 1/M^2$ and 
\be
\frac{d\sigma_T}{dM^2}\sim \frac{1}{(M^2+Q^2)^2}\,\label{ajm}
\ee
follow. Thus, the above cross section can be interpreted as the total 
high-energy cross section of target proton and aligned jet fluctuation of 
the photon, i.e., of two soft hadronic objects. Therefore, a similar elastic 
cross section is expected, $\sigma^D_T\sim\sigma_T$ (cf. 
Fig.~\ref{fig:ajm}b). The resulting diffractive structure function, as 
defined in the previous chapter, reads 
\be
F_2^{D(3)}(\xi,\beta,Q^2)\sim \frac{\beta}{\xi}\,.\label{ajmf2d}
\ee

It is interesting that the very simple arguments outlined above result in an 
expression for $F_2^{D(3)}$ that captures two important features of the HERA 
data: the leading-twist nature of diffraction and the approximate $1/\xi$ 
behaviour. The main problems of the model are the precise energy 
dependence, which is measured to be somewhat steeper than $1/\xi$, and the 
limit $\beta\to 0$, where a constant or even rising behaviour of 
$F_2^{D(3)}$ is observed. 

The above discussion shows, in very simple terms, what was also one of the 
main qualitative results of the more technical treatment of the last 
chapter: diffractive electroproduction is based on a soft hadronic 
high-energy cross section. Such cross sections are very successfully 
described within the framework of pomeron exchange~\cite{dl}. Let us recall 
the basic underlying concepts (see~\cite{col} for a detailed discussion 
or~\cite{fro} for a brief introduction). 

\begin{figure}[ht]
\begin{center}
\vspace*{.2cm}
\parbox[b]{6.7cm}{\psfig{width=6.7cm,file=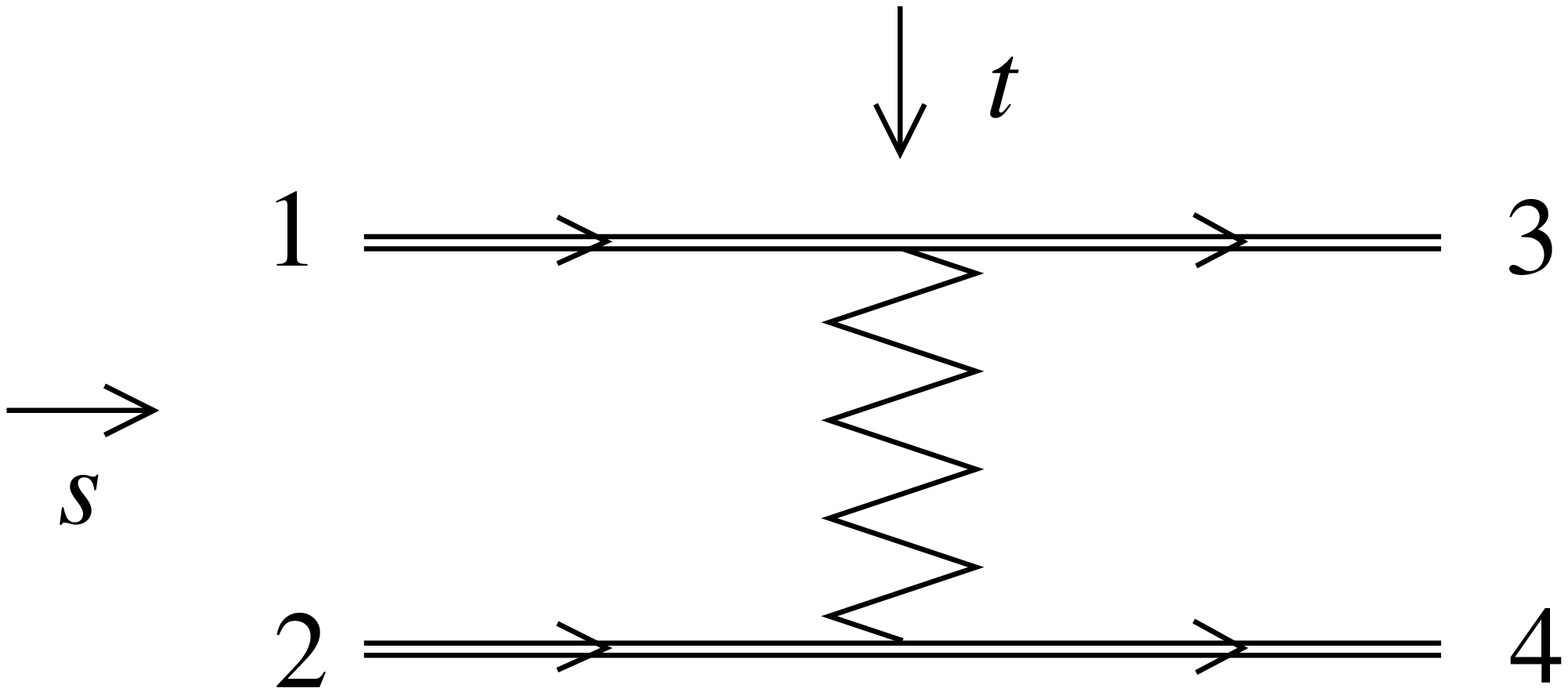}}\\
\end{center}
\refstepcounter{figure}
\label{fig:reg}
{\bf Figure \ref{fig:reg}:} Scattering process $12\to 34$ via reggeon 
exchange.
\end{figure}

Using analyticity and crossing symmetry, the amplitude $T_{12\to 34}(s,t)$, 
depicted in Fig.~\ref{fig:reg}, can be related to the amplitude $T_{1\bar{3} 
\to\bar{2}4}(s',t')$, where $s'=t$, $t'=s$, and bared numbers denote 
antiparticles. The partial wave expansion for this crossed amplitude reads 
\be
T_{1\bar{3}\to\bar{2}4}(s',t')=\sum_{l=0}^{\infty}(2l+1)a_l(s')P_l
(\cos\theta)\,,
\ee
where $\theta$ is the centre-of-mass frame scattering angle, which is a 
function of $s',t'$ and the particle masses, and $P_l$ are Legendre 
polynomials. Let the two functions $a_\eta(l,t)$ with $\eta=+1$ and 
$\eta=-1$ be the analytic continuations to complex $l$ of the two sequences 
$\{a_l(t),\,\,l=0,2,4,...\}$ and $\{a_l(t),\,\,l=1,3,5,...\}$. In the 
simplest non-trivial case, the only singularity of $a_\eta(l,t)$ is a single 
$t$-dependent pole at $l=\alpha(t)$. It can then be shown that, in the limit 
$s\to\infty$,
\be
T_{12\to 34}(s,t)=\beta_{13}(t)\beta_{24}(t)\,\zeta_\eta(\alpha(t))\,
\left(\frac{s}{s_0}\right)^{\alpha(t)}\,,\label{rf}
\ee
where $s_0$ is an arbitrary scale factor, $\beta_{13}$ and $\beta_{24}$ are 
two unknown functions of $t$, and 
\be
\zeta_\eta(\alpha(t))=\frac{1+\eta e^{-i\pi\alpha(t)}}{\sin \pi\alpha(t)}
\label{sifa}
\ee
is the signature factor, depending on the signature $\eta$ of the relevant 
Regge trajectory $\alpha(t)$. If $a_\eta(l,t)$ has a more complicated 
analytic structure, the rightmost singularity in the $l$ plane dominates 
the behaviour at large $s$. 

Within the present context, the essential predictions of the asymptotic 
expression Eq.~(\ref{rf}) are the power-like energy dependence 
$s^{\alpha(t)}$ and the factorization of the unknown $t$ dependence into 
the two vertex factors $\beta_{13}$ and $\beta_{24}$. This last feature, 
which underlies the graphic representation of reggeon exchange in 
Fig.~\ref{fig:reg}, is relevant if the same Regge trajectory governs 
different scattering processes. Note also that, for positive $t=s'$ and 
integer $l$, $\alpha(t)$ describes the positions of poles of the physical 
amplitude $T_{1\bar{3}\to\bar{2}4}(s',t')$. Such poles are expected whenever 
an on-shell particle with appropriate mass $m^2=s'$ and angular momentum $l$ 
can be created in the collision of 1 and $\bar{3}$. Indeed, most Regge 
trajectories pass through known physical states with mass $m^2=t$ and 
angular momentum $\alpha(t)$. 

The Froissart bound~\cite{froi} on the high-energy growth of total cross 
sections, 
\be
\sigma_{tot}\le\frac{\pi}{m_\pi^2}\ln^2(s/s_0)\,,
\ee
where $m_\pi$ is the pion mass and $s_0$ is an unknown scale factor, implies 
that $\alpha(0)\le 1$ for all Regge trajectories. However, it was observed 
early on that a very good fit to $pp$ and $p\bar{p}$ cross sections, which 
were measured to rise at high energy, could be obtained by assuming the 
dominance of a single pole with $\alpha(0)>1$~\cite{cgm}. The corresponding 
trajectory, originally introduced with $\alpha(0)=1$~\cite{cf}, is known as 
the pomeron trajectory (cf. the Pomeranchuk condition of~\cite{pom}). 

Donnachie and Landshoff demonstrated that a large set of different hadronic 
cross sections can be fitted with an intercept $\ap(0)=1.08$~\cite{dl}. 
In spite of the power-like growth of Eq.~(\ref{rf}), the predicted cross 
sections are so small that the Froissart bound is not violated below the 
Planck scale. It is then argued that unitarity is not a serious problem at 
all realistic energies. However, one should keep in mind that all following 
analyses based on a pomeron trajectory with $\alpha(0)>1$ are, strictly 
speaking, not self-consistent in the framework of Regge theory. In fact, it 
is likely that the rightmost singularity in the complex $l$ plane is not a 
single pole but a cut, in which case many of the following results would 
have to be reconsidered. 

\begin{figure}[ht]
\begin{center}
\vspace*{.2cm}
\parbox[b]{6cm}{\psfig{width=6cm,file=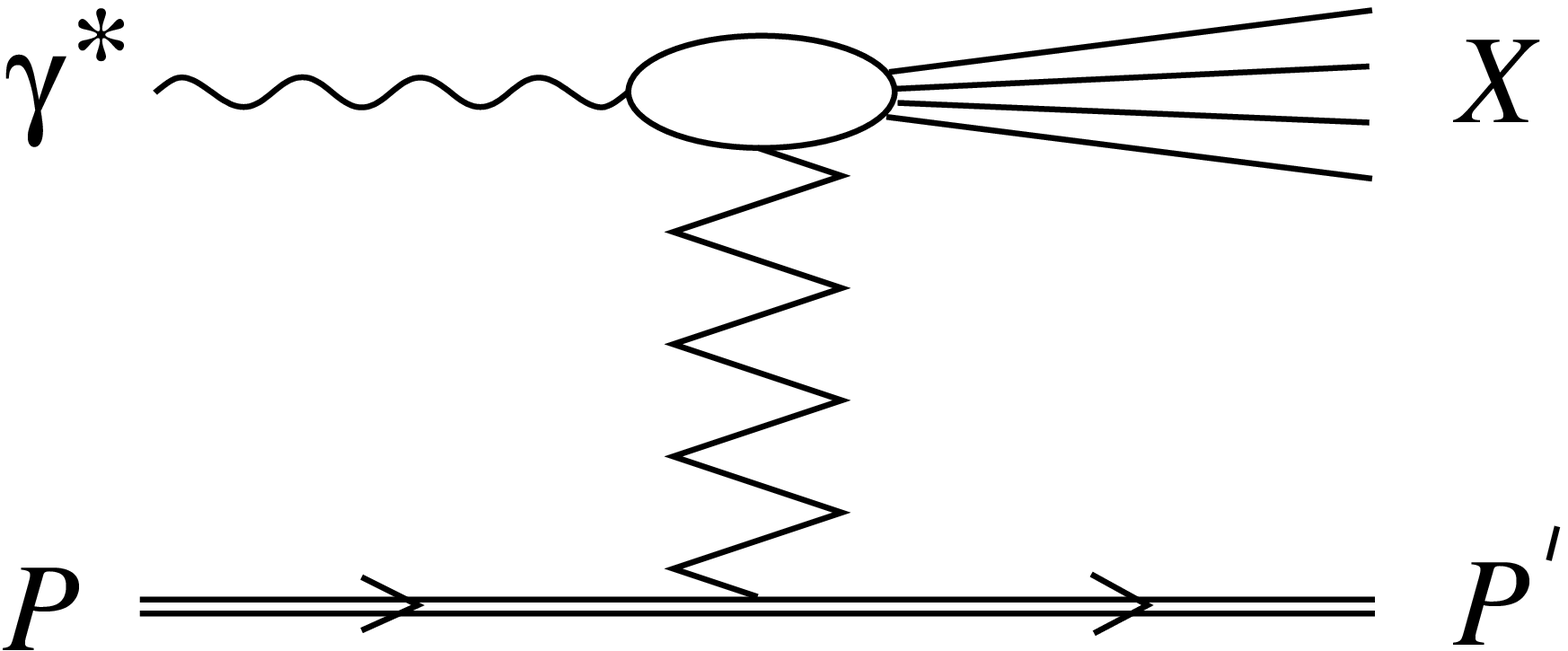}}\\
\end{center}
\refstepcounter{figure}
\label{fig:pe}
{\bf Figure \ref{fig:pe}:} Diffractive electroproduction via pomeron 
exchange. 
\end{figure}

Given the universal success of the Donnachie-Landshoff pomeron, it is 
natural to apply the same trajectory to the quasi-elastic amplitude of 
diffractive electroproduction. Thus, the diagram in Fig.~\ref{fig:ajm}b is 
interpreted in terms of pomeron exchange (cf. Fig.~\ref{fig:pe}). In this 
situation, the soft energy dependence of the pomeron and the measured $t$ 
dependence of the proton-proton-pomeron vertex are naturally combined with 
the aligned jet model prediction of Eq.~(\ref{ajm}). Ignoring the vertex 
factor from the upper part of Fig.~\ref{fig:pe}, one can write 
\be
\frac{d\sigma_T}{dt\,dM^2}\sim \frac{\beta_{pp}^2(t)\,s^{\,2\,(\ap(t)-1)}}
{(M^2+Q^2)^2}\,.
\ee
The non-trivial energy dependence represents a clear improvement compared 
to Eqs.~(\ref{ajm}) and (\ref{ajmf2d}). However, our ignorance of the upper 
vertex prevents an unambiguous prediction of the $M^2$ distribution. In 
fact, the obtained suppression at large $M^2$, which corresponds to the 
region of small $\beta$, appears to be the main qualitative problem of the 
presented model. Furthermore, the treatment of the hard scale $Q^2$ is 
rather na\"\i ve in view of the impressive successes of QCD perturbation 
theory in the description of inclusive structure functions. 

As is shown in the next section, a better understanding of the $M^2$ 
dependence is possible in the framework of Regge theory. The obtained 
results allow for an interpretation of the upper vertex in 
Fig.~\ref{fig:pe} in terms of a $\gamma^*$-pomeron collision and thus for 
the application of QCD perturbation theory.

\section{Pomeron structure function}\label{sect:psf}
It is the purpose of this section to explain the interpretation of 
Fig.~\ref{fig:pe} in terms of a $\gamma^*$-pomeron collision, thereby 
providing the background for the ensuing partonic treatment of the pomeron. 
Before doing so, it is helpful to recall Mueller's generalization of the 
optical theorem and the resulting triple-pomeron interpretation of single 
diffractive dissociation. 

\begin{figure}[ht]
\begin{center}
\vspace*{.1cm}
\parbox[b]{15cm}{\psfig{width=15cm,file=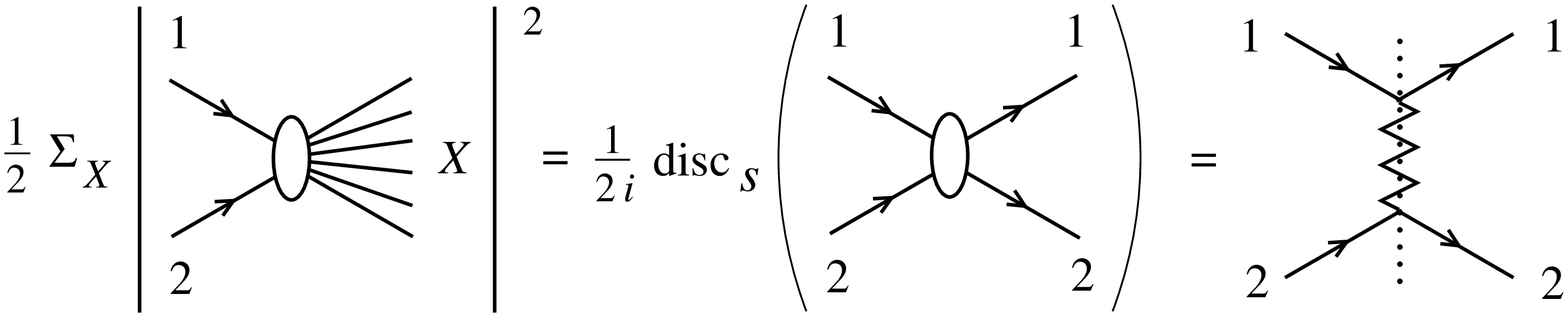}}\\
\vspace*{-.1cm}
\end{center}
\refstepcounter{figure}
\label{fig:ot}
{\bf Figure \ref{fig:ot}:} Graphic representation of the optical theorem for
a two-particle total cross section. In the rightmost diagram, which is 
relevant in the pomeron-dominated high-energy limit, the cut is represented 
by a dotted line. 
\end{figure}

The optical theorem relates the two-particle total cross section to the 
imaginary part of the forward amplitude. This is illustrated by the first 
equality in Fig.~\ref{fig:ot}. Here $\sqrt{s}$ is the centre-of-mass energy 
of particles 12, i.e., the mass of the system $X$, and the discontinuity 
across the real axis is defined by 
\be
\mbox{disc}_s A(s)= A(s+i\epsilon)-A(s-i\epsilon)\,.
\ee
The last equality in Fig.~\ref{fig:ot} illustrates the high-energy limit, 
which is dominated by pomeron exchange. 

\begin{figure}[ht]
\begin{center}
\vspace*{.1cm}
\parbox[b]{15.6cm}{\psfig{width=15.6cm,file=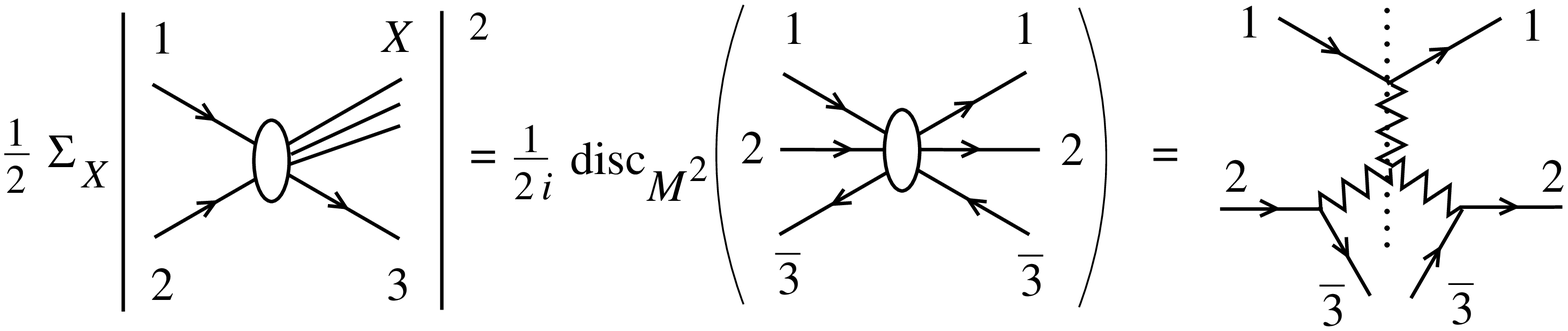}}\\
\vspace*{-.2cm}
\end{center}
\refstepcounter{figure}
\label{fig:mot}
{\bf Figure \ref{fig:mot}:} Graphic representation of Mueller's 
generalization of the optical theorem. In the rightmost diagram, relevant 
for $s/M^2\to\infty$ and $M^2\to\infty$, the cut is only through the 
pomeron coupled to particle 1.
\vspace*{-.1cm}
\end{figure}

Mueller's generalization~\cite{mue} (see also~\cite{col}), which is 
illustrated by the first equality in Fig.~\ref{fig:mot}, relates a 
one-particle inclusive cross section to a six-particle amplitude. The 
derivation uses crossing symmetry to reinterpret the outgoing particle 
3 as the incoming antiparticle $\bar{3}$. In very much the same way as for 
the usual optical theorem, completeness of the sum over $X$ and unitarity 
are employed to relate the inclusive cross section to the discontinuity of 
the amplitude. Note, however, that in Fig.~\ref{fig:mot} the discontinuity 
is taken in $M^2$, which is the squared mass of the system $X$ and different 
from the variable $s$ characterizing the original process with incoming 
particles 1 and 2. 

The rightmost diagram in Fig.~\ref{fig:mot} is obtained in the double 
limit $s/M^2\to\infty$ and $M^2\to\infty$~\cite{tar} (see also~\cite{col}). 
If $s\gg M^2$ and $M^2$ is much larger than any of the other variables, the 
process $12\to3X$ is dominated by pomeron exchange, and the amplitude 
receives a factor $\zeta_\eta(\alpha(t))\,(s/M^2)^{\ap(t)}$. The appearance 
of the ratio $s/M^2$ is a non-trivial result of the relevant kinematics. The 
remaining $M^2$ dependence of the amplitude is given by the cut that is left 
after the two pomerons coupled to particles 2 and 3 are factorized. If the 
high-energy limit of this amplitude, which corresponds to $M^2\to\infty$, is 
again dominated by pomeron exchange, a factor $\zeta_\eta(\alpha(0))\, 
(M^2)^{\ap(0)}$ results. The appearance of these three `pomeron 
propagators' is illustrated by the three zigzag lines on the r.h. side of 
Fig.~\ref{fig:mot}. The corresponding cut amplitude reads 
\be
\mbox{disc}\,T=\left|\beta_{23}(t)\,\zeta_\eta(\alpha(t))\,\left(\frac{s}
{M^2}\right)^{\ap(t)}\right|^2 \, \mbox{disc}\left(G(t)\,\zeta_\eta
(\alpha(0))\,\left(\frac{M^2}{s_0}\right)^{\ap(0)} \beta_{11}(0)\right)\,,
\label{tpa}
\ee
where $G(t)$ is the unknown triple-pomeron vertex. Since the $t$ dependence 
of $G(t)$ is measured to be weak and since, for the pomeron, $\eta=+1$, the 
approximations $G(t)\simeq G(0)$ and $|\zeta_\eta|\simeq 1$ can be used at 
small $|t|$. The resulting cross section for the process $12\to3X$ is given 
by 
\be
\frac{d\sigma}{dt\,dM^2}=\frac{1}{16\pi^2 M^2s_0}|\beta_{23}(t)|^2
\left(\frac{s}{M^2}\right)^{\,2\,(\ap(t)-1)}\,|G(0)\beta_{11}(0)|\,\left(
\frac{M^2}{s_0}\right)^{\ap(0)-1}\,.\label{dstm}
\ee

This expression for the cross section suggests the following interpretation
\cite{kt}. Incoming particle 2 radiates a pomeron carrying a fraction 
$\xi=M^2/s$ of its momentum. Then, this pomeron collides with particle 1 
producing the diffractive state $X$ with mass $M$. According to this 
interpretation, Eq.~(\ref{dstm}) can be rewritten as 
\be
\frac{d\sigma}{dt\,d\xi}=f_{\pom}(\xi,t)\sigma_{1\pom}(M^2)\,,\label{dstxi}
\ee
where 
\be
f_{\pom}(\xi,t)=\frac{1}{16\pi^2s_0}|\beta_{23}(t)|^2\,\xi^{\,1-2\ap(t)}
\ee
is the pomeron flux factor, characterizing the probability for the 
transition $2\to3\pom$, and
\be
\sigma_{1\pom}(M^2)=|G(0)\beta_{11}(0)|\,\left(\frac{M^2}{s_0}
\right)^{\ap(0)-1}
\ee
is the total cross section for the collision of the pomeron and particle 1. 
Since the pomeron is not a real particle, the decomposition of 
Eq.~(\ref{dstm}) into pomeron flux and total cross section is ambiguous. 
The definitions given here correspond to interpreting $\beta_{23}(t)$ and 
$\zeta_\eta(\alpha(t))\,(s/M^2)^{\ap}$ in Eq.~(\ref{tpa}) as 
pomeron-particle-particle vertex and pomeron propagator respectively. The 
rest of the diagram corresponds to the pomeron-particle scattering 
amplitude, which is used to define the pomeron-particle cross section. 

In their seminal paper on diffractive jet production in hadron-hadron 
scattering, Ingelman and Schlein exploited this picture to predict the rate 
of high-$p_\perp$ jets within the diffractive final state $X$~\cite{is}. 
They suggested that, according to the parton model, hard processes in the 
collision of particle 1 and pomeron can be described in terms of a 
convolution of two parton distributions and a hard partonic cross section. 
The idea of introducing a parton distribution for the pomeron proved 
to be very successful phenomenologically and was widely used in subsequent 
analyses of hard diffractive processes. 

\begin{figure}[ht]
\begin{center}
\vspace*{-.3cm}
\parbox[b]{6cm}{\psfig{width=6cm,file=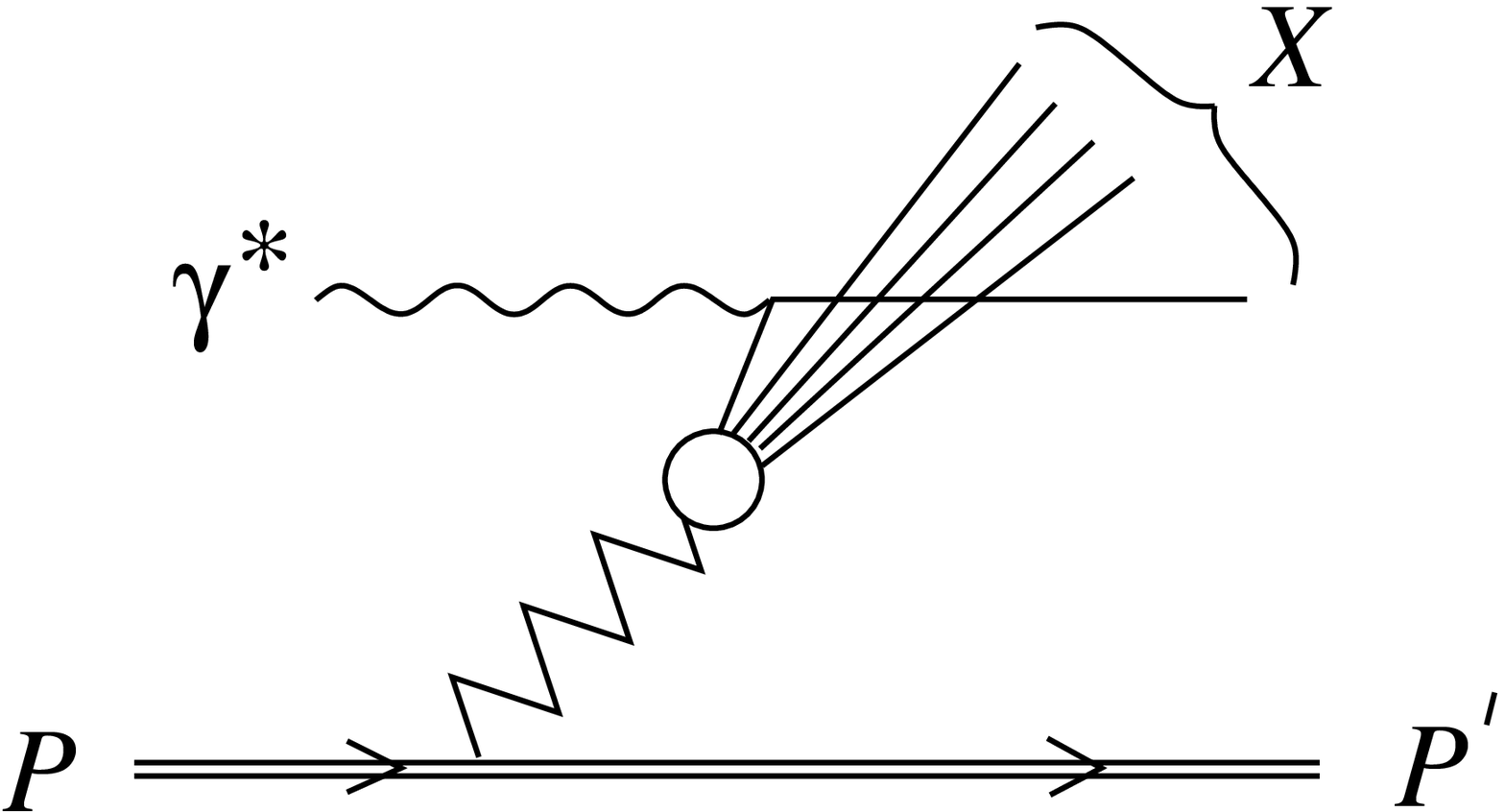}}\\
\vspace*{-.1cm}\end{center}
\refstepcounter{figure}
\label{fig:psf}
{\bf Figure \ref{fig:psf}:} Diffractive DIS via the pomeron structure 
function.
\end{figure}

Donnachie and Landshoff applied this idea to the case of diffractive 
DIS~\cite{dl1,dl2}, where particle 1 is a highly virtual photon and can 
therefore be treated in perturbation theory. In this approach, HERA 
diffraction at the level of a few per cent of the total DIS cross section 
was predicted as early as 1987~\cite{dl1}. In contrast to Ingelman and 
Schlein, who focussed on the gluon distribution of the pomeron, Donnachie 
and Landshoff introduced a quark distribution, which can be directly probed 
by the virtual photon. The resulting physical picture of diffraction is 
shown in Fig.~\ref{fig:psf}. 

In complete analogy to Eq.~(\ref{dstxi}), the diffractive cross section can 
be written as
\be
\frac{d\sigma_T}{dt\,d\xi}=f_{\pom}(\xi,t)\,\sigma_T^{\gamma^*\pom}\,,
\label{dsdtdxi}
\ee
where, according to the conventional parton model, the photon-pomeron cross 
section is given in terms of the quark distribution $q_{\pom}(\beta,Q^2)$ 
of the pomeron, $\sigma_T^{\gamma^*\pom}=(\pi e^2/Q^2)\,2\,\beta q_{\pom}
(\beta,Q^2)$. Here $\beta=x/\xi$ is the fraction of the pomeron momentum 
carried by the struck quark, and the factor 2 is introduced to account for 
the antiquark contribution. The resulting diffractive structure function 
reads
\be
F_2^{D(4)}(x,Q^2,\xi,t)=f_{\pom}(\xi,t)\,2\beta q_{\pom}(\beta,Q^2)\,,
\label{f2dp}
\ee
and a non-trivial $Q^2$ dependence is naturally expected on the basis of the 
Altarelli-Parisi evolution of the quark distribution. 

The above normalization of pomeron flux and pomeron-particle cross section 
is consistent with~\cite{col}. However, other normalizations are also 
frequently used (compare the conventions of~\cite{is,dl1,dl2,kt,beta}).

\section{Diffractive parton distributions}\label{sect:dpd}
Loosely speaking, the analysis of diffraction in terms of diffractive 
parton distributions is equivalent to the analysis in terms of the 
\nopagebreak
partonic pomeron of Ingelman and Schlein, `minus its Regge content' 
\cite{sop}. Diffractive parton distributions provide a 
\pagebreak[0]
direct, perturbative-QCD-based approach to the hard part of the process, 
without introducing any less well established non-perturbative concepts. 

The basic theoretical ideas are due to Trentadue and Veneziano, who 
proposed to parametrize semi-inclusive hard processes in terms of 
`fracture functions'~\cite{vt}, and to Berera and Soper, who defined 
similar quantities for the case of hard diffraction~\cite{bs} and coined 
the term `diffractive parton distributions'. The following discussion is 
limited to the latter, more specialized framework. 

In short, diffractive parton distributions are conditional probabilities. 
A diffractive parton distribution $df^D_i(y,\xi,t)/d\xi\,dt$ describes the 
probability of finding, in a fast moving proton, a parton $i$ with momentum 
fraction $y$, under the additional requirement that the proton 
remains intact while being scattered with invariant momentum transfer $t$ 
and losing a small fraction $\xi$ of its longitudinal momentum. Thus, the 
corresponding $\gamma^*p$ cross section can be written as~\cite{bs1} 
\be
\frac{d\sigma(x,Q^2,\xi,t)^{\gamma^*p\to p'X}}{d\xi\,dt}=\sum_i\int_x^\xi 
dy\,\hat{\sigma}(x,Q^2,y)^{\gamma^*i}\left(\frac{df^D_i(y,\xi,t)}{d\xi\,dt}
\right)\, ,\label{sx}
\ee
where $\hat{\sigma}(x,Q^2,y)^{\gamma^*i}$ is the total cross section for 
the scattering of a virtual photon characterized by $x$ and $Q^2$ and 
a parton of type $i$ carrying a fraction $y$ of the proton momentum. The 
above factorization formula holds in the limit $Q^2\to\infty$ with $x$, 
$\xi$ and $t$ fixed. 

At leading order and in the case of transverse photon polarization, only 
the quark distribution contributes. For one quark flavour with one unit of 
electric charge, the well-known partonic cross section reads 
\be
\hat{\sigma}_T(x,Q^2,y)^{\gamma^*q}=\frac{\pi e^2}{Q^2}\,\delta(1-y/x)\,,
\ee
giving rise to the diffractive cross section 
\be
\frac{d\sigma(x,Q^2,\xi,t)^{\gamma^*p\to p'X}}{d\xi\,dt}=\frac{2 \pi e^2}
{Q^2}\,\frac{x\,df^D_q(x,\xi,t)}{d\xi\,dt}\, ,
\ee
where the factor $2$ is introduced to account for the antiquark 
contribution. 

The main difference to fracture functions is the requirement of a specific 
$t$ transferred to the final state hadron. Since fracture functions are 
defined to include the $t$ integration, a non-negligible contribution to 
the relevant cross section arises from processes where the outgoing hadron, 
in this case the proton, belongs to the current fragmentation region. This 
contribution complicates the $Q^2$ dependence of fracture functions, but 
not of diffractive parton distributions. Even if the $t$ integration is 
performed, the above problem does not arise in diffractive kinematics, where 
$\xi$ is small and the production of a very energetic forward proton in 
the fragmentation of the current is strongly suppressed. 

It is now obvious that Eqs.~(\ref{dsdtdxi}) and (\ref{f2dp}) of the last 
section form a special case of the present more general framework. There, 
Regge phenomenology and the assumption of a partonic pomeron lead to the 
specific form 
\be
x\,\frac{df^D_q(x,\xi,t)}{d\xi\,dt}=f_{\pom}(\xi,t)\,\beta q_{\pom}(\beta)
\ee
for the diffractive quark distribution. However, the factorizing dependence 
on $\beta=x/\xi$ and the expression for the pomeron flux factor $f_{\pom}
(\xi,t)$ given in the last section have not been derived in the framework of 
QCD. 

The operator definition of diffractive parton distributions can be obtained 
as follows~\cite{bs1}. Consider first a complex scalar field $\phi(x)$, 
which can be written as 
\be
\phi(x)=\int d\tilde{k}\left[\hat{a}(k)e^{-ikx}+\hat{b}^\dagger(k)
e^{ikx}\right]\quad,\quad d\tilde{k}=\frac{d^3\vec{k}}{(2\pi)^3 2k_0}\,,
\ee
where $\hat{a}$ and $\hat{b}$ are the annihilation operators of particles 
and antiparticles respectively. In this field theory, the number of particles 
with momentum in the interval $d^3\vec{k}$ is measured by the operator 
\be
d\hat{N}=d\tilde{k}\,\hat{a}^\dagger(k)\hat{a}(k)\,.
\ee
The conventional inclusive distribution $f(y)$ of partons of the field 
$\phi$ is given by the number of particles in a momentum interval 
$dk_+=P_+dy$, normalized to the size of the interval $dy$. Note that, in 
this section, the co-ordinate system is such that the proton and photon 
momenta have large plus and minus components respectively. This is 
customary for the discussion of DIS in the Breit frame and contrasts with 
the remainder of this review, where the proton rest frame is emphasized and, 
correspondingly, the photon momentum is defined to have a large plus 
component. 

The particle number is measured in the state of a hadron with momentum $P$, 
and integration over all transverse momenta $k_\perp$ is assumed. Thus, the 
explicit formula reads
\be
(2\pi)^3\,2P_0\,\delta^3(\vec{P}'-\vec{P})\,f(y)\,dy=\int d^2k_\perp\,
\langle P'|\frac{d\hat{N}}{d^2k_\perp}|P\rangle\,,
\ee
where the prefactor $(2\pi)^3\,2P_0\,\delta^3(\vec{P}'-\vec{P})$ is a result 
of the normalization of the hadronic state $|P\rangle$,
\be
\langle P'|P\rangle=(2\pi)^3\,2P_0\,\delta^3(\vec{P}'-\vec{P})\,.
\ee
It can then be shown that, in terms of the fundamental field $\phi$, the 
distribution of scalar partons reads
\be
f(y)=\frac{y P_+}{4\pi}\int dx_- e^{-iy P_+x_-/2}\langle P|
\phi^\dagger(0,x_-,0_\perp)\phi(0,0,0_\perp)|P\rangle\,.\label{spd}
\ee
In the case of a gauge theory, this definition has to be supplemented with 
a link operator
\be
U_{x_-,0}=P\exp\left(-\frac{i}{2}\int_0^{x_-}A_+(0,y_-,0_\perp)
dy_-\right)\,,
\ee
connecting the two scalar field operators. 

Before generalizing this definition of the parton distribution to the 
diffractive case, it is convenient to rewrite it in the form
\be
f(y)=\frac{y P+}{4\pi}\int dx_- e^{-iy P_+x_-/2}\sum_X\langle P|
\phi^\dagger(0,x_-,0_\perp)U_{x_-,\infty}|X\rangle\,\langle X|U_{\infty,0}
\,\phi(0,0,0_\perp)|P\rangle\,.\label{odi}
\ee
In the diffractive case, the operators describing the creation and 
annihilation of the parton are the same. However, the proton is required to 
appear in the final state carrying momentum $P'$. Thus, the above definition
is changed to
\bea
\frac{df^D_q(y,\xi,t)}{d\xi\,dt}&=&\frac{y P_+}{64\pi^3}\int dx_- 
e^{-iy P_+x_-/2}\label{od}\\
&&\times\sum_X\langle P|\phi^\dagger(0,x_-,0_\perp)U_{x_-,\infty}|
P',X\rangle\,\langle P',X|U_{\infty,0}\,\phi(0,0,0_\perp)|P\rangle\,,
\nonumber
\eea
where $\langle P',X|$ denotes the outgoing state, the only restriction on 
which is the presence of a scattered proton with momentum $P'$. 

The above formulae for inclusive and diffractive distributions of scalar 
partons are generalized to the case of spinor quarks by the substitution 
\be
\phi^\dagger(0,x_-,0_\perp)\phi(0,0,0_\perp)\,\longrightarrow
\frac{1}{2y P_+}\bar{\psi}(0,x_-,0_\perp)\gamma_+\psi(0,0,0_\perp)\,,
\ee
and to the case of gluons by the substitution
\be
\phi^\dagger(0,x_-,0_\perp)\phi(0,0,0_\perp)\,\longrightarrow\frac{1}
{(y P_+)^2}F^\dagger(0,x_-,0_\perp)^{+\mu}F(0,0,0_\perp)_\mu{}^+\,.
\label{gsub}
\ee

The operator expressions appearing in the definitions of both the inclusive 
and diffractive parton distributions are ultraviolet divergent. They are 
conveniently renormalized with the $\overline{\mbox{MS}}$ prescription, 
which introduces the scale $\mu$ as a further argument. The distribution 
functions then read $f(x,\mu^2)$ and $df^D(x,\xi,t,\mu^2)/d\xi dt$. 

Accordingly, Eq.~(\ref{sx}) has to be read in the $\overline{\mbox{MS}}$ 
scheme, with a $\mu$ dependence appearing both in the parton distributions 
and in the partonic cross sections. The claim that Eq.~(\ref{sx}) holds to 
all orders implies that these $\mu$ dependences cancel, as is well known 
in the case of conventional parton distributions. Since the partonic cross 
sections are the same in both cases, the diffractive distributions obey 
the usual Altarelli-Parisi evolution equations, 
\be
\frac{d}{d(\ln\mu^2)}\,\frac{df^D_i(x,\xi,t,\mu^2)}{d\xi\,dt}=\sum_j 
\int_x^\xi\frac{dy}{y}P_{ij}(x/y)\frac{df^D_j(y,\xi,t,\mu^2)}{d\xi\,dt}\,. 
\ee
with the ordinary splitting functions $P_{ij}(x/y)$.

Clearly, this is equivalent to the assertion that, in the operator 
definition of Eq.~(\ref{od}), the ultraviolet divergences are independent 
of the final state proton $P'$. If this is the case, the Altarelli-Parisi 
evolution of the distribution functions follows from the operator 
definitions exactly as in the inclusive case of Eq.~(\ref{odi}). 

Thus, for the analysis of diffractive DIS, it is essential to gain 
confidence in the validity of the factorization formula Eq.~(\ref{sx}). 
Berera and Soper first pointed out~\cite{bs1} that such a factorization 
proof could be designed along the lines of related results for other QCD 
processes~\cite{fact} (see~\cite{cssr} for a review). 

Using Mueller's method of cut vertices~\cite{cv}, Grazzini, Trentadue and 
Veneziano~\cite{gtv} proved, in the framework of a simple scalar model, that 
the above factorization property holds for `extended fracture functions'. 
These objects differ from fracture functions in that they are differential 
in the momentum transfer $t$. They are thus equivalent to the diffractive 
parton distributions discussed here. 

Collins~\cite{cfa} showed that factorization holds in full QCD by 
generalizing the essential step of dealing with soft gluon interactions, 
which lies at the heart of many previous factorization proofs~\cite{cssr}, 
to the case of diffractive DIS. A brief summary of the essential arguments 
is presented below. 

\begin{figure}[ht]
\begin{center}
\vspace*{.2cm}
\parbox[b]{7cm}{\psfig{width=7cm,file=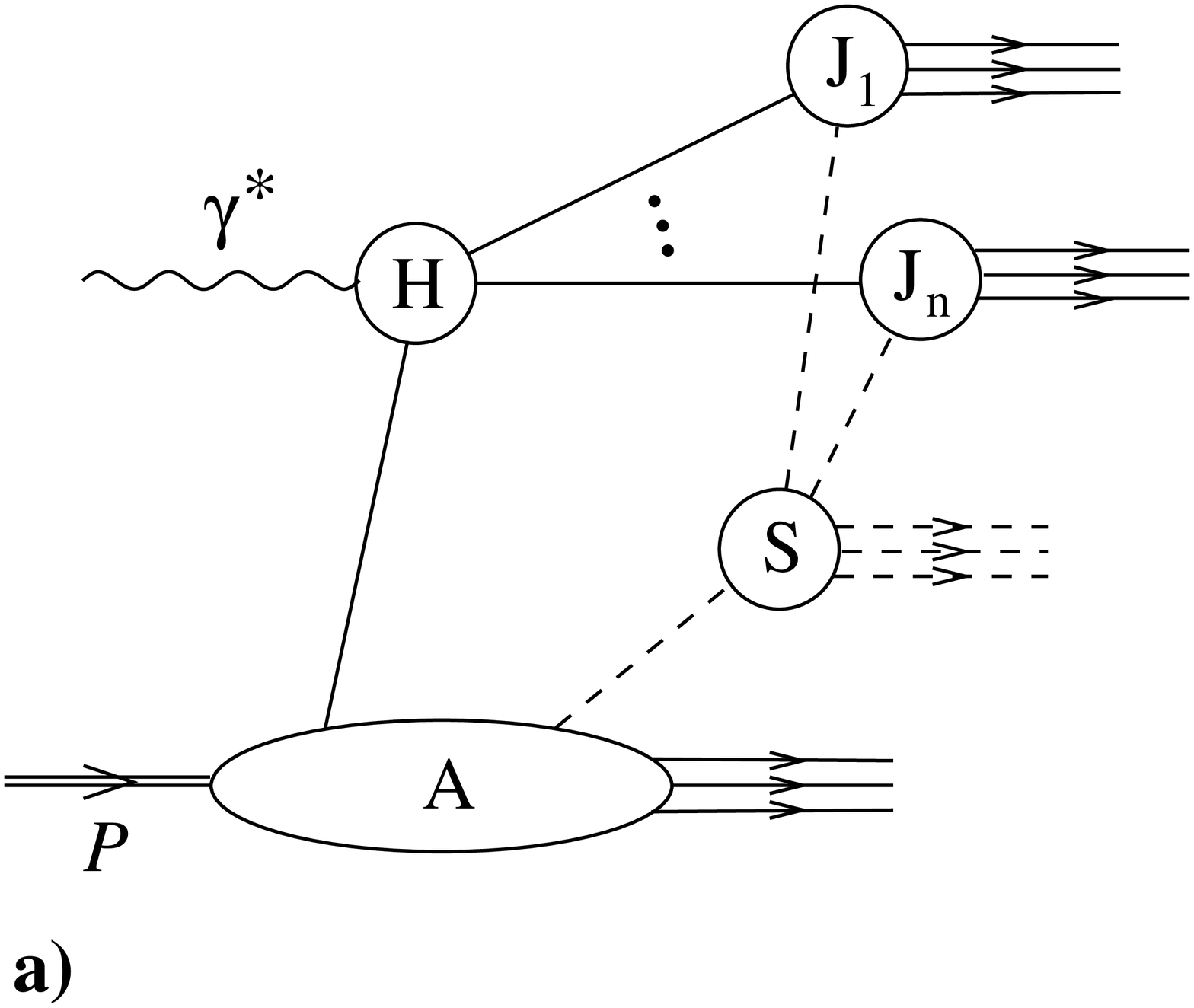}}\hspace{1cm}
\parbox[b]{7cm}{\psfig{width=7cm,file=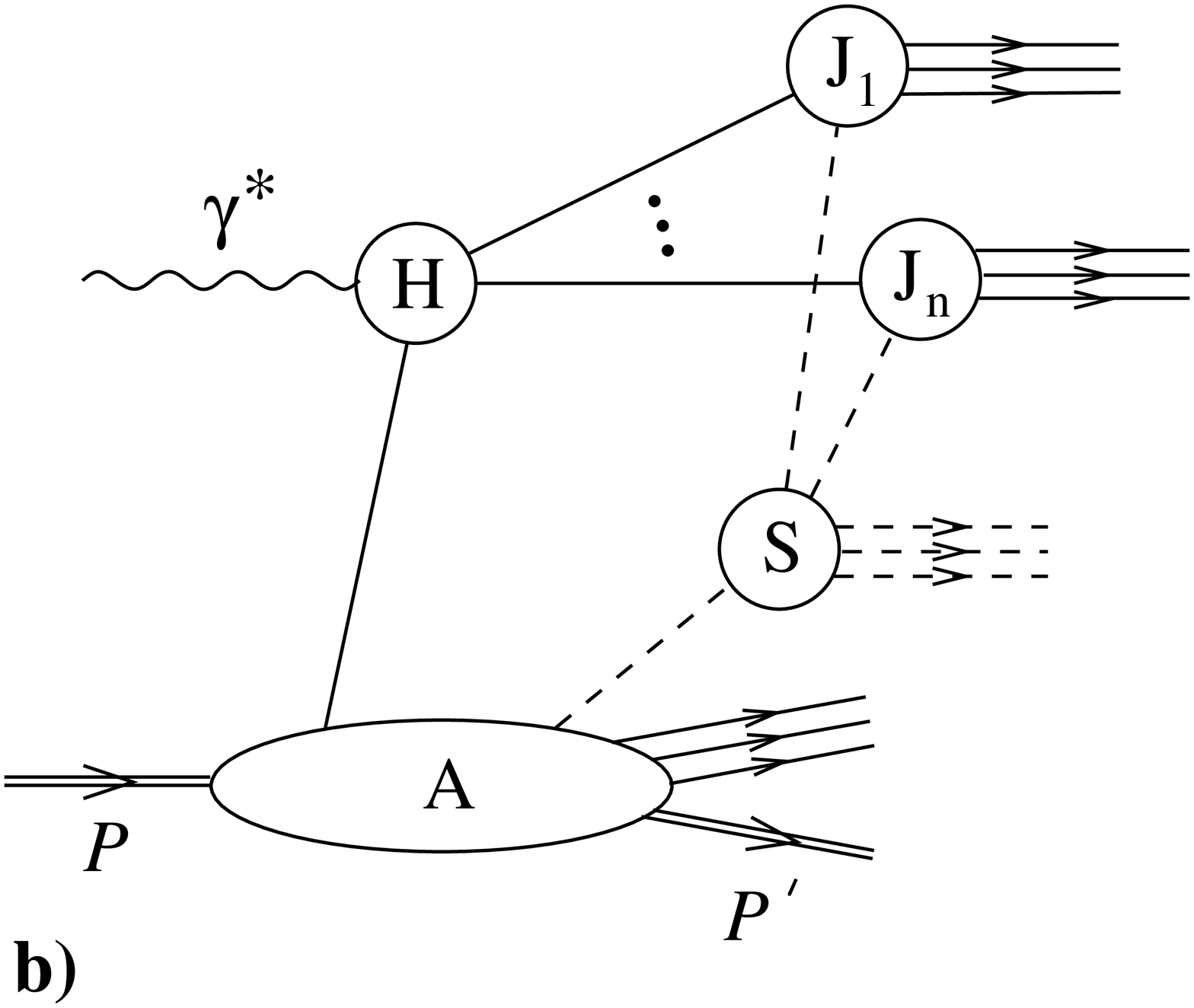}}\\
\end{center}
\refstepcounter{figure}
\label{fig:fact}
{\bf Figure \ref{fig:fact}:} Leading regions in inclusive (a) and 
diffractive (b) DIS (cf.~\cite{cfa}).
\end{figure}

To begin with, recall the main ideas of the factorization proof for 
inclusive DIS. It is based on the dominance of contributions from so-called 
leading regions, shown in Fig.~\ref{fig:fact}a. Here the hard subgraph is 
denoted by H, the subgraphs with momenta collinear to $P$ and to the 
produced jets are denoted by A and J${}_1\cdots$ J${}_n$ respectively, and 
the soft subgraph is denoted by S. The analysis is performed in the Breit 
frame, where lines in H have typical virtuality $Q^2$, lines in A and 
J${}_1\cdots$ J${}_n$ have small virtualities but may have large 
longitudinal momentum components of order $Q$, and all components of momenta 
in S are small compared to $Q$. 

The rationale behind the discussion in terms of leading regions is the 
realization that the momentum integrations are dominated by pinch 
singularities, i.e., poles of the propagators that can not be avoided by 
deforming the integration contours. In the limit of large $Q^2$, such 
singularities are associated with the leading regions shown in 
Fig.~\ref{fig:fact}. By power counting, only one hard line may connect the 
A and H subgraphs, while arbitrarily many soft (dashed) lines may connect 
the soft subgraph with other parts of the diagram. 

An essential part of the factorization proof is the demonstration that the 
soft subgraph can be factored out. More specifically, it has to be shown 
that the soft lines are not important for the subgraphs H and J${}_1\cdots$ 
J${}_n$, so that a perturbative hard cross section with free partons in the 
final state can be used. This is achieved using Ward identities, which, 
however, can only be applied in the region where all components of the 
soft momenta are small and of comparable size. Difficulties arising in the 
region where one component of a soft momentum is much smaller than the 
other components can be solved by appropriately deforming the integration 
contour. 

\begin{figure}[ht]
\begin{center}
\vspace*{.2cm}
\parbox[b]{6.3cm}{\psfig{width=6.3cm,file=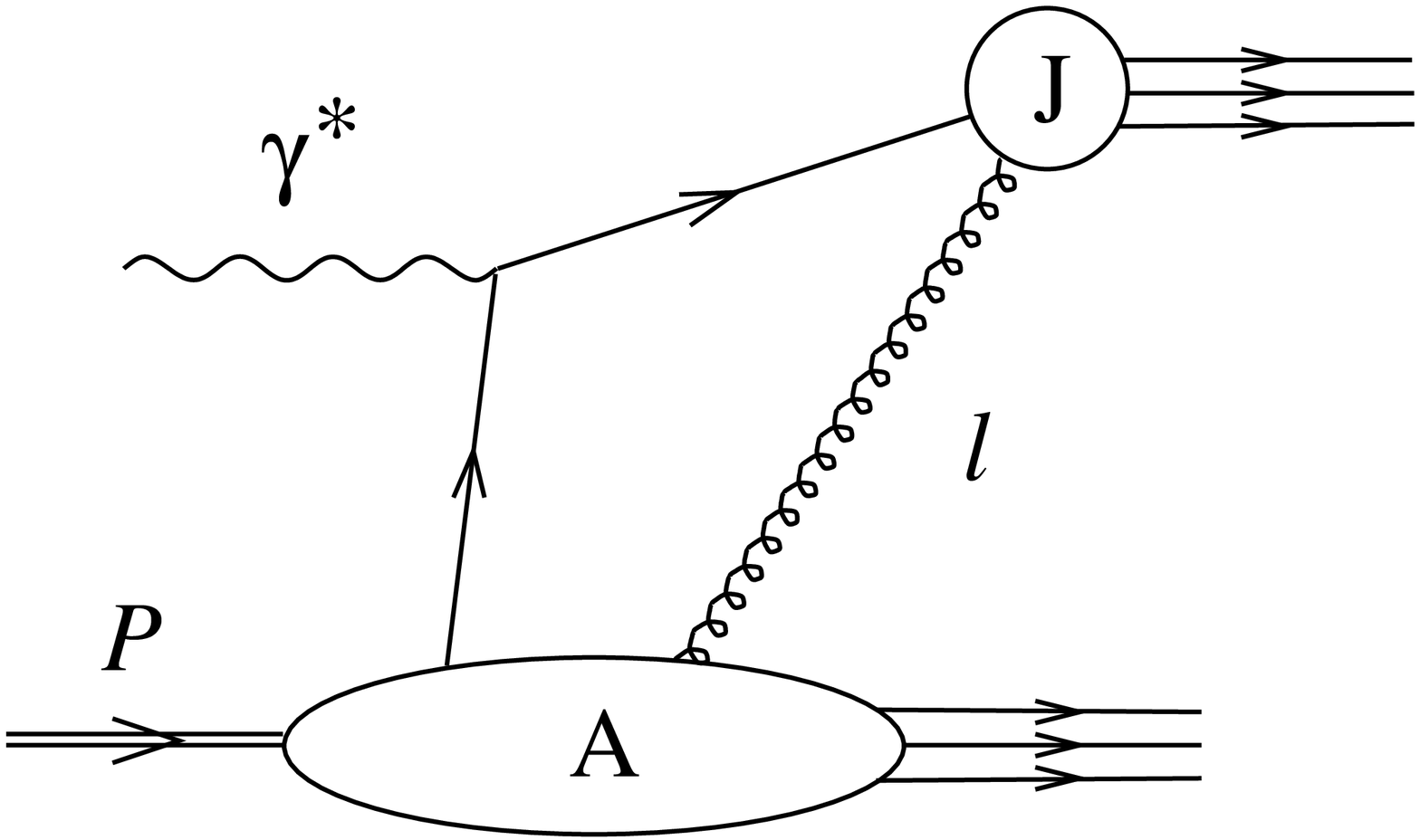}}
\end{center}
\refstepcounter{figure}
\label{fig:og}
{\bf Figure \ref{fig:og}:} Leading order process in DIS with additional 
exchange of a soft gluon between the target and current jet subgraphs 
(cf.~\cite{cfa}). 
\end{figure}

To see this in more detail, consider the particularly simple diagram of 
Fig.~\ref{fig:og}, where there is only one current jet, and a single soft 
gluon connects the corresponding jet subgraph with subgraph A. Let the soft 
gluon with momentum $l$ couple to outgoing particles with momenta 
$k_{\mbox{\scriptsize J}}$ and $k_{\mbox{\scriptsize A}}$ in the subgraphs 
J and A respectively. The particle propagators 
\be
\frac{i}{(k_{\mbox{\scriptsize J}}-l)^2-m^2+i\epsilon}\qquad\mbox{and}\qquad
\frac{i}{(k_{\mbox{\scriptsize A}}+l)^2-m^2+i\epsilon}
\ee
attached to the gluon vertices produce poles in the complex $l_+$ and $l_-$ 
planes that lie above and below the real axis respectively. This is also 
true for further $l$ dependent propagators in J and A since, due to 
completeness, final state interactions can be disregarded in both subgraphs. 
Thus, regions where either $l_+$ or $l_-$ are too small can always be 
avoided by deforming the integration contour. This takes us back to the 
genuinely soft region where all components of $l$ are small and of comparable 
size, and where Ward identities can be used to factor out the soft subgraph. 

It is precisely this part of the factorization proof that is affected in the 
case of diffraction. The requirement of a final state proton leads to the 
presence of final state interactions in subgraph A of Fig.~\ref{fig:og} 
(as shown in Fig.~\ref{fig:fact}b) that can not be neglected. Since both 
initial and final state interactions appear in A, poles on both sides of the 
$l_-$ plane exist, and the contour can no longer be deformed to avoid the 
dangerous region of too small $l_-$. However, Collins was able to 
demonstrate~\cite{cfa} that deformations of the $l_+$ contour can instead be 
used to show that this dangerous region does not produce a non-factorizing 
contribution. The argument is based on a detailed analysis of the pole 
structure in subgraphs J${}_1\cdots$ J${}_n$ (or the single jet subgraph J 
in the simple example above). It leads to the conclusion that, exactly as in 
the inclusive case, the soft subgraph can be factored out. The result is a 
convolution of the calculable hard part with diffractive parton 
distributions. 

Note that diffractive factorization does not hold in the case of 
hadron-hadron collisions. A general argument for this was given in 
\cite{cfs}, and the effect was also found in the model calculation of 
\cite{bs}. The reason is that, in contrast to the DIS case, the use of 
completeness in the final state can not be avoided in the factorization 
proof for hadron-hadron collisions. This completeness is lost if a final 
state proton with a given momentum is required, and a breakdown of the 
factorization theorem results.

\section{Target rest frame point of view}\label{sect:trf}
In this section, the connection between the target rest frame point of 
view, used in the semiclassical approach, and the Breit frame point of 
view, relevant for the two previous sections, is established. In particular, 
the consistency of the semiclassical approach with the concept of 
diffractive parton distributions and with the factorization formulae of 
the last section are demonstrated. 

Even before the all-orders factorization proofs of~\cite{gtv} and 
\cite{cfa}, it was suggested that diffractive factorization could be 
understood in the semiclassical picture in the proton rest frame 
\cite{sopa}. This was then explicitly shown in the leading order analysis 
of \cite{h}, on which the present section is largely based. Calculating the 
cross section with the methods of Chapter~\ref{sect:sc}, a result is 
obtained that can be written as a convolution of a partonic cross section 
and a diffractive parton distribution. Within the semiclassical model, this 
diffractive parton distribution is explicitly given in terms of integrals 
of non-Abelian eikonal factors in the background field. 

To be more specific, it is explicitly shown that the amplitude contains 
two fundamental parts: the usual hard scattering amplitude of a partonic 
process, and the amplitude for soft eikonal interactions with the external 
colour field. The latter part is determined by the scattering of one of the 
partons from the photon wave function. This parton has to have small 
transverse momentum and has to carry a relatively small fraction of the 
longitudinal photon momentum. In a frame where the proton is fast, this 
parton can be interpreted as a parton from the diffractive structure 
function. 

The special role played by the soft parton in the photon wave function 
was also discussed in~\cite{nz,pt1,wue,bhma} in the framework of two 
gluon exchange. However, the present approach has the two following 
advantages: firstly, by identifying the hard part as a standard 
photon-parton scattering cross section, the necessity for non-covariant 
photon wave function calculations is removed. Secondly, once it is 
established that the main contribution comes from the soft region, 
non-perturbative effects are expected to become important. The eikonal 
approximation provides a simple, self-consistent model for this 
non-perturbative region. 

\begin{figure}[ht]
\begin{center}
\vspace*{.2cm}
\parbox[b]{9cm}{\psfig{width=9cm,file=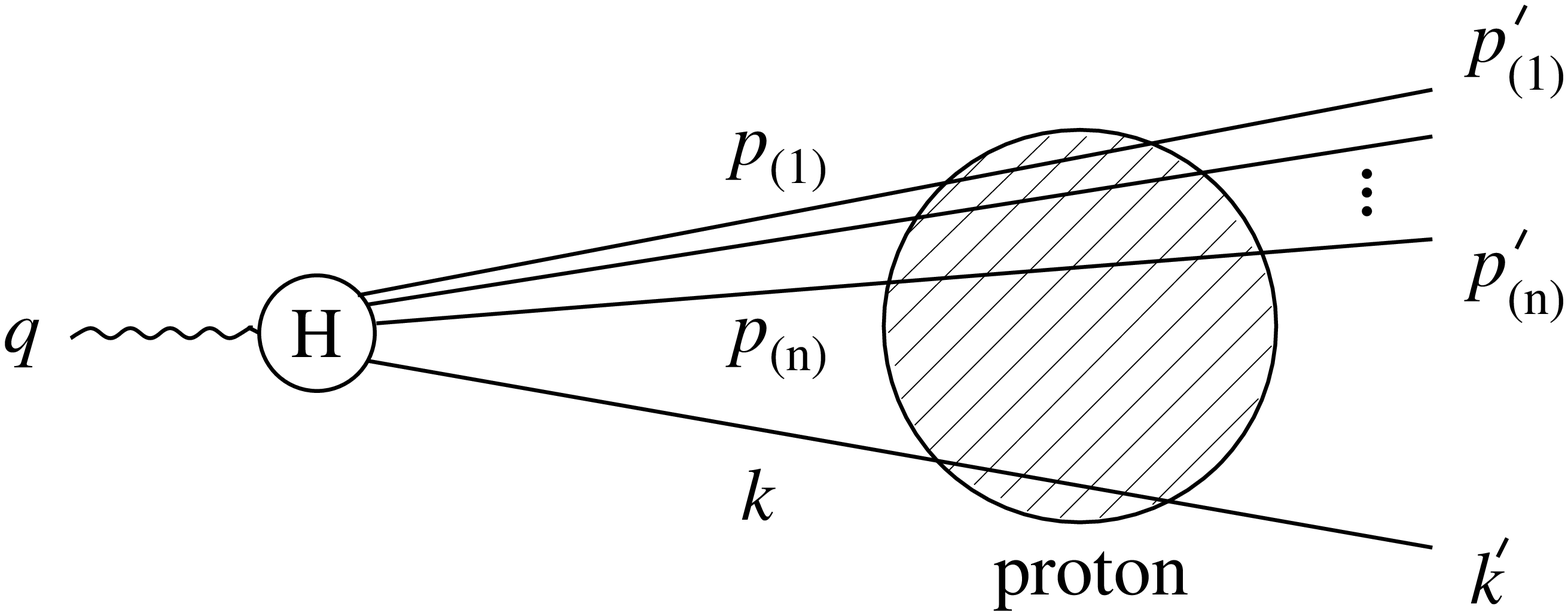}}
\end{center}
\refstepcounter{figure}
\label{fig:dpd}
{\bf Figure \ref{fig:dpd}:} Hard diffractive process in the proton rest frame. 
The soft parton with momentum $k$ is responsible for the leading twist 
behaviour of the cross section. 
\end{figure}

The explicit calculation closely follows the calculation of 
Chapter~\ref{sect:sc}. It is convenient to start with the particularly 
simple case of scalar partons. The kinematic situation is shown 
symbolically in Fig.~\ref{fig:dpd}. The process is split into two parts, 
the hard amplitude for the transition of the photon into a virtual partonic 
state and the scattering of this state off the external field. 

To keep the amplitude for the first part (denoted by H) hard, the 
transverse momenta $p_{(j)\perp}'$ $(j=1...n)$ are required to be large, 
i.e., $\sim Q$. The momentum $k_\perp$ is small, i.e., $\sim\Lambda$, and 
the corresponding parton carries only a small fraction ($\sim\Lambda^2/ 
Q^2$) of the longitudinal photon momentum in the proton rest frame. While 
the hardness condition for particles 1 through $n$ is introduced 
``by hand'', simply to make the process tractable, the softness of the last 
particle follows automatically from the requirement of leading twist 
diffraction. 

The cross section for the scattering off a soft external field reads 
\be
d\sigma=\frac{1}{2q_0}\,|T|^2\,2\pi\delta(q_0-q_0')\,dX^{(n+1)}
\,,\quad\mbox{where}\quad q'=k'+\mbox{$\sum$}p_{(j)}'\, .\label{cs1}
\ee
All momenta are given in the proton rest frame, $T$ is the amplitude 
corresponding to Fig.~\ref{fig:dpd}, and $dX^{(n+1)}$ is the usual phase 
space element for $n+1$ particles. 

According to Eq.~(\ref{qs}), each of the particles interacts with the 
external field via the effective vertex 
\be
V(p',p)=2\pi\delta(p_0'-p_0)\,2p_0\,\left[\tilde{U}(p_\perp'-p_\perp)
-(2\pi)^2\delta^2(p_\perp'-p_\perp)\right]\,.\label{hev}
\ee
The amplitude $T$ can now be built from the hard amplitude $T_H$ (symbolized 
by H in Fig.~\ref{fig:dpd}) together with the effective vertices defined by 
Eq.~(\ref{hev}) and the appropriate propagators. It is convenient to 
consider first the amplitude $T'$, which is defined like $T$ but 
without the transverse-space $\delta$-function terms of Eq.~(\ref{hev}), 
\[
i\,2\pi\delta(q_0-q_0')\,T'=\int T_H\,\prod_j\!\left(\frac{i}{p_{(j)}^2}
\,2\pi\delta(p_{(j)0}'-p_{(j)0})\,2p_{(j)0}\,\tilde{U}(p_{(j)\perp}'-
p_{(j)\perp})\,\frac{d^4p_{(j)}}{(2\pi)^4}\right) 
\]
\be
\hspace*{5cm}\times\left(\frac{i}{k^2}\,2\pi\delta(k_0'-k_0)\,2k_0\,
\tilde{U}(k_\perp'-k_\perp)\right)\, .\label{amp1}
\ee
In this equation, colour indices have been suppressed. Notice also that some 
of the produced partons are antiparticles. The corresponding matrices $U$ 
have to be replaced by $U^\dagger$. To keep the notation simple, this is 
not shown explicitly. 

The integrations over the light-cone components $p_{(j)+}$ can be performed 
using the appropriate energy $\delta$-functions. After that, the 
$p_{(j)-}$ integrations are performed by picking up the poles of the 
propagators $1/p_{(j)}^2$. The result is 
\be
T'=\int T_H\,\prod_j\!\left(\tilde{U}(p_{(j)\perp}'-p_{(j)\perp})\,\frac{d^2
p_{(j)\perp}}{(2\pi)^2}\right) \frac{2k_0}{k^2}\,\tilde{U}(k_\perp'-k_\perp)
\, .\label{amp2}
\ee
Here poles associated with the $p_{(j)-}$ dependence of $T_H$ have been 
disregarded since their contribution is cancelled by diagrams where part of 
the hard interaction occurs after the scattering off the external field. 

Next, a change of integration variables is performed, 
\be
d^2p_{(n)\perp}\,\to\, d^2k_\perp\,.
\ee
Since the external field is assumed to be soft, it can only transfer 
transverse momenta of order $\Lambda$, i.e., $p_{(j)\perp}\simeq 
p_{(j)\perp}'$ for all $j$. In general, the amplitude $T_H$ will be 
dominated by the hard momenta of order $Q$. Therefore, it can be assumed 
that $T_H$ is constant if the momenta $p_{(j)\perp}$ vary on a scale 
$\Lambda$. In this approximation, the integrations over $p_{(j)\perp}$ 
$(j=1...n-1)$ can be performed in Eq.~(\ref{amp2}), resulting in 
$\delta$-functions in impact parameter space. These manipulations give 
the result 
\be
T'=\int\frac{d^2k_\perp}{(2\pi)^2}\,\frac{2k_0}{k^2}\,T_H
\int_{x_\perp,y_\perp}\bigg(\prod_jU(x_\perp)\bigg)U(x_\perp+y_\perp)
\,e^{-ix_\perp\Delta_\perp-iy_\perp(k_\perp'-k_\perp)}\,\, ,\label{amp3}
\ee
where $\Delta$ is the total momentum transferred from the proton to the 
diffractive system and, in particular, $\Delta_\perp=k_\perp'+\sum 
p_{(j)\perp}'$. It is intuitively clear that the relative proximity of the 
high-$p_\perp$ partons in impact parameter space leads to the corresponding 
eikonal factors being evaluated at the same position $x_\perp$. 

Now, the colour structure of the amplitude will be considered in more 
detail. Spelling out all the colour indices and introducing explicitly the 
colour singlet projector $P$, the relevant part of the amplitude reads: 
\be
T_{colour}'=T_H^{a_1...a_nb}\,\bigg(\prod_jU(x_\perp)
\bigg)^{a_1'...a_n'}_{a_1...a_n}\,U(x_\perp+
y_\perp)^{b'}_b\,P_{a_1'...a_n'b'}\, .
\ee
Using the fact that $T_H$ is an invariant tensor in colour space, and 
introducing the function $W$ defined in Eq.~(\ref{wdef}), the following 
formula is obtained: 
\be
T_{colour}=T_{colour}'-T_H^{a_1...a_nb}P_{a_1...a_nb}=
T_H^{a_1...a_nb}\,W^{b'}_b P_{a_1...a_nb'}\, .\label{tcol}
\ee
Here the first equality states the explicit relation between $T'$ and the 
true amplitude $T$, where the trivial contribution of zeroth order in $A$ 
has been subtracted. This subtraction corresponds to the unit matrix on the 
r.h. side of Eq.~(\ref{wdef}). 

For colour covariance reasons
\be
T_H^{a_1...a_nb} P_{a_1...a_nb'}=\mbox{const.}\,\times\,\delta^b_{b'}\, .
\label{const}
\ee
Since the photon is colour neutral, the following equality holds: 
\be
T_H^{a_1...a_nb}\,T^*_{H\,a_1...a_nb}=|\,T_H^{a_1...a_nb} P_{a_1...a_nb}\,
|^2=|\mbox{const.}|^2N_c^2\, .\label{const1}
\ee
Here the partons are assumed to be in the fundamental representation of 
the colour group $SU(N_c)$. Combining Eq.~(\ref{tcol}) with 
Eqs.~(\ref{const}) and (\ref{const1}), it becomes clear that the colour 
structure of the hard part decouples from the eikonal factors, 
\be
|T_{colour}|^2=\frac{1}{N_c}\,|\,\mbox{tr}[W]\,|^2\,\,|T_H|^2\, .\label{tc}
\ee
The hard part will be interpreted in terms of an incoming small-$k_\perp$ 
parton that collides with the virtual photon to produce the outgoing partons 
1 through $n$. Therefore, a factor $1/N_c$ for initial state colour 
averaging is included in the definition of $|T_H|^2$. 

In the expression for the cross section, the two functions $W$ appear in the 
combination 
\be
\Big(\mbox{tr}[W_{x_\perp}(y_\perp)]\Big)\Big(\mbox{tr}[W_{x_\perp'}
(y_\perp')]\Big)^*\,e^{-i(x_\perp-x_\perp')\Delta_\perp}\, ,\label{ww}
\ee
with independent integrations over $x_\perp,x_\perp',y_\perp$ and 
$y_\perp'$. If the external field is sufficiently smooth, the functions $W$ 
vary only slowly with $x_\perp$ and $x_\perp'$. Therefore, after 
integration over $x_\perp$ and $x_\perp'$, the expression in Eq.~(\ref{ww}) 
produces an approximate $\delta$-function in $\Delta_\perp$. Furthermore, 
it is assumed that the measurement is sufficiently inclusive, i.e., the 
hard momenta $p_{(j)_\perp}'$ are not resolved on a soft scale $\Lambda$. 
This corresponds to a $\Delta_\perp$-integration, which gives an 
approximate $\delta$-function in $x_\perp\!-x_\perp'$. Since the expression 
in Eq.~(\ref{ww}) will always appear under $x_\perp,\,x_\perp'$ and 
$\Delta_\perp$ integration, the above considerations justify the 
substitution 
\be
e^{-i(x_\perp-x_\perp')\Delta_\perp}\to (2\pi)^2\,\delta^2(x_\perp-x_\perp')
\,\delta^2(\Delta_\perp)\, .\label{subs}
\ee
Combining Eqs.~(\ref{cs1}), (\ref{amp3}) and (\ref{tc}), the following 
formula for the cross section results, 
\be
d\sigma\!=\frac{1}{2q_0}\int |T_H|^2 \int_{x_\perp}\left|\int\frac{d^2
k_\perp}{(2\pi)^2}\,\frac{\mbox{tr}[\tilde{W}_{x_\perp}(k_\perp'\!\!\!-\!
k_\perp)]}{\sqrt{N_c}\,k^2}\right|^2\!\!(2k_0)^2(2\pi)^3\,\delta^2\!\left(
\mbox{$\sum$}p_{(j)\perp}\right)\delta(q_0\!\!-\!q_0')\,dX^{(n+1)}.
\label{cs2}
\ee
Note that the soft momentum $k_\perp'$ has been neglected in the transverse 
$\delta$-function. 

To finally establish the parton model interpretation of diffraction, the 
hard partonic cross section based on $|T_H|^2$ has to be identified in 
Eq.~(\ref{cs2}). Consider the process
\be
\gamma^*(q)+q(yP)\to q(p_{(1)}')+\dots+q(p_{(n)}')\, ,\label{pp}
\ee
where the photon collides with a parton carrying a fraction $y$ of the 
proton momentum and produces $n$ high-$p_\perp$ final state partons. The 
cross section is approximately given by
\be
d\hat{\sigma}(y)=\frac{1}{2(\hat{s}+Q^2)}\,|T_H|^2\,(2\pi)^4\delta^4(q-k-
\mbox{$\sum$}p_{(j)}')\,dX^{(n)}\, ,\label{pcs}
\ee
where $\hat{s}=(\sum p_{(j)}')^2$ and the quantities $|T_H|^2$ and $k$ are 
the same as in the previous discussion. Equation~(\ref{pcs}) is not exact 
for several reasons. On the one hand, $|T_H|^2$ is defined in terms of the 
unprimed momenta $p_{(j)}$, which differ slightly from $p_{(j)}'$. On the 
other hand, the vector $k$ is slightly off shell and has, in general, a 
non-zero transverse component. However, both effects correspond to 
$\Lambda/Q$ corrections, where $Q$ stands generically for the hard scales 
that dominate $T_H$. 

Using Eq.~(\ref{pcs}), the cross section of Eq.~(\ref{cs2}) can now be 
rewritten as 
\be 
d\sigma=\int dk_-\int\frac{\hat{s}+Q^2}{2\pi q_0}\,(2k_0)^2\,d\hat{\sigma}
(y)\int_{x_\perp}\left|\int\frac{d^2k_\perp}{(2\pi)^2}\,\frac{\mbox{tr}
[\tilde{W}_{x_\perp}(k_\perp'\!\!\!-\!k_\perp)]}{\sqrt{N_c}\,k^2}\right|^2
\frac{d^3k'}{(2\pi)^3\,2k_0'}\, .\label{cs3}
\ee
The light-cone component $k_-$ is given by $-k_-=yP_-=ym_{\pro}$. Note 
that the minus sign in this formula comes from the interpretation of the 
parton with momentum $k$ as an incoming particle in Eq.~(\ref{pcs}). This 
is, in fact, the crucial point of the whole calculation: due to the 
off-shellness of $k$, the corresponding parton can be interpreted as an 
incoming particle in both the process of Eq.~(\ref{pp}) and in the soft 
scattering process off the external field. The latter process, where an 
almost on-shell parton with momentum $k$ scatters softly off the external 
field changing its momentum to the on-shell vector $k'$, is most easily 
described in the proton rest frame. By contrast, the natural frame for the 
hard part of the diagram is the Breit frame or a similar frame. In such a 
frame, the $k_-$ component is large and negative, so that the above parton 
can be interpreted as an almost on-shell particle with momentum $-k$, 
colliding head-on with the virtual photon. 

Substituting the variables $y$ and $\xi$ for $k_-$ and $k_3'$, the cross 
section, Eq.~(\ref{cs3}), takes the form 
\be
\frac{d\sigma}{d\xi}=\int_x^\xi dy\,\hat{\sigma}(y)\left(
\frac{df^D_s(y,\xi)}{d\xi}\right)\, ,\label{sx1}
\ee
where the diffractive parton distribution for scalars is 
\be
\frac{df^D_s(y,\xi)}{d\xi}=\frac{1}{\xi^2}\left(\frac{\beta}{1-\beta}
\right)\int\frac{d^2k_\perp'(k_\perp'^2)^2}{(2\pi)^4N_c}\int_{x_\perp}
\left|\int\frac{d^2k_\perp}{(2\pi)^2}\,\frac{\mbox{tr}[\tilde{W}_{x_\perp}
(k_\perp'\!\!\!-\!k_\perp)]}{k_\perp'^2\beta+k_\perp^2(1-\beta)}\right|^2\,.
\label{fx}
\ee
This result is in complete agreement with the concepts described in the last 
section. In contrast to the formulae presented there, the above parton 
distribution is inclusive in $t$. However, as will be seen in an example 
calculation in Sect.~\ref{sect:scd}, parton distributions differential in 
$t$ can also be obtained in this formalism. 

For an external colour field that is smooth on a soft scale $\Lambda$ and 
confined to a region of approximate size $1/\Lambda$, the function 
$\mbox{tr}[W_{x_\perp}(y_\perp)]$ is also smooth and vanishes at 
$y_\perp=0$ together with its first derivative. From this it can be derived 
that the $k_\perp$ and $k_\perp'$ integrations in Eq.~(\ref{fx}) are 
dominated by the soft scale. This justifies, a posteriori, the softness 
assumption for one of the partons used in the derivation. 

The qualitative result is that the eikonal scattering of this soft parton 
off the proton field determines the diffractive parton distribution. The 
hard part of the photon evolution can be explicitly separated and expressed 
in terms of a standard cross section for photon-parton collisions. 

Having worked out the kinematics in the simple scalar case, it is 
straightforward to extend the calculation to realistic quarks and gluons. 
The introduction of spinor or vector partons does not affect the 
calculations leading to the generic expression in Eq.~(\ref{sx1}). However, 
those parts of the calculation responsible for the specific form of 
Eq.~(\ref{fx}) have to be changed if the soft parton is a spinor or vector 
particle. 

Referring the reader to Appendix~\ref{sect:dqgd} for details, only the 
final formulae for diffractive quark and gluon distributions in the 
semiclassical approach are presented below. They read 
\be
\frac{df^D_q(y,\xi)}{d\xi}=\frac{2}{\xi^2}\int
\frac{d^2k_\perp'(k_\perp'^2)}{(2\pi)^4N_c}\int_{x_\perp}\left|\int
\frac{d^2k_\perp}{(2\pi)^2}\,\frac{k_\perp\mbox{tr}[\tilde{W}_{x_\perp}
(k_\perp'\!\!\!-\!k_\perp)]}{k_\perp'^2\beta+k_\perp^2(1-\beta)}\right|^2
\,,\label{fxsp}
\ee
for the case of a realistic spinor quark, and
\be
\frac{df^D_g(y,\xi)}{d\xi}=\frac{1}{\xi^2}\left(\frac{\beta}{1-\beta}\right)
\int\!\frac{d^2k_\perp'(k_\perp'^2)^2}{(2\pi)^4\,
(N_c^2-1)}\int_{x_\perp}\left|\int\frac{d^2k_\perp}{(2\pi)^2}\,
\frac{\mbox{tr}[\tilde{W}_{x_\perp}^{\cal A}(k_\perp'\!\!\!-\!k_\perp)]\,
t^{ij}}{k_\perp'^2\beta+k_\perp^2(1-\beta)}\right|^2\, ,\label{fxg}
\ee
with 
\be
t^{ij}=\delta^{ij}+\frac{2k_\perp^ik_\perp^j}{k_\perp'^2}\left(
\frac{1-\beta}{\beta}\right)\, ,
\ee
for the case of a gluon. It was checked explicitly that these 
distribution functions together with the appropriate partonic cross 
sections reproduce the results of Sects.~\ref{sect:qq} and \ref{sect:hfs}. 
Henceforth, the simplified notation $dq(y,\xi)/d\xi$ and $dg(y,\xi)/d\xi$ 
will be used for the diffractive quark and gluon distribution calculated 
above.

\begin{figure}[ht]
\begin{center}
\vspace*{-.5cm}
\parbox[b]{15cm}{\psfig{width=15cm,file=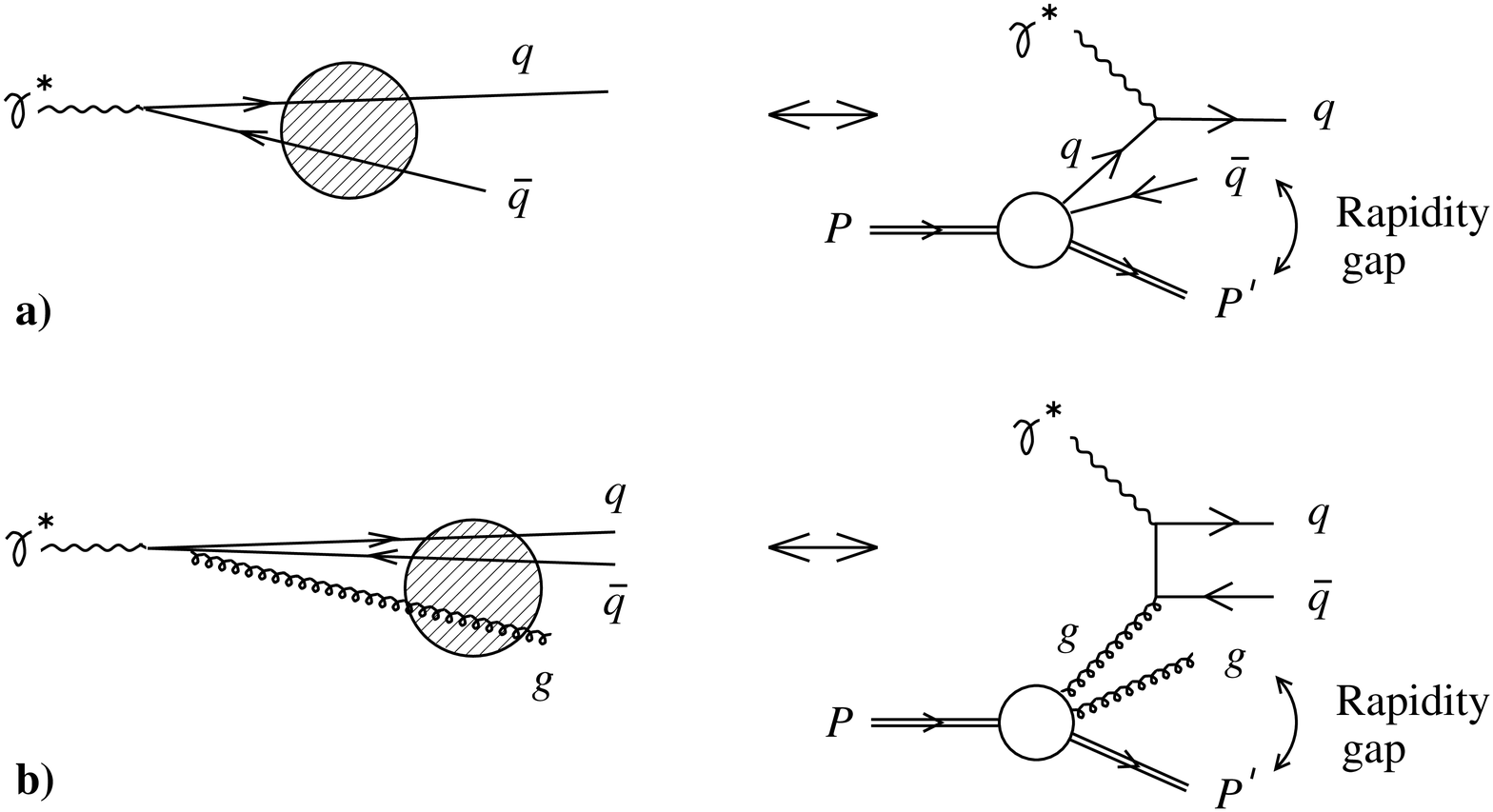}}\\
\end{center}
\refstepcounter{figure}
\label{fig:f2d}
{\bf Figure \ref{fig:f2d}:} Diffractive DIS in the proton rest frame (left) 
and the Breit frame (right); asymmetric quark fluctuations correspond to 
diffractive quark scattering, asymmetric gluon fluctuations to diffractive 
boson-gluon fusion. 
\end{figure}

The application of these diffractive quark and gluon distributions to 
diffractive DIS is summarized in an intuitive way in Fig.~\ref{fig:f2d}. 
On the l.h. side, the two lowest-order processes, $q\bar{q}$ and 
$q\bar{q}g$ state production, are shown from the target rest frame point of 
view. On the r.h. side, the same two processes are shown from the Breit 
frame point of view, which is conventionally used to discuss the partonic 
interpretation of DIS. The essence of the calculations presented in the 
present section is the identification of these two viewpoints, as a result 
of which explicit formulae for the diffractive parton distributions, 
expressed in terms of the target colour field, are obtained. 

Note that it is incorrect to interpret the r.h. side of Fig.~\ref{fig:f2d} 
as a two-step process, where a colour-neutral cluster is first emitted by 
the proton and then probed by the virtual photon. If this was the case, 
the two-gluon or two-quark cluster relevant in this calculation would 
necessarily lead to parton distributions symmetric in $\beta$ and 
$1-\beta$. A counter example to this is provided by the model distributions 
derived in Sect.~\ref{sect:lh}. 

Thus, the idea of a `pre-formed' colour neutral cluster, as it is usually 
associated with the pomeron structure function, is not supported by the 
present calculation. 

\begin{figure}[t]
\begin{center}
\vspace*{-.5cm}
\parbox[b]{13.7cm}{\psfig{width=13.7cm,file=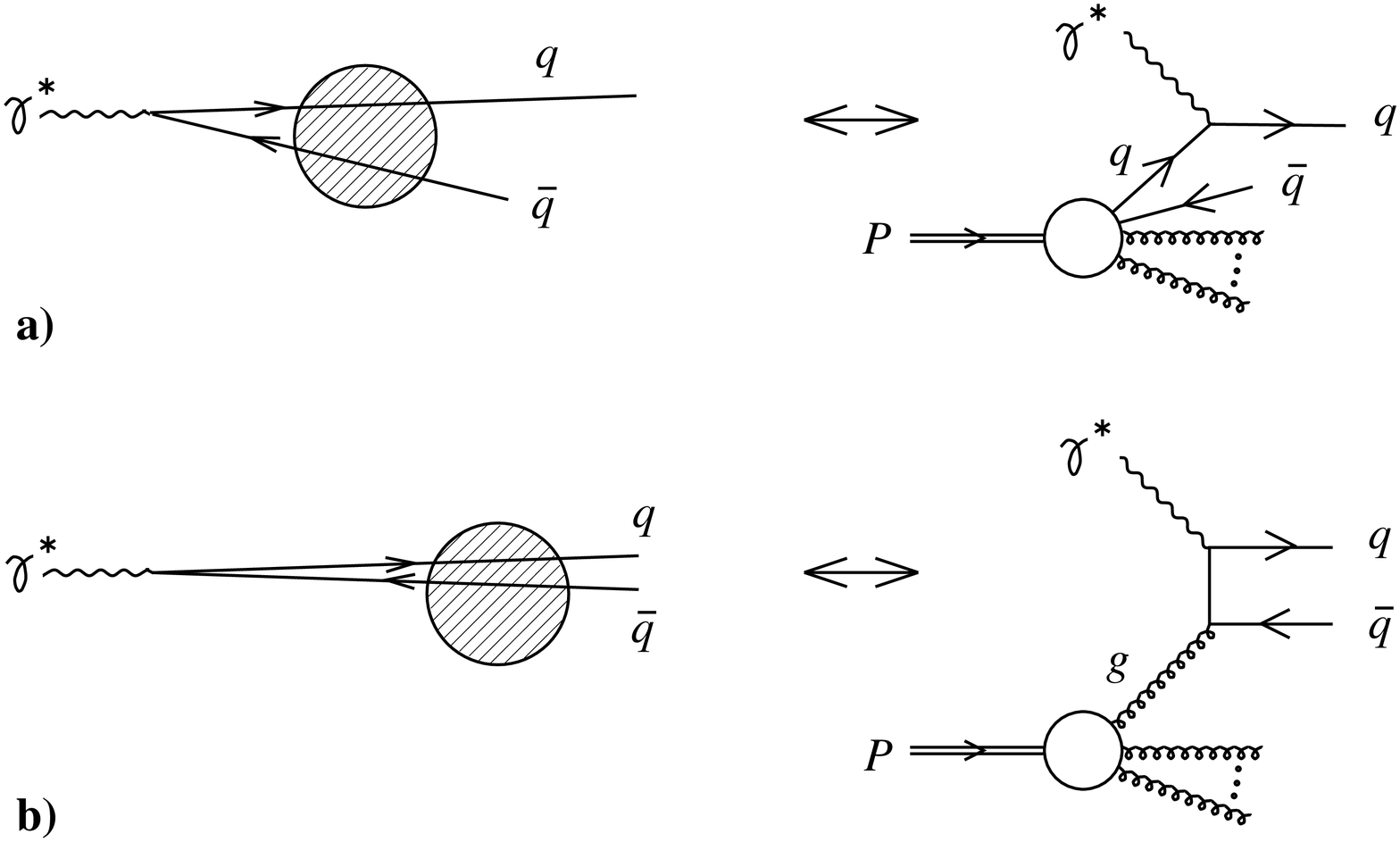}}\\
\end{center}
\refstepcounter{figure}
\label{fig:f2}
{\bf Figure \ref{fig:f2}:} Inclusive DIS in the proton rest frame (left) 
and the Breit frame (right); asymmetric fluctuations correspond to quark 
scattering (a), symmetric fluctuations to boson-gluon fusion (b). 
\end{figure}

At this point, it is appropriate to add a brief discussion of the target 
rest frame vs Breit frame interpretations of inclusive DIS~\cite{bgh}. The 
leading order semiclassical calculation of inclusive DIS, which amounts 
essentially to inclusive $q\bar{q}$ pair production off an external field, 
was given in Sect.~\ref{sect:qq}. It was shown that, in contrast to the 
diffractive case, both asymmetric and symmetric $q\bar{q}$ configurations 
contribute to the leading twist cross section. As explained in more detail 
in Appendix~\ref{sect:isf}, these configurations correspond, in the 
parton model, to leading order quark scattering, testing the inclusive 
quark distribution, and boson-gluon fusion, testing the inclusive gluon 
distribution (cf.~Fig~\ref{fig:f2}). The symmetric configurations have a 
small transverse size and test directly the one-gluon component of the 
target colour field. This is the reason why, in the semiclassical 
framework, the leading order calculation is already sensitive to the gluon 
distribution of the target. 

The inclusive gluon distribution plays a very special role. In contrast 
to both the inclusive quark distribution and the diffractive quark and 
gluon distributions, it is only sensitive to the short distance structure 
of the proton field, and it is enhanced by an explicit factor $1/\alpha_s$ 
(see Appendix~\ref{sect:isf}). As a result, the dominance of the inclusive 
over the diffractive DIS cross section, which is of fundamental importance 
for the successful phenomenological analysis of \cite{bgh} (cf. 
Sect.~\ref{sect:edsf}), emerges. 

\newpage

\mychapter{Two Gluon Exchange}\label{sect:tge}
So far, inclusive diffraction, as parametrized, e.g., by the diffractive 
structure function $F_2^D$, was at the centre of interest of this review. 
It was argued that, for inclusive processes, the underlying colour singlet 
exchange is soft, and two corresponding approaches, the semiclassical 
framework and the pomeron picture, were described in some detail. 

In perturbative QCD, the simplest possibility of realizing colour singlet 
exchange is via two $t$ channel gluons. In fact, the colour singlet 
exchange in certain more exclusive diffractive processes is, with 
varying degree of rigour, argued to be governed by a hard scale. In such 
cases, two gluon exchange dominates. Some of these processes are 
discussed in the present chapter. Finally, attempts to approach the whole 
diffractive cross section in two-gluon models are described.

\section{Elastic meson production}\label{sect:emp}
Elastic meson electroproduction is the first diffractive process that was 
claimed to be calculable in perturbative QCD~\cite{rys,bro}. It has since 
been considered by many authors, and a fair degree of understanding has been 
achieved as far as the perturbative calculability and the factorization of 
the relevant non-perturbative parton distributions and meson wave functions 
are concerned. 

To begin with, consider the electroproduction of a heavy $q\bar{q}$ bound 
state off a given classical colour field. This calculation represents an 
alternative derivation Ryskin's celebrated result~\cite{rys} for elastic 
$J/\psi$ production. 

\begin{figure}[ht]
\begin{center}
\vspace*{.2cm}
\parbox[b]{9cm}{\psfig{width=9cm,file=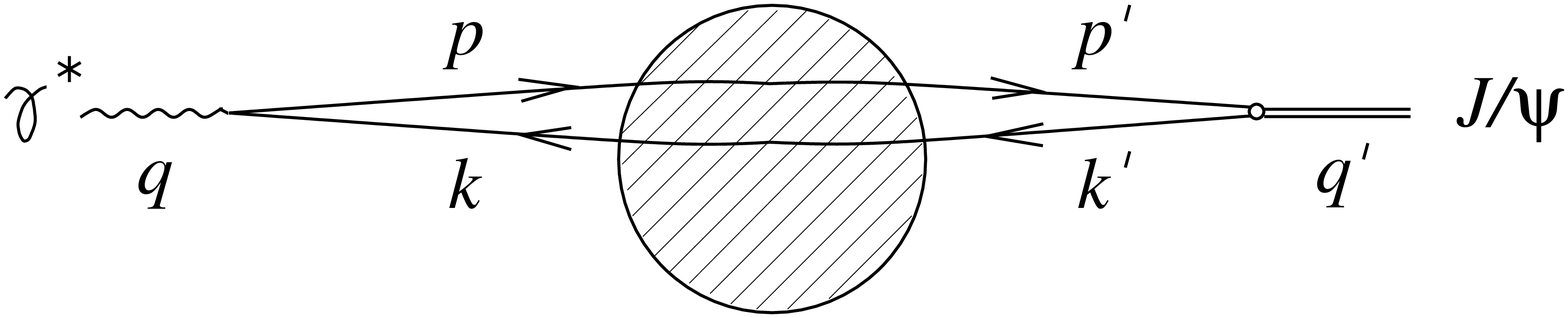}}
\end{center}
\refstepcounter{figure}
\label{fig:jpsi}
{\bf Figure \ref{fig:jpsi}:} Leading order amplitude for the elastic 
production of a $J/\Psi$ particle off an external colour field. 
\end{figure}

The relevant amplitude is shown in Fig.~\ref{fig:jpsi}. In the 
non-relativistic limit, the two outgoing quarks are on-shell, and each 
carries half of the $J/\psi$ momentum. Thus, the two quark propagators with 
momenta $p'=k'=q'/2$ and the $J/\psi$ vertex are replaced with the 
projection operator $g_J\epsilons_J(\ks'+m)$. Here 
\be
g_J^2=\frac{3\Gamma^J_{ee}m_J}{64\pi\aem^2}\,,
\ee
$\Gamma^J_{ee}$ is the electronic decay width of the $J/\psi$ particle, 
$m_J=2m$ is its mass, and $\epsilon_J$ its polarization vector~\cite{bj}. 

Using the notation and calculational technique of Sect.~\ref{sect:qq}, the 
amplitude of Fig.~\ref{fig:jpsi} can now be written as 
\be
i2\pi\delta(q_0'-q_0)T_{q\bar{q}}=ie_cg_J\!\!\int\!\!
\frac{d^4k}{(2\pi)^4}\mbox{tr}\bigg[\epsilons_J(\ks'+m)V_q(p',p)
\frac{i}{\ppl-m}\epsilons_\gamma(q)\frac{i}{-\ks-m}V_{\bar{q}}(k,k')
\bigg],\label{tjqq}
\ee
where $e_c=(2/3)e$ is the electric charge of the charm quark. The Dirac 
structure is simplified employing the identities
\be
-g_J\epsilons_J(\ks'+m)=\frac{g_J}{2m}(\ks'-m)\epsilons_J(\ks'+m)=\frac{g_J}
{2m}\,\sum_rv_r(k')\bar{v}_r(k')\,\epsilons_J\,\sum_su_s(p')\bar{u}_s(p')
\ee
as well as Eqs.~(\ref{prop1}) and (\ref{prop2}). The further calculation 
proceeds along the lines of Sect.~\ref{sect:qq} using, in particular, the 
quark scattering vertices Eqs.~(\ref{v1}) and (\ref{v2}). Adding the two 
contributions where only the quark or only the antiquark is scattered to 
Eq.~(\ref{tjqq}), the full amplitude $T=T_{q\bar{q}}+T_q+T_{\bar{q}}$ takes 
the form
\bea
\!\!\!\!\!\!T&\!=\!&  -{ie_cg_J\over 8\pi^2m}\ q_+
  \int d^2k_{\perp}\ {\alpha(1-\alpha)\over N^2 + k_{\perp}^2}\sum_{r's'}
  \ [\bar{u}_{s'}(\bar{p})\epsilons_\gamma(q)v_{r'}(\bar{k})]\,
    [\bar{v}_{r'}(\bar{k'})\epsilons_J(q)u_{s'}(\bar{p'})]\label{tjpsi}
\\
\!\!\!\!\!\!\!\!&&\hspace*{.2cm}\times \left[\tilde{U}(p_\perp'-p_\perp)
\tilde{U}^\dagger(k_\perp-k_\perp')-(2\pi)^4\delta^2(p_\perp'-p_\perp)
\delta^2(k_\perp'-k_\perp)\right]\,.\nonumber
\eea
Note that this is very similar to Eq.~(\ref{tqbq}) with the $U$ matrix 
structure replaced according to Eq.~(\ref{uudd}). The main difference is 
that the produced quarks are projected onto the $J/\psi$ state. 

Since both $Q$ and $m$ are considered to be hard scales while $U$ and 
$U^\dagger$ are governed by the soft hadronic scale $\Lambda$, the integrand 
in Eq.~(\ref{tjpsi}) can be expanded in powers of the soft momentum 
$k_\perp$. The leading power of the amplitude is given by the first 
non-vanishing term. In the case of forward production, $p_\perp'=k_\perp'= 
0$, the dependence on the external colour field takes the form 
\bea
&&\int d^2k_\perp k_\perp^2 \mbox{tr}\left[\tilde{U}(p_\perp'-p_\perp)
\tilde{U}^\dagger(k_\perp-k_\perp')-(2\pi)^4\delta^2(p_\perp'-p_\perp)
\delta^2(k_\perp'-k_\perp)\right]\nonumber
\\
&=&\int d^2k_\perp k_\perp^2 \int_{x_\perp}\int_{y_\perp}\mbox{tr}\left[
U(x_\perp)\,U^\dagger(y_\perp)\,-\,1\right]\,e^{ik_\perp(y_\perp-x_\perp)}
\nonumber\\
&=&-(2\pi)^2\partial_{y_\perp}^2\int_{x_\perp}\mbox{tr}W_{x_\perp}(y_\perp)
\Big|_{y_\perp=0}\,\,.\label{ddw}
\eea

\begin{figure}[ht]
\begin{center}
\vspace*{.2cm}
\parbox[b]{9cm}{\psfig{width=9cm,file=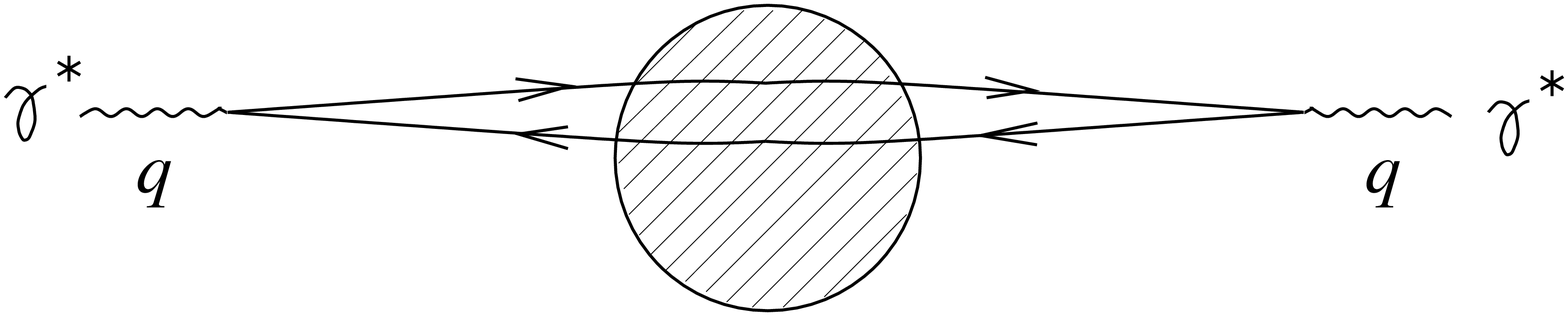}}
\end{center}
\refstepcounter{figure}
\label{fig:comp}
{\bf Figure \ref{fig:comp}:} The Compton scattering amplitude within the 
semiclassical approach. 
\end{figure}

Now, the crucial observation is that precisely the same dependence on the 
external field is present in the amplitude for forward Compton scattering 
shown in Fig.~\ref{fig:comp}. In the case of longitudinal photon 
polarization, the transverse size of the $q\bar{q}$ pair is always small, 
and the target field enters only via the second derivative of $W$ that 
appears in Eq.~(\ref{ddw}). 

Thus, comparing with the parton model result for longitudinal photon 
scattering, this derivative can be identified in terms of the gluon 
distribution of the target proton \cite{bhmpt}, 
\be
-\partial_{y_\perp}^2\int_{x_\perp}\mbox{tr}W_{x_\perp}(y_\perp)
\Big|_{y_\perp=0}=2\pi^2\alpha_sxg(x)\,.\label{gd}
\ee
Using this relation\footnote{Note 
that this is also consistent with the result for longitudinal 
photon scattering given in Eq.~(\ref{slt}) since
\be
-\partial_{y_\perp}^2\int_{x_\perp}\mbox{tr}W_{x_\perp}(y_\perp)
\Big|_{y_\perp=0}=\int_{x_\perp}\mbox{tr}\left[\partial_{y_\perp}W_{x_\perp}
(0)\,\partial_{y_\perp}W_{x_\perp}^\dagger(0)\right]\,,
\ee
which follows from unitarity of the $U$ matrices. 
} 
and Eq.~(\ref{tjpsi}), the amplitudes for the forward production of 
transversely and longitudinally polarized $J/\psi$ mesons by transversely 
and longitudinally polarized virtual photons are obtained. To go from 
amplitudes for scattering off an external field to usual covariant 
amplitudes, a factor $2m_{\pro}$ has to be introduced. Under the additional 
assumption $Q^2\gg m_J^2$, the covariant amplitudes for longitudinal and 
transverse polarization read 
\be
T_L=-i64\pi^2\alpha_sg_Je\,(xg(x))\frac{s}{3Q^3}\quad,\quad T_T=
\frac{m_J}{Q}T_L\,,\label{tltt}
\ee
where $\sqrt{s}$ is the centre-of-mass energy of the $\gamma^*p$ collision. 

It is not surprising that the gluon distribution of Eq.~(\ref{gd}), 
calculated according to Fig.~\ref{fig:comp}, shows no scaling violations and 
only the trivial Bremsstrahlungs energy dependence $\sim 1/x$. The reason 
for this is the softness assumptions of the semiclassical calculation. 
Firstly, the eikonal approximation implies that all longitudinal modes of 
the external field are much softer than the photon energy. Secondly, the 
reduction of the field dependence to a transverse derivative is only 
justified if the scales governing the quark loop, i.e., $Q^2$ in the case 
of Fig.~\ref{fig:comp} and $Q^2$ and $m^2$ in the case of 
Fig.~\ref{fig:jpsi}, are harder than the transverse structure of $W$. 
These two approximations, evidently valid for a given soft field, are also 
justified for a dynamical target governed by QCD as long as only leading 
logarithmic accuracy in both $1/x$ and $Q^2$ is required. Thus, a 
non-trivial dependence on $1/x$ and $Q^2$ can be reintroduced into 
Eq.~(\ref{tltt}) via the measured gluon distribution, keeping in mind that 
the result is only valid at double-leading-log accuracy. 

Note that this is precisely what was claimed in the original two-gluon 
exchange calculation of~\cite{rys}. Note also that, in distinction from 
this presentation, the calculation of~\cite{rys} obtains the gluon 
distribution by coupling the $t$ channel gluons to the quarks of the target
and identifying the logarithmic integral over the transverse gluon momentum 
as the logarithm accompanying the usual quark-to-gluon splitting function. 

\begin{figure}[ht]
\begin{center}
\vspace*{.2cm}
\parbox[b]{10cm}{\psfig{width=5cm,file=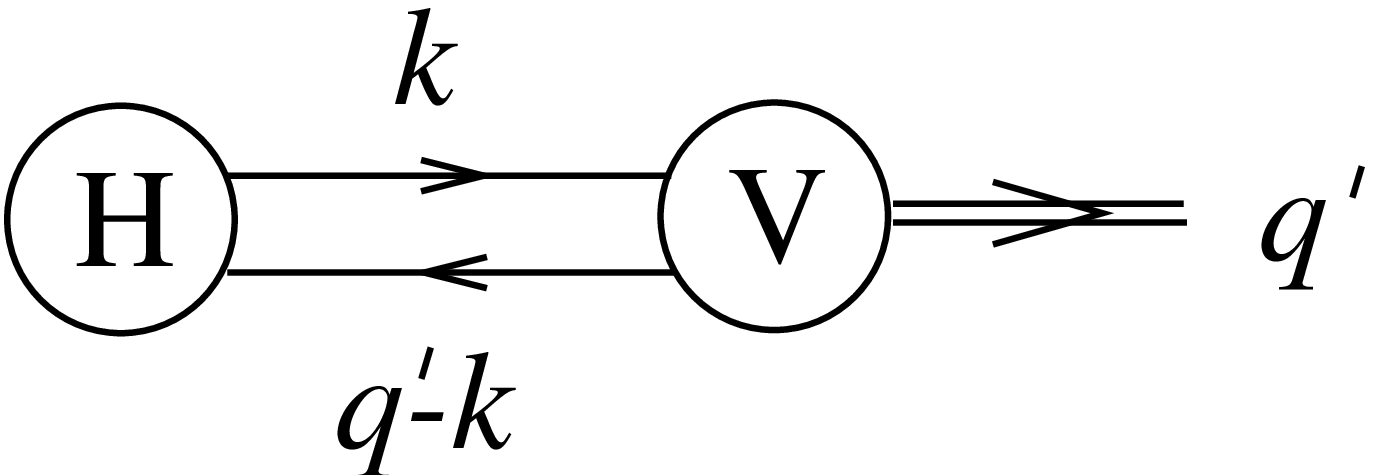}}
\end{center}
\refstepcounter{figure}
\label{fig:hs}
{\bf Figure \ref{fig:hs}:} Generic diagram for meson production in a hard 
process. 
\end{figure}

The first essential extension of the above fundamental result is related to 
the treatment of the bound state produced. Brodsky et al.~\cite{bro} showed 
that, at least for longitudinal photon polarization, a perturbative 
calculation is still possible in the case of light, non-perturbative bound 
states like the $\rho$ meson. The calculation is based on the concept of 
the light-cone wave function of this meson. Referring the reader 
to~\cite{cz} for a detailed review, a brief description of the main ideas 
involved is given below (cf.~\cite{pire}). For this purpose, consider the 
generic diagram for the production of a light meson in a hard QCD process 
given in Fig.~\ref{fig:hs}. 

Assume that, as shown in this figure, all diagrams can be cut across two 
quark-lines, the constituent quarks of the meson, in such a way as to 
separate the hard process H from the soft meson formation vertex V, which 
is defined to include the propagators. The amplitude can be written as 
\be
T=\int d^4k\,T_H(k)V(k)=\int_0^1 dz\,T_H(z)\,\frac{q'_+}{2}\int dk_-d^2
k_\perp V(k)=\int_0^1 dz\,T_H(z)\phi(z)\,, 
\ee
where $z=k_+/q_+'$, and the last equality is simply the definition of the 
light-cone wave function $\phi$ of the meson. The two crucial 
observations leading to the first of these equalities are the approximate 
$k_-$ and $k_\perp$ independence of $T_H$ and the restriction of the $z$ 
integration to the interval from 0 to 1. The first is the result of the hard 
scale that dominates $T_H$, the second follows from the analytic structure 
of $V$. In QCD, the $k_\perp$ integration implicit in $\varphi$ usually 
has an UV divergence due to gluon exchange between the quarks. Therefore, 
one should really read $\phi=\phi(z,\mu^2)$, where the cutoff $\mu^2$ is of 
the order of the hard scale that governs $T_H$. At higher orders in 
$\alpha_s$, the hard amplitude $T_H$ develops a matching IR cutoff 
dependence. 

A more rigorous definition of the light-cone wave function, required, in 
particular, for the discussion of higher order corrections, can be given 
in the operator language. Without going into further detail, note that
the above wave function $\phi$ satisfies the relation 
\be
\langle\mbox{meson}(q')|\varphi^\dagger(y)\varphi(-y)|0\rangle=\int_0^1 dz
\,e^{i(1-2z)(q'y)}\phi(z)\,,
\ee
where $y^2=0$ and $\varphi$ is the field operator, for simplicity a scalar, 
corresponding to the particle content of the meson. 

The case of the $\rho$ meson is more complicated because it is a 
vector particle built from spin-(1/2) constituents. The matrix element 
relevant for the exclusive electroproduction of a longitudinally polarized 
vector meson reads 
\be
\langle\rho(q')|\bar{\psi}(y)\gamma^\mu\psi(-y)|0\rangle=
\int_0^1 dz\,e^{i(1-2z)(q'y)}\left[q'^\mu(\epsilon\!\cdot\!y) f_a(z)+
\epsilon^\mu f_b(z)+y^\mu(\epsilon\!\cdot\!y)f_c(z)\right],
\ee
where $\epsilon$ is the polarization vector of the meson and $f_{a,b,c}$ are 
leading twist $q\bar{q}$ distribution functions (see \cite{ball} for more 
details on these quantities and related issues). After the convolution with 
the appropriate hard production amplitude, only the combination 
\be
\phi_\rho(z)=f_b(z)+\frac{i}{2}\frac{\partial}{\partial z}f_a(z)
\ee
survives. Under the simplifying assumption that the $\rho$ is built from 
one flavour of quarks with one unit of electric charge (which does not 
affect the final result if normalized by the decay width) the longitudinal 
amplitude reads~\cite{bro,broc}
\be
T_L=-i4\pi^2\alpha_s e\,(xg(x))\frac{s}{3Q^3m_\rho}\int_0^1 dz
\frac{\phi_\rho(z)}{z(1-z)}\,.\label{tlr}
\ee
The underlying calculation of the hard scattering process is analogous to 
the case of the $J/\psi$ discussed above. Although the shape of the 
light-cone wave function $\phi_\rho$ is not known, its normalization can 
be related to the electronic width $\Gamma^\rho_{ee}$ of the $\rho$ via the 
relations 
\be
\int_0^1 dz\,\phi_\rho(z) = \sqrt{2}\,m_\rho f_\rho\qquad,\qquad 
\Gamma^\rho_{ee}=\frac{8\pi\aem^2f_\rho^2}{3m_\rho}\,.\label{rn}
\ee
If the non-relativistic approximation used in the $J/\psi$ case was to 
be applied here, a wave function $\phi_\rho$ proportional to 
$\delta(z-1/2)$ would result. In this case, Eqs.~(\ref{tlr}) and (\ref{rn}) 
would give an unambiguous prediction for the $\rho$ production amplitude 
in terms of its electronic width, precisely as for the $J/\psi$ amplitude. 

Note that, in the transverse case, the hard amplitude generates a $z$ 
dependence which is even more singular at the endpoints than the factor 
$1/z(1-z)$ of Eq.~(\ref{tlr}). While the expected fall-off of the wave 
function at the endpoints is sufficient to render Eq.~(\ref{tlr}) finite, 
this is probably not true in the transverse case. Thus, perturbative 
calculability in the sense of the $J/\psi$ and the longitudinal $\rho$ 
amplitudes can not be claimed. 

Since the gluon distribution is defined by the imaginary part of the 
forward Compton scattering amplitude, the above result for longitudinal 
vector meson production, $T_L\sim ixg(x)$, is, strictly speaking, only the 
imaginary part of the full amplitude. Assuming that the energy dependence 
is given by $xg(x)\sim (1/x)^\alpha$, where the intercept $\alpha$ belongs 
to an even signature trajectory, the real part follows from the signature 
factor of Eq.~(\ref{sifa}). If $\alpha-1$ is small, the ratio of real and 
imaginary part is approximately $(\pi/2)(\alpha-1)$, and the real part 
correction can be introduced into Eq.~(\ref{tlr}) via the 
substitution~\cite{bro,rrml}
\be
ixg(x)\quad\to\quad ixg(x)+\frac{\pi}{2}\,\frac{\partial(xg(x))}
{\partial\ln(1/x)}\,. 
\ee

The discussion of vector meson production given above was limited to the 
double-leading-log approximation as far as the colour singlet exchange in 
the $t$ channel is concerned. To go beyond this approximation, the concept 
of `non-forward' or `off-diagonal' parton distributions, introduced some 
time ago (see~\cite{mea,dit} and refs. therein) and discussed by Ji~\cite{ji} 
and Radyushkin~\cite{rad} in the present context, has to be used. Recent 
reviews of these quantities and their evolution, which interpolates between 
the Altarelli-Parisi and the Brodsky-Lepage evolution equations, can be 
found in~\cite{gm}. 

\begin{figure}[ht]
\begin{center}
\vspace*{.2cm}
\parbox[b]{9cm}{\psfig{width=9cm,file=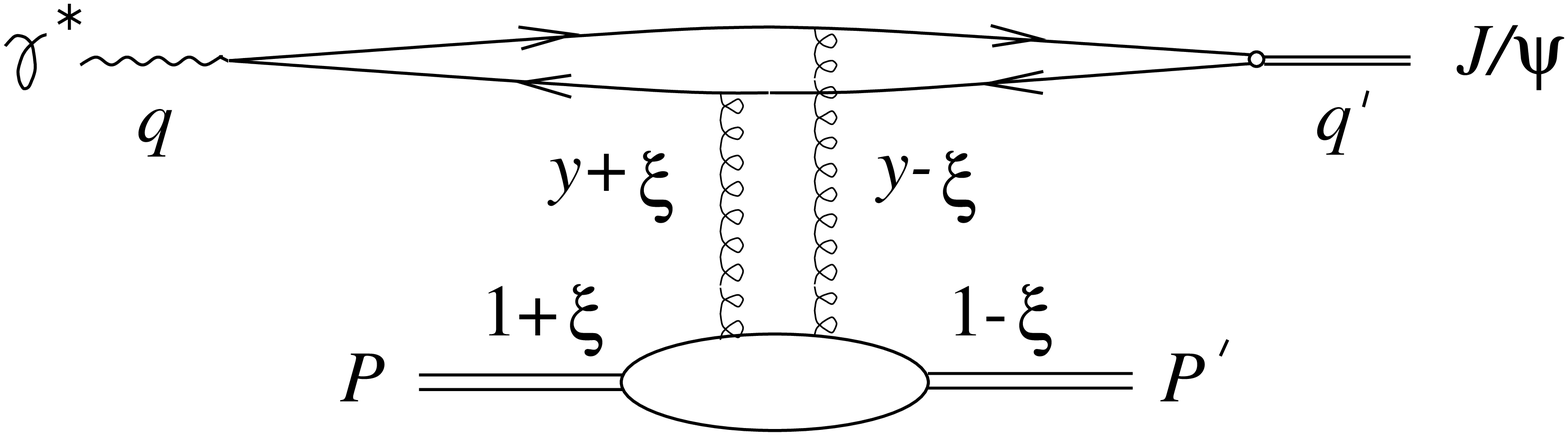}}
\end{center}
\refstepcounter{figure}
\label{fig:jpsi1}
{\bf Figure \ref{fig:jpsi1}:} Elastic $J/\psi$ production (three further 
diagrams with the gluons connected in different ways have to be added). The 
two gluon lines and the incoming and outgoing proton are labelled by their 
respective fractions of the plus component of $\bar{P}\equiv(P+P')/2$. 
\end{figure}

Recall first that the semiclassical viewpoint of Figs.~\ref{fig:jpsi} and 
\ref{fig:comp} is equivalent to two gluon exchange as long as the 
transverse size of the energetic $q\bar{q}$ state is small. So far, the 
recoil of the target in longitudinal direction has been neglected. However, 
such a recoil is evidently required by the kinematics. For what follows, it 
is convenient to use a frame where $q_-$ is the large component of the 
photon momentum. In Fig.~\ref{fig:jpsi1}, the exchanged gluons and the 
incoming and outgoing proton with momenta $P$ and $P'$ are labelled by their 
respective fractions of the plus component of $\bar{P}\equiv(P+P')/2$. If 
$\Delta$ is the momentum transferred by the proton, $\xi\bar{P}_+=\Delta_+ 
/2$. The variable $y$ is an integration variable in the gluon loop. 

The lower part of the diagram in Fig.~\ref{fig:jpsi1} is a generalization 
of the conventional gluon distribution (cf. Eqs.~(\ref{spd}) and 
(\ref{gsub})). It can be described by the non-forward gluon distribution 
\be
H_g(y,\xi,t)=\frac{1}{4\pi y\bar{P}_+}\int dx_- e^{-iy\bar{P}_+x_-/2}\langle 
P'|F^\dagger(0,x_-,0_\perp)^{+\mu}F(0,0,0_\perp)_\mu{}^+|P\rangle\,.
\label{nfpd}
\ee
Recently, it has been shown~\cite{dg} that, as in the case of conventional 
parton distributions~\cite{jaf}, no time ordering of the operators in 
Eq.~(\ref{nfpd}) is required. 

The description of elastic meson production in terms of non-forward parton 
distributions is superior to the double-leading-log approach of 
\cite{rys,bro} since $\alpha_s$ corrections to the hard amplitude, meson 
wave function and parton distribution function can, at least in principle, 
be systematically calculated. However, the direct relation to the measured 
conventional gluon distribution is lost. A new non-perturbative quantity, 
the non-forward gluon distribution, is introduced, which has to be measured 
and the evolution of which has to be tested -- a very complicated problem 
given the uncertainties of the experiment and of the meson wave functions 
involved. 

Over the recent years, the theory of non-forward parton distributions has 
developed into an active research field in its own rights, a detailed 
account of which is beyond the scope of this paper (see, however,~\cite{gm} 
for recent reviews). Important issues include the investigation of different 
models for non-forward distribution functions~\cite{mod}, helicity-flip 
distributions~\cite{heli}, the further study of non-forward evolution 
equations~\cite{evol}, and possibilities of predicting the non-forward from 
the forward distribution functions~\cite{fnf}. The latter suggestion relies 
on the observation that, at sufficiently high $Q^2$, the non-trivial $\xi$ 
dependence (cf.~Fig.~\ref{fig:jpsi1} and Eq.~(\ref{nfpd})) is largely 
determined by the $Q^2$ evolution. 

Furthermore, following the basic results of~\cite{rys,bro}, a number of 
interesting phenomenological analyses of meson production have appeared. 
The analyses of \cite{fks} and \cite{nem} focus, among other issues, on the 
effects of the meson wave functions. More details of these approaches will 
be given in Sect.~\ref{sect:mp}, when experimental results are discussed. 
The effects of Fermi motion and quark off-shellness have recently also been 
discussed in~\cite{cro}. Other interesting topics discussed in the 
literature include the form of the energy dependence~\cite{edep}, shadowing 
effects~\cite{shad}, Sudakov suppression~\cite{suda}, effects of 
polarization~\cite{pola}, the $t$ slope~\cite{nem,tslo}, and the intrinsic 
transverse momentum of the hadron~\cite{itra}. Vector meson production at 
large momentum transfer ($|t|\gg \Lambda^2$) is discussed in~\cite{fr,lart}. 

Note especially the proposal of~\cite{mrt} to approach both transversely 
and longitudinally polarized diffractive $\rho$ production on the basis of 
open $q\bar{q}$ production combined with the idea of parton-hadron duality. 

Vector meson electroproduction was also investigated in the framework of 
the model of the stochastic vacuum, which emphasizes the non-perturbative 
aspects of the process \cite{dgkp,dkp,rue}. The crucial observation is that, 
at realistic $Q^2$, the asymptotic hard regime has not yet set in, and the 
typical transverse size of the $q\bar{q}$ fluctuation scattering off the 
target is not small. More details on this approach are found in 
Sect.~\ref{sect:sv}, as well as in Sects.~\ref{sect:edsf} 
and~\ref{sect:mp}. 

It should be emphasized that the above list of interesting subjects and 
related papers is in no way complete.

\section{Factorization}\label{sect:fact}
Having discussed the leading order results for the cases of heavy vector 
meson production and light vector meson production with longitudinal 
polarization, the next logical step is to ask whether the systematic 
calculation of higher order corrections is feasible. For this, it is 
necessary to understand the factorization properties of the hard amplitude 
and the two non-perturbative objects involved, i.e., the meson wave function 
and the non-forward gluon distribution. In this section, a brief discussion 
of the general proof by Collins, Frankfurt and Strikman~\cite{cfs1} is 
given. In addition, the essential role played by gauge invariance is 
explained in the framework of a simple model~\cite{hl}. This illustrates, 
in a particularly intuitive target rest frame approach, the physical 
mechanism underlying factorization properties in the small-$x$ limit. 

\begin{figure}[ht]
\begin{center}
\vspace*{.2cm}
\parbox[b]{6.5cm}{\psfig{width=6.5cm,file=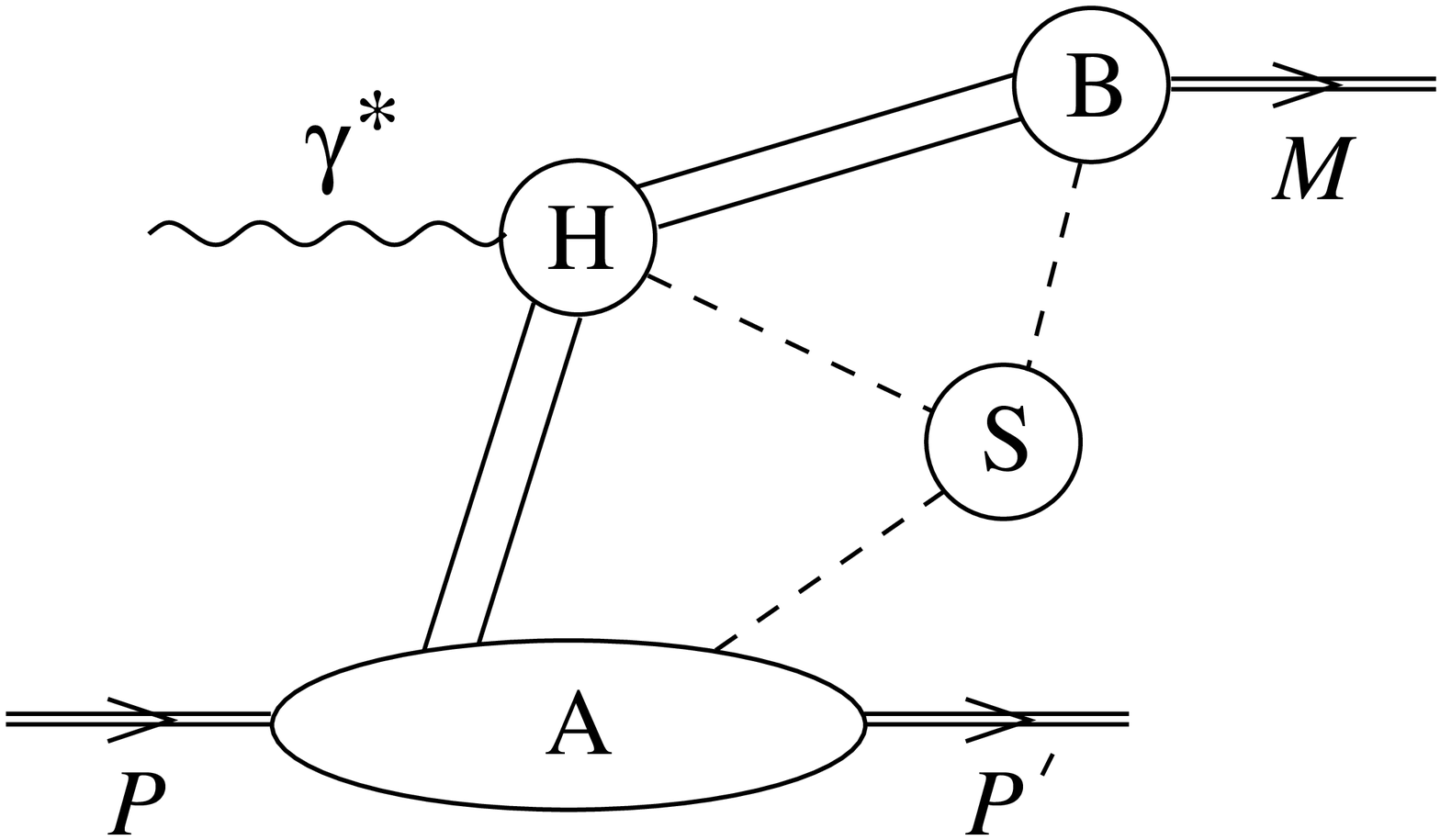}}
\end{center}
\refstepcounter{figure}
\label{fig:lrm}
{\bf Figure \ref{fig:lrm}:} Leading regions in elastic meson 
production (cf.~\cite{cfs1}). 
\end{figure}

The analysis of~\cite{cfs1} is based on the method of leading regions 
discussed previously in Sect.~\ref{sect:dpd}. The leading regions relevant 
for elastic meson production are shown in Fig.~\ref{fig:lrm}. The main line 
of reasoning is analogous to the factorization proofs for inclusive hard 
scattering. As before, H is the hard subgraph, A is the subgraph with 
momenta collinear with the incoming and outgoing proton, and S is the soft 
subgraph. Subgraph B contains only lines which are collinear with the 
produced meson. 

Factorization means that, to leading order in $1/Q$, the amplitude 
corresponding to the process in Fig.~\ref{fig:lrm} can be written as 
\be
T=\int_0^1dz\int dy\,H(y,x/2)\,T_H(Q^2,y/x,z,\mu^2)\,\phi_V(z,\mu^2)\,,
\label{vma}
\ee
where $T_H$ is the hard scattering amplitude, $\phi$ is the light-cone 
wave function of the vector meson produced, and $H$ is the non-forward 
parton distribution of the proton. It could, for example, be the 
non-forward gluon distribution $H_g$ of the last section. The variable 
$x=\xbj$ is the usual DIS Bjorken variable. Equation~(\ref{vma}) implies 
that the soft subgraph in Fig.~\ref{fig:lrm} is irrelevant, and only two 
parton lines, i.e., the minimal number required by the exchanged quantum 
numbers, connect the hard subgraph H with both A and B. 

According to~\cite{cfs1}, the essential steps of the prove include the 
demonstration that the leading regions are indeed given by 
Fig.~\ref{fig:lrm}, the derivation of a power counting formula showing 
that H is connected to A and B by the minimal number of lines, and the 
demonstration of factorization of the soft subgraph using Ward identities. 

An important complication compared to conventional factorization proofs 
results from the fact that final state interactions are important for both 
the outgoing proton and the produced meson. Thus, the soft gluons of S have 
to be factorized from both of these subgraphs at the same time. 

Furthermore, the so-called endpoint regions for the two quark lines 
connecting H and B have to be analyzed in detail. Endpoint regions are 
those kinematic domains where one of the two quarks carries almost all of 
the meson's longitudinal momentum. They are thus endpoints of the $z$ 
integration in Eq.~(\ref{vma}). An essential result of~\cite{cfs1} is 
the suppression of these endpoint regions to all orders of perturbation 
theory in the case of longitudinal photon polarization. 

In the case of transverse polarization, it was shown that the amplitude is 
suppressed by a power of $Q$ relative to the longitudinal case. However, the 
endpoints are not suppressed relative to the intermediate $z$ region, and 
therefore the factorization formula Eq.~(\ref{vma}) can not be established 
in the transverse case. 

A discussion of helicity and transversity parton distributions, measurable 
via the polarization of the produced meson, is also contained in~\cite{cfs1}. 
This will not be reproduced here. More recently, factorization 
proofs similar to~\cite{cfs1} have been given for the process of deeply 
virtual Compton scattering~\cite{csp}, where the situation is simpler since 
no soft meson wave function has to be factorized. Note also that the 
factorization proofs of~\cite{cfs1,csp}, which use Breit frame kinematics, 
do not rely on the limit of small $x$. 

\begin{figure}[ht]
\begin{center}
\vspace*{.2cm}
\parbox[b]{13.3cm}{\psfig{width=13.3cm,file=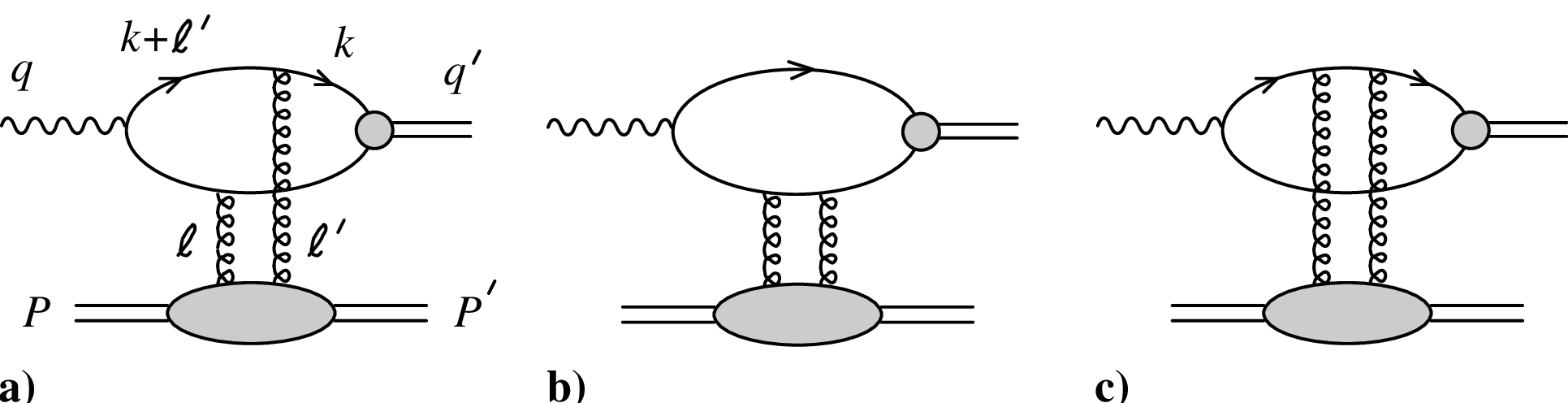}}
\end{center}
\refstepcounter{figure}
\label{fig:lo}
{\bf Figure \ref{fig:lo}:} The leading amplitude for a point-like meson 
vertex. 
\end{figure}

However, it is instructive to see in a particularly simple situation how 
factorization works specifically at small $x$, from the point of view of the 
target rest frame commonly used for the description of small-$x$ processes 
\cite{hl}. For this purpose, consider a very energetic scalar photon that 
scatters off a hadronic target producing a scalar meson built from two 
scalar quarks (see Fig.~\ref{fig:lo}). The quarks are coupled to the photon 
and the meson by point-like scalar vertices $ie$ and $i\lambda$, where $e$ 
and $\lambda$ have dimension of mass. The coupling of the gluons to the 
scalar quarks is given by $-ig\,(r_\mu+r_\mu')$, where $r$ and $r'$ are the 
momenta of the directed quark lines, and $g$ is the strong gauge coupling. 

Under quite general conditions \cite{hl}, the gluon momenta satisfy the 
relations \linebreak 
$\ell_+,\,\ell_+'\ll q_+,\quad\,\ell_-,\,\ell_-'\ll P_-$ and $\ell ^2\sim 
\ell^{\prime 2} \sim-\ell_\perp^2$. Then the lower bubble in 
Fig.~\ref{fig:lo} effectively has the structure 
\be
F^{\mu\nu}(\ell,\ell',P)\simeq\delta(P_-\ell_+)\,F(\ell_\perp^2)\,P^\mu 
P^\nu\,,\label{fd}
\ee
which is defined to include both gluon propagators and all colour factors. 
A similar expression was found by Cheng and Wu~\cite{cw} in a tree model 
for the lower bubble. 

Assume that $F$ restricts the gluon momentum to be soft, $\ell_\perp^2\ll 
Q^2$. In the high-energy limit, it suffices to calculate 
\be
M=\int\frac{d^4\ell}{(2\pi)^4}T^{\mu\nu}F_{\mu\nu}\simeq\int
\frac{d^4\ell}{4(2\pi)^4}T_{++}F_{--}\,,\label{it}
\ee
where 
\be
T^{\mu\nu}=T^{\mu\nu}(\ell,\ell',q)=T^{\mu\nu}_a+T^{\mu\nu}_b+T^{\mu\nu}_c
\label{tmn}
\ee
is the sum of the upper parts of the diagrams in Fig.~\ref{fig:lo}. 

Note that, because of the symmetry of $F_{\mu\nu}$ with respect to the two 
gluon lines, the amplitude $T^{\mu\nu}$ of Eq.~(\ref{tmn}) is used instead 
of the properly-symmetrized upper amplitude 
\be
T_{\mbox{\scriptsize sym}}^{\mu\nu}(\ell,\ell',q)=\frac{1}{2}[T^{\mu\nu}
(\ell,\ell',q)+ T^{\nu\mu}(-\ell',-\ell,q)]\,.
\ee
The two exchanged gluons together form a colour singlet and so the 
symmetrized amplitude $T_{\mbox{\scriptsize sym}}^{\mu\nu}$ satisfies the 
same Ward identity as for two photons, 
\be
T_{\mbox{\scriptsize sym}}^{\mu\nu}(\ell,\ell',q)\ell_\mu \ell_\nu'=0\,.
\label{w1}
\ee
Writing this equation in light-cone components and setting $\ell_\perp=
\ell_\perp'$, as appropriate for forward production, it follows that, for 
the relevant small values of $\ell _-$, $\ell '_-$, $\ell_+$ and $\ell_+'$, 
\be
T_{\mbox{\scriptsize sym,}++} \sim\ell_\perp^2\,
\ee
in the limit $\ell_\perp^2\to 0$. Here the fact that the tensor 
$T_{\mbox{\scriptsize sym}}^{\mu\nu}$, which is built from $\ell'$, $\ell$ 
and $q$, has no large minus components has been used. The $\ell_-$ 
integration makes this equation hold also for the original, unsymmetrized 
amplitude, 
\be
\int d\ell_-T_{++} \sim\ell_\perp^2\,.\label{w2}
\ee
This is the crucial feature of the two-gluon amplitude that will simplify 
the calculation and lead to the factorizing result below. 

Consider first the contribution from diagram a) of Fig.~\ref{fig:lo} to 
the $\ell_-$ integral of $T_{++}$, which is required in Eq.~(\ref{it}), 
\be
\int d\ell_-T_{a,++}=-4eg^2q_+\int\frac{d^4k}{(2\pi)^3}\,\frac{z(1-z)}
{N^2+(k_\perp+\ell_\perp)^2}\,\frac{i\lambda}{k^2(q'-k)^2}\,.
\label{ta}
\ee
Here $N^2=z(1-z)Q^2$, $z=k_+/q_+$ and the condition $\ell_+=0$, enforced 
by the $\delta$-function in Eq.~(\ref{fd}), has been anticipated. 

Now $\int d\ell_-T_{b,++}$ and $\int d\ell_-T_{c,++}$ each carry no 
$\ell_\perp$ dependence. So, to ensure the validity of Eq.~(\ref{w2}), the 
sum of the three diagrams must be 
\be
\int d\ell_-T_{++}=4eg^2q_+\int\frac{d^4k}{(2\pi)^3}z(1-z){\cal N}
\frac{i\lambda}{k^2(q'-k)^2}\,,
\label{intit}
\ee
where
\be
{\cal N}=\Big [\frac{1}{N^2+k_\perp^2}-\frac{1}
{N^2+(k_\perp+\ell_\perp)^2}\Big ]\sim \frac{\ell_\perp^2}
{(N^2+k_\perp^2)^2}\,.\label{intitt}
\ee
Note the $1/Q^4$ behaviour obtained after a cancellation of $1/Q^2$ 
contributions from the individual diagrams. This cancellation, which is 
closely related to the well-known effect of colour transparency~\cite{ct}, 
has been discussed in~\cite{ad} in the framework of vector meson 
electroproduction. 

Introduce the $k_\perp$ dependent light-cone wave function of the meson 
\be
\phi(z,k_\perp^2)=-\frac{iq'_+}{2}\int dk_-dk_+\,
\frac{i\lambda}{(2\pi)^4k^2(q'-k)^2}\,\delta (k_+-zq'_+).\label{wv}
\ee
The final result following from Eqs.~(\ref{it}) and (\ref{intit}) is a 
convolution of the production amplitude of two on-shell quarks and the 
light-cone wave function:
\be
M=ieg^2s\left(\int\frac{d^2\ell_\perp}{2(2\pi)^3}\ell_\perp^2F(
\ell_\perp^2)\right)\int dz\int d^2k_\perp\frac{z(1-z)}{(N^2+k_\perp^2)^2}
\phi(z,k_\perp^2)\,.\label{lot}
\ee
This corresponds to the $O(\ell_\perp^2)$ term in the Taylor expansion
of the contribution from Fig.~\ref{fig:lo}a, given in Eq.~(\ref{ta}). 

\begin{figure}[ht]
\begin{center}
\vspace*{.2cm}
\parbox[b]{4.8cm}{\psfig{width=4.8cm,file=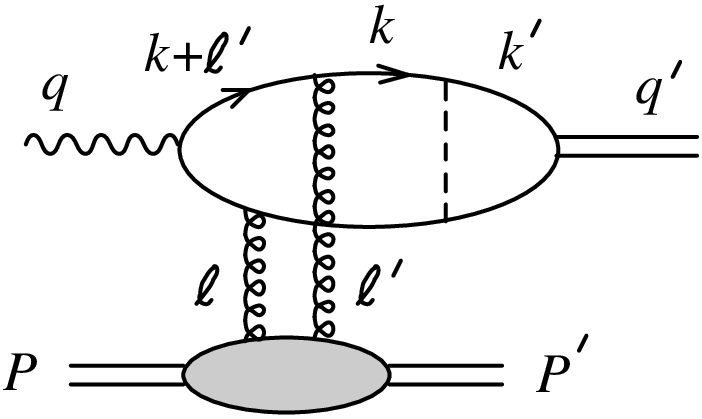}}
\end{center}
\refstepcounter{figure}
\label{fig:nlo}
{\bf Figure \ref{fig:nlo}:} Diagram for meson production with the vertex 
modelled by scalar particle exchange.
\end{figure}

At leading order, factorization of the meson wave function was trivial 
since the point-like quark-quark-meson vertex $V(k^2,(q'-k)^2)=i\lambda$ 
was necessarily located to the right of the all other interactions. To 
see how factorization comes about in the simplest non-trivial situation, 
consider the vector meson vertex 
\be
V(k^2,(q'-k)^2)=\int\frac{d^4k'}{(2\pi)^4}\,\frac{i\lambda\lambda'^2}{k'^2
(q'-k')^2(k-k')^2}\,,\label{tf2v}
\ee
which corresponds to the triangle on the r.h. side of Fig.~\ref{fig:nlo}. 
Here, the dashed line denotes a colourless scalar coupled to the scalar 
quarks with coupling strength $\lambda'$. 

\begin{figure}[ht]
\begin{center}
\vspace*{-.2cm}
\parbox[b]{13.3cm}{\psfig{width=13.3cm,file=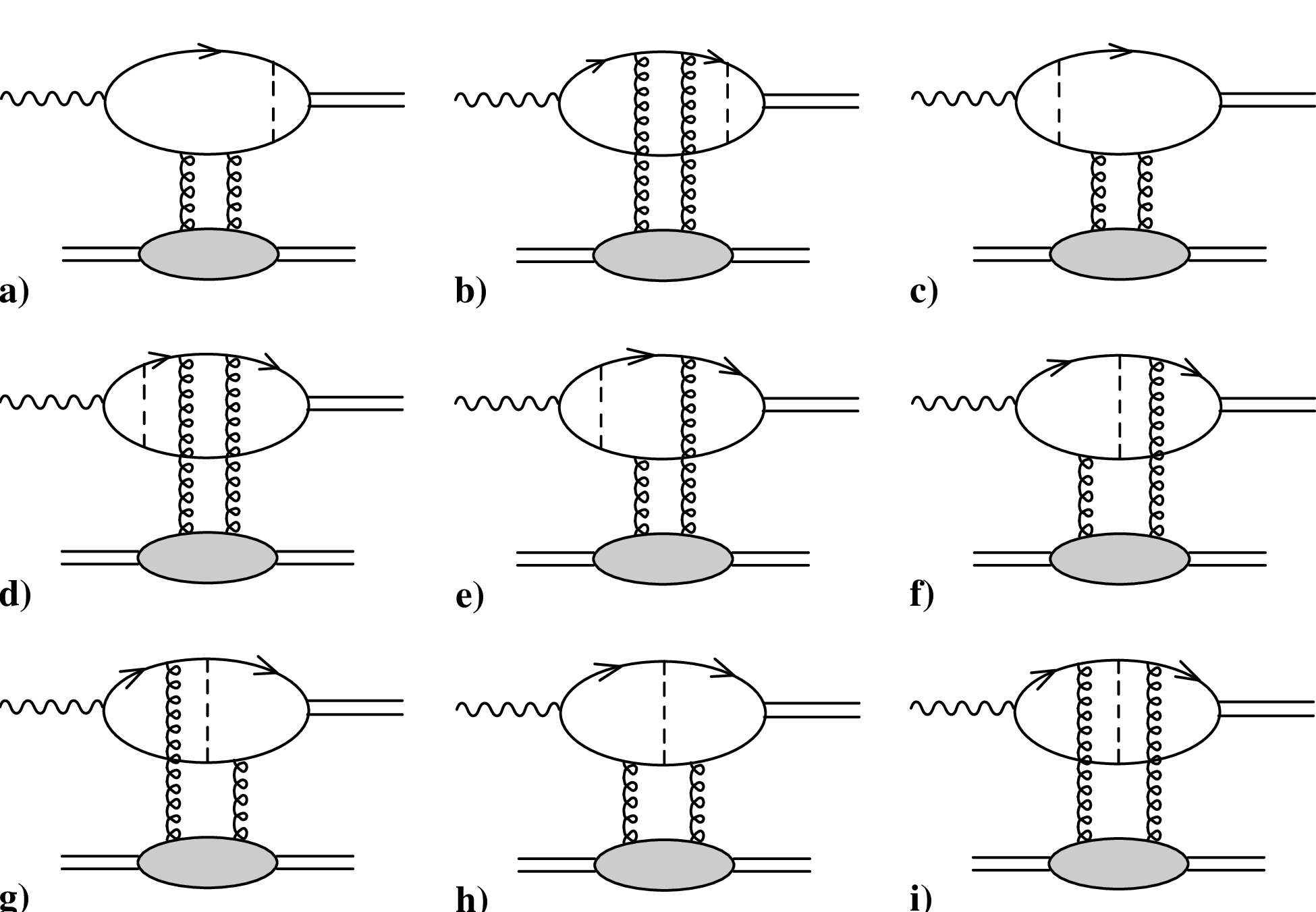}}
\end{center}
\refstepcounter{figure}
\label{fig:rest}
{\bf Figure \ref{fig:rest}:} The remaining diagrams contributing to meson 
production within the above simple model for the meson wave function.
\end{figure}

The diagram of Fig.~\ref{fig:nlo} by itself gives no consistent description 
of meson production since it lacks gauge invariance. This problem is not 
cured by just adding the two diagrams \ref{fig:lo}b) and c) with the blob 
replaced by the vertex $V$. It is necessary to include all the diagrams 
shown in Fig.~\ref{fig:rest}.

The same gauge invariance arguments that lead to Eq.~(\ref{w2}) apply to 
the sum of all the diagrams in Figs.~\ref{fig:nlo} and \ref{fig:rest}. 
Therefore, the complete result for $T_{++}$, which is now defined by the 
sum of the upper parts of all these diagrams, can be obtained by extracting 
the $\ell_\perp^2$ term at leading order in the energy and $Q^2$. Such a 
term, with a power behaviour $\sim \ell_\perp^2/Q^4$, is obtained from 
the diagram in Fig.~\ref{fig:nlo} (replace $i\lambda$ in Eq.~(\ref{ta}) 
with the vertex $V$ of Eq.~(\ref{tf2v})) by expanding around $\ell_\perp=0$. 
It can be demonstrated that none of the other diagrams gives rise to such a 
leading-order $\ell_\perp^2$ contribution (see~\cite{hl} for more details). 

The complete answer is given by the $\ell_\perp^2$ term from the Taylor 
expansion of Eq.~(\ref{ta}). The amplitude $M$ is precisely the one of 
Eqs.~(\ref{lot}) and (\ref{wv}), with $i\lambda$ substituted by $V$ of 
Eq.~(\ref{tf2v}). The correctness of this simple factorizing result has also 
been checked by explicitly calculating all diagrams of Fig.~\ref{fig:rest}. 

The above simple model calculation can be summarized as follows. The 
complete result contains leading contributions from diagrams that cannot 
be factorized into quark-pair production and meson formation. However, the 
answer to the calculation can be anticipated by looking only at one 
particular factorizing diagram. The reason for this simplification is gauge 
invariance. In the dominant region, where the transverse momentum 
$\ell_\perp$ of the two $t$-channel gluons is small, gauge invariance 
requires the complete quark part of the amplitude to be proportional to 
$\ell_\perp^2$. The leading $\ell_\perp^2$ dependence comes exclusively 
from one diagram. Thus, the complete answer can be obtained from this 
particular diagram, which has the property of factorizing explicitly if the 
two quark lines are cut. The resulting amplitude can be written in a 
factorized form.

\section{Charm and high-$p_\perp$ jets}\label{sect:ccpt}
In the two previous sections, the exclusive production of vector mesons was 
described as an example of a diffractive process with hard colour singlet 
exchange. As a different possibility of keeping the colour singlet exchange 
in diffraction hard, the diffractive production of heavy quarks~\cite{ccnz, 
ccle,cclo,ccd} and of high-$p_\perp$ jets~\cite{nz,di,blw,pt1} was 
considered by many authors. However, as will become clear from the 
discussion below, both processes can be associated with either soft or 
hard colour singlet exchange, and it is necessary to distinguish the two 
mechanisms carefully~\cite{bhmcc,bhmpt}. The semiclassical approach 
provides a very convenient framework for this analysis. 

One might expect the hard scale, provided by the transverse momentum of the
jets, to ensure the applicability of perturbation theory. Indeed, the 
production of final states containing only two high-$p_\perp$ jets can be 
described by perturbative two-gluon exchange~\cite{tge,nztge}. This process 
has been studied in detail by several groups and higher-order corrections 
have already partially been considered~\cite{nz,di,blw,pt1}. 

Below, the two simplest configurations, \qq\ and  \qqg, are discussed 
following \cite{bhmpt}. In both cases, diffractive processes are obtained 
by projecting onto the colour singlet configuration of the final state 
partons. Although the discussion focusses on high-$p_\perp$ jets, all 
qualitative results carry over to the case of diffractive charm 
production~\cite{bhmcc}. Technically, it is not important whether the hard 
scale in the diffractive final state is $p_\perp^2$ or $m_c^2$. One 
has simply to replace the high-$p_\perp$ $q\bar{q}$ jets with $c\bar{c}$ 
jets, whose transverse momentum will automatically be $\sim m_c$. However, 
there is a clear phenomenological difference. On the one hand, $m_c$ is 
fixed and not very large, while the hard scale $p_\perp$ can, at least 
in principle, be arbitrarily high. On the other hand, charm production 
is simpler to analyse since it does not require the identification of jets. 

Consider the production of a diffractive \qq\ final state. Using the 
results of Sect.~\ref{sect:qq}, for the transversely polarized photon one 
easily finds
\be
\left.{d\sigma_T\over dt\,d\alpha\,dp_\perp'^2}\right|_{t=0}={\sum_q e_q^2 
 \aem\over 2N_c(2\pi)^6}(\alpha^2+(1\!-\!\alpha)^2)\left|
 \int_{x_{\perp}}\int d^2p_\perp \frac{p_\perp \mbox{tr}\tilde{W}(p_\perp'-
 p_\perp)}{N^2+p_{\perp}^2}\right|^2\!\!\! .\label{hse}
\ee
There are two essential differences compared to Eq.~(\ref{dst}): firstly, 
the colour trace is taken at the amplitude level to ensure colour 
neutrality of the $q\bar{q}$ state; secondly, two independent $x_\perp$ 
integrations are applied to the two factors $W$ and $W^\dagger$. This is 
the result of Eq.~(\ref{hse}) being differential in $t$ at $t=0$, in 
contrast to Eq.~(\ref{dst}), where the $t$ integration has been performed. 

The cross section for large transverse momenta is calculated by expanding 
the integrand around $p_\perp=p_\perp'$, as exercised in Sect.~\ref{sect:qq} 
for inclusive electroproduction in the longitudinal case. Note however that, 
due to the colour singlet condition, the first two terms 
(cf.~Eqs.~(\ref{ex1}) and (\ref{ex2})) do not contribute, so that the 
leading contribution comes from the third term, which is proportional to the 
second derivative of $W$ at the origin (compare the discussion of 
diffraction in Sect.~\ref{sect:qq}). Even higher terms of the Taylor series 
give rise to contributions suppressed by powers of $p_\perp'^2$, thus 
demonstrating the dominance of the short distance behaviour of 
$\mbox{tr}W_{x_{\perp}}(y_{\perp})$. The leading order result reads 
\be
\left.{d\sigma_T \over dt d\alpha dp'^2_{\perp}}\right|_{t=0}=
{\sum_q e_q^2 \aem\over 384\pi^2} (\alpha^2 + (1-\alpha)^2)
\left|\left({\partial\over \partial p'_{\perp}}\right)^2
{p'_{\perp}\over N^2 + p'^2_{\perp}}\right|^2\left|\partial_{y_\perp}^2
\int_{x_\perp}\mbox{tr}W_{x_\perp}(0)\right|^2\, .\label{xqq}
\ee
As the derivation illustrates, this cross section describes the interaction
of a small \qq\ pair with the proton. Hence, it is perturbative or hard.
According to Eq.~(\ref{gd}), the cross section Eq.~(\ref{xqq}) is 
proportional to the square of the gluon distribution. In order to obtain 
the $t$ integrated cross section, one has to multiply Eq.~(\ref{xqq}) by the 
constant 
\be
C=\left(\int\frac{d\sigma}{dt}dt\right)\bigg/\left(\frac{d\sigma}{dt}
\bigg|_{t\simeq 0}\right)\sim \Lambda^2\, ,
\ee
where $\Lambda$ is a typical hadronic scale. The resulting cross section
integrated down to the transverse momentum 
$p'^2_{\perp,\mbox{\footnotesize cut}}$ yields a contribution to the 
diffractive structure function $F_2^D$ which is suppressed by 
$\Lambda^2/p'^2_{\perp,\mbox{\footnotesize cut}}$. 

As discussed in Sect.~\ref{sect:hfs}, a leading twist diffractive cross 
section for jets with \mbox{$p_{\perp}\sim Q$} requires at least three 
partons in the final state, one of which has to have low transverse 
momentum. It can be written as a convolution of ordinary partonic cross 
sections with diffractive parton distributions. In the case of 
high-$p_{\perp}$ quark jets there is an additional wee gluon. The partonic 
process is then boson-gluon fusion, and the cross section 
\be\label{bgfu}
{d\sigma_T\over d\xi dp'^2_{\perp}} = \int_x^\xi dy 
{d\hat{\sigma}_T^{\gamma^* g\rightarrow q\bar{q}}(y,p_{\perp}') 
 \over dp'^2_{\perp}} {dg(y,\xi)\over d\xi}
\ee
involves the diffractive gluon distribution of Eq.~(\ref{fxg}). 

In addition to boson-gluon fusion, the QCD Compton process can also produce
high-$p_{\perp}$ jets. In this case either the quark or the antiquark
is the wee parton. The corresponding cross section 
\be
{d\sigma_T\over d\xi dp'^2_{\perp}}=\int_x^\xi dy {d\hat{\sigma}_T(y,
p'_{\perp})^{\gamma^*q\rightarrow gq} \over dp'^2_{\perp}}{dq(y,\xi)\over 
d\xi}\,,\label{com}
\ee
involves the diffractive quark distribution of Eq.~(\ref{fxsp}). An analogous 
relation holds in the antiquark case.

The cross sections of Eqs.~(\ref{bgfu}), (\ref{com}) for diffractive 
boson-gluon fusion and diffractive Compton scattering can be evaluated 
along the lines described in~\cite{bhmcc}. In the leading-$\ln(1/x)$ 
approximation, one obtains for the longitudinal and transverse boson-gluon 
fusion cross sections 
\begin{eqnarray}
\frac{d\sigma_L}{d\alpha dp_\perp'^2}&=&\frac{\Sigma_q e_q^2\aem
\alpha_s}{2\pi^3}\, \frac{[\alpha(1-\alpha)]^2Q^2p_\perp'^2}{(N^2+
p_\perp'^2)^4}\, \ln(1/x) h_{\cal A}, \label{eq:slg}
\\ \nn
\frac{d\sigma_T}{d\alpha dp_\perp'^2}&=&\frac{\Sigma_q e_q^2\aem
\alpha_s}{16\pi^3}\,\frac{(\alpha^2+(1\!-\!\alpha)^2)\,(p_\perp'^4+
N^4)}{(N^2+p_\perp'^2)^4}\,\ln(1/x) h_{\cal A} \,,\label{eq:stg}
\\
h_{\cal A} &=&\int_{y_\perp}\int_{x_\perp}\frac{\left|\mbox{tr}
W^{\cal A}_{x_\perp}(y_\perp)\right|^2}{y_\perp^4}\,.\label{ha}
\end{eqnarray}
Similarly, one finds for the QCD-Compton cross sections
\begin{eqnarray}
\frac{d\sigma_L}{d\alpha dp_\perp'^2}&=&\frac{16 \Sigma_q e_q^2\aem
\alpha_s}{27\pi^3} \frac{Q^2}{[\alpha(1-\alpha)] \hat{Q}^6}\, h_{\cal F}, 
\label{eq:slq}
\\ \nonumber\\
\frac{d\sigma_T}{d\alpha dp_\perp'^2}&=&\frac{4 \Sigma_q
  e_q^2\aem\alpha_s}{27 \pi^3 \hat{Q}^6 p_\perp'^2} 
\left[ \hat{Q}^4 - 2Q^2(\hat{Q}^2+Q^2) +
  \frac{\hat{Q}^4+Q^4}{\alpha(1\!-\!\alpha)} \right]
\,h_{\cal F}\, , \label{eq:stq}\\
&&\qquad \hat{Q}^2=Q^2 + {p'^2_{\perp}\over \alpha(1-\alpha)}\, ,
\end{eqnarray}
where the constant $h_{\cal F}$ is defined analogously to Eq.~(\ref{ha}) 
but with the $U$ matrices in the fundamental representation. 

Comparing Eqs.~(\ref{eq:slg}), (\ref{eq:stg}) with Eqs.~(\ref{eq:slq}), 
(\ref{eq:stq}), it is apparent that the configurations with a wee gluon are 
enhanced by $\ln(1/x)$ at small $x$ relative to those with a wee quark or 
antiquark. The origin of this enhancement can be understood as follows. 
Eqs.~(\ref{bgfu}) and (\ref{com}) provide cross sections differential in 
$\xi$, $p_\perp'^2$ and $y$ or, equivalently, in  $\xi$, $p_\perp'^2$ and 
$\alpha$, since ${y=x \left[Q^2 + p'^2_{\perp}/\alpha(1\!-\!\alpha)\right] 
/Q^2}$. The above $p_\perp$-spectra are obtained after performing the 
$\xi$-integration from $y$ to some fixed $\xi_0 \ll 1$. The integral is
dominated by $\xi \gg y$, i.e., $beta\ll 1$, where one has $y dg/d\xi \sim
1/\xi$. The $\xi$-integration then yields a factor $\ln(1/x) +
\mbox{constant}$. In contrast, the diffractive quark distribution behaves as
$y dq/d\xi \sim y/\xi^2$, and consequently the $\xi$-integration only yields 
a constant. 

Simple arguments concerning colour, outlined in~\cite{bhmcc}, suggest an 
additional large suppression of the wee fermion contributions due to colour 
factors ($h_{\cal A} \gg h_{\cal F}$). This is in qualitative agreement 
with model calculations of Sects.~\ref{sect:scd} and \ref{sect:lh}, which 
predict the diffractive gluon distribution to be much larger than the quark 
distribution. As a result, it can be claimed that configurations with a wee 
gluon dominate over those with a wee fermion in the small-$x$ region 
relevant to diffraction. The latter shall be ignored from now on. 

The differential cross section for the leading order \qq\ fluctuation can be 
calculated as described above (cf. Eq.~(\ref{xqq})). The longitudinal and 
transverse cross sections are 
\begin{eqnarray}
\frac{d\sigma_L}{d\alpha dp'^{2}_\perp}&=&\frac{2\Sigma_q e_q^2\aem
\alpha_s^2\pi^2[\xi G(\xi)]^2C}{3}\, \frac{[\alpha(1-\alpha)]^2Q^2 (a^2-
p'^{2}_\perp)^2}{(N^2+p'^{2}_\perp)^6}\,,\label{slg} 
\\ \nonumber\\
\frac{d\sigma_T}{d\alpha dp'^{2}_\perp}&=&\frac{2 \Sigma_q e_q^2\aem
\alpha_s^2\pi^2[\xi G(\xi)]^2C}{3} \,\frac{(\alpha^2+(1-\alpha)^2)
p'^{2}_\perp N^4}{(N^2+p'^{2}_\perp)^6}\, .\label{eq:stqqg}
\end{eqnarray}
\noindent Identical differential distributions have been found for two-gluon
exchange in leading order~\cite{nz}. One can easily see that the region of 
$\alpha$ close to 0 or 1 dominates, and that high-\pp  configurations are 
unlikely. This statement will now be quantified. 

The quantitative differences between the \qq\ and \qqg\ configurations are 
particularly  pronounced in the integrated cross section with a lower cut on 
the  transverse  momentum of the quarks. Since the overall normalization of 
the contributions is uncertain (it is inherently non-perturbative), the 
shape in $p_{\perp}'^2$ of each configuration is compared.  Consider the 
quantity
\be
\sigma(p'^2_{\perp,\mbox{\footnotesize cut}})= 
\int_{p'^2_{\perp,\mbox{\scriptsize cut}}}^{\infty} {dp_\perp'^2}
\int_{0}^{1} d\alpha
\frac{d\sigma}{dp_\perp'^2 d\alpha}\,,
\ee
which is the fraction of events remaining above a certain minimum 
$p_\perp'^2$. The integrand here is obtained by adding the contributions 
from longitudinal and transverse photons. Figure~\ref{fig:psq} shows the 
dependence of the corresponding event fraction on the lower limit, 
$p'^2_{\perp,\mbox{\footnotesize cut}}$. Each curve is normalized to its 
value at $p'^2_{\perp,\mbox{\footnotesize cut}} = 5 $~GeV$^2$. One can see 
that the spectrum for the \qqg\ configuration is much harder than that for 
the \qq\ configuration. This is expected since in boson-gluon fusion 
$p_\perp$ is distributed logarithmically between the soft scale and $Q$ 
thus resulting in a significant high-$p_\perp$ tail above 
$p'_{\perp,\mbox{\footnotesize cut}}$.

\gnufig{psq}{ The fraction of diffractive events with $p'^2_{\perp}$ above 
$p'^2_{\perp,\mbox{\footnotesize cut}}$ for $Q^2$ of 10 GeV$^2$ and 100
GeV$^2$ (lower and upper curve in each pair).}{
\setlength{\unitlength}{0.1bp}
\begin{picture}(4320,2592)(0,0)
\includegraphics{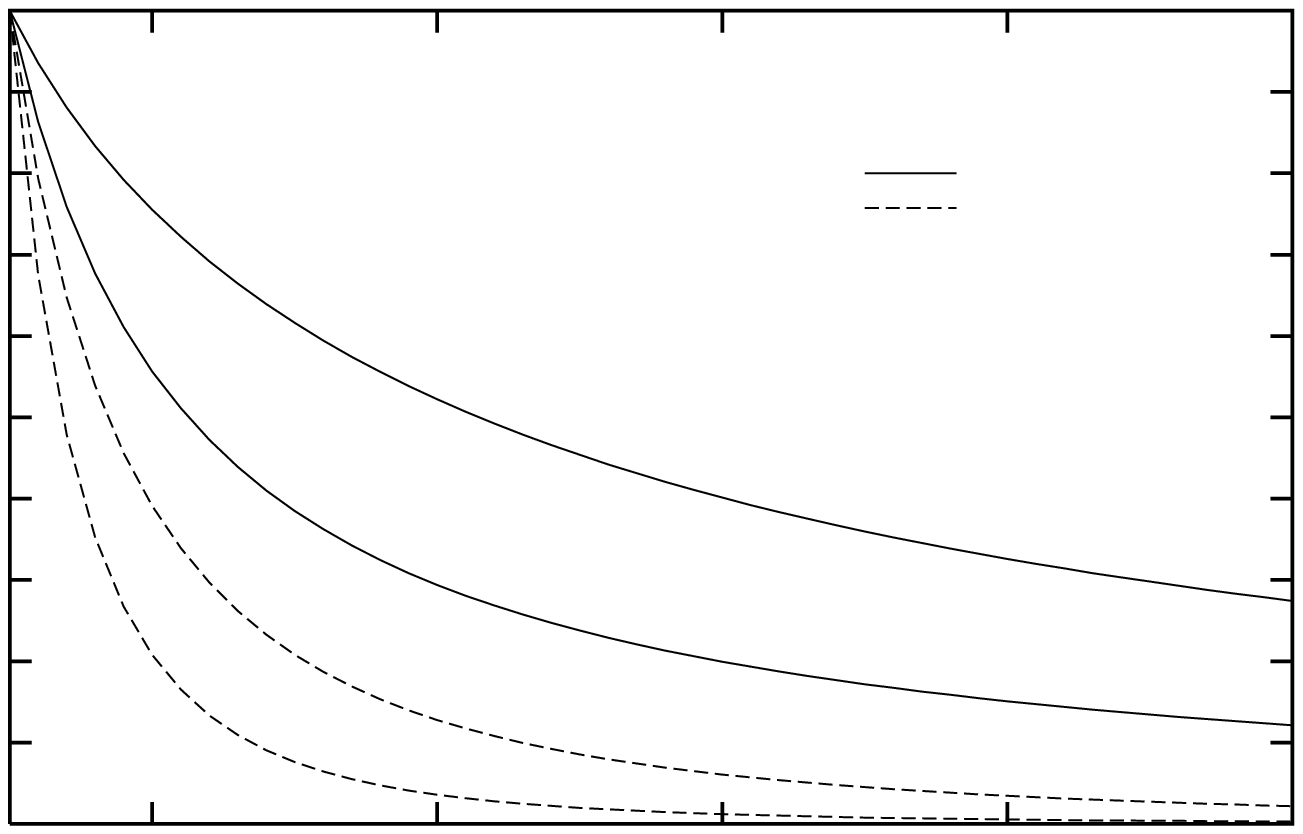}
\put(2876,1924){\makebox(0,0)[r]{ }}
\put(2876,2024){\makebox(0,0)[r]{ }}
\put(2679,1907){\makebox(0,0)[l]{$q \bar{q} ~~ $}}
\put(2679,2024){\makebox(0,0)[l]{$q \bar{q} g $}}
\put(2023,-118){\makebox(0,0)[l]{$p'^2_{\perp,\mbox{cut}} ~(\mbox{GeV}^2)$}}
\put(100,1321){%
\makebox(0,0)[b]{\shortstack{Fraction  of Events}}%
}
\put(4157,50){\makebox(0,0){50}}
\put(3336,50){\makebox(0,0){40}}
\put(2515,50){\makebox(0,0){30}}
\put(1694,50){\makebox(0,0){20}}
\put(873,50){\makebox(0,0){10}}
\put(413,2492){\makebox(0,0)[r]{1}}
\put(413,2258){\makebox(0,0)[r]{0.9}}
\put(413,2024){\makebox(0,0)[r]{0.8}}
\put(413,1789){\makebox(0,0)[r]{0.7}}
\put(413,1555){\makebox(0,0)[r]{0.6}}
\put(413,1321){\makebox(0,0)[r]{0.5}}
\put(413,1087){\makebox(0,0)[r]{0.4}}
\put(413,853){\makebox(0,0)[r]{0.3}}
\put(413,618){\makebox(0,0)[r]{0.2}}
\put(413,384){\makebox(0,0)[r]{0.1}}
\put(413,150){\makebox(0,0)[r]{0}}
\end{picture}
}

Let $M_j$ be the invariant mass of the two-jet system in diffractive events 
containing two high-$p_\perp$ jets in the diffractive final state. The 
measurement of this observable provides, in principle, a clean distinction 
between \qq\ final states, where $M_j^2=M^2$, and \qqg\ final states, where 
$M_j^2<M^2$. In practice, however, this requires the contribution of the 
wee gluon to the diffractive mass, which is responsible for the difference 
between $M^2$ and $M_j^2$, to be sufficiently large. To quantify the 
expectation within the semiclassical approach, consider the transverse 
photon contribution to the differential diffractive cross section 
$d\sigma/dM^2dM_j^2$. 

In the case of a \qq\ final state this cross section can be obtained 
directly from Eq.~(\ref{eq:stqqg}),
\be
\frac{d\sigma_T}{dM^2dM_j^2}=\Sigma_q e_q^2\aem\alpha_s^2
\pi^2[\xi G(\xi)]^2C\,\delta(M^2-M_j^2)\,\frac{16M^2Q^4\sqrt{1-\kappa}}
{3(M^2+Q^2)^6\kappa}\,.\label{mqq}
\ee
Here the $\delta$-function setting $M^2=M_j^2$ is only precise up to 
hadronization effects, which are expected to be of the order of the hadronic 
scale. The dependence on the transverse momentum cutoff enters via the 
variable $\kappa=4p'^2_{\perp,\mbox{\footnotesize cut}}/M_j^2$.

By contrast, the mass distribution for diffractive processes with 
three-particle final states is not peaked at $M_j^2=M^2$. Concentrating, as 
before, on the transverse photon polarization and on the contribution from 
the diffractive gluon distribution, the following formula can be derived 
from Eq.~(\ref{bgfu})
\be
\frac{d\sigma_T}{dM^2dM_j^2}=2\pi\Sigma_q e_q^2\aem
\alpha_s\,y^2\frac{dg(y,\xi)}{d\xi}\,\,\frac{Q^4+M_j^4}{(Q^2+M_j^2)^5}\left[
2\,\mbox{Arctanh}\sqrt{1-\kappa}-\sqrt{1-\kappa}\,\right]\,.\label{mqqg}
\ee
Explicit results can be obtained by using model calculations for the 
diffractive gluon distribution (cf. Chapter~\ref{sect:mod}) or utilizing a 
simple parametrization (cf. the numerical predictions of 
\cite{bhmcc,bhmpt}.) 

Note that the $\ln(1/x)$-enhancement present in Eq.~(\ref{eq:stg}) can be 
recovered if the $M^2$ integration is performed in Eq.~(\ref{mqqg}). The 
origin of this enhancement is the integration measure $dM^2/M^2$, which 
appears since $y^2(dg(y,\xi)/d\xi)\sim 1/M^2$ for $M^2$ sufficiently large. 
The main contribution to the total cross section comes from the region where 
$M^2$ is significantly larger than $M_j^2$. Thus it is clear, even without 
detailed calculations, that the distribution of Eq.~(\ref{mqqg}) differs 
qualitatively from Eq.~(\ref{mqq}). With sufficient statistics, a 
determination of the relative weight of soft colour singlet exchange, 
relevant in the \qqg\ case, and hard colour singlet exchange, relevant in 
the \qq\ case, should be feasible.

The above calculations can be summarized as follows. The diffractive 
production of a \qq\ final state with high $p_\perp$ or with charm quarks 
proceeds via hard colour singlet exchange. The description in the 
semiclassical framework reproduces the two-gluon exchange calculations. 
By contrast, high-$p_\perp$ jets or charm in \qqg\ final states are 
predominantly produced via boson-gluon fusion. The colour neutralization 
mechanism is soft, and the cross section is proportional to the diffractive 
gluon distribution. The boson-gluon fusion mechanism is distinguished by 
a much harder $p_\perp$ spectrum and a diffractive mass that is, on 
average, much larger than the invariant mass of the two-jet system. 

The energy dependence of the process could be very different for the above 
two mechanisms. For hard colour singlet exchange, a steep rise is expected 
from the known small-$x$ behaviour of the gluon distribution. For soft 
colour singlet exchange, the $\xi$ dependence is expected to be less steep. 

In Ref.~\cite{ccle}, the importance of higher order $\alpha_s$ corrections 
to diffractive charm production was estimated in the framework of two-gluon 
exchange. A sizeable enhancement of the cross section was found. Clearly, 
the $c\bar{c}g$ final state is part of these corrections. However, the 
above discussion shows that this final state is dominated by the 
region where the gluon is soft. In this case, the $c\bar{c}g$ contribution 
is not perturbatively calculable and can not be considered an $\alpha_s$ 
correction to the hard $c\bar{c}$ process. Thus, the systematic 
calculability of corrections to hard $c\bar{c}$ or jet production is an 
interesting open problem, which may be related to the problem of defining 
`exclusive' jet production at higher orders.

\section{Inclusive diffraction}\label{sect:id} 
There are a number of attempts by several authors to approach the bulk of 
the diffractive DIS data, as described by the structure function $F_2^D$, 
from the perspective of two gluon exchange in the $t$ channel. The degree 
to which perturbation theory is taken seriously varies significantly in 
the different investigations discussed below. 

In what can be possibly called the most modest approach, two gluons with an 
appropriate form-factor-like coupling to the proton are used as a simple 
model for colour singlet exchange, even if this exchange is believed to be 
non-perturbative. In the context of the diffractive electroproduction cross 
section at HERA, two gluon exchange calculations were performed in 
\cite{wf}, where the importance of the soft region was pointed out, and 
the result was parametrized in terms of an effective two gluon form factor 
of the proton. 

Furthermore, the two gluon calculation was used to describe both diffractive 
and inclusive cross sections in terms of the $q\bar{q}$ component of the 
light-cone wave function of the virtual photon and of $\sigma(\rho)$, the 
cross section for a $q\bar{q}$ pair to interact with the hadronic target
\cite{wf}. If the $q\bar{q}$ component dominates, which is, however, a 
non-trivial assumption, this approach allows one to link the diffractive 
cross section at $t=0$ to the inclusive DIS cross section via the optical 
theorem. 

In Ref.~\cite{di}, the diffractive structure function was discussed on the 
basis of the exchange of two gluons with non-perturbative propagators in the 
sense of the Landshoff-Nachtmann model~\cite{ln}. Further analyses extend 
the two gluon exchange calculations to include the $q\bar{q}g$ component of 
the incoming photon~\cite{pt1,wue,nztge}. 

Going one step further in the direction of perturbation theory, the known 
relation between the $q\bar{q}$ cross section $\sigma(\rho)$ and the 
inclusive gluon distribution~\cite{fms}, 
\be
\sigma(\rho)=\frac{\pi^2}{3}\alpha_s[xg(x)]\rho^2+{\cal O}
(\rho^4)\,,\label{srgd}
\ee
may be employed for the calculation of diffractive cross sections. This is 
similar to what was discussed in Sects.~\ref{sect:emp} and \ref{sect:ccpt} 
in the case of meson production and high-$p_\perp$ jet or charm production 
respectively. However, such a relation to the gluon distribution, employed, 
e.g., in~\cite{nz,nzchi}, is problematic since the diffractive structure 
function is dominated by large transverse sizes of the $q\bar{q}$ 
fluctuation of the photon, where Eq.~(\ref{srgd}) is not valid. It was 
emphasized in~\cite{glm} that small, perturbative values of $\rho$ become 
more important if the analysis is restricted to small diffractive masses. 

In spite of the evident problems with the applicability of perturbation 
theory to the diffractive structure function, it is still interesting to 
take the perturbative approach even further, calculating the BFKL 
leading-log corrections~\cite{bfkl} to the two gluon exchange amplitude. A 
possible formal justification of such a treatment can be obtained by 
considering diffractive DIS off a heavy quark-antiquark state, a so-called 
onium, which provides a perturbative scale in addition to the $Q^2$ of the 
virtual photon. The focus is clearly on the energy dependence of the cross 
section, which so far can not be quantitatively described by 
non-perturbative QCD based methods. 

A detailed discussion of the BFKL technique of summing leading logarithms 
in the high-energy limit of perturbative QCD amplitudes is beyond the scope 
of the present review. The essential qualitative result, relevant for the 
following brief overview, concerns the scattering of two small colour 
dipoles at very high centre-of-mass energy $\sqrt{s}$. It states that all 
corrections of the form $\alpha_s\ln s$ to the above process, which proceeds 
via two gluon exchange at leading order, can be summed. This results 
in a power-like growth of the amplitude with $s$. One may think of the 
energy logarithms as being associated with gluonic ladders, although the 
ladder topology does not exhaust all relevant diagrams. 

More recently, a colour dipole picture of the BFKL amplitude has been 
developed~\cite{nz,mcd,mu,nzcd}. In this picture, each of the colliding 
colour dipoles radiates gluons, thus creating new colour dipoles with smaller 
energy, which are the source of further gluon radiation. Eventually, two 
dipoles, one from each of the two colliding cascades, interact via simple 
two gluon exchange. The equivalence with the original BFKL technique has 
been established for the most fundamental, but not for all relevant 
applications. In particular, the role of the large $N_c$ limit, which 
is used in addition to the leading logarithmic approximation~\cite{mcd,mu}, 
is not yet fully understood. 

For illustration, the particularly simple formula for the total cross 
section of two onia with radius $R$ and mass $m$ is given (see, 
e.g.,~\cite{mu}). It reads 
\be
\sigma(s)=16\pi^2R^2\alpha_s^2\frac{(s/m^2)^{(\ap-1)}}{\sqrt{(7/2)
\alpha_sN_c\zeta(3)\ln(s/m)}}\,,
\ee
where $\ap=1+(4N_c\alpha_s/\pi)\ln 2$ is the intercept of the BFKL pomeron. 
The reader is referred to the original papers~\cite{bfkl} and to the modern 
introductory text~\cite{fr} for further details. 

The most straightforward application of the above perturbative techniques 
to diffraction at HERA relies on postulating the exchange of a BFKL 
pomeron between target proton and partonic photon fluctuation. For forward 
diffraction, the conventional BFKL amplitude has to be introduced between, 
say, the $q\bar{q}$ fluctuation of the photon and a quark of the proton. 
More generally, the BFKL amplitude at non-zero $t$ is needed. Corresponding 
formulae appear as a bye-product in the more general investigations of 
\cite{bwtr,blwtr}. In the framework of the colour dipole approach, they 
are discussed in~\cite{gnz,bp}. Clearly, the method provides the desired 
rise of the cross section with energy or, equivalently, of $F_2^D$ with 
$1/\xi$. However, with the simple asymptotic BFKL result, the rise appears 
to be far too strong (cf. Sect.~\ref{sect:edsf} more details). 

\begin{figure}[ht]
\begin{center}
\vspace*{.2cm}
\parbox[b]{10cm}{\psfig{width=10cm,file=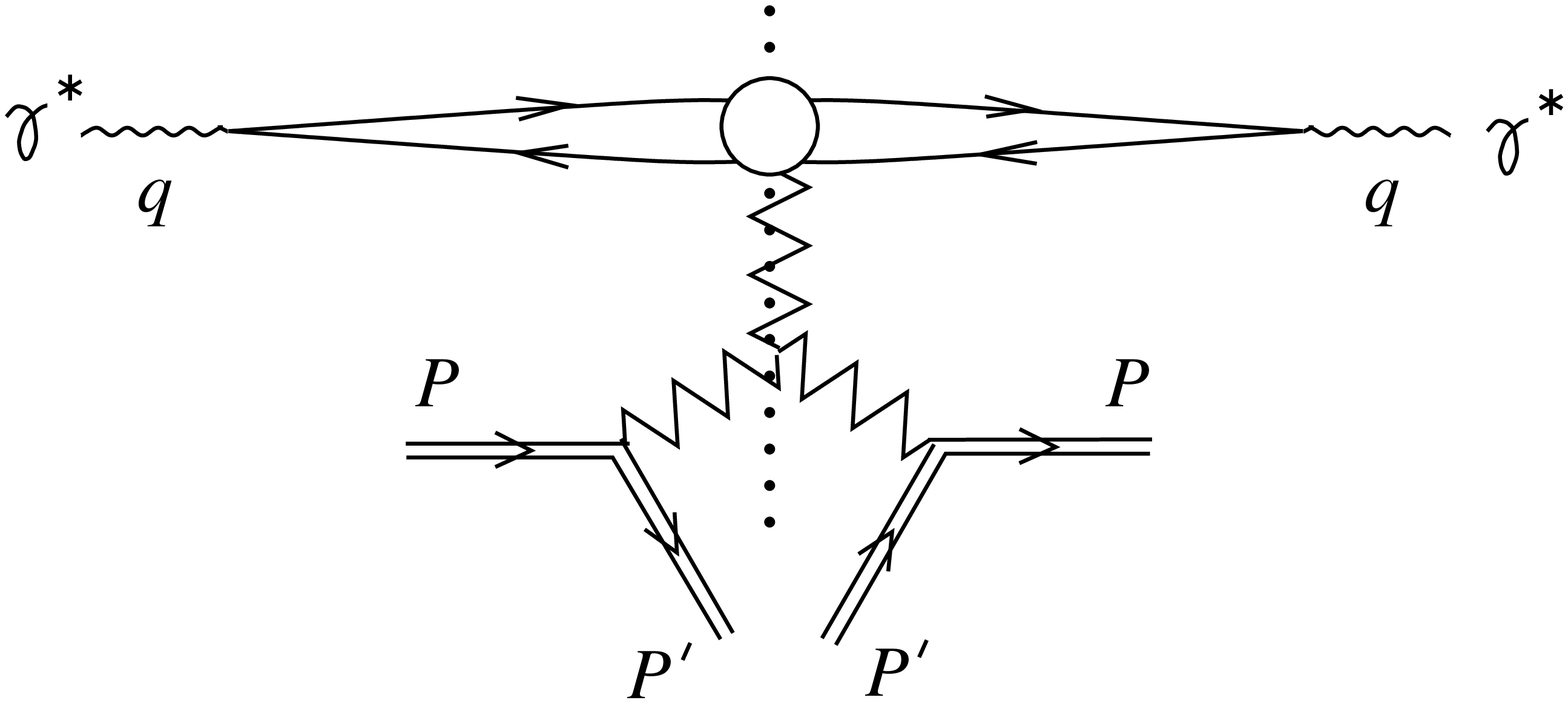}}
\end{center}
\refstepcounter{figure}
\label{fig:tp}
{\bf Figure \ref{fig:tp}:} Triple pomeron vertex in diffractive 
electroproduction of large masses. 
\end{figure}

It is a further, even more challenging theoretical problem to consider the 
double limit $W^2\gg M^2\gg Q^2$, which corresponds to the simultaneous 
small-$\xi$ and small-$\beta$ limit of $F_2^D$. The relevant process, 
illustrated in Fig.~\ref{fig:tp}, is similar to large-mass soft diffraction, 
discussed in Sect.~\ref{sect:psf}. Several authors proposed to 
use perturbative BFKL techniques for the calculation of this process of 
large-mass electroproduction. The obvious idea is to use the knowledge of 
the perturbative BFKL amplitude and to interpret the three pomerons in 
Fig.~\ref{fig:tp} as BFKL pomerons. In particular, the cut pomeron in 
the upper part of the diagram corresponds to gluonic radiation being 
responsible for the large diffractive mass created in the process. The 
dominance of gluons in the large mass region is well-known from fixed order 
perturbative calculations. 

In the framework of conventional BFKL summation, the triple pomeron vertex 
in diffractive electroproduction was considered in~\cite{bwtr,blwtr}. 
It was found that the cut pomeron in Fig.~\ref{fig:tp} is more complicated 
than the usual gluonic ladder. Four-gluon states were found to contribute 
to the leading amplitude. 

The triple pomeron vertex was derived in the colour dipole approach and 
used for the calculation of large-mass diffractive electroproduction 
in~\cite{bptp,gnztp}. As expected, an enhancement of the cross section 
both in the limit of large $W^2$ and large $M^2$ is obtained. More details 
of the phenomenological analysis will be given in Sect.~\ref{sect:edsf}. 

In spite of the impressive theoretical work discussed above, the 
perturbative calculability of the triple pomeron vertex appearing in 
large-mass hard diffraction remains questionable. On the one hand, it is not 
clear how far down the gluon ladder the influence of the hard scale $Q^2$, 
introduced by the $\gamma^*$, extends. On the other hand, it is difficult to 
justify the ad-hoc introduction of a hard scale at the bottom of the 
diagram, as advocated in~\cite{bptp}, since the proton is a soft hadronic 
object.

Note finally that next-to-leading order results in the BFKL framework 
have recently become available (see~\cite{nlo} and refs. therein). For 
realistic values of $\alpha_s$, the obtained corrections are larger than 
the leading order contributions and of opposite sign, thus complicating 
the theoretical status of the perturbative energy dependence even 
further. However, according to the very recent result of~\cite{blm}, the 
use of BLM scale setting improves the situation dramatically. 

\newpage

\mychapter{Models for the Colour Field of the Proton}\label{sect:mod}
In this chapter, three different models for the procedure of averaging over 
the target colour field configurations are described. Using the formulae of 
Sect.~\ref{sect:trf}, these models give rise to predictions for the 
diffractive quark and gluon distribution of the target hadron and thus for 
diffractive electroproduction cross sections.

\section{Small colour dipole}\label{sect:scd}
A particularly simple model based entirely on perturbation theory has 
recently been suggested by Hautmann, Kunszt and Soper~\cite{hks}. The 
authors study diffraction as quasi-elastic scattering off a special target 
photon that couples to only one flavour of very massive ($M \gg \Lambda$) 
quarks. The large quark mass justifies a completely perturbative treatment 
of the target and the diffractive system. In this situation, the required 
$t$ channel colour singlet exchange is realized by two gluons coupling to 
the massive quark loop of the target. The analysis of \cite{hks} is based 
on the operator definitions of diffractive parton distributions 
(Eqs.~(\ref{od})--(\ref{gsub})), evaluated in an explicit two gluon exchange 
calculation. 

Here, the results of~\cite{hks} will be derived in the semiclassical 
framework, following Appendix B of~\cite{bgh}. In the 
semiclassical approach, the two $t$ channel gluons are understood to be 
radiated by the massive quark loop and are treated as the colour field 
generating tr$W$tr$W^\dagger$. The semiclassical calculation proceeds as 
follows. 

Equations (\ref{fxsp}) and (\ref{fxg}) have the structure 
\be
\frac{df_i^D}{d\xi}=F_i\left[\int_{\x}\mbox{tr}W_{\x}\,\mbox{tr}
W_{\x}^\dagger\right]\,,
\ee
where $F_i$ (with $i=q,g$) is a linear functional depending on $\int\mbox{tr}
W_{\x}(\y)\,\mbox{tr}W_{\x}^\dagger(\y')$, interpreted as a function of $\y$ 
and $\y'$. To be differential in $t$, one simply writes 
\be
\frac{df_i^D}{d\xi\,dt}=\frac{1}{4\pi}F_i\left[\int_{\x}\int_{\x'}\mbox{tr}
W_{\x}\,\mbox{tr}W_{\x'}^\dagger e^{iq_\perp(\x'-\x)}\right]\,,
\ee
with $q_\perp^2=-t$.

The field responsible for tr$W$ is created by a small colour dipole which, 
in turn, is created by the special photon that models the target. At leading 
order in perturbation theory, the colour field of a static quark is 
analogous to its electrostatic Coulomb field. The field of a quark 
travelling on the light cone in $x_-$ direction at transverse position 
$0_\perp$ has therefore the following line integral along the $x_+$ 
direction, 
\be
-\frac{ig}{2}\int A_-\,dx_+\,=\,-ig^2\int\frac{d^2k_\perp}{(2\pi)^2}\,\cdot
\,\frac{e^{ik_\perp\x}}{k_\perp^2}\,\,\,.
\ee
It is exactly this type of line integral that appears in the exponents of 
the non-Abelian phase factors $U$ and $U^\dagger$ that form $W$ 
(cf.~Eq.~(\ref{um})). A straightforward calculation shows that the function 
tr$W$ produced by a dipole consisting of a quark at $\rho_\perp$ and an 
antiquark at $0_\perp$ reads 
\bea
\mbox{tr}W_{\x}(\y)&=&-\frac{g^4(N_c^2-1)T_R}{2}\left[\int_{k_\perp,
k_\perp'}\frac{\left(1-e^{-ik_\perp\rho_\perp}\right)\left(1-e^{-ik_\perp'
\rho_\perp}\right)}{(2\pi)^4 k_\perp^2 k_\perp'^2}\right]
\nn
\nn
&&\times \left(1-e^{ik_\perp\y}\right)\left(1-e^{ik_\perp'\y}\right)\,
e^{i(k_\perp+k_\perp')\x}\,,\label{trw}
\eea
where $T_F=1/2$ and $T_A=N_c$ have to be used for the fundamental and 
adjoint representation respectively. 

The final formulae for the diffractive 
parton distributions of the target are obtained after 
integrating over the transverse sizes of the colour dipoles with a weight 
given by the $q\bar{q}$ wave functions of the incoming and outgoing target 
photon. They read
\bea
\frac{df_i^D}{d\xi\,dt}&\!\!=\!\!&\int dz\,d^2\rho_\perp \int dz'\,d^2
\rho_\perp'\,\frac{1}{4\pi}F_i\left[\int_{\x}\int_{\x'}\mbox{tr}W_{\x}\,
\mbox{tr}W_{\x'}^\dagger e^{iq_\perp(\x'-\x)}\right]\label{faf}
\\
\nn
&&\times \frac{1}{2}\sum_{\epsilon,\epsilon'}\left[\psi_\gamma^*(z,
\rho_\perp,p_\perp',\epsilon_\perp')\,\psi_\gamma(z,\rho_\perp,0_\perp,
\epsilon_\perp)\right]\left[\psi_\gamma^*(z,'\rho_\perp',p_\perp',
\epsilon_\perp')\,\psi_\gamma(z',\rho_\perp',0_\perp,\epsilon_\perp)
\right]\,,\nonumber
\eea
where tr$W_{\x}(\y)$ is produced by the field of a quark at $\rho_\perp$ and 
an antiquark at $0_\perp$, and tr$W_{\x'}(\y')$ is produced by the field of 
a quark at $\rho_\perp'$ and an antiquark at $0_\perp$, as detailed in 
Eq.~(\ref{trw}).

The wave function $\psi_\gamma(z,\rho_\perp,0_\perp,\epsilon_\perp)$ 
characterizes the amplitude for the fluctuation of the incoming target 
photon with polarization $\epsilon$ and transverse momentum $0_\perp$ into 
a $q\bar{q}$ pair with momentum fractions $z$ and $1-z$ and transverse 
separation $\rho_\perp$. Similarly, the wave function $\psi_\gamma^*(z,
\rho_\perp,p_\perp',\epsilon'_\perp)$ characterizes the amplitude for the 
recombination of this $q\bar{q}$ pair into a photon with polarization 
$\epsilon'$ and transverse momentum $p_\perp'=-q_\perp$. The summation over 
the helicities of the intermediate quark states, which are conserved by the 
high-energy gluonic interaction, is implicit. 

The required product of photon wave functions can be calculated following 
the lines of~\cite{bhm} and using the matrix elements of 
Appendix~\ref{sect:me}. It reads explicitly 
\vspace*{.3cm}
\be
\hspace*{-9cm}\psi_\gamma^*(z,\rho_\perp,p_\perp',\epsilon_\perp')\,
\psi_\gamma(z,\rho_\perp,0_\perp,\epsilon_\perp)\label{swv}
\ee
\[
\hspace*{2cm}=\frac{N_ce^2e_q^2}{2(2\pi)^5}\int_{k_\perp,k_\perp'}
\mbox{tr}\Phi^\dagger(z,k_\perp',M,\epsilon_\perp')\Phi(z,k_\perp,M,
\epsilon_\perp)e^{i\rho_\perp(k_\perp'-k_\perp+zp_\perp')}\,,
\]
where the notation of~\cite{hks},
\be
\Phi(z,k_\perp,M,\epsilon_\perp)=\frac{1}{(k_\perp^2+M^2)}\left[
\,(1-z)\,\epsilon_\perp\cdot\sigma\,k_\perp\cdot\sigma-z\,k_\perp\cdot\sigma
\,\epsilon_\perp\cdot\sigma+iM\,\epsilon_\perp\cdot\sigma\,\right]\,,
\ee
has been used, $M$ is the quark mass, and $\sigma_{1,2}$ are the first two 
Pauli matrices. Note that for $p_\perp'=0$, the average of the diagonal 
elements ($\epsilon_\perp=\epsilon_\perp'$) in Eq.~(\ref{swv}) reproduces 
the well-known formula for the square of the photon wave function~\cite{wf}. 

Inserting Eq.~(\ref{swv}) into Eq.~(\ref{faf}) and introducing 
explicitly the required functionals $F_i$ specified by Eqs.~(\ref{fxsp}) and 
(\ref{fxg}), the formulae of~\cite{hks} for diffractive quark and gluon 
distribution are exactly reproduced. For the lengthy final expressions the 
reader is referred to the original paper, where a number of plots, based on 
the numerical evaluation of these formulae, is also given. Qualitatively, 
the behaviour can be summarized as follows. The quark distribution 
$\beta (df_q^D\,/\,d\xi\,dt)$ falls off like $\beta$ as $\beta\to 0$ and 
approaches a constant value, which is small compared to intermediate 
$\beta$ points, as $\beta\to 1$. The gluon distribution $\beta (df_g^D\,/\, 
d\xi\,dt)$ approaches a sizeable constant as $\beta\to 0$ and a small 
constant as $\beta\to 1$. For $t=0$, the behaviour at $\beta\to 1$ changes 
-- quark and gluon distributions vanish approximately as $(1-\beta)$ and 
$(1-\beta)^2$. The overall normalization of the gluon distribution is found 
to be much larger than that of the quark distribution. 

Even though a real proton is very different from a small colour dipole, 
it would certainly be interesting to perform a phenomenological analysis 
on the basis of the above model.

\section{Large hadron}\label{sect:lh}
In this section, the colour field averaging procedure is described for 
the case of a very large hadronic target, where a quantitative treatment 
becomes possible under minimal additional assumptions. The following 
discussion is based on~\cite{hw,bgh}, where the large hadron model was 
developed and applied to a combined analysis of both diffractive and 
inclusive structure functions (see also Sect.~\ref{sect:trf} and 
Appendix~\ref{sect:isf} of this review). 

McLerran and Venugopalan observed that the large size of a hadronic target, 
realized, e.g., in an extremely heavy nucleus, introduces a new hard scale 
into the process of DIS~\cite{mv}. From the target rest frame point of 
view, this means that the typical transverse size of partonic 
fluctuations of the virtual photon remains small~\cite{hw}, thus justifying 
the perturbative treatment of the photon wave function in the semiclassical 
calculation. 

The basic arguments underlying this important result are best explained 
in the simple case of a longitudinally polarized photon coupled to 
scalar quarks with one unit of electric charge. As far as the 
$Q^2$-behaviour of the total $\gamma^*p$ cross section is concerned, this 
is analogous to the standard partonic process where a transverse photon 
couples to spinor quarks~\cite{bd}. 

In analogy to~\cite{wf}, the longitudinal cross section can be written
as
\begin{equation}
\sigma_L=\int d^2\rho_\perp \sigma(\rho)
  W_L(\rho)\,,\label{conv}
\end{equation}
with the square of the wave function of the virtual photon given by 
\begin{equation}
W_L(\rho)=\frac{3\aem}
  {4\pi^2}\int d\alpha\,N^2
K_0^2(N\rho)\,.\label{wl}
\end{equation}
Here $\rho=|\rho_\perp|$ is the transverse size of the
$q\bar{q}$ pair, $\alpha$ is the longitudinal momentum fraction of the
photon carried by the quark, $N^2=\alpha(1\!-\!\alpha)Q^2$, and $K_0$ 
is a modified Bessel function. Note that, in contrast to~\cite{wf}, $W_L$ 
is defined to include the integration over $\alpha$. 

Within the semiclassical approach, the dipole cross section $\sigma(\rho)$ 
is given by
\begin{equation}
\sigma(\rho)=\frac{2}{3}\int d^2x_\perp
\mbox{tr}\left[\mbox{\bf 1}-U(x_\perp)
U^\dagger(x_\perp+\rho_\perp)\right]\,,\label{sc}
\end{equation}
but it is convenient to formulate the following arguments in terms of the 
more general quantity $\sigma(\rho)$. 

\begin{figure}[ht]
\begin{center}
\vspace*{.2cm}
\parbox[b]{8cm}{\psfig{width=8cm,file=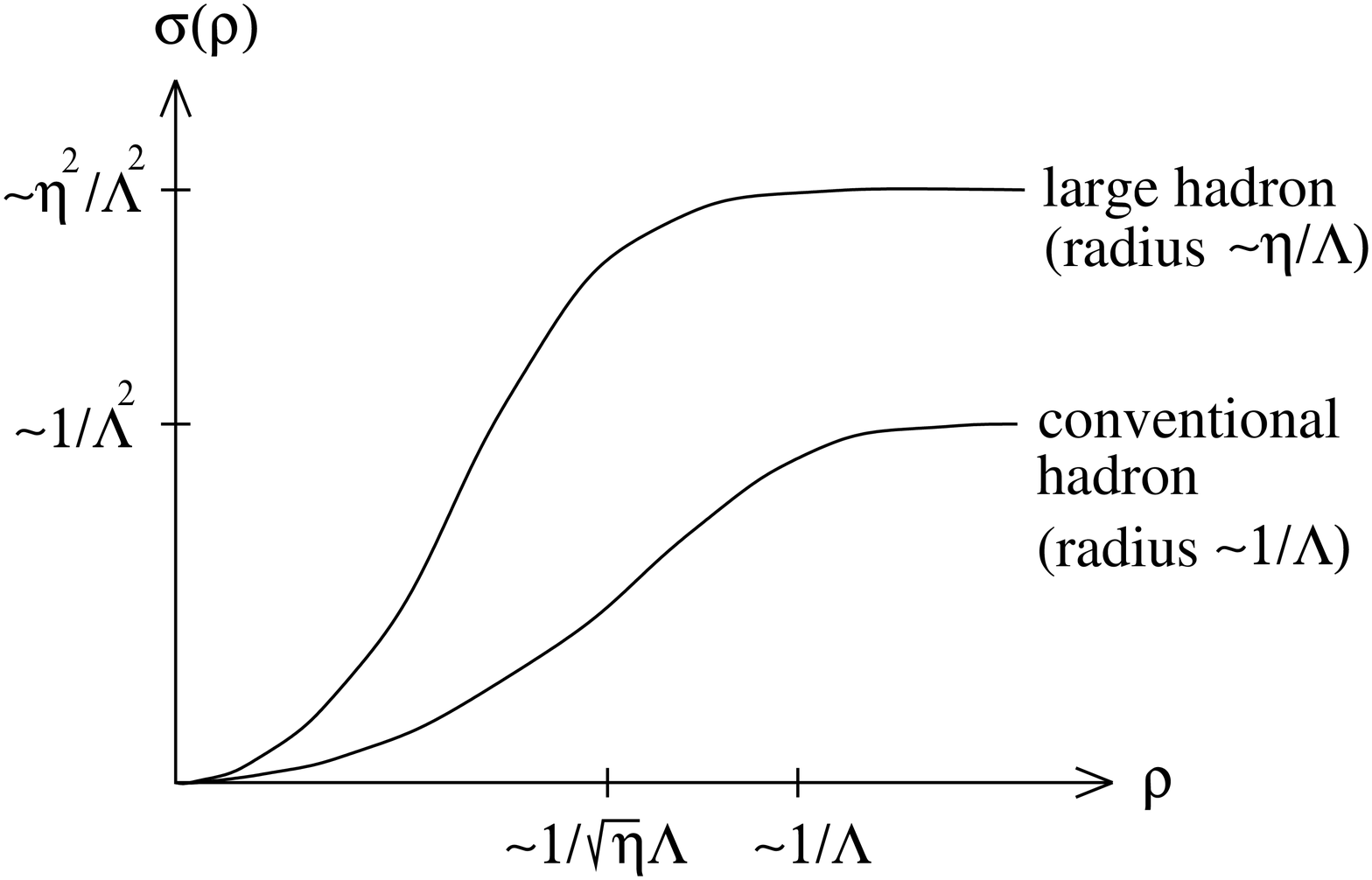}}
\end{center}
\refstepcounter{figure}
\label{fig:sigma}
{\bf Figure \ref{fig:sigma}:} Qualitative behaviour of the function 
$\sigma(\rho)$.
\end{figure}

The functional form of $\sigma(\rho)$ is shown qualitatively in
Fig.~\ref{fig:sigma}. For conventional hadrons of size $\sim 1/\Lambda$
(where $\Lambda\sim\Lambda_{\mbox{\footnotesize QCD}}$), its typical 
features are the quadratic rise at small $\rho$ ($\sigma(\rho)\sim \rho^2$ 
with a proportionality constant ${\cal O}(1)$) and the saturation at 
$\sigma(\rho) \sim 1/\Lambda^2$, which occurs at $\rho\sim 1/\Lambda$. 
Consider now the idealized case of a very large target of size 
$\eta/\Lambda$ with $\eta\gg 1$ ($\eta\sim A^{1/3}$ for a nucleus). It is 
easy to see that at small $\rho$ the functional behaviour is given by 
$\sigma(\rho)\sim\eta^3\rho^2$ while saturation has to occur at $\sigma( 
\rho)\sim\eta^2/\Lambda^2$ for geometrical reasons. It follows that the 
change from quadratic rise to constant behaviour takes place at $\rho \sim 
1/\sqrt{\eta}\Lambda$, i.e., at smaller $\rho$ than for conventional 
targets. 

{}From the above behaviour of $\sigma(\rho)$, the dominance of small 
transverse distances in the convolution integral of Eq.~(\ref{conv}) will 
now be derived. For this purpose, a better understanding of the function 
$W_L(\rho)$ is necessary. Recalling that $K_0(x)\sim\ln(1/x)$ for $x\ll 1$ 
while being exponentially suppressed for $x\gg 1$, it is easy to see that 
$W_L(\rho)\sim Q^2\ln^2(1/\rho^2Q^2)$ for $\rho\ll 1/Q$ and $W_L(\rho)\sim 
1/\rho^4Q^2$ for $\rho\gg 1/Q$. Here numerical constants and non-leading 
terms have been suppressed.

Under the assumption $\Lambda^2\ll \eta\Lambda^2\ll Q^2$, the integral
in Eq.~(\ref{conv}) can now be estimated by decomposing it into three
regions with qualitatively different behaviour of the functions
$W_L(\rho)$ and $\sigma(\rho)$,
\begin{equation}
\sigma_L=\sigma_L^{\ri}+\sigma_L^{\ri\!\ri}+\sigma_L^{\ri\!\ri\!\ri}=\left(
\int_0^{1/Q^2}+\int_{1/Q^2}^{1/\eta \Lambda^2}+
\int_{1/\eta \Lambda^2}^{\infty}\right)\pi d\rho^2
\sigma(\rho)W_L(\rho)\,.
\end{equation}
Of the three contributions 
\begin{eqnarray}
\sigma_L^{\ri}\qquad\sim &\displaystyle\int_0^{1/Q^2}
d\rho^2\,\eta^3\rho^2\,
\,Q^2\ln^2(1/\rho^2Q^2)&\sim\quad\frac{\eta^3}{Q^2}
\nonumber \\ \nonumber\\
\sigma_L^{\ri\!\ri}\qquad\sim &
\displaystyle\int_{1/Q^2}^{1/\eta \Lambda^2}d
\rho^2\,\eta^3\rho^2\,\frac{1}{\rho^4Q^2}&
\sim\quad\frac{\eta^3}{Q^2}\ln(Q^2/
\eta\Lambda^2)
\\ \nonumber\\
\sigma_L^{\ri\!\ri\!\ri}\qquad\sim &\displaystyle
\int_{1/\eta \Lambda^2}^{\infty}d\rho^2\,\frac{\eta^2}
{\Lambda^2}\,\frac{1}
{\rho^4Q^2}&\sim\quad\frac{\eta^3}{Q^2}\nonumber
\end{eqnarray}
the second one dominates, giving the total cross section 
\begin{equation}
\sigma_L\sim\frac{\eta^3}{Q^2}\ln(Q^2/\eta\Lambda^2)\,.
\end{equation}
It is crucial that the third integral is dominated by contributions
from its lower limit. Therefore, the overall result is not sensitive
to values of $\rho$ that are larger than $1/\sqrt{\eta}\Lambda$. Phrased 
differently, for sufficiently large targets the transverse size of the
$q\bar{q}$ component of the photon wave function stays perturbative.

This result can be carried over to the realistic case of a transverse 
photon and spinor quarks, which is obtained simply by substituting 
$K_0^2(\rho N)$ with $2[\alpha^2+(1\!-\!\alpha)^2]K_1^2(\rho N)$ in 
Eq.~(\ref{wl}). From the above derivation, one can also expect the result 
to hold for the transverse size of the $q\bar{q}g$ component of the photon 
wave function, both in the case of inclusive and diffractive scattering. 

Note that this does not imply a complete reduction to perturbation theory 
since the long distance which the partonic fluctuation travels in the 
target compensates for its small transverse size, thus requiring the 
eikonalization of gluon exchange. 

\begin{figure}[ht]
\begin{center}
\vspace*{.2cm}
\parbox[b]{12cm}{\psfig{width=9cm,file=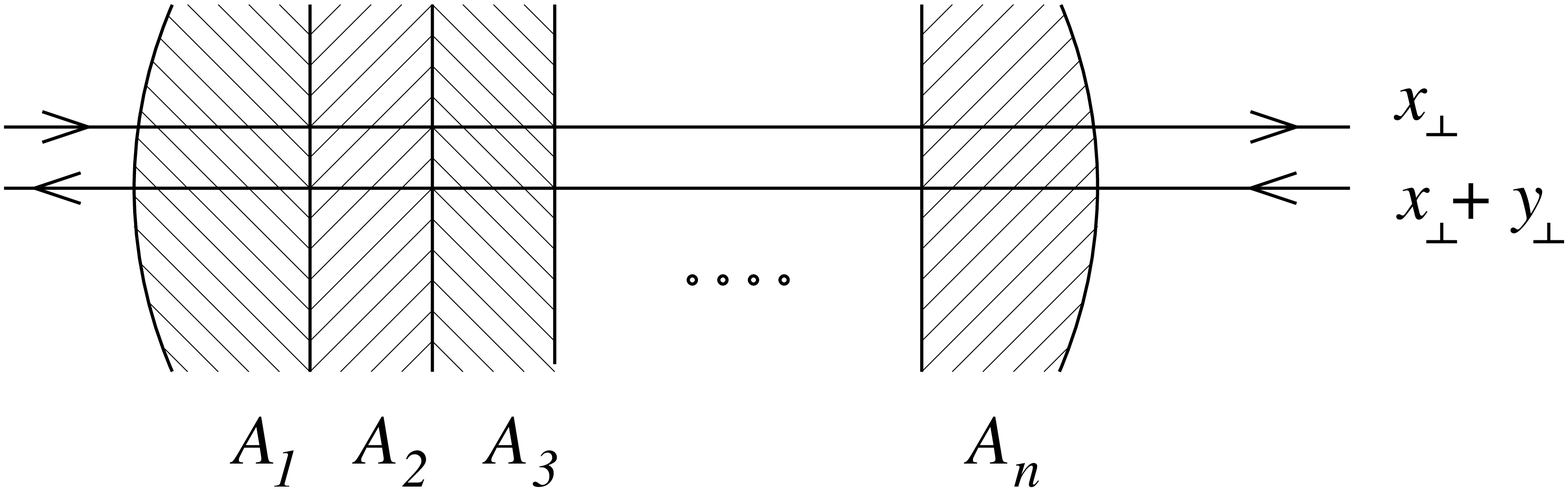}}
\end{center}
\refstepcounter{figure}
\label{fig:lt}
{\bf Figure \ref{fig:lt}:} Colour dipole travelling through a large
hadronic target.
\end{figure}

Within this framework, it is natural to introduce the additional assumption 
that the gluonic fields encountered by the partonic probe in distant regions 
of the target are not correlated (cf.~\cite{jkmw} and the somewhat 
simplified discussion in~\cite{hw}). Thus, one arrives at the situation 
depicted in Fig.~\ref{fig:lt}, where a colour dipole passes a large number 
of regions, each one of size $\sim 1/\Lambda$, with mutually uncorrelated 
colour fields $A_1$ ... $A_n$.

Consider the fundamental quantity $W_{x_\perp}(y_\perp)_{ij}[A]\, 
W^\dagger_{x_\perp}(y_\perp')_{kl}[A]$ which, after specifying the 
required representation and appropriately contracting the colour indices 
$ijkl$, enters the formulae for inclusive and diffractive parton 
distributions (cf. Sect.~\ref{sect:trf} and Appendix~\ref{sect:isf}). 
According to Eq.~(\ref{wdef}), this quantity is the sum of four terms, the 
most complicated of which involves four $U$ matrices, 
\vspace*{.2cm}
\bea
&&\hspace*{-1.3cm}
\left\{U_{x_\perp}[A]\,U^\dagger_{x_\perp+y_\perp}[A]\right\}_{ij}\left\{
U_{x_\perp+y_\perp'}[A]\,U^\dagger_{x_\perp}[A]\right\}_{kl}\label{um1}
\\
&&\hspace*{3cm}
=\left\{U_{x_\perp}[A_n]\cdots U_{x_\perp}[A_1]\,\,\,
U^\dagger_{x_\perp+y_\perp}[A_1]\cdots U^\dagger_{x_\perp+y_\perp}[A_n]
\right\}_{ij}
\nonumber
\\
&&\hspace*{3cm}\vspace*{.2cm}
\times\left\{U_{x_\perp+y_\perp'}[A_n]\cdots U_{x_\perp+y_\perp'}[A_1]\,\,\,
U^\dagger_{x_\perp}[A_1]\cdots U^\dagger_{x_\perp}[A_n]\right\}_{kl}\,.
\nonumber
\eea
The crucial assumption that the fields in regions $1$ ... $n$ are 
uncorrelated is implemented by writing the integral over all field 
configurations as
\be
\int_{A}=\int_{A_1}\cdots\int_{A_n}\,\,,\label{int}
\ee
i.e., as a product of independent integrals. Here the appropriate weighting 
provided by the target wave functional is implicit in the symbol $\int_A$. 

Under the integration specified by Eq.~(\ref{int}), the $U$ matrices on the 
r.h. side of Eq.~(\ref{um1}) can be rearranged to give the result 
\be
\hspace*{-6.5cm}\int_A
\left\{U_{x_\perp}[A]\,U^\dagger_{x_\perp+y_\perp}[A]\right\}_{ij}\left\{
U_{x_\perp+y_\perp'}[A]\,U^\dagger_{x_\perp}[A]\right\}_{kl}\label{ro}
\ee
\[
\hspace*{2cm}
=\int_{A_1}\cdots\int_{A_n}
\left\{U_{x_\perp}[A_1]\,\,U^\dagger_{x_\perp+y_\perp}[A_1]\cdots 
U_{x_\perp}[A_n]\,\,U^\dagger_{x_\perp+y_\perp}[A_n]\right\}_{ij}
\]
\[\hspace*{4cm}
\times\left\{U_{x_\perp+y_\perp'}[A_n]\,\,U^\dagger_{x_\perp}[A_n]\cdots
U_{x_\perp+y_\perp'}[A_1]\,\,U^\dagger_{x_\perp}[A_1]\right\}_{kl}\,.
\]
To see this, observe that the $A_1$ integration acts on the integrand
$\{U_{x_\perp}[A_1]\,U^\dagger_{x_\perp+y_\perp}[A_1]\}_{i'j'}$ $\{
U_{x_\perp+y_\perp'}[A_1]\,U^\dagger_{x_\perp}[A_1]\}_{k'l'}$ transforming 
it into an invariant colour tensor with the indices $i'j'k'l'$. The 
neighbouring matrices $U_{x_\perp}[A_2]$ and $U^\dagger_{x_\perp}[A_2]$ can 
now be commuted through this tensor structure in such a way that the 
expression $\{U_{x_\perp}[A_2]\,U^\dagger_{x_\perp+y_\perp}[A_2]\}_{i''j''}$ 
$\{U_{x_\perp+y_\perp'}[A_2]\,U^\dagger_{x_\perp}[A_2]\}_{k''l''}$ emerges. 
Subsequently, the $A_2$ integration transforms this expression into an 
invariant tensor with indices $i''j''k''l''$. Repeating this 
argument, one eventually arrives at the structure displayed on the r.h. side 
of Eq.~(\ref{ro}).

To evaluate Eq.~(\ref{ro}) further, observe that it represents a contraction 
of $n$ identical tensors 
\be
F_{ijkl}=\int_{A_m} \{U_{x_\perp}[A_m]\,U^\dagger_{x_\perp+y_\perp}[A_m] 
\}_{ij}\,\{U_{x_\perp+y_\perp'}[A_m]\,U^\dagger_{x_\perp}[A_m]\}_{kl}\,, 
\label{ft}
\ee
where the index $m$ refers to any one of the regions $1$ ... $n$ into which 
the target is subdivided. At this point, the smallness of the transverse 
separations $y_\perp$ and $y_\perp'$, enforced by the large size of the 
target, is used. In fact, for a target of geometrical size $\sim n/\Lambda$ 
(where $n\gg 1$), the relevant transverse distances are bounded by 
$y^2\sim y'^2\sim 1/n\Lambda^2$. 

Assuming that size and $\x$ dependence of typical field configurations $A_m$ 
are characterized by the scale $\Lambda$, it follows that the products 
$U_{\x}U^\dagger_{\x+\y}$ and $U_{\x+\y'}U^\dagger_{\x}$ are close to unit
matrices for all relevant $\y$ and $\y'$. Therefore, it is justified to 
write 
\be
U_{x_\perp}[A_m]\,U^\dagger_{x_\perp+y_\perp}[A_m]=\exp\big\{iT^af^a(\x,\y) 
[A_m]\big\}\,,\label{exp}
\ee
where $T^a$ are the conventional group generators and $f^a$ are functions 
of $\x$ and $\y$ and functionals of $A_m$. Equation~(\ref{exp}) and its $\y'$ 
analogue are expanded around $\y=\y'=0$ (which corresponds to $f^a(\x,0)=0$) 
and inserted into Eq.~(\ref{ft}). At leading non-trivial order, the result 
reads 
\be
F_{ijkl}=\delta_{ij}\delta_{kl}\left(1-\frac{1}{2}\gamma C_R(y^2+
y'^2)\right)+\gamma(y_\perp y_\perp')T^a_{ij}T^a_{kl}\,,\label{ftf}
\ee
where $C_R$ is the Casimir number of the relevant representation 
($C_R=C_{F,A}$) and the constant $\gamma$ is defined by 
\be
\int_A f^a(\x,\y)f^b(\x,\y') = \gamma \delta^{ab} (\y\y')+{\cal O}(y^2y'^2)
\,.\label{ga}
\ee
Note that the absence of terms linear in $f^a$ and the simple structure on 
the r.h. side of Eq.~(\ref{ga}) are enforced by colour covariance and 
transverse space covariance. The absence of an explicit $\x$ dependence is 
a consequence of the homogeneity that is assumed to hold over the large 
transverse size of the target. Neglecting boundary effects, the $\x$ 
integration is accounted for by multiplying the final result with a 
parameter $\Omega\sim n^2/\Lambda^2$ that characterizes the geometrical 
cross section of the target. 

Substituting the $n$ tensors $F_{ijkl}$ on the r.h. side of Eq.~(\ref{ro}) 
by the expression given in Eq.~(\ref{ftf}) and contracting the colour 
indices as appropriate for the inclusive and diffractive case respectively, 
one obtains, in the large-$N_c$ limit, 
\be
\int_A\left\{U_{x_\perp}U^\dagger_{x_\perp+y_\perp}\right\}_{ij}\left\{
U_{x_\perp+y_\perp'}U^\dagger_{x_\perp}\right\}_{ji}=d_R\left[1-\frac{1}{2}
\gamma C_R(y_\perp-y_\perp')^2\right]^n\,,\label{uu1}
\ee
\be
\int_A\left\{U_{x_\perp}U^\dagger_{x_\perp+y_\perp}\right\}_{ii}\left\{
U_{x_\perp+y_\perp'}U^\dagger_{x_\perp}\right\}_{jj}=d_R^2\left[1-\frac{1}{2}
\gamma C_R(y_\perp^2+y_\perp'^2)\right]^n\,,\label{uu2}
\ee
where $d_R$ is the dimension of the representation. 

Since $n$ is assumed to be large and the typical values of $y^2$ and $y'^2$ 
do not exceed $1/n\Lambda^2$, the formula $(1-x/n)^n\simeq\exp[-x]$ can be 
applied to the r.h. sides of Eqs.~(\ref{uu1}) and (\ref{uu2}). Furthermore, 
contributions proportional to $\{U_{\x}U^\dagger_{\x+\y}\}_{ij}\delta_{kl}
\,,$ $\delta_{ij}\{U_{\x+\y'}U^\dagger_{\x}\}_{kl}$ and $\delta_{ij}
\delta_{kl}$ have to be added to obtain the complete expression for 
$W_{x_\perp}(y_\perp)_{ij}\,W^\dagger_{x_\perp}(y_\perp')_{kl}$. The 
corresponding calculations are straightforward and the result reads 
\begin{eqnarray}
\int_{x_\perp}\int_A\mbox{tr}\left(W_{x_\perp}(y_\perp)W^\dagger_{x_\perp}
(y_\perp')\right)&=&\Omega d_R\left[1-e^{-a_Ry^2}-e^{-a_Ry'^2}+e^{-a_R(y_\perp-
y_\perp')^2}\right]\,,\label{ww1}
\\
\int_{x_\perp}\int_A\mbox{tr}W_{x_\perp}(y_\perp)\mbox{tr}
W^\dagger_{x_\perp}(y_\perp')&=&\Omega d_R^2\left[1-e^{-a_Ry^2}\right]\,
\left[1-e^{-a_Ry'^2}\right]\,,\label{ww2}
\end{eqnarray}
where $a_R=n\gamma C_R/2$ plays the role of a saturation scale. 

The above calculation, performed at large $N_c$ and for the case of a large 
target subdivided into many uncorrelated regions, has no immediate 
application to realistic experiments. However, it provides a set of 
non-perturbative inclusive and diffractive parton distributions which are 
highly constrained with respect to each other. For the purpose of a 
phenomenological analysis, it is convenient to consider $\Omega$ and 
$a\equiv n\gamma N_c/4$ as new fundamental parameters, giving rise to the 
following formulae for the basic hadronic quantities required in 
Sect.~\ref{sect:trf} and Appendix~\ref{sect:isf}, 
\begin{eqnarray}
\int_{x_\perp}\int_A\mbox{tr}\left(W^{\cal F}_{x_\perp}(y_\perp)
W^{{\cal F}\dagger}_{x_\perp}(y_\perp')\right)&=&\Omega N_c\left[1-e^{-ay^2}-
e^{-ay'^2}+e^{-a(y_\perp-y_\perp')^2}\right]\label{ww0}\,,
\\
\frac{1}{N_c}\int_{x_\perp}\int_A\mbox{tr}W^{\cal F}_{x_\perp}(y_\perp)
\mbox{tr}W^{{\cal F}\dagger}_{x_\perp}(y_\perp')&=&\Omega N_c\left[1-
e^{-ay^2}\right]\,\left[1-e^{-ay'^2}\right]\,,\label{wwf}
\\
\frac{1}{N_c^2}\int_{x_\perp}\int_A\mbox{tr}W^{\cal A}_{x_\perp}(y_\perp)
\mbox{tr}W^{{\cal A}\dagger}_{x_\perp}(y_\perp')&=&\Omega N_c^2\left[1-
e^{-2ay^2}\right]\,\left[1-e^{-2ay'^2}\right]\,.\label{wwa}
\end{eqnarray}
Here the indices ${\cal F}$ and ${\cal A}$ stand for the fundamental and 
adjoint representation. A similar, Glauber type $y^2$ dependence has been 
recently used in the analyses of~\cite{gbw,gbw1}. Note that according to 
Eqs.~(\ref{ww0})--(\ref{wwa}) the diffractive structure function is not 
suppressed by a colour factor relative to the inclusive structure function, 
as originally suggested in~\cite{bh1}. 

Using Eq.~(\ref{wwf}) in the generic semiclassical formula Eq.~(\ref{fxsp}) 
for the diffractive quark distribution, the explicit result 
\be
{dq(\beta,\xi)\over d\xi}={a\Omega N_c(1-\beta)\over 2\pi^3\xi^2} 
h_q(\beta)\;, 
\ee
is obtained. Here $h_q(\beta)$ is an integral over two Feynman-type 
parameters,
\be
h_q(\beta) = 4 \int_0^\infty dxdx'\frac{\left(\frac{\displaystyle 
\sb+x}{\displaystyle (1-\beta+(\sb+x)^2)^2}\right)\,
\left(\frac{\displaystyle \sb+x'}{\displaystyle (1-\beta+( 
\sb+x')^2)^2}\right)}
{(x+x')\sb + (1-\beta)\left({\displaystyle x\over \displaystyle 
\sb+x}+{\displaystyle x'\over \displaystyle \sb+x'} 
\right)}\;, 
\ee
which can be evaluated analytically at $\beta=0$ and $\beta=1$ yielding 
$h_q(0)=1/2$ and $h_q(1)=3\pi^2/8-2$. 

\begin{figure}[t]
\begin{center}
\parbox[b]{7.5cm}{\psfig{file=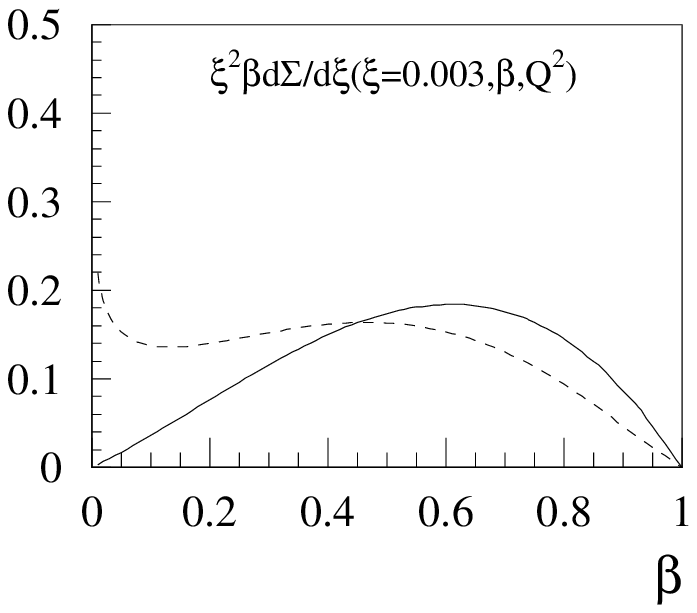,width=7.5cm}}
\parbox[b]{7.5cm}{\psfig{file=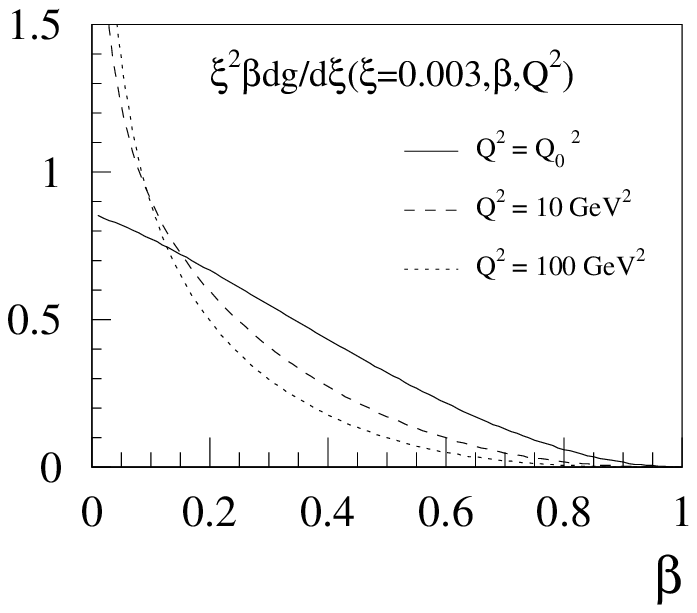,width=7.5cm}}\\
\end{center}
\refstepcounter{figure}
\label{fig:difpdf}
{\bf Figure \ref{fig:difpdf}:}
Diffractive quark and gluon distributions in the large hadron model at the 
initial scale $Q_0^2$ and after $Q^2$ evolution.
\end{figure}

Analogously, Eqs.~(\ref{wwa}) and (\ref{fxg}) give rise to an explicit 
formula for the diffractive gluon distribution. It reads 
\be
{dg(\beta,\xi)\over d\xi}={a\Omega N_c^2(1-\beta)^2\over 2\pi^3 \xi^2 
\beta} h_g(\beta)\;, 
\ee
where $h_g(\beta)$ is given by the two-dimensional integral
\be
h_g(\beta) = 2\int_0^\infty dxdx'\frac{
\left(\frac{\displaystyle 1-\beta+3(1+x)^2\beta}{\displaystyle (1+x)^2(1- 
\beta+(1+x)^2\beta)^2}\right)
\left(\frac{\displaystyle 1-\beta+3(1+x')^2\beta}{\displaystyle (1+x')^2
(1-\beta+(1+x')^2\beta)^2}\right)}
{(x+x')\beta + (1-\beta)\left({\displaystyle x\over\displaystyle 1+x}+{
\displaystyle x'\over\displaystyle 1+x'}\right)}\;.
\ee
This integral is easily evaluated for $\beta=0$ and $\beta=1$ yielding 
$h_g(0)=4\ln 2$ and $h_g(1)=45\pi^2/32-17/2$. For general $\beta$, 
$h_q(\beta)$ and $h_g(\beta)$ can be evaluated numerically. The results can 
be inferred from the solid curves in Fig.~\ref{fig:difpdf}, where the 
distribution $d\Sigma/d\xi=6dq/d\xi$ is displayed to account for the 3 
generations of light quarks and antiquarks. The total normalization, the 
value of $\xi$, and the $Q^2$ evolution given in Fig.~\ref{fig:difpdf} are 
not relevant for the present section and will be discussed in the context 
of the phenomenological analysis of Sect.~\ref{sect:edsf}. 

The diffractive distributions displayed in Fig.~\ref{fig:difpdf} are 
multiplied by $\beta$ and thus reflect the distribution of momentum carried 
by the partons. The quark distribution is peaked around $\beta \simeq 
0.65 $, thus being harder than the distribution $\beta (1-\beta)$ suggested 
in~\cite{dl1}. It vanishes like $\beta$ for $\beta\to 0$ and 
like $(1-\beta)$ at large $\beta$; the gluon distribution $\beta 
d g/d \xi$, on the other hand, approaches a constant for 
$\beta \to 0$ and falls off like $(1-\beta)^2$ at large $\beta$. 
This asymptotic behaviour in the small- and large-$\beta$ region is in 
agreement with the results obtained in the perturbative approach of 
\cite{hks} at $t=0$. In spite of the $(1-\beta)^2$ behaviour, gluons 
remain important even at large $\beta$, simply due to the large total 
normalization of this distribution.

Very recently, a closely related discussion of diffractive and inclusive 
structure functions has been given in \cite{kml}. The authors focus on the 
process where the target hadron remains intact, discussed in the first part 
of Sect.~\ref{sect:av}, and its relation to inclusive DIS. Technically, the 
results obtained are very similar to those of the large hadron model as 
described above \cite{bgh}. It is also emphasized in \cite{kml} that the 
results can be generalized to the case of conventional hadrons if the 
assumption of a Gaussian distribution of colour sources of \cite{mv} is 
correct.

\section{Stochastic vacuum}\label{sect:sv}
A further fundamentally non-perturbative approach to high-energy hadronic 
processes, which has recently been applied to diffractive structure 
functions by Ramirez~\cite{ram}, is based on the model of the stochastic 
vacuum. The model was originally developed by Dosch and Simonov in Euclidean 
field theory~\cite{ds}. A detailed description of the stochastic vacuum 
approach to high-energy scattering, introduced originally in~\cite{dk} and 
closely related to the eikonal approach of \cite{nac}, can be found 
in~\cite{dfk} (see also \cite{nare} for a comprehensive review). The method 
was first applied to diffractive DIS in~\cite{dgkp}, where vector 
meson production processes were considered. Here, only a brief description 
of the main underlying ideas will be given. 

The fundamental assumption underlying the model of the stochastic vacuum 
of~\cite{ds} is that of a convergent cumulant expansion for the vacuum 
expectation value of path ordered products of field operators. To calculate 
the average $\langle...\rangle$ of the path ordered exponential 
\be
P\exp\left(\int_0^t\hat{O}(s)ds\right)=\sum_{n=0}^\infty\,\,\int_0^tds_1 
\int_0^{s_1}ds_2...\int_0^{s_{n-1}}ds_n\,\hat{O}(s_1)...\hat{O}(s_n)\,,
\ee
the so-called path ordered cumulants $((...))$, defined by 
\vspace*{-.5cm}

\hspace*{-1cm}
\parbox{16cm}{
\bea
\langle 1\rangle &=& ((1))\nonumber\\
\langle 1,2 \rangle &=& ((1,2))+((1))((2))\nonumber\\
\langle 1,2,3 \rangle &=& ((1,2,3))+((1))((2,3))+((1,2))((3))+((1))((2))
((3))\nonumber\\
\langle 1,2,3,4 \rangle &=& ((1,2,3,4))+\hspace{2.5cm}\cdots\hspace{2.5cm}
+((1))((2))((3))((4))\,,\nonumber
\eea}
\hspace{-4cm}\hfill\parbox{2cm}{\bea \eea}
\vspace*{-.3cm}

\noindent
are introduced. An expansion in the above cumulants can be applied to 
the Wegner-Wilson loop in a non-Abelian gauge theory. The supposition that 
the cumulants are decreasing sufficiently fast with increasing distance 
between the operators leads to the area law in the purely gluonic case. 
Neglecting all cumulants higher than quadratic in the fields amounts to the 
assumption of a Gaussian process, where all higher correlators can be 
obtained from the two-point Green's function. All of the above is naturally 
formulated in Euclidean space. The two-point correlator or, more precisely, 
its analytic continuation to Minkowski space, is the fundamental object in 
applications to high-energy scattering. 

In the treatment of diffractive DIS, the stochastic vacuum approach 
describes the hadron as well as the virtual photon in terms of fast partons 
moving in opposite directions. The colour field facilitating the 
interaction is soft in the centre-of-mass frame of the $\gamma^*p$ 
collision (cf. the general discussion of soft hadronic processes 
in~\cite{nac}). In this situation, the partons from both sides interact 
with the field in an eikonalized way, and the actual model of the 
stochastic vacuum is used to evaluate the correlation function of the 
resulting oppositely directed light-like Wegner-Wilson lines. 

\begin{figure}[t]
\begin{center}
\parbox[b]{11cm}{\psfig{file=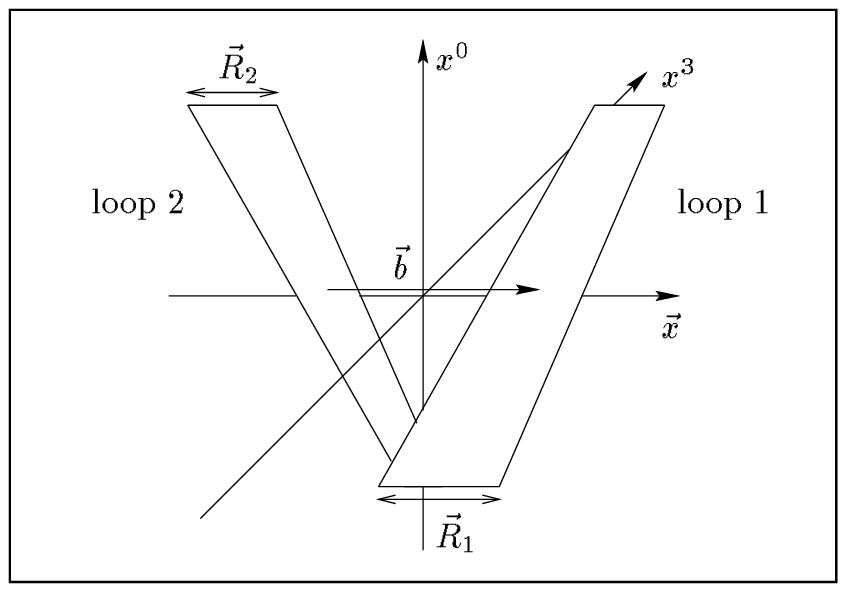,width=11cm}}
\end{center}
\refstepcounter{figure}
\label{fig:wls}
{\bf Figure \ref{fig:wls}:} Wegner-Wilson loops formed by the paths of 
quarks and antiquarks inside two dipoles. The impact parameter $\vec{b}$ 
connects the centres of the two loops, while $\vec{R}_1$ and $\vec{R}_2$ 
point from the quark to the antiquark line of each dipole. All three are 
vectors in the transverse plane of the collision (figure from~\cite{dr}). 
\end{figure}

In the simplest case, where the photon fluctuates into a $q\bar{q}$ pair 
and the hadron is modelled as a quark-diquark system, the interaction 
amplitude of two Wegner-Wilson loops has to be calculated (see 
Fig.~\ref{fig:wls}). Introducing the notation 
\be
W=P\exp\left[-ig\int_{\partial S}A_\mu(z)\,dz^\mu\right]\label{wdef1}
\ee
for the path ordered integral around the loop $\partial S$, which is 
extended to light-like infinity in both directions, this amplitude can be 
written as 
\be
J(\vec{b},\vec{R}_1,\vec{R}_2)=\left\langle\frac{1}{N_c}\mbox{tr}(W_1(
\vec{b}/2,\vec{R}_1)-1)\,\frac{1}{N_c}\mbox{tr}(W_2(-\vec{b}/2,\vec{R}_2) 
-1)\right\rangle_A\,.\label{jdef}
\ee
The notation of this section follows the original papers reviewed (see, 
e.g.,~\cite{dgkp,dfk}), which means that the function defined in 
Eq.~(\ref{wdef1}) differs by a unit matrix from the closely related 
function $W$ used previously in the context of the semiclassical approach. 
The brackets $\langle...\rangle_A$ denote the vacuum expectation value or, 
in the functional language, the integration over all colour field 
configurations. 

The line integrals in Eq.~(\ref{jdef}) are transformed into surface 
integrals with the help of the non-Abelian Stokes theorem 
\be
P\exp\left[\int_{\partial S}-igA_\mu(z)\,dz_\mu\right]=P_S\exp\left[
\int_S-igF_{\mu\nu}(z,\omega)d\Sigma^{\mu\nu}(z)\right]\,.\label{sto}
\ee
Here $F_{\mu\nu}(z,\omega)$ are the field strength tensors 
$F_{\mu\nu}(z)$, parallel-transported to a common reference point $\omega$. 
The operator $P_S$ denotes an appropriate surface ordering of these matrix 
valued tensors (see~\cite{dfk} and refs. therein for further details). The 
surface $S$ that is chosen for each of the two Wegner-Wilson loops is the 
upper side of the pyramid with the loop as base and the origin of the 
co-ordinate system as apex (see Fig.~\ref{fig:wls}). The reference point 
$\omega$ is also chosen to be at the origin. 

The interaction amplitude for two Wegner-Wilson loops, in 
its explicit form of Eq.~(\ref{jdef}) or expressed through surface 
integrals according to Eq.~(\ref{sto}), can not be evaluated by direct 
application of the cumulant expansion methods discussed above. The problem 
is that those methods are adopted for one path ordered integral, in contrast 
to the two path (or surface) ordered integrals required for $J(\vec{b}, 
\vec{R}_1,\vec{R}_2)$. In Ref.~\cite{dk} it was suggested to apply the 
Gaussian factorization hypothesis directly to products of the fields 
$F_{\mu\nu}^a(z,\omega)$, 
\be
\langle F(1)...F(2n)\rangle = \sum_{\mbox{\footnotesize all pairings}} 
\langle F(i_1)F(j_1)\rangle ...\langle F(i_n)F(j_n)\rangle\,,
\ee
where the single argument of $F$ stands for Lorentz and colour indices and 
space-time co-ordinates. Note, however, that this factorization assumption 
is not equivalent to the original approach of \cite{ds} (cf. the discussion 
in~\cite{krae,nare}). 

Equation~(\ref{jdef}) can now be evaluated if the fundamental correlator 
\be
\left\langle g^2 F^c_{\mu\nu}(x,\omega)F^d_{\sigma\rho}(y,\omega)
\right\rangle_A
\ee
is known. Under the further assumption that this correlator depends neither 
on the path used for transporting the field to $\omega$ nor on the 
position of the reference point $\omega$ itself, the most general form reads 
\bea
\left\langle g^2 F^c_{\mu\nu}(x,\omega)F^d_{\rho\sigma}(y,\omega)
\right\rangle_A&\!\!\!\!=\!\!\!\!&\frac{\delta^{cd}\langle g^2FF\rangle}{12
(N_c^2-1)}\bigg\{\kappa(g_{\mu\rho}g_{\nu\sigma}-g_{\mu\sigma}g_{\nu\rho})
D(z^2/a^2)\label{corr}\\
\nonumber\\
&&+(1\!-\!\kappa)\frac{1}{2}\left[\partial_\mu(z_\rho g_{\nu\sigma}\!-\!
z_\sigma g_{\nu\rho})+\partial_\nu(z_\sigma g_{\mu\rho}\!-\!z_\rho 
g_{\mu\sigma})\right]D_1(z^2/a^2)\bigg\}\,,\nonumber
\eea
where $D$ and $D_1$ are, a priori, two independent functions, and $\langle 
g^2FF\rangle$ is the gluon condensate. Note that the model of the stochastic 
vacuum is formulated in Euclidean field theory and the analytic continuation 
to Minkowski space is non-trivial. The intricacies of this process and, in 
particular, the constraints it imposes on the shape of the functions $D$ 
and $D_1$ will not be discussed here. For completeness, the correlation 
functions used, e.g., in~\cite{dgkp} are given: 
\be
D_1(z^2/a^2)=D(z^2/a^2)=\frac{27\pi^4}{4}i\int\frac{d^4k}{(2\pi)^4}\,
\frac{k^2}{(k^2-(3\pi/8)^2)^4}\,e^{-ik\cdot z/a}\,.
\ee
Detailed information about the parameters entering Eq.~(\ref{corr}) and the 
functional form of $D$ and $D_1$ is available from low energy hadronic 
physics as well as from lattice calculations (see, e.g., the recent results 
of \cite{meg}). Given these low-energy parameters, a large number of soft 
hadronic high-energy scattering processes is successfully described by the 
model of the stochastic vacuum. For the present review, it is sufficient to 
state that, within the present model, the behaviour of the correlator in 
Eq.~(\ref{corr}) is quantitatively known. 

Thus, the method for evaluating the fundamental dipole amplitude of 
Eq.~(\ref{jdef}) is now established. The amplitude vanishes for small 
dipoles, as expected from colour transparency, and grows linearly for large 
dipoles, as suggested by the geometric picture of string-string scattering. 

The calculation of diffractive processes is straightforward as soon as a 
specific quark-diquark wave function of the hadronic target is chosen. The 
procedure of folding the dipole amplitude with the virtual photon and the 
hadron wave functions introduces a certain model dependence on the hadronic 
side. 

An essential difference between the present framework and the semiclassical 
approach discussed above is the frame of reference where the model for the 
hadronic target is formulated. From the perspective of the virtual photon, 
the situations are similar: the $q\bar{q}$ fluctuation scatters off a soft 
colour field. However, in the semiclassical approach (cf., e.g., 
Sect.~\ref{sect:lh}), this field is modelled on the basis of ideas about 
hadron colour fields in the rest frame of the hadron. By contrast, 
the stochastic vacuum model uses an intermediate frame, where both 
projectile and target are fast (e.g., the centre-of-mass frame), as the 
natural frame for the colour field. The target hadron is described in terms 
of quarks interacting with this field in an eikonalized way, and the field, 
as seen by the projectile, is a result of the distribution of these 
partons and of the gluon dynamics in the stochastic vacuum model. 

As an interesting consequence of the above discussion, the conditions 
required for the eikonal approximation to work are different in the two 
approaches. While, in the semiclassical model, the partons from the photon 
wave function have to be fast in the target rest frame, the stochastic 
vacuum approach requires them to be fast in, say, the centre-of-mass frame. 
The latter is a far more stringent condition, which can be numerically 
important in phenomenological applications~\cite{rue,ram}. 

The reader is referred to Sections~\ref{sect:edsf} and \ref{sect:mp} for a 
brief discussion of recent applications of the presented model to 
diffractive electroproduction data.

\newpage

\mychapter{Recent Experimental Results}\label{sect:exp}
In this chapter, a brief discussion of experimental results in diffractive 
electroproduction, focussing in particular on the diffractive structure 
function, on specific features of the diffractively produced final 
state, and on exclusive meson production is given. The discussion is aimed 
at the demonstration of interesting features of different theoretical models 
and calculational approaches in the light of the available data (see 
also~\cite{exr} for recent reviews). From an experimentalist's perspective, 
what follows can only be considered a very brief and incomplete overview.

\section{Diffractive structure function}\label{sect:edsf}
As already explained in Chapter~\ref{sect:basic}, the observation that 
diffractive DIS is of leading twist and the ensuing measurement of 
diffractive structure functions lie at the heart of increased recent 
interest in the field. While the basic experimental facts were stated at 
the beginning of this review, the present section supplies some additional 
details concerning $F_2^D$ and compares observations with theoretical 
ideas. 

To begin with, a brief discussion of the most recent and most precise data 
on $F_2^D$, produced by the H1 and ZEUS collaborations~\cite{nh1,nzeus}, 
is given. As mentioned previously, the most precise analyses have to be based 
on measurements where the scattered proton is not tagged. In Ref.~\cite{nh1}, 
this problem is handled by starting from a cross section
\be
\frac{d^{\,5}\sigma_{ep\to eXY}}{dx\,dQ^2\,d\xi\,dt\,dM_Y^2}\,,
\ee
where, by definition, the two clusters $X$ and $Y$ are separated by 
the largest rapidity gap in the hadronic final state of the event. 
This definition allows for the possibility that the proton breaks up into a 
final state $Y$ with mass square $P'^2=M_Y^2$ (cf.~Fig.~\ref{fig:dep}). A 
structure function $F_2^{D(3)}$ is then defined, as in Sect.~\ref{sect:dsf}, 
after integration over $t$ and $M_Y^2$. The event selection of~\cite{nh1} 
is based on the requirement that $X$ is fully contained in the main 
detector while $Y$ passes unobserved into the beam pipe. This implies that 
$M_Y\lsim 1.6\,\mbox{GeV}$ and $|t|\lsim 1\,\mbox{GeV}^2$, thus 
approximately specifying the relevant integration region for $M_Y^2$ and 
$t$. A rapidity cut ensures that $X$ is well separated from $Y$. 

The most recent ZEUS analysis~\cite{nzeus} is based on the so-called 
$M_X$ method, introduced in~\cite{subt}. Similarly to what was discussed 
above, it is assumed that $X$ is contained in the detector while $Y$ 
escapes down the beam pipe. Thus, $M_X$ is simply defined to be the full 
hadronic mass contained in the main detector components. In contrast to 
the H1 analysis, the diffractive cross section, integrated over all $t$ 
and $M_Y$, is defined by subtracting from this full cross section the 
non-diffractive contribution. The subtraction is based on the ansatz 
\be
\frac{d{\cal N}}{d\,\ln M_X^2}=D+c\,\exp(b\,\ln M_X^2)\,\label{zfit}
\ee
for the event distribution in the region of not too large masses. Here $D$ 
is the diffractive contribution and the second term is the non-diffractive 
contribution. The exponential $\ln M_X^2$ dependence is expected as a result 
of the Poisson distribution of particles emitted between current and target 
jet regions in non-diffractive DIS. This leads to an exponential 
distribution of rapidity gaps between the detector limit and the most 
forward detected particles and therefore to the above $M_X$ dependence. 
The diffractive contribution is not taken from the fit result for $D$ but is 
determined by subtracting from the observed number of events the 
non-diffractive contribution that corresponds to the fit values of $b$ 
and $c$. 

For more details concerning the data analysis and the experimental results 
themselves the reader is referred to the original papers~\cite{nh1,nzeus} 
and to the figures in the remainder of this section, where theoretical 
models are compared with measured values of $F_2^D$. It can be seen from 
the direct comparison of H1 and ZEUS data found in~\cite{nzeus} that the 
two methods give similar results, although the H1 values for $F_2^{D(3)}$ 
have a tendency towards a faster $Q^2$ rise for any given $\beta$. 

The $M_X$ method of ZEUS has the advantage of subtracting events which
happen to have a rapidity gap in spite of non-singlet colour exchange. It 
also allows for diffractive events with relatively forward particles, which 
are excluded by the rapidity cut of H1~\cite{er}. However, the subtraction 
of the non-diffractive background also introduces new uncertainties, in 
particular the dependence on the region in $M_X$ where the fit according to 
Eq.~(\ref{zfit}) is performed. 

Note also that a measurement of $F_2^{D(4)}$, based on the use of the 
leading proton spectrometer of ZEUS, was reported in~\cite{f2d4}. The 
results for $F_2^{D(3)}$, obtained by explicit $t$ integration, are probably 
the cleanest ones from both the theoretical and experimental perspectives. 
They agree with the latest $F_2^{D(3)}$ measurements discussed above, but 
at the moment they are, unfortunately, less precise because of limited 
statistics. The observed $t$ dependence of $F_2^{D(4)}$ can be parametrized 
as $e^{bt}$, with $b=7.2\pm 1.1(\mbox{stat.})\mst{+0.7}{-0.9}(\mbox{syst.})
\,\mbox{GeV}^{-2}$.

\subsubsection*{Leading twist analysis}
As mentioned previously, the leading twist behaviour of small-$x$ 
diffraction is one of its most striking features. For this reason, and in 
keeping with the general perspective of this review, it is convenient to 
start the discussion with calculations, such as the semiclassical approach, 
that focus on the leading twist nature of the process. 

Detailed numerical predictions exhibiting this behaviour were first made in 
\cite{wf} based on simple aligned jet model calculations with soft two 
gluon exchange. Further theoretical and numerical developments by these and 
other authors~\cite{nz,nztge,gnz}, focussing, in particular, on 
high-mass diffraction and higher twist effects, are discussed below. 

Data analyses based on the idea of parton distributions of the pomeron 
\cite{is} and their $Q^2$ evolution were performed early on by many authors 
\cite{pa,gs}. The most recent H1 publication~\cite{nh1} includes such a 
partonic analysis as well. 

\begin{figure}[t]
\begin{center}
\vspace*{.2cm}
\parbox[b]{14cm}{\psfig{width=14cm,file=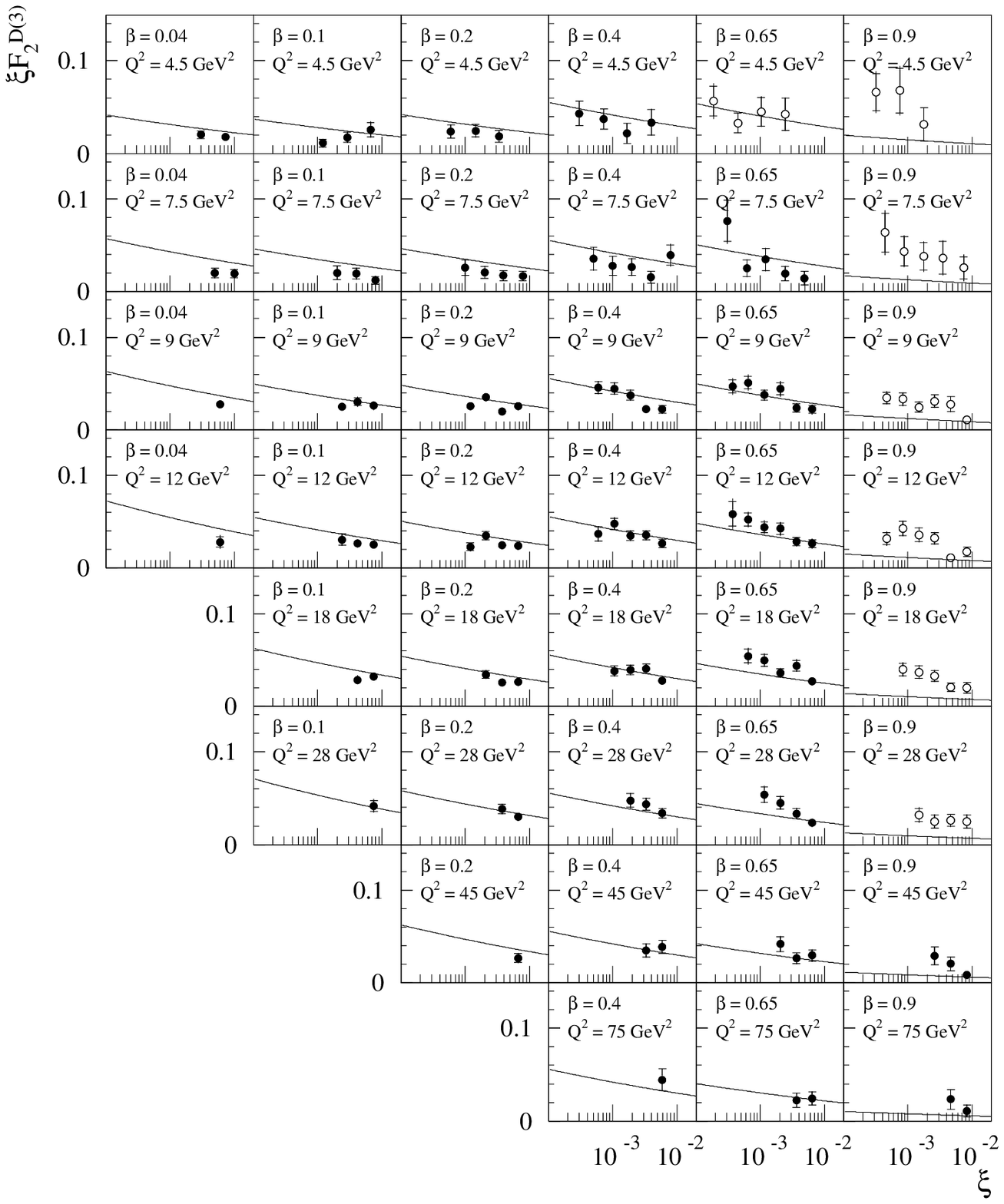}}
\end{center}\vspace*{-.3cm}
\refstepcounter{figure}
\label{fig:f2dh1}
{\bf Figure \ref{fig:f2dh1}:} 
The structure function $F_2^{D(3)}(\xi,\beta,Q^2)$ computed in the 
semiclassical approach with H1 data from~\protect\cite{nh1}. The open data 
points correspond to $M^2 \leq 4$~GeV$^2$ and are not included in the fit. 
\end{figure}

\begin{figure}[t]
\begin{center}
\vspace*{.2cm}
\parbox[b]{9cm}{\psfig{width=9cm,file=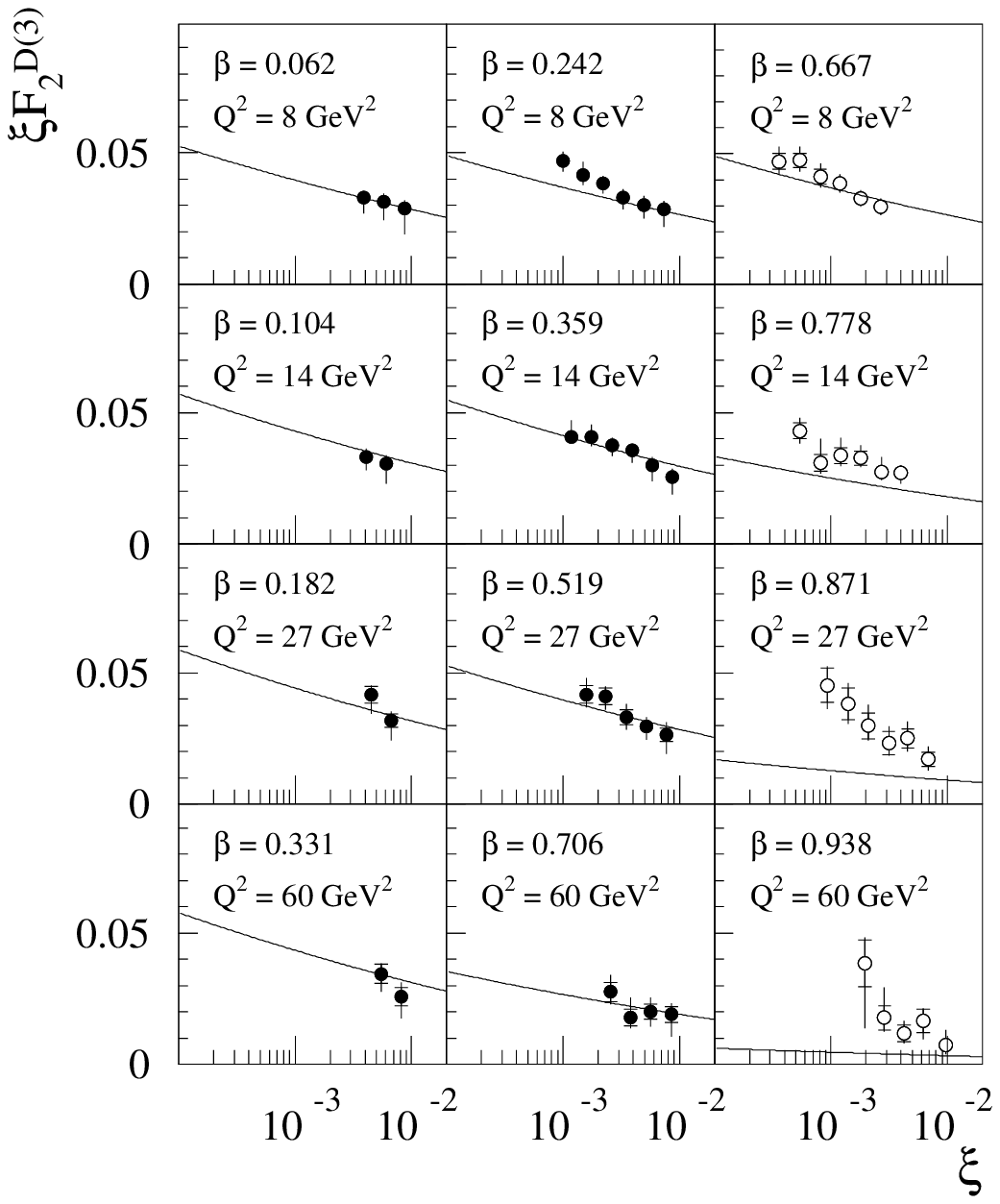}}
\end{center}\vspace*{-.3cm}
\refstepcounter{figure}
\label{fig:f2dzeus}
{\bf Figure \ref{fig:f2dzeus}:} 
The structure function $F_2^{D(3)}(\xi,\beta,Q^2)$ computed in the 
semiclassical approach with ZEUS data from~\protect\cite{nzeus}. The open 
data points correspond to $M^2 \leq 4$~GeV$^2$ and are not included in the 
fit.
\end{figure}

\begin{figure}[t]
\begin{center}
\parbox[c]{5.5cm}{\psfig{file=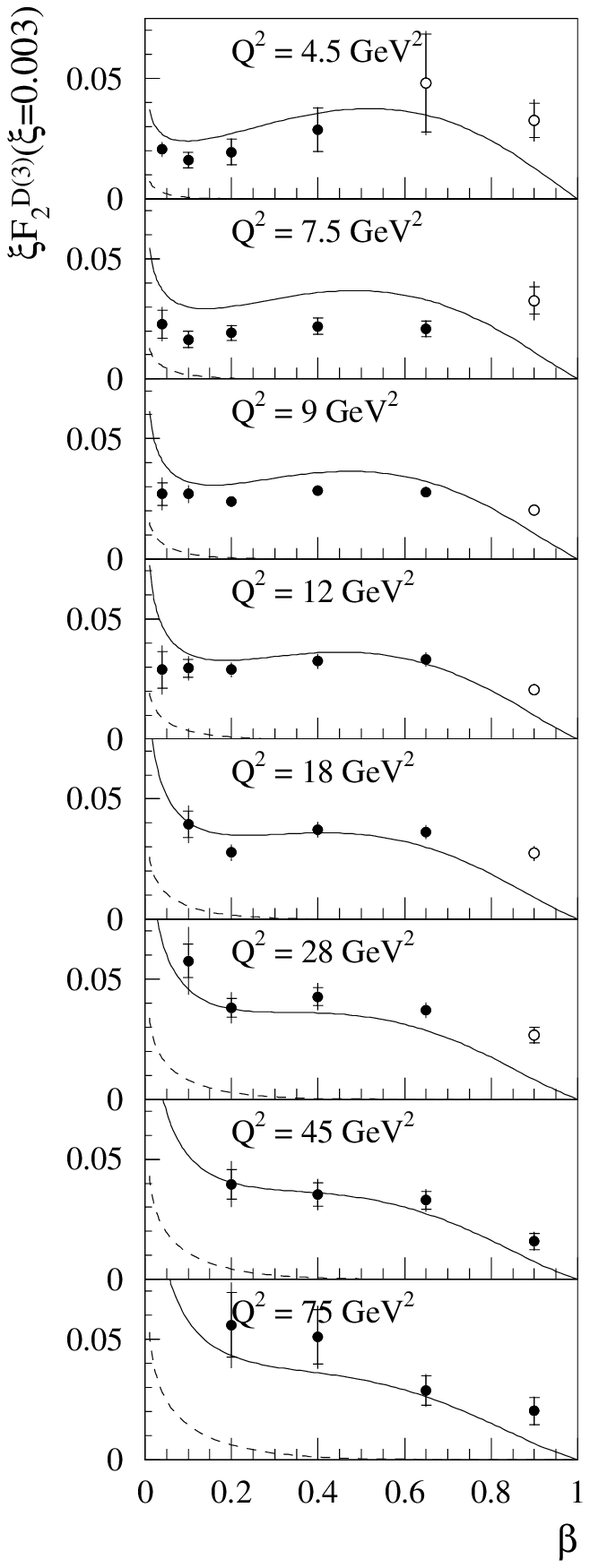,width=5.5cm} }\parbox[c]{1.8cm}
{\hspace{1.8cm}}\parbox[c]{5.5cm}{\psfig{file=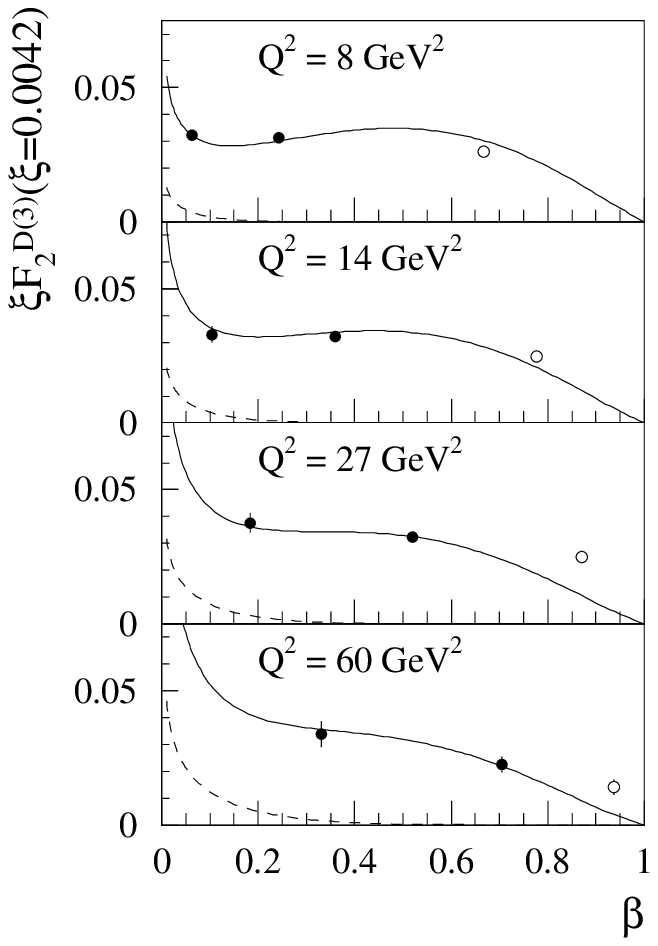,width=5.5cm} }\\
\vspace{-.25cm}
\end{center}
\refstepcounter{figure}
\label{fig:difq2}
{\bf Figure \ref{fig:difq2}:}
Dependence of the diffractive structure function $F_2^{D(3)}$ on $\beta$
and $Q^2$, compared to data from H1 (left) and ZEUS 
(right)~\protect\cite{nh1,nzeus}. Open data points 
correspond to $M^2\leq 4$~GeV$^2$. The charm content of the 
structure function is indicated as a dashed line.
\end{figure}

Here, a detailed description of the approach of~\cite{bgh} will be given, 
which is based on diffractive parton distributions in the sense of 
Berera and Soper~\cite{bs}, calculated in the semiclassical approach with 
a specific model for the averaging over all target colour fields 
(cf.~Sect.~\ref{sect:lh}). As can be seen from Figs.~\ref{fig:f2dh1} to 
\ref{fig:difq2}, a satisfactory description of $F_2^{D(3)}$ is achieved 
in this framework. 

In the presented approach, the diffractive distributions of 
Sect.~\ref{sect:lh} and the inclusive distributions of 
Appendix~\ref{sect:isf} are used as non-perturbative input at some small 
scale $Q_0^2$. They are evolved to higher $Q^2$ using the leading-order 
DGLAP equations~\cite{dglap}. The non-perturbative parameters of the 
model as well as the scale $Q_0^2$ are then determined from a combined 
analysis of experimental data on inclusive and diffractive structure 
functions. 

At first sight, the semiclassical description of parton distribution 
functions always predicts an energy dependence corresponding to a classical 
bremsstrahlung spectrum: $q(x),g(x)\sim 1/x$. However, one expects, in a 
more complete treatment, a non-trivial energy dependence to be induced 
since the field averaging procedure encompasses more and more modes of the 
proton field with increasing energy of the probe (cf. the discussion at the 
end of Sect.~\ref{sect:av}). This energy dependence is parametrized in the 
form of a soft, logarithmic growth of the normalization of diffractive and 
inclusive parton distributions with the collision energy $\sim 1/x$, 
consistent with the unitarity bound. As a result, the additional parameter 
$L$ is introduced into the formulae of both Sect.~\ref{sect:lh} and 
Appendix~\ref{sect:isf}, where the overall normalization factor $\Omega$ 
has to be replaced according to 
\begin{equation}
\Omega \to \Omega \left(L - \ln x \right)^2.
\end{equation}

The following expressions for the diffractive parton distributions are 
obtained, 
\begin{eqnarray}
\frac{dq\left(\beta,\xi,Q_0^2\right)}{d\xi} & = & \frac{a\Omega N_c 
(1-\beta) 
\left(L - \ln \xi\right)^2}{2\pi^3\xi^2} h_q(\beta)\; , \label{qdinp} \\
\frac{dg\left(\beta,
\xi,Q_0^2\right)}{d\xi} & = & \frac{a\Omega N_c^2 (1-\beta)^2 
\left(L - \ln \xi\right)^2}{2\pi^3 \beta \xi^2} h_g(\beta)\; \label{gdinp},
\end{eqnarray}
where the functions $h_{q,g}(\beta)$ are defined in Sect.~\ref{sect:lh}. 
Corresponding expressions for the inclusive input distributions of quarks 
and gluons are given in Appendix~\ref{sect:isf}. 

Thus, the input distributions depend on $a$, $\Omega$, $L$, and on the scale 
$Q_0^2$ at which these distributions are used as a boundary condition for 
the leading-order DGLAP evolution\footnote{\samepage 
Note that the two variables $a$ 
and $\Omega$ can not be combined into one since the inclusive quark 
distribution depends on them in a more complicated way.}. At this  order, 
the measured structure function $F_2$ coincides with the transverse 
structure function. In defining structure functions and parton 
distributions, all three light quark flavours are assumed to yield the same 
contribution, such that the singlet quark distribution is 
simply six times the quark distribution defined above. Valence quarks are 
absent in the semiclassical approach. Charm quarks are treated entirely as 
massive quarks in the fixed flavour number scheme~\cite{ffns}. Thus, 
$n_f=3$ in the DGLAP splitting functions, and only gluon and singlet quark 
distributions are evolved. 

The structure functions $F_2$ and $F_2^{D(3)}$ are then given 
by the singlet quark distribution and a massive 
charm quark contribution due to boson-gluon fusion. Explicit formulae 
can, for example, be found in~\cite{gs}. For the numerical studies, the 
values $\Lambda_{{\rm LO},n_f=3}= 144$~MeV ($\alpha_s(M_Z)=0.118$), 
$m_c=1.5$~GeV, $m_b=4.5$~GeV are used, and the massive charm quark 
contribution is evaluated for a renormalization and factorization scale 
$\mu_c = 2 m_c$.  

The resulting structure functions can be compared with HERA data on the 
inclusive structure function $F_2(x,Q^2)$~\cite{incl} and on the 
diffractive structure function $F_2^{D(3)}(\xi,\beta,Q^2)$~\cite{nh1,nzeus}. 
These data sets from the H1 and ZEUS experiments are used to determine 
the unknown parameters of the model. The following selection criteria are 
applied to the data: $x\leq 0.01$ and $\xi \leq 0.01$ are needed to justify 
the semiclassical description of the proton; with $Q_0^2$ being a fit 
parameter, a sufficiently large minimum $Q^2=2$~GeV$^2$ is required to 
avoid that the data selection is influenced by the current value of 
$Q_0^2$; finally $M^2 > 4$~GeV$^2$ is required in the diffractive case to 
justify the leading-twist analysis. 

The optimum set of model parameters is determined from a minimization of 
the total $\chi^2$ (based on statistical errors only) of the selected 
data. As a result, 
\vspace*{-.4cm}

\parbox{13cm}{
\begin{eqnarray}
Q_0^2 & =& 1.23 \; {\rm GeV}^2\; , \nonumber\\
L &=& 8.16 \; , \nonumber\\
\Omega &=& (712\;   {\rm MeV})^{-2} \; , \nonumber\\
a & = & \left( 74.5\;  {\rm MeV} \right)^2 \; .\nonumber
\end{eqnarray}}
\hfill\parbox{2cm}{\bea\label{fitpar}\eea}
\vspace*{-.4cm}

\noindent
All parameters are given with a precision which allows reproduction of the 
plots, but which is inappropriate with respect to the crudeness of the 
model. The distributions obtained with these fitted parameters yield a 
good qualitative description of all data on inclusive and diffractive DIS 
at small $x$, as illustrated in Figs.~\ref{fig:f2i},~\ref{fig:f2dh1} 
and~\ref{fig:f2dzeus}.  The starting scale $Q_0^2$ is in the region where 
one would expect the transition between perturbative and non-perturbative 
dynamics to take place; the two other dimensionful parameters $\Omega L^2$ 
and $a$ are both of the order of typical hadronic scales. 

The approach fails to reproduce the data on $F_2^{D(3)}$ for low $M^2$. 
This might indicate the importance of higher twist contributions in this 
region (see, e.g.,~\cite{bekw}). It is interesting to note that a breakdown 
of the leading twist description is also observed for inclusive structure 
functions~\cite{mrst2}, where it occurs for similar invariant hadronic 
masses, namely $W^2\lsim 4$~GeV$^2$. 

The perturbative evolution of inclusive and diffractive structure functions 
is driven by the gluon distribution, which is considerably larger 
than the singlet quark distribution in both cases. With the parameters 
obtained above, it turns out that the inclusive gluon distribution is about 
twice as large as the singlet quark distribution. By contrast, the relative 
magnitude and the $\beta$ dependence of the diffractive distributions are 
completely independent of the model parameters. Moreover, their absolute 
normalization is, up to the slowly varying factor $1/\alpha_s(Q_0^2)$, 
closely tied to the normalization of the inclusive gluon distribution. 

In spite of the $(1-\beta)^2$ behaviour, gluons remain important even at 
large $\beta$, simply due to the large total normalization of this 
distribution. As a result, the quark distribution does not change with 
increasing $Q^2$ for $\beta\simeq 0.5$ and is only slowly decreasing for 
larger values of $\beta$. 

The dependence of the diffractive structure function on $\beta$ and $Q^2$ is 
illustrated in Fig.~\ref{fig:difq2}, where the predictions are compared with 
experimental data~\cite{nh1,nzeus} at fixed $\xi=0.003$ (H1) and $\xi= 
0.0042$ (ZEUS). The underlying diffractive parton distributions $dg/d\xi$ 
and $d\Sigma/d\xi=6dq/d\xi$ at input scale $Q_0^2$ and their $Q^2$ 
evolution are shown in Fig.~\ref{fig:difpdf}. 

It is an essential feature of the above semiclassical analysis that the rise 
of $F_2(x,Q^2)$ and of $F_2^{D(3)}(\xi,\beta,Q^2)$ at small $x$ and $\xi$ 
has the same, {\it non-perturbative} origin in the energy dependence of 
the average over soft field configurations in the proton. With increasing 
$Q^2$, the $x$ dependence is enhanced by perturbative evolution in the 
case of the inclusive structure function while, in the diffractive case, 
the $\xi$ dependence remains unchanged.

\subsubsection*{Importance of higher twist}
\begin{figure}[p]
\begin{center}
\vspace*{.2cm}
\parbox[b]{14cm}{\psfig{width=14cm,file=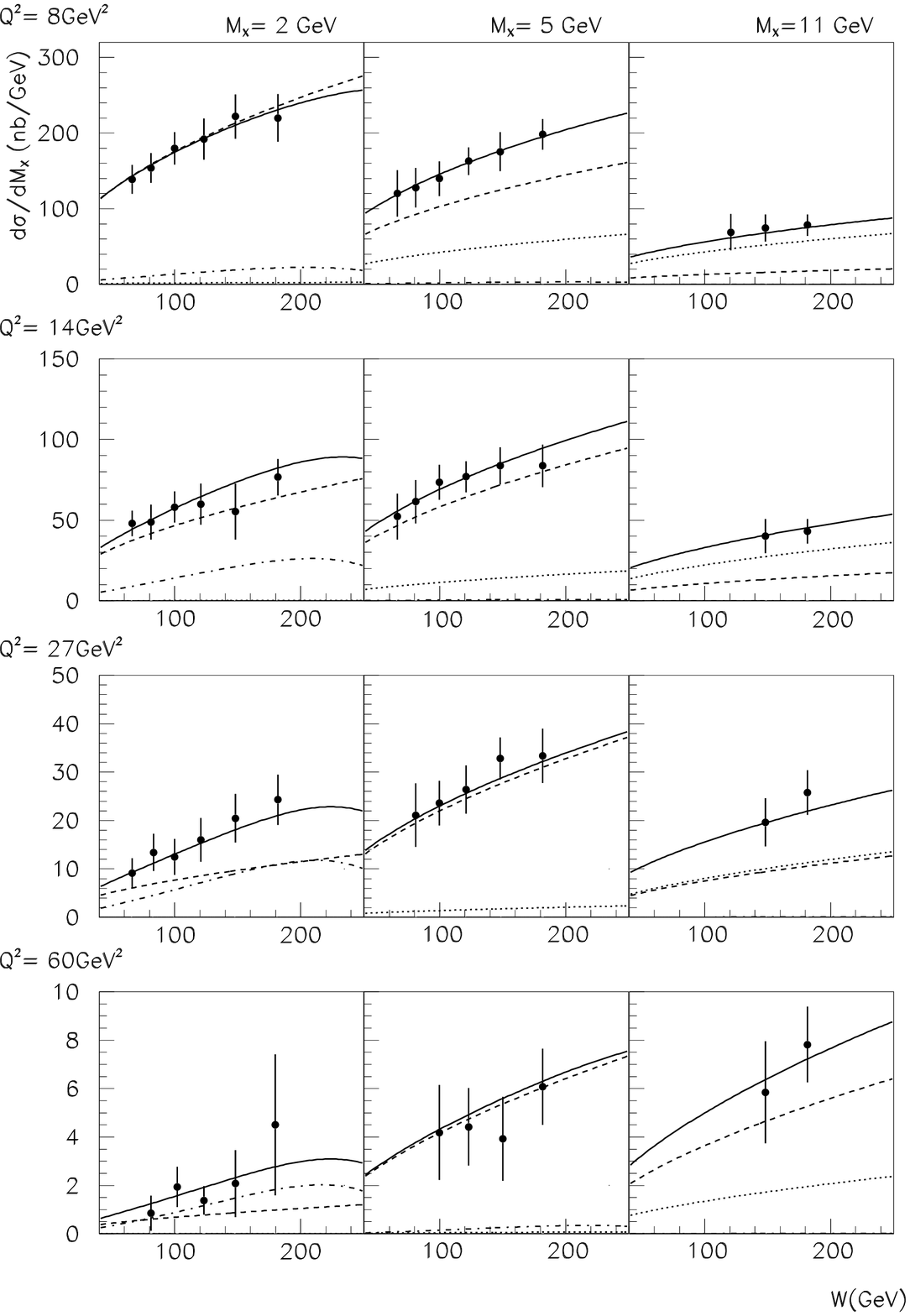}}
\end{center}
\refstepcounter{figure}
\label{fig:zba}
{\bf Figure \ref{fig:zba}:} 
Fit to ZEUS data for $F_2^{D(3)}(\xi,\beta,Q^2)$ by Bartels et al. 
\cite{bekw}. Solid line: total result; dashed line: $F_{q\bar{q}}^T$; 
dotted line: $F_{q\bar{q}g}^T$; dashed-dotted line: $\Delta F_{q\bar{q}}^L$
(data from \cite{kow}, see also \cite{nzeus}).
\end{figure}

\begin{figure}[p]
\begin{center}
\vspace*{.2cm}
\parbox[b]{14cm}{\psfig{width=14cm,file=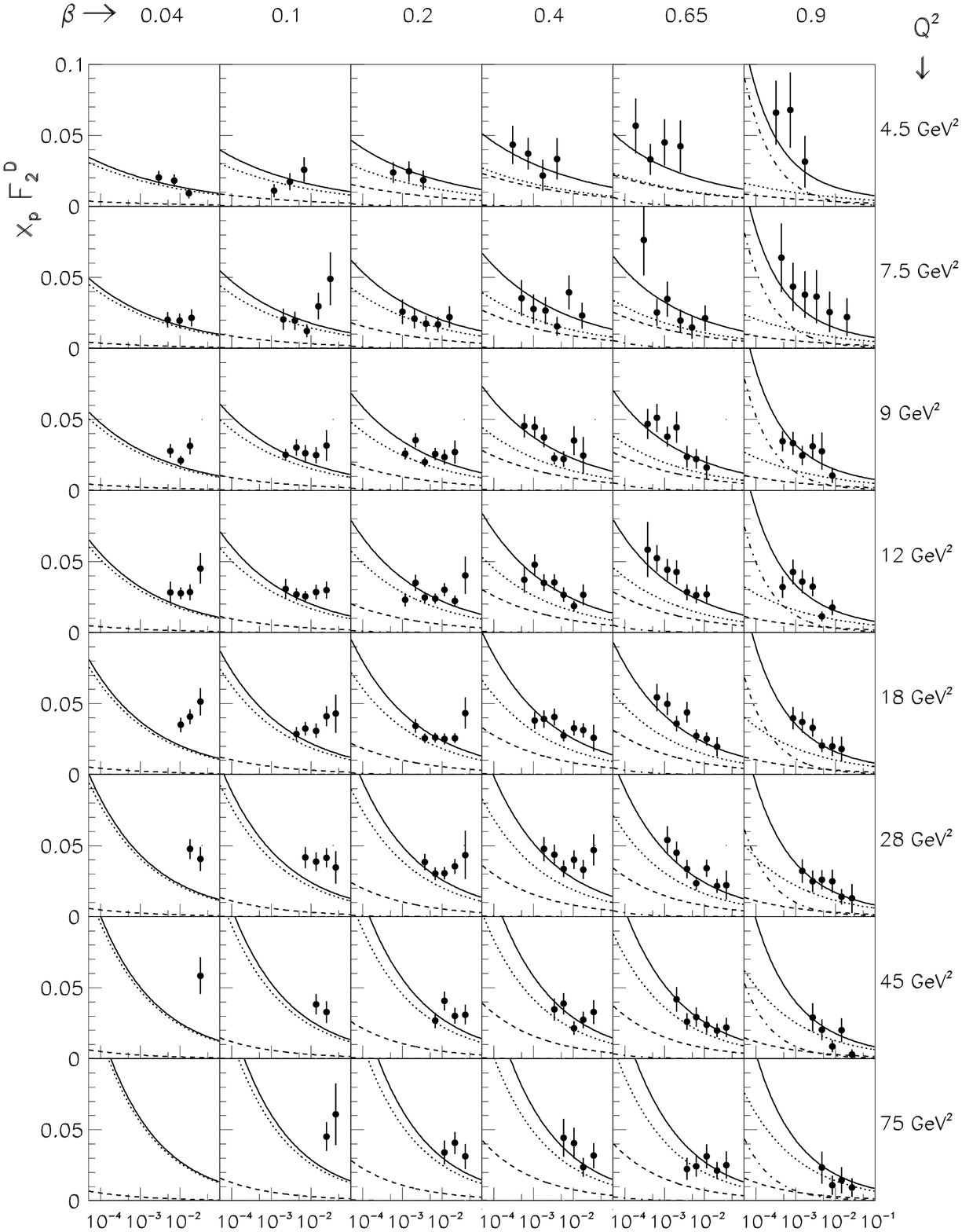}}
\end{center}
\refstepcounter{figure}
\label{fig:h1ba}
{\bf Figure \ref{fig:h1ba}:} 
Fit to H1 data for $F_2^{D(3)}(\xi,\beta,Q^2)$ by Bartels et al. 
\cite{bekw}. Solid line: total result; dashed line: $F_{q\bar{q}}^T$; 
dotted line: $F_{q\bar{q}g}^T$; dashed-dotted line: $\Delta F_{q\bar{q}}^L$ 
(data from \cite{nh1}).
\end{figure}

To illustrate the importance of higher twist contributions, consider the 
parametrization of Bartels et al.~\cite{bekw} (cf.~Figs.~\ref{fig:zba} and 
\ref{fig:h1ba}). In this analysis, four different contributions to the 
diffractive structure function $F_2^{D(3)}$ are considered. These are 
$F_{q\bar{q}}^T$ and $F_{q\bar{q}g}^T$ -- the leading twist contributions 
of $q\bar{q}$ and $q\bar{q}g$ components of the transverse photon wave 
function -- supplemented by $\Delta F_{q\bar{q}}^L$ and $\Delta 
F_{q\bar{q}}^T$ -- the higher twist contributions of the $q\bar{q}$ 
component of longitudinal and transverse photon wave function. 

As has already been explained in Sect.~\ref{sect:qq}, there is no leading 
twist contribution to diffraction from the $q\bar{q}$ component of the 
longitudinal photon. Nevertheless, this component is important because of 
its dominance in the region $\beta\to 1$. 

The approach of~\cite{bekw} is to write down generic expressions for the 
above four contributions to the structure function. These expressions 
incorporate the qualitative knowledge about $\beta$, $\xi$ and $Q^2$ 
dependence of each contribution. This knowledge is based on the two gluon 
exchange calculations of~\cite{pt1,wue} and is largely in agreement with what 
can be learned from the semiclassical treatment presented in this review. 
For example, the leading twist contributions are expected to have a softer 
$1/\xi$ behaviour than the higher twist contributions, which are expected to 
behave as $(\xi g(\xi))^2$ (cf. Sect.~\ref{sect:ccpt}). Furthermore, the 
longitudinal $q\bar{q}$ component is the only one known not to vanish at 
$t=0$ and $\beta\to 1$. 

The relative weights of the four contributions, which are, in principle, 
determined by photon wave function and structure of the target, are allowed 
to vary in the fit of~\cite{bekw}. The results of the two 
independent fits to ZEUS and H1 data (cf. Figs.~\ref{fig:zba} and 
\ref{fig:h1ba}) are in good agreement with physical 
expectations. Note especially the dominance of $\Delta F_{q\bar{q}}^L$ at 
$\beta\to 1$ and of $F_{q\bar{q}g}^T$ at $\beta\to 0$. Note also the 
steep $\xi$ dependence of $\Delta F_{q\bar{q}}^L$, the need of which is 
visible, in particular, in the top right-hand bins of Fig.~\ref{fig:h1ba}. 
Thus, the data clearly has room for the higher twist contributions 
introduced in~\cite{bekw}. It would certainly be desirable to include 
such contributions in the leading twist analysis of~\cite{bgh}, with its 
consistent treatment of the $Q^2$ evolution. An obvious problem to be 
solved before such a development can be realized is the $\xi$ dependence 
of leading and higher twist terms. 

As an interesting feature of the data, the authors of \cite{bekw} notice 
that the H1 results can be described by two different sets of parameters: 
one in which $F_{q\bar{q}g}^T$ is enhanced at large $\beta$ (corresponding 
to Fig.~\ref{fig:h1ba}) and a second one which is more similar to the ZEUS 
fit. The solution with enhanced $q\bar{q}g$ contribution can be interpreted 
as the analogue, within the framework of this parametrization, of the 
singular gluon proposal of H1 \cite{nh1}. 

As can be seen from the QCD analysis in \cite{nh1}, the H1 data exhibit 
some preference for a gluon distribution that is large for high $\beta$, 
such that for low values of $Q^2$ most of the pomeron's momentum is carried 
by a single gluon. This singular gluon proposal, which has some similarity 
with the na\"\i ve boson-gluon fusion model of~\cite{bh1}, is hard to 
justify theoretically. In particular, the validity of the partonic leading 
twist picture at $\beta\sim 1$ and not too large $Q^2$ is questionable. 
Given the successful analysis of~\cite{bgh} and the alternative, less 
singular parametrizations in~\cite{nh1,bekw}, a singular diffractive gluon 
distribution is not unavoidable with the present data. 

Note that the QCD analysis of \cite{nh1} has the important advantage of 
explicitly dealing with the Reggeon contribution, which, although 
suppressed in the high-energy limit, represents an important correction 
in the upper region of the $\xi$ range. A Reggeon contribution is also 
included in the recent, more detailed analysis of H1 data~\cite{br} in the 
framework of the parametrization of~\cite{bekw}. For a discussion of 
Reggeon exchange in diffractive DIS the reader is referred, e.g., to 
\cite{regg}. 

\begin{figure}[t]
\begin{center}
\vspace*{-.3cm}
\parbox[b]{8cm}{\psfig{width=8cm,file=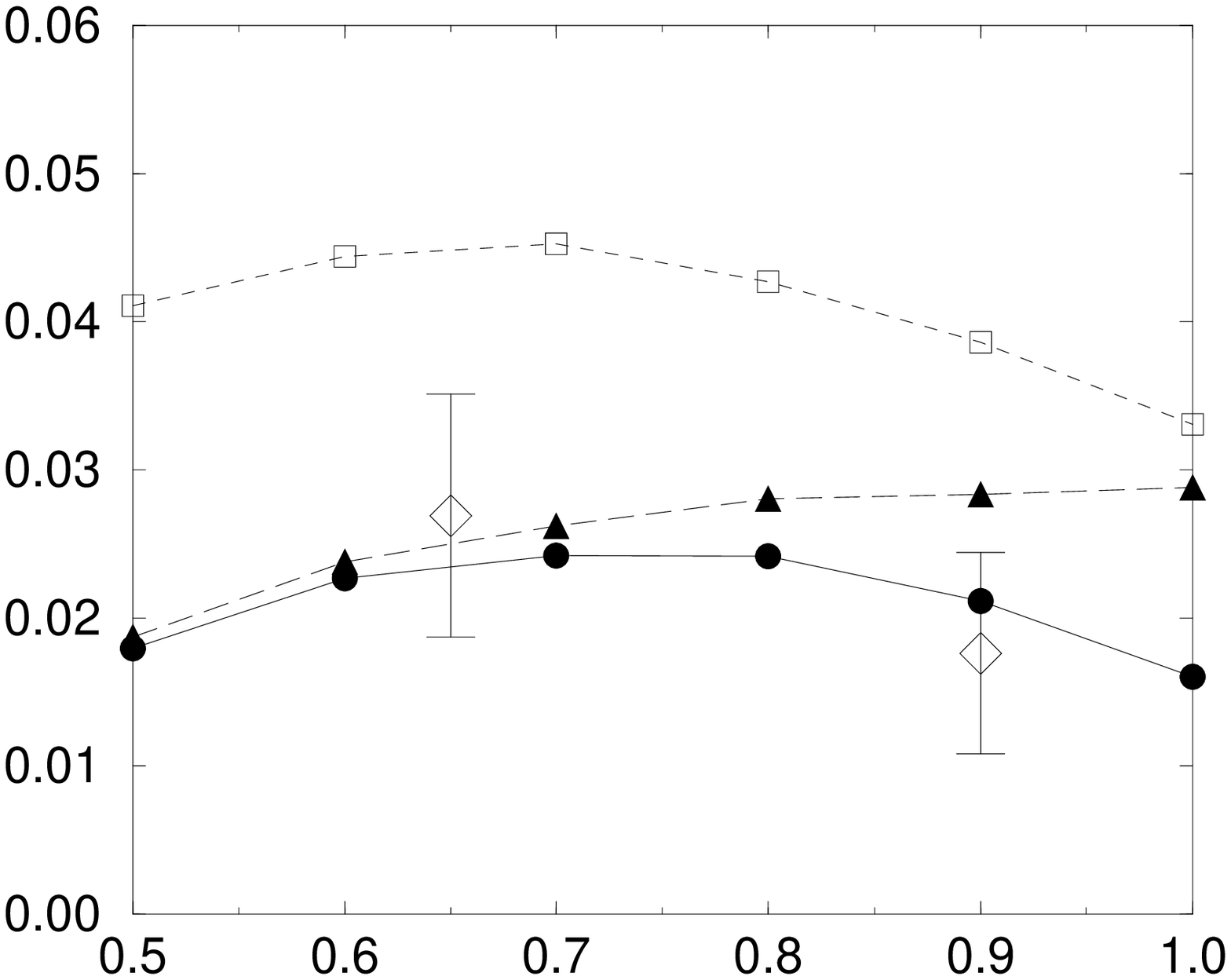}}
\put(-110,0){$\beta$}
\end{center}
\refstepcounter{figure}
\label{fig:ram}
{\bf Figure \ref{fig:ram}:} 
Diffractive structure functions at $Q^2=12\,\mbox{GeV}^2$ and $x=0.0075$ in 
the model of the stochastic vacuum, calculated by Ramirez \cite{ram}, 
compared to H1 data from~\cite{nh1}: $\xi F_2^{D(3)}(\xi,\beta,Q^2)$ and 
$\xi F_T^{D(3)} (\xi,\beta,Q^2)$ in the modified model (long-dashed and 
solid lines respectively); $\xi F_2^{D(3)}(\xi,\beta,Q^2)$ in the model 
without modifications (short-dashed line); figure from \cite{ram}.
\end{figure}

A further interesting analysis, based on the model of the stochastic vacuum 
(cf. Sect.~\ref{sect:sv}) and including both leading and higher twist 
contributions, was reported by Ramirez~\cite{ram}. As an example, results 
for $F_2^{D(3)}$ and for the transverse contribution $F_T^{D(3)}$ at 
$Q^2=12\,\mbox{GeV}^2$ and $x=0.0075$ are shown in Fig.~\ref{fig:ram}. 

The presented calculation is limited to the $q\bar{q}$ fluctuation of the 
photon and does not include Altarelli-Parisi evolution. Therefore, the 
author limits the discussion of data to the high-$\beta$ region, where 
gluon radiation effects are not yet dominant. The model as used in 
previous analyses tends to overshoot the data. Modifications proposed by 
Rueter (see~\cite{rue,ruet}) subtract integration regions where one of 
the quarks is too slow for the eikonal approach to work, and where the 
transverse size of the pair is too small for the non-perturbative model to 
be applicable. The modified model gives a good description of the data 
(cf. Fig.~\ref{fig:ram}). 

It is particularly interesting to see the absolute magnitude of the 
longitudinal structure function at large $\beta$ as a prediction of an 
explicit calculation. Furthermore, the model shows a non-vanishing 
contribution from the $q\bar{q}$ component at $\beta\to 1$. This is an 
improvement compared to the treatment of~\cite{bgh} which is related to the 
better description of the dynamics of the outgoing proton at non-zero $t$ 
in the calculation of~\cite{ram}. 

Note that the above approach allows, in principle, for the inclusion of 
explicit $q\bar{q}g$ contributions of the photon wave function as well as 
for the all orders resummation of $\ln Q^2$ terms, in a way similar 
to~\cite{bgh}. It would be very interesting to see the results of such an 
extended analysis. 

The analysis of Bertini et al.~\cite{bert} is devoted to the influence of 
higher twist terms on the $Q^2$ dependence of diffractive DIS. It is 
found that, similar to what is known about the longitudinal $q\bar{q}$ 
contribution, the higher twist terms in the transverse cross section are 
short distance dominated and are therefore perturbatively calculable. These 
contributions affect the $Q^2$ dependence at $\beta\to 1$ and should be 
subtracted from the cross section before a conventional DGLAP analysis 
of the large $\beta$ region is performed. 

The data analyses of \cite{bpr,mpr} are based on the dipole picture of the 
BFKL pomeron discussed briefly in Sect.~\ref{sect:id}. The diffractive 
structure function is described as a sum of two components: the elastic 
component, where the $q\bar{q}$ fluctuation of the virtual photon scatters 
off the proton \cite{bp}, and the inelastic component, where, based on the 
original $q\bar{q}$ fluctuation, a multi-gluon state develops, giving rise 
to high-mass diffraction  \cite{bptp}. While the first contribution is 
modelled by BFKL pomeron exchange in the $t$ channel, the second one 
involves the triple-pomeron coupling and a cut gluon ladder representing 
the final state. 

In Ref.~\cite{bpr}, a combined analysis of both inclusive and diffractive 
structure functions was performed. The inclusive structure function was 
fitted on the basis of a colour dipole BFKL calculation, which, however, 
contains a number of free parameters and a smaller pomeron intercept than 
na\"\i ve perturbation theory would suggest. Introducing further parameters 
for the normalization of the elastic and inelastic components discussed 
above, qualitative agreement with both the diffractive and inclusive DIS 
data was achieved. 

A further phenomenological investigation of diffraction on the basis of 
colour dipole BFKL was performed in \cite{mpr}, where a good 7 parameter 
fit to the diffractive structure function was presented. It is certainly 
a challenging problem to gain a better understanding of the dynamics, thus 
reducing the number of required parameters. In particular, the 
applicability of perturbative methods needs further investigation. 

Very recently, an analysis of diffraction has appeared~\cite{gbw1} that is 
based on the previous inclusive DIS fit with a Glauber type model for the 
dipole cross section $\sigma(\rho)$~\cite{gbw}. Motivated by the idea 
of saturation (compare the discussion in \cite{musa}), an energy or $x$ 
dependence of $\sigma(\rho)$ is introduced according to the formula 
\be
\sigma(x,\rho)=\sigma_0\left\{1-\exp\left(-\frac{\rho^2}{4R_0^2(x)}\right)
\right\}\,,
\ee
where the function $R_0(x)$ vanishes as $x\to 0$. The authors find a similar 
energy dependence of the diffractive and inclusive cross section. Two new 
parameters required for the analysis of diffraction are the diffractive 
slope and the fixed coupling constant $\alpha_s$. The very successful 
description of $F_2^D$ is based on the $q\bar{q}$ and $q\bar{q}g$ 
fluctuations of the virtual photon. It includes higher twist terms but no 
Altarelli-Parisi evolution. 

In the different approaches discussed above, separate predictions for 
transverse and longitudinal photon cross sections can be derived. 
Corresponding measurements, which are, however, very difficult, would 
be immensely important in gaining a better understanding of the underlying 
colour singlet exchange. An interesting possibility of obtaining the 
required polarization information is the measurement and analysis of the 
distribution in the azimuthal angle, which, at non-zero $t$, characterizes 
the relative position of the leptonic and hadronic plane of a diffractive 
event \cite{azi}.

\section{Final states}

A further tool for studying the mechanism of diffraction is provided by 
the details of the hadronic final states. In this section, the focus is on 
events with a diffractive mass of the order of $Q^2$ or larger. Diffractive 
production of single particles forms the subject of the next section. 

The first topic is the production of jets with high transverse momentum 
relative to the $\gamma^*p$ collision axis within the diffractively 
produced final state. Such diffractive dijet events were reported soon 
after the beginning of the investigation of diffraction at HERA~\cite{zpt1}. 
Given the kinematic constraints and the available number of events, 
a direct analysis of jet cross sections in diffractive DIS proved difficult. 
Therefore, event shape observables such as thrust were first considered 
by both collaborations~\cite{thr}. One of the main results was the 
alignment of the diffractive state along the $\gamma^*p$ axis and, 
furthermore, the presence of significant transverse momentum components. 
If this is to be interpreted in terms of diffractive parton distributions, 
a large gluonic component seems to be required. 

\begin{figure}[t]
\begin{center}
\vspace*{.2cm}
\parbox[b]{10cm}{\psfig{width=10cm,file=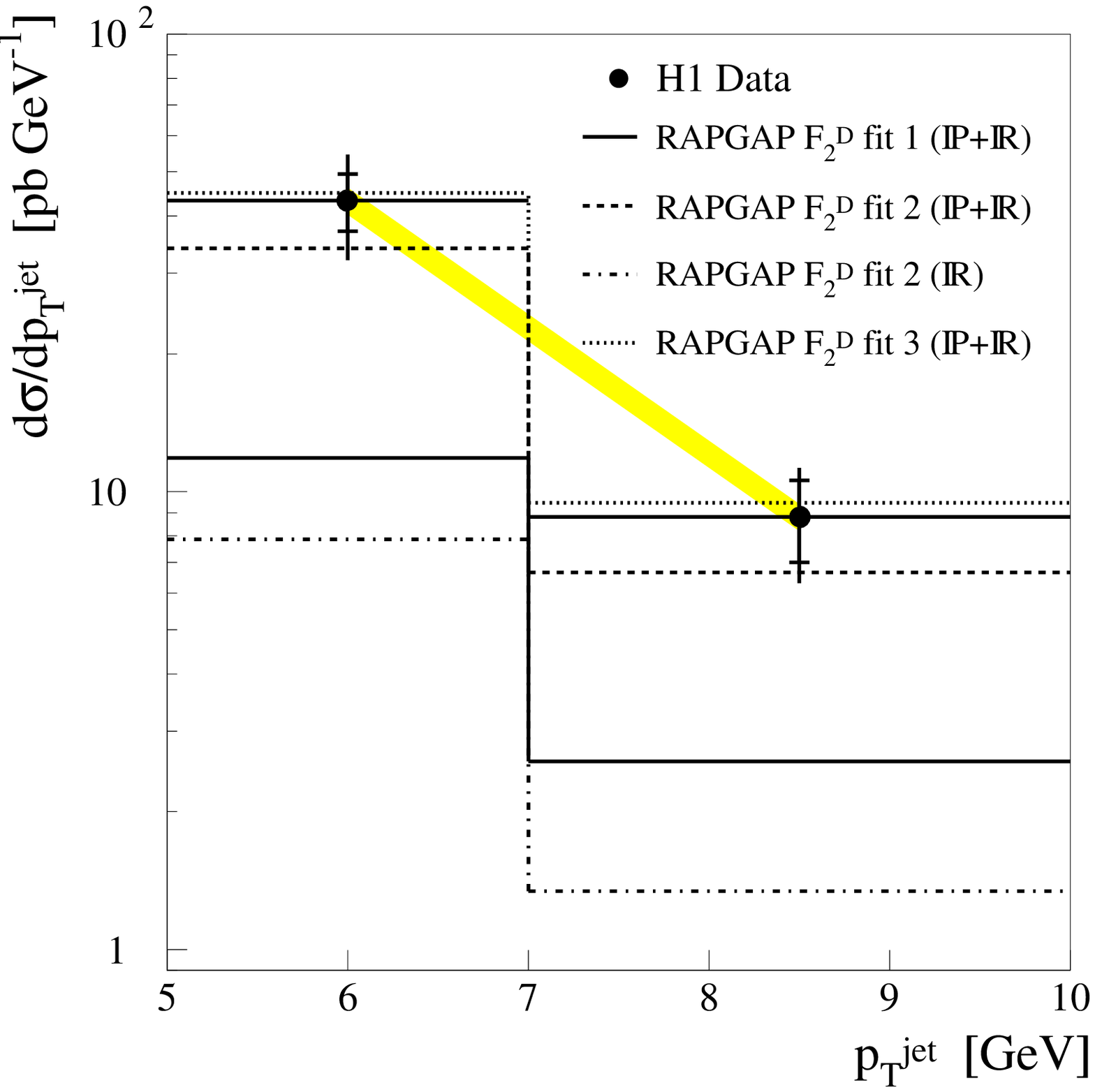}}
\end{center}
\refstepcounter{figure}
\label{fig:pt}
{\bf Figure \ref{fig:pt}:} Differential cross section in transverse jet 
momentum for events where the diffractively produced hadronic state contains 
two jets. Photon virtualities vary in the range $7.5<Q^2<80$ GeV${}^2$. The 
data are compared to predictions of RAPGAP Monte Carlo models with leading 
order pomeron and reggeon parton distributions: quarks only (fit 1), `flat' 
gluon dominated (fit 2) and `peaked' gluon dominated (fit 3). The figure 
is from~\cite{h1pt}, where further details can be found. 
\end{figure}

A direct analysis of diffractive dijet events has recently been reported 
by the H1 collaboration~\cite{h1pt}. In the following, this analysis is 
discussed in more detail since, for such measurements, a qualitative 
comparison with different theoretical models is simpler and less affected 
by hadronization effects. As an illustration, one of the distributions 
published in~\cite{h1pt} is reproduced in Fig.~\ref{fig:pt}. 

As explained in Sect.~\ref{sect:ccpt}, jets in the diffractive final state 
can be the result of either boson-gluon fusion, with the gluon coming from 
the diffractive gluon distribution of the proton, or of so-called exclusive 
dijet production, where the jets are associated with the pure $q\bar{q}$ 
component of the photon wave function. The former case, viewed in the 
target rest frame, is associated with the $q\bar{q}g$ component of the 
photon wave function, where the gluon has small $p_\perp$ and the colour 
singlet exchange in the $t$ channel is soft. 

In Fig.~\ref{fig:pt}, the data points are compared to predictions based on 
the boson-gluon fusion scenario. They are well described if the diffractive 
parton distributions are dominated by the gluon, a situation realized, e.g., 
in the semiclassical calculation based on the colour field model of 
Sect.~\ref{sect:lh}. It would certainly be interesting to substitute the 
diffractive parton distributions underlying the Monte Carlo results of 
Fig.~\ref{fig:pt} (see, e.g.,~\cite{jung} for a recent review of Monte 
Carlo generators used for HERA diffraction) with the model distributions of 
Chapter~\ref{sect:mod}. However, given the precision of the data, which is 
at present consistent with the two very different gluon distributions of 
fit 2 and fit 3, it appears unlikely that any gluon dominated model can be 
ruled out. 

The most important qualitative conclusion to be drawn is that the data 
favours boson-gluon fusion rather than exclusive dijet production. As 
discussed in Sect.~\ref{sect:ccpt}, the exclusive process has a much 
steeper $p_\perp$ distribution, which would be reflected in the quality of 
the fit in Fig.~\ref{fig:pt}. This conclusion is made even clearer by the 
distribution of the variable $z^{\mbox{\scriptsize jets}}_{\pom}$, which 
characterizes the fraction of the proton's momentum loss absorbed by the 
jets. In exclusive jet production, $z^{\mbox{\scriptsize jets}}_{\pom}=1$, 
which, even taking into account the smearing caused by hadronization, is in 
sharp contrast to the data. Thus, the observations suggest that only a 
small fraction of the observed dijet events are related to exclusive jet 
production with hard colour singlet exchange. It would be important to try 
to quantify this statement. 

A further interesting aspect of the diffractive final state is the presence 
of open charm. So far, no published results by either the H1 or ZEUS 
collaboration are available on this subject. However, preliminary data have 
recently been presented by both collaborations~\cite{h1c,zeusc}. Cross 
section measurements for the production of $D^{*\pm}$ mesons in diffractive 
events are reported. It is found in~\cite{zeusc} that both $W^2$ and $Q^2$ 
dependences of diffractively produced open charm as well as the fraction 
of charmed events are consistent with the observations made in inclusive 
DIS. This is in agreement with the semiclassical scenario of~\cite{bhmcc} 
or, equivalently, with a boson-gluon fusion scenario based on gluon 
dominated diffractive parton distributions. 

It should finally be mentioned that a number of more inclusive diffractive 
final state analyses focussing, in particular, on charged particle 
distributions~\cite{cpd}, transverse energy flow~\cite{tef} and the 
multiplicity structure~\cite{ms} were reported by both collaborations. 
Although a detailed discussion of the obtained results can not be given here, 
it is important to note that all observations seem to be consistent with 
the physical picture of a partonic pomeron. Clearly, this is also in accord 
with the more general concept of diffractive parton distributions. As 
expected, particle distributions and energy flow support the by now 
well-established picture of colour singlet exchange with the target. In 
agreement with the jet oriented analyses, a significant gluonic component 
is required by the data. It was emphasized by H1 (see~\cite{cpd,ms}) that a 
relatively hard diffractive gluon distribution is favoured. 

As already mentioned in the Introduction, an approach describing 
diffractive DIS by the assumption of soft colour interactions in the final 
state was proposed by Edin, Ingelman and Rathsman in~\cite{eir}. The 
idea of soft colour exchange is similar to the na\"\i ve boson-gluon fusion 
model of \cite{bh1}, but the Monte-Carlo based implementation is quite 
different. After the hard scattering, the physical state is given by a 
number of quarks and gluons -- those created in the scattering process and 
those from the proton remnant. It is assumed that, at this point, colour 
exchange can take place between each pair of colour charges, the probability 
of which is described by a certain phenomenological parameter $R$. This 
changes the colour topology and leads, in certain cases, to colour singlet 
subsystems, giving rise to rapidity gaps in the final state. The number 
of gap events initially increases with $R$, but saturates or even decreases 
at larger values of this parameter since further colour exchanges can 
destroy neutral clusters previously created. 

A more detailed description of the soft colour interaction model can be 
found in \cite{eir1}, where the consistency of the approach with the 
large transverse energy flow in the proton hemisphere, a feature of HERA 
data that appears to be `orthogonal' to the rapidity gaps, is demonstrated. 
In a more recent development \cite{rath}, the colour exchange is described 
on the level of colour strings rather than on the partonic level discussed 
above. Furthermore, $e^+e^-$ data is included in the analysis. In conclusion 
it can be said that an impressive agreement with many features of hadronic 
final states in small-$x$ DIS has been achieved within the framework of 
soft colour interactions. It is an interesting question in how far the 
perturbatively known suppression of interactions with small colour-singlet 
objects, i.e., colour transparency, is consistent with this approach.

\section{Meson production}\label{sect:mp}
Exclusive meson production in DIS is, both theoretically and experimentally, 
a very interesting and active area that certainly deserves more space than 
is devoted to it in the present review. The main theme of the previous 
chapters was the semiclassical approach, the applicability of which to 
meson production processes is limited by the lack of a genuine prediction 
for the energy dependence of the amplitude. Although, as discussed in 
Sect.~\ref{sect:emp}, Ryskin's double-leading-log result is reproduced, two 
essential problems in going beyond it, the dynamics of the energy growth 
and details of the meson wave function, have not been addressed. 
Nevertheless, this review would be incomplete without at least a brief 
discussion of the most important new experimental results and their 
implications for theory.

Recent analyses of exclusive vector meson electroproduction have been 
published by ZEUS for $\rho$ and $J/\psi$ mesons~\cite{zrho} and by H1 for 
$\rho$ mesons~\cite{h1r}. New preliminary results on exclusive $J/\psi$ 
production were also reported by H1~\cite{h1j}, with previous measurements 
published in~\cite{h1ro}. Agreement is observed between the most recent 
results of the two experiments~\cite{h1r}.

\begin{figure}[t]
\begin{center}
\vspace*{-.1cm}
\parbox[b]{14cm}{\psfig{width=14cm,file=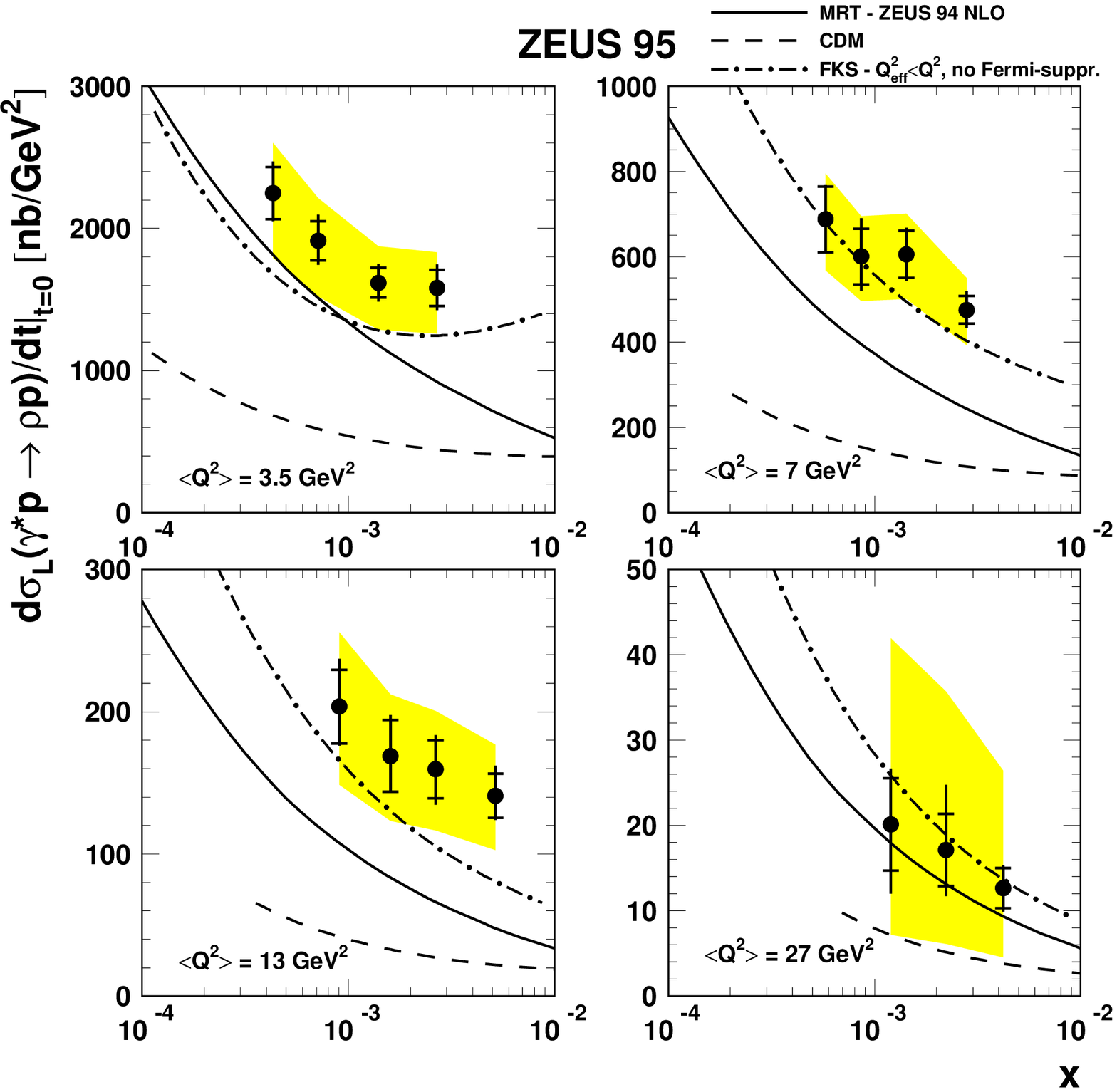}}
\end{center}
\refstepcounter{figure}
\label{fig:rho}
{\bf Figure \ref{fig:rho}:} Forward longitudinal $\rho$ electroproduction 
cross section $d\sigma/dt|_{t=0}$ as measured by the ZEUS collaboration 
\cite{zrho}. Shaded areas indicate normalization uncertainties in addition 
to the error bars shown. The curves are based on calculations by Martin, 
Ryskin and Teubner~\cite{mrt} (solid line), by Frankfurt, Koepf and 
Strikman~\cite{fks} (dashed-dotted line) and on colour dipole model 
calculations of Nemchik et al.~\cite{nem} (dashed line). The figure is 
from~\cite{zrho}, where further details can be found. 
\end{figure}

For illustration, the ZEUS results for a particularly interesting quantity, 
the forward longitudinal $\rho$ meson cross section, are shown in 
Fig.~\ref{fig:rho}. As discussed in Sects.~\ref{sect:emp} and 
\ref{sect:fact}, this quantity can, at least in principle, be calculated in 
perturbation theory. If the asymptotic form of the meson wave function 
$\phi_\rho(z)\sim z(1-z)$ is used, the amplitude of Eq.~(\ref{tlr}) gives 
rise to the cross section formula~\cite{bro,broc} 
\be
\frac{d\sigma}{dt}\,\bigg|_{t=0}=\frac{12\pi^3\Gamma^\rho_{ee}m_\rho
\alpha_s^2(Q^2)[xg(x,Q^2)]^2}{\aem Q^6}\,.\label{rcs}
\ee

The main qualitative predictions of this formula are in good agreement 
with experimental results. Taking into account the anomalous dimension of 
the gluon distribution, the observed $Q^2$ dependence of roughly $Q^{-5}$ 
is consistent with Eq.~(\ref{rcs}). The data indicate that the energy 
dependence becomes more pronounced with growing $Q^2$, as expected from the 
square of the gluon distribution. However, taking the above formula at 
face value and using, e.g., the MRS(A$'$) gluon distribution~\cite{mrsa}, 
the absolute normalization of the cross section comes out too high. 

Several improvements on the result of~\cite{bro} were discussed in 
\cite{fks}. In particular, it was found that the Fermi motion of the quarks 
in the vector meson leads to a suppression factor in the cross section. 
Furthermore, in comparing the recent ZEUS data with the perturbative 
calculation, the gluon distribution is evaluated at a scale 
$Q_{\mbox{\scriptsize eff}}^2<Q^2$. The new scale is determined from the 
average transverse size of the $q\bar{q}$ pair, which turns out to be 
larger in the $\rho$ production process than, say, in longitudinal 
electroproduction. Such a rescaling is certainly legitimate in view of the 
leading logarithmic nature of the approach. It leads to a further reduction 
of the cross section. While both suppression effects together produce 
results that are too small , either one of them gives a reasonable fit (see 
Fig.~\ref{fig:rho} and \cite{zrho}). Note also the claim \cite{hood} that, 
at least in the $J/\psi$ case, Fermi motion effects might be considerably 
smaller than expected. 

The approach of~\cite{mrt}, briefly mentioned in Sect.~\ref{sect:emp}, also 
gives a reasonable fit to the $\rho$ production data. Both the results of 
FKS~\cite{fks} and MRT~\cite{mrt} shown in Fig.~\ref{fig:rho} are based on 
the ZEUS~94~NLO gluon distribution, which, in the relevant parameter 
region, is not too different from, e.g., MRS(A$'$)~\cite{mrsa}. Note that 
all calculations which derive their total normalization from the 
identification of the two gluon exchange amplitude with the gluon 
distribution are affected by the large experimental uncertainties of the 
latter. 

The results of~\cite{nem}, also displayed in Fig.~\ref{fig:rho}, are based 
on calculations that emphasize the non-perturbative aspects of the process. 
The interaction of the $q\bar{q}$ pair with the target is described in 
terms of the colour dipole cross section $\sigma(\rho,\nu)$, where $\nu$ is 
the photon energy. The amplitude for the production of a $q\bar{q}$ pair 
of transverse size $\rho$ is convoluted with a meson wave function with 
non-trivial $\rho$ dependence. In the relevant region, where $Q^2$ is not 
yet very large, these non-perturbative effects and the contributions from 
$\sigma(\rho,\nu)$ at relatively large $\rho$ are important. It is proposed 
that the energy and $Q^2$ dependence of the measured cross section will 
allow for the `scanning' of the dipole cross section $\sigma(\rho,\nu)$ as a 
function of $\rho$ (cf.~\cite{kop}) and provide evidence for BFKL dynamics. 

The violation of $s$ channel helicity conservation is found to be small. 
However, evidence for a helicity flip amplitude on the level of $8\pm3\%$ 
of the non-flip amplitude was reported in~\cite{h1r}. A particularly 
interesting feature of the data is the relatively small ratio $R=\sigma_L/ 
\sigma_T$ of longitudinal and transverse cross sections, which reaches the 
value $R\simeq 3$ at $Q^2\simeq 20\,\mbox{GeV}^2$~\cite{h1r}. This is in 
sharp contrast to a na\"\i ve extrapolation of the perturbative result of 
Eq.~(\ref{tltt}), leading one to expect $R\sim Q^2/m_\rho^2$. Even though 
the transverse cross section is not perturbatively calculable in the case of 
light vector mesons, one would still expect a linear growth of $R$ with 
$Q^2$, which is not favoured by the data~\cite{zrho,h1r}. 

In the stochastic vacuum approach of \cite{dgkp,dkp,rue}, the supposition 
that, at presently accessible values of $Q^2$, the perturbative regime is 
not yet reached is taken even more seriously. The analysis is based on a 
fundamentally nonperturbative model of the interaction with the target 
hadron (cf. Sect.~\ref{sect:sv}) and is thus ideally suited for studying the 
transition region to photoproduction, as well as the transverse cross 
section, where QCD factorization theorems fail. Clearly, the approach 
depends on the model for the $\rho$ meson wave function, but certain 
observables, such as the ratio of longitudinal and transverse cross sections 
or elastic slopes, represent relatively robust predictions. In particular, 
the ratio of longitudinal and transverse $\rho$ meson production cross 
sections is well described \cite{dkp}. 

In its original form, the model of the stochastic vacuum predicts constant 
high-energy cross sections. This limits the applicability of the model in 
the HERA regime, where the energy growth of meson electroproduction cross 
sections can not be neglected. Recently, the model has been 
extended~\cite{rue} by introducing a phenomenological energy dependence 
based on two different pomerons coupled to small and large dipoles, very 
much in the spirit of~\cite{tp}. Furthermore, for small dipole 
configurations, a perturbative two gluon exchange contribution was added. 
This allowed for a good description of both photo- and electroproduction of 
vector mesons and of $F_2(x,Q^2)$ in the HERA regime. 

The presently available data can also be described in the generalized 
vector dominance approach of \cite{sss}. In contrast to the calculations 
discussed so far, this approach predicts both longitudinal and transverse 
vector meson production cross sections to fall like $\sim 1/Q^4$ at 
asymptotically large virtualities. 

Note also the recent perturbative model calculation of~\cite{cro}, which 
finds an asymptotic behaviour of $R=\sigma_L/\sigma_T$ that is 
qualitatively different from the expected linear $Q^2$ growth. The 
calculations of~\cite{mrt,cro,sss} reproduce the experimentally observed 
flat behaviour of $R$ at high $Q^2$. 

The most striking feature of the measured $J/\psi$ production cross 
sections is the strong energy growth $\sim(W^2)^{0.4}$. Its approximate 
$Q^2$ independence up to $Q^2\sim 13\,\mbox{GeV}^2$ suggests that, in the 
present data, the charm quark mass rather than the photon virtuality 
play the dominant role in making the colour singlet exchange hard. 
The flavour-symmetric ratio $\sigma(J/\psi)/\sigma(\rho)=8/9$ is not yet 
reached at $Q^2=13\,\mbox{GeV}^2$. 

Fitting the $t$ dependence of $\rho$ and $J/\psi$ electroproduction cross 
sections with the function $e^{bt}$, values of $b\sim 8\,\mbox{GeV}^{-2}$ 
and $b\sim 5\,\mbox{GeV}^{-2}$ respectively were obtained~\cite{zrho}. 

The exclusive electroproduction of excitations of both the $\rho$ and the 
$J/\psi$ meson have also received much theoretical interest. However, 
experimental data on this subject is only beginning to emerge \cite{exc}. 

\newpage

\mychapter{Conclusions}\label{sect:conc} 
In the above review, different methods employed in the treatment of 
diffractive electroproduction processes were put into context. 

The semiclassical approach, which was chosen as the starting point of the 
discussion, is well suited to develop a basic understanding of the 
underlying physical effects. In this approach, non-diffractive and 
diffractive DIS are treated along parallel lines. Diffraction occurs if, 
after scattering off the target, the partonic fluctuation of the virtual 
photon emerges in a colour singlet state. Altarelli-Parisi evolution of 
both the diffractive and inclusive structure function is related to the 
presence of higher Fock states of the photon. These fluctuations have a 
small transverse size and do not affect the soft colour exchange with the 
target. The application of the semiclassical approach to experimental 
results is particularly simple if the approach is used to derive both 
diffractive and inclusive parton distributions at some small input scale. 
In this case, the analysis of all higher-$Q^2$ data proceeds with standard 
perturbative methods. Different models for the underlying colour fields can 
be compared to diffractive and inclusive structure function data in a very 
direct way. 

If the scattering amplitude of the partonic fluctuation of the photon and 
the target grows with energy, then, in the semiclassical framework, this 
growth enters both the diffractive and inclusive input distributions in 
precisely the same way. Such an energy growth is expected to be generated 
by the process of averaging over all relevant colour field configurations. 
However, no explicit non-perturbative calculation exists so that, at 
present, the energy dependence remains one of the most challenging aspects 
of the method. 

The idea of the pomeron structure function combines methods of Regge theory 
with the partonic description of hard scattering processes in QCD. In the 
treatment of diffractive electroproduction, both the semiclassical approach 
and the pomeron structure function are equivalent as far as the hard part 
of the process, i.e., the $Q^2$ evolution, is concerned. In this respect, 
they represent different realizations of the more general and less 
predictive concept of diffractive parton distributions. However, as far as 
the colour singlet exchange mechanism is concerned, the pomeron structure 
function idea is very different in that it assumes the existence of a 
`pre-formed' pomeron state, off which the virtual photon scatters. By 
contrast, in the semiclassical approach the diffractive character of an 
event is only determined after the photon fluctuation has passed the 
target. No pre-formed colour singlet object is required. From this point of 
view, the semiclassical description of diffraction shows similarity to the 
idea of probabilistic soft colour exchange added on top of standard Monte 
Carlo models. However, in contrast to this method, the semiclassical 
calculation keeps track of the relevant transverse distances in the parton 
cascade, ascribing soft colour exchange only to large size configurations. 

In an apparently quite different approach, the $t$ channel colour singlet 
exchange, characteristic of diffractive processes, is realized by the 
exchange of two $t$ channel gluons. This is justified for certain specific 
final states, like longitudinally polarized vector mesons or two-jet 
systems, where the colour singlet exchange is governed by a hard scale. 
From the semiclassical perspective, this effect manifests itself in the 
small transverse size of the photon fluctuation at the moment of passing 
the external colour field. Thus, only the small distance structure of the 
field is tested, which corresponds to the two gluon exchange amplitude 
discussed above. 

The fundamental advantage of two gluon exchange calculations is the 
understanding of the energy dependence which, one may hope, can be achieved 
in this framework by the resummation of large logarithms. However, leading 
order resummation predicts a far too steep energy growth, and the recently 
obtained next-to-leading correction is very large, thus calling the whole 
method into question. It is also possible that the mechanism responsible 
for the energy dependence in soft processes is fundamentally different 
from the above perturbative ideas. 

Nevertheless, leading logarithmic two gluon calculations were successful 
in relating certain diffractive cross sections to the conventional gluon 
distribution, the energy growth of which is measured. Going beyond leading 
logarithmic accuracy, the non-forward gluon distribution was introduced. 
This is a promising tool for the further study of diffractive processes 
with hard colour singlet exchange. 

On the experimental side, much attention has been devoted to the diffractive 
structure function, which represents one of the most interesting new 
observables in small-$x$ DIS. Leading twist analyses, based on the concept 
of diffractive parton distributions, are successful in describing the bulk 
of the data. It has been demonstrated that a simple large hadron model for 
the target colour fields allows for the derivation of a consistent set of 
diffractive and inclusive parton distributions in the semiclassical 
framework. Other colour field models, such as a perturbative dipole field 
and the model of the stochastic vacuum, promise a successful data analysis 
as well. It can be hoped that future work with the data will expose the 
qualitative differences among the available models, thus advancing our 
understanding of the proton bound state and non-perturbative QCD dynamics. 

The available data shows the importance of higher twist contributions in 
the small-mass region of diffractive structure functions. A consistent 
theoretical description of both leading and higher twist effects has yet to 
be developed. Furthermore, the energy dependence of the diffractive 
structure function and, even more importantly, its relation to the energy 
dependence of the inclusive structure function are interesting unsolved 
problems. While the na\"\i ve summation of logarithms appears to be 
disfavoured, no new standard framework has emerged. The proposal of an 
identical, non-perturbative energy dependence of both diffractive and 
inclusive cross sections at some small virtuality $Q_0^2$ is 
phenomenologically successful. This energy growth is reflected in the $\xi$ 
dependence of the diffractive structure function, which is not altered by 
the perturbative $Q^2$ evolution. 

The detailed investigation of diffractive final states provides a further 
broad field where different approaches can be tested. Here, considerable 
improvements on present data are expected in the near future. So far, the 
analyses performed emphasize the dominance of the soft colour exchange 
mechanism and the importance of the diffractive gluon distribution. 
Diffractive meson production is, on a qualitative level, well described by 
perturbative QCD wherever the dominance of the hard scale is established. 
However, large uncertainties remain, and non-perturbative contributions are 
important even at relatively high values of $Q^2$. In particular, the ratio 
of longitudinal and transverse $\rho$ meson production cross sections is 
far smaller than suggested by simple perturbative estimates. A better 
understanding of the dynamics of meson formation appears to be required. 

In summary, it is certainly fair to say that diffractive electroproduction 
at small $x$ proved a very rich and interesting field. Its investigation 
over the recent years has lead, in many different ways, to the improvement 
of our understanding of QCD dynamics on the interface of perturbative and 
non-perturbative physics. However, many important problems, such as the 
energy dependence and the systematic treatment of higher twist 
contributions, remain at present unsolved.

\subsubsection*{Acknowledgements}
I am deeply indebted to W. Buchm\"uller, who introduced me to the subject 
of diffraction at HERA, for our very interesting and fruitful collaboration 
and for his continuing support of my work over the recent years. I am also 
very grateful to S.J.~Brodsky, T.~Gehrmann, P.V.~Landshoff, M.F.~McDermott 
and H.~Weigert, with whom I had the privilege to work on several of the 
topics described in this review and from whose knowledge I have greatly 
benefited. Furthermore, I would like to thank J.~Bartels, M.~Beneke, 
E.R.~Berger, J.C.~Collins, M.~Diehl, J.R.~Forshaw, L.~Frankfurt, 
G.~Ingelman, B.~Kopeliovich, L.~McLerran, A.H.~Mueller, D.E.~Soper, 
M.~Strikman, T.~Teubner, B.R.~Webber and M.~W\"usthoff for their many 
valuable comments and discussions. Finally, I am especially grateful to 
H.G.~Dosch and O.~Nachtmann for their encouragement, for numerous 
interesting discussions, and for their comments on the final version of 
the manuscript.

\newpage
\begin{appendix}

\noindent{\Huge\bf \underline{Appendix}}
\mychapter{Derivation of Eikonal Formulae}\label{sect:eik}
A particularly simple way to derive the eikonal formulae of 
Sect.~\ref{sect:ef} is based on the direct summation of all Feynman 
diagrams of the type shown in Fig.~\ref{fig:gsu} in the high-energy limit. 
In the case of scalar quarks, the one gluon exchange contribution to the 
$S$-matrix element reads
\be
S_1(p',p)=-ig(p_\mu+p_\mu')\int d^4x A^\mu(x)e^{ix(p'-p)}\,.\label{s1}
\ee

Working in a covariant gauge and in the target rest frame, all components 
of $A$ are expected to be of the same order of magnitude. If, in the 
high-energy limit, $p_+$ and $p_+'$ are the large momentum components, the 
approximation $(p_\mu+p'_\mu)A^\mu\simeq(p_++p'_+)A_-/2$ can be made. 
Assuming that the $x$ dependence of $A$ is soft, one can write
\be
\int dx_-e^{ix_-(p'_+-p_+)/2}A_-(x_+,x_-,x_\perp)\simeq4\pi\,\delta(p_+'-
p_+)\,A_-(x_+,x_\perp)\,,\label{df}
\ee
where the $x_-$ dependence of $A$ has been suppressed on the r.h. side 
since it is irrelevant for the process. More precisely, one has to think of 
the incoming and outgoing particles as wave packets, localized, say, at 
$x_-\simeq 0$, which is the $x_-$ position where $A$ has to be evaluated. 
Equation~(\ref{s1}) takes the form 
\be
S_1(p',p)=2\pi\,\delta(p_+'-p_+)\,2p_+\int d^2x_\perp e^{-ix_\perp(
p_\perp'-p_\perp)}\left(-\frac{ig}{2}\int dx_+A_-(x_+,x_\perp)\right)\,,
\ee
which, within the approximation $k_+\,\delta(k_+)\simeq k_0\,\delta(k_0)$, 
is precisely the first term of the expansion of Eq.~(\ref{qs}) in powers of 
$A$. 

The $n$ gluon exchange contribution to the $S$-matrix element can be 
written as 
\bea
S_n(p',p)&=&\prod_{i=2}^n\left\{-\frac{ig}{2}\int d^4x^{(i)}
\int\frac{d^4p^{(i-1)}}{(2\pi)^4}\left(p^{(i)}_++p^{(i-1)}_+\right)A_-( 
x^{(i)})\frac{i\,e^{\,i\,x^{(i)}\,(p^{(i)}-p^{(i-1)})}}{(p^{(i-1)})^2+i 
\epsilon}\right\}
\nonumber\\
\nonumber\\
&&\times\Bigg\{-\frac{ig}{2}\int d^4x^{(1)}\left(p^{(1)}_++p^{(0)}_+\right)
A_-(x^{(1)})e^{\,i\,x^{(1)}\,(p^{(1)}-p^{(0)})}\Bigg\}\,,\label{sn}
\eea
where $x^{(i)}$ is the space-time variable at vertex $i$, the momentum of 
the quark line between vertex $i$ and vertex $i\!+\!1$ is denoted by 
$p^{(i)}$, and the initial and final momenta are $p=p^{(0)}$ and $p'= 
p^{(n)}$. With Eq.~(\ref{df}), all $p_+$ integrations in Eq.~(\ref{sn}) 
become trivial, and all momentum plus components become identical with 
$p^{(0)}_+$. Integrations over the momentum minus components are performed 
using the identity 
\be
\int dp_-\frac{e^{-ip_-(y_+-z_+)/2}}{p_+p_--p_\perp^2+i\epsilon}=-\theta
(y_+-z_+)\frac{2\pi i}{p_+}\,.\label{pmi}
\ee
The resulting step functions translate into path ordering of the matrix 
valued fields $A$ along the plus direction, 
\be 
\int dx^{(1)}_+...\,dx^{(n)}_+\theta(x^{(2)}_+-x^{(1)}_+)...\,\theta(
x^{(n)}_+-x^{(n-1)}_+)\{\cdots\}=\frac{1}{n!}\int dx^{(1)}_+...\,
dx^{(n)}_+\,P\{\cdots\}\,.
\ee

Now that all dependence on the intermediate transverse momenta 
$p^{(1)}_\perp...\,p^{(n-1)}_\perp$ originating in the propagators has 
disappeared, the corresponding integrations are straightforward. They 
produce $\delta$-functions in transverse co-ordinate space, 
\be
\int d^2p_\perp e^{ip_\perp(y_\perp-z_\perp)}=(2\pi)^2\,\delta^2(y_\perp-
z_\perp)\,.\label{pti}
\ee

Making use of Eqs.~(\ref{pmi}) -- (\ref{pti}), the expression for the $n$th 
order contribution to the $S$-matrix element finally takes the form 
\be
\hspace*{-.3cm}
S_n(p',p)=2\pi\,\delta(p_+'-p_+)\,2p_+\left\{\int d^2x_\perp e^{-ix_\perp
(p_\perp'-p_\perp)}\right\}\frac{1}{n!}\,P\left(-\frac{ig}{2}\int dx_+A_-( 
x_+,x_\perp)\right)^n\,.\label{snf}
\ee
Since this is precisely the $n$th term of the expansion of Eq.~(\ref{qs}) 
in powers of $A$, the derivation of the eikonal formula in the scalar case 
is now complete. 

To obtain the eikonal formula for spinor quarks, Eq.~(\ref{qss}), write 
down the analogue of Eq.~(\ref{sn}) using the appropriate expressions for 
quark propagator and quark-quark-gluon vertex. Applying the identity of 
Eq.~(\ref{prop2}) to each of the quark propagators, neglecting the 
$\gamma_+$ term, which is suppressed in the high-energy limit, and making 
use of the relation 
\be
\bar{u}_{s'}(k')(-ig\As)u_s(k)\simeq -igk_+A_-\,\delta_{ss'}\,,\label{qeik}
\ee
valid if $k_+\simeq k_+'$ are the big components, the exact structure of 
Eq.~(\ref{sn}) is recovered. The only difference is an additional 
overall factor $\delta_{ss'}$, corresponding to the conservation of the 
quark helicity. The further calculation, leading to the analogue of 
Eq.~(\ref{snf}), is unchanged, and the eikonal formula follows. 

The further generalization to the gluon case, Eq.~(\ref{gss}), proceeds 
along the same lines. Begin by writing down the analogue of Eq.~(\ref{sn}), 
i.e., the $S$-matrix element for Fig.~\ref{fig:gsu} with a gluon line 
instead of the quark line. The expression contains conventional three-gluon 
vertices and gluon propagators in Feynman gauge. Contributions with 
four-gluon vertices are suppressed in the high-energy limit. The colour 
structure is best treated by introducing matrices 
\be
A_{\cal A}=A^a(T^a_{\cal A})^{bc}=-iA^af^{abc}\,,
\ee
where $f^{abc}$ are the usual structure constants appearing in the 
three-gluon vertex. Now, the product of matrices $A$ in Eq.~(\ref{sn}) 
is, in the gluonic case, simply replaced by an identical product of adjoint 
representation matrices $A_{\cal A}$. 

Next, the $g^{\mu\nu}$ tensor of each gluon propagator $-ig^{\mu\nu}/k^2$ 
is decomposed according to the identity 
\be
g^{\mu\nu}=\left(\,\sum_{i=1}^2e_{(i)}^\mu e_{(i)}^\nu+\frac{m^\mu k^\nu}
{(mk)}+\frac{k^\mu m^\nu}{(mk)}-\frac{m^\mu m^\nu}{(mk)^2}k^2\right)\,,
\label{gten}
\ee
where $m$ is a light-like vector with a non-zero minus component, 
$m=(0,2,0_\perp)$, and the polarization vectors are defined by $e k=
e m =0$ and $e^2=-1$. An explicit choice is given by 
\be
e_{(i)}=\Big(\,0,\,\frac{2(k_\perp\epsilon_{(i)\perp})}{k_+},\,
\epsilon_{(i)\perp}\,\Big)\,,
\ee
where the transverse basis $\epsilon_{(1)\perp}=(1,0)$ and 
$\epsilon_{(2)\perp}=(0,1)$ has been used. If, in the high-energy limit, 
$g^{\mu\nu}$ appears between two three-gluon vertices, the last three terms 
of Eq.~(\ref{gten}) can be neglected. Note that, for the second and third 
term, this is a non-trivial statement since the vector $k$ in the numerator 
could, in principle, compensate for the suppression by the $k_+$ in the 
denominator. However, this is prevented by the gauge invariance of the 
three gluon vertex. 

Thus, the analogue of Eq.~(\ref{sn}) is written down with three-gluon 
vertices and propagators proportional to 
\be
g^{\mu\nu}\simeq\sum_{i=1}^2e_{(i)}^\mu e_{(i)}^\nu\,.
\ee
Now, each of the three-gluon vertices $V_{\mu\nu\sigma}(-k',k,k'\!-\!k)$ 
(where all momenta are incoming and colour indices are suppressed) appears 
between two transverse polarization vectors and simplifies according to 
\be
e_{(i')}^\mu(k')\,V_{\mu\nu\sigma}(-k',k,k'\!-\!k)\,A_{\cal A}^\sigma\,
e_{(i)}^\nu(k)\,\simeq -igk_+A_{{\cal A}\,-}\delta_{ii'}\,,
\ee
where $A_{\cal A}$ is the external field. This is similar to what was found 
in the quark case in Eq.~(\ref{qeik}), with the difference that helicity 
conservation is replaced by polarization conservation. Therefore, as 
before, the structure of Eq.~(\ref{sn}) is recovered, but with the external 
field in the adjoint representation and with an additional polarization 
conserving $\delta$-function. The further calculation, leading to the 
analogue of Eq.~(\ref{snf}), is unchanged, and the eikonal formula in the 
gluonic case follows. 

\newpage

\mychapter{Spinor Matrix Elements}\label{sect:me}
In this appendix, the spinor matrix elements of the type $\bar{u}_{s'}(k') 
\epsilons(q)u_s(k)$, required,  e.g.,  for the calculation the transition 
from virtual photon to $q\bar{q}$ pair, are listed. 

Using light-cone components for vectors, $a=(a_+,a_-,a_\perp)$, and 
orienting the photon momentum along the positive $z$-axis, $q=(q_+, 
-Q^2/q_+,0_\perp)$, the longitudinal and transverse polarization vectors 
can be defined as 
\be
\epsilon_L=(q_+/Q,Q/q_+,0_\perp)\quad,\quad\epsilon_{\pm 1}=(0,0,
\epsilon_\perp(\pm))\,,
\ee
where $\epsilon_\perp(\pm)=(1,\pm i)/\sqrt{2}$. 

The Dirac representation of $\gamma$ matrices and the conventions of 
\cite{nb} for Dirac spinors are used. Introducing the matrix 
\be
\varepsilon=\left(\begin{array}{rr} 0 & 1 \\ -1 & 0 \end{array}\right)\,.
\ee
and the two-component spinors
\be
\chi_{\frac{1}{2}}=\left(\begin{array}{c} 1 \\ 0 \end{array}\right)\quad,
\quad\chi_{-\frac{1}{2}}=\left(\begin{array}{c} 0 \\ 1 \end{array}\right)\,,
\ee
the two independent Dirac spinors, with $s=\pm 1/2$, for negative and 
positive frequency solutions can be explicitly written as 
\be
u_s(k)=\sqrt{k_0+m}\left( \begin{array}{c} \chi_s\\[-.3cm] \\ 
\frac{\ds \vec{\sigma}
\vec{k}}{\ds k_0+m}\chi_s \end{array} \right)\quad,\quad
v_s(k)=-\sqrt{k_0+m}\left( \begin{array}{c} \frac{\ds \vec{\sigma}\vec{k}}
{\ds k_0+m}\varepsilon\chi_s \\[-.3cm] 
\\ \varepsilon\chi_s \end{array} \right)\,,
\ee
where $\vec{\sigma}$ is the vector formed by the three Pauli matrices. 

The result for longitudinal photon polarization can be obtained from the 
relation 
\be 
\bar{u}_{s'}(k')\gamma_0 u_s(k) = \delta_{s's}2\sqrt{k_0k_0'}\,,\label{lp}
\ee
which holds at leading order in the high-energy expansion. From this, the 
matrix element $\bar{u}_{s'}(k')\epsilons_L(q)u_s(k)$ is obtained using the 
gauge invariance of the $q\bar{q}$-photon vertex, $\bar{u}_{s'}(k')\qs 
u_s(k) = 0$. Since the quark mass and transverse momentum do not enter the 
leading order relation, Eq.~(\ref{lp}), they may be neglected so that the 
relation $v_s(k)=-u_{-s}(k)$ holds. Thus, the complete result takes the 
form 
\[
\bar{u}_{s'}(k')\epsilon_L u_s(k) = \bar{v}_{s'}(k')\epsilon_L v_s(k)=\!-\!
\bar{v}_{-s'}(k')\epsilon_L u_s(k)=\!-\!\bar{u}_{-s'}(k')\epsilon_L u_s(k)= 
\delta_{s's}2Q\sqrt{\alpha(1\!\!-\!\!\alpha)}\,,
\]
\be
\vspace*{-.3cm}
\ee
where $\alpha=k_0/q_0$ and $\epsilon_L=\epsilon_L(q)$. 

Neglecting terms suppressed in the high-energy limit, the relevant matrix 
elements for transverse photon polarization read 
\bea
\!\!\!\!\! 
\bar{u}_{s'}(k') \epsilons_{+1\,/\,-1}(q) u_s(k)&=&-\sqrt{\frac{2}{k_0k_0'}}
\left(\begin{array}{c|c} k_{\perp+}k_0'\quad\big/\quad k_{\perp-}'k_0 &
                         m(k_0'-k_0)\quad\big/\quad 0 \\[.2cm] \hline\\[-.3cm]
                         0\quad\big/\quad m(k_0-k_0') &
                         k_{\perp+}'k_0\quad\big/\quad k_{\perp-}k_0' 
\end{array}\right)_{s's}\label{ma1}\\ \nonumber\\ \nonumber\\
\!\!\!\!\! 
\bar{v}_{s'}(k') \epsilons_{+1\,/\,-1}(q) v_s(k)&=&-\sqrt{\frac{2}{k_0k_0'}}
\left(\begin{array}{c|c} k_{\perp+}'k_0\quad\big/\quad k_{\perp-}k_0' &
                         0\quad\big/\quad m(k_0'-k_0)\\[.2cm] \hline\\[-.3cm]
                         m(k_0-k_0')\quad\big/\quad 0 &
                         k_{\perp+}k_0'\quad\big/\quad k_{\perp-}'k_0 
\end{array}\right)_{s's}\\ \nonumber\\ \nonumber\\
\!\!\!\!\! 
\bar{u}_{s'}(k') \epsilons_{+1\,/\,-1}(q) v_s(k)&=&+\sqrt{\frac{2}{k_0k_0'}}
\left(\begin{array}{c|c} \!\!\!-\!m(k_0+k_0')\quad\big/\quad 0 &
                         k_{\perp+}k_0'\quad\big/\quad k_{\perp-}'k_0
                         \\[.2cm] \hline\\[-.3cm]
                         k_{\perp+}'k_0\quad\big/\quad k_{\perp-}k_0' &
                         0 \quad\big/\quad m(k_0'+k_0)
\end{array}\right)_{s's}\\ \nonumber\\ \nonumber\\
\!\!\!\!\! 
\bar{v}_{s'}(k') \epsilons_{+1\,/\,-1}(q) u_s(k)&=&+\sqrt{\frac{2}{k_0k_0'}}
\left(\begin{array}{c|c} 0\!\quad\big/\quad \!\!\!\!-\!\!m(k_0+k_0') &
                         k_{\perp+}'k_0\quad\big/\quad k_{\perp-} k_0'
                         \\[.2cm] \hline\\[-.3cm]
                         k_{\perp+}k_0'\quad\big/\quad k_{\perp-}'k_0 &
                         m(k_0'+k_0)\!\quad\big/\quad 0
\end{array}\right)_{s's}\!\!\!\!\!\!\!\!\!\:.\label{ma4}
\eea
Here the notation $k_{\perp\pm}=k_1\pm ik_2$ is used, and the four entries 
(11), (12), (21) and (22) of the two-by-two matrices on the r.h. sides 
of the above equations correspond to the combinations $(s's)=(+ 
\frac{1}{2}\,+\frac{1}{2})\,,\,\,\, (+\frac{1}{2}\,-\frac{1}{2})\,,\,\,\, 
(-\frac{1}{2}\,+\frac{1}{2})$ and $(-\frac{1}{2}\,-\frac{1}{2})$. For each 
of these four entries, the expression before and after the oblique stroke 
`$\,/\,$' corresponds to positive and negative photon polarization 
respectively. For example, 
\be
\bar{u}_{+\frac{1}{2}}(k') \epsilons_{+1}(q) u_{-\frac{1}{2}}(k)
=-\sqrt{\frac{2}{k_0k_0'}}\,\,\,m(k_0'-k_0)\,.
\ee

The calculation of the transition from virtual photon to $q\bar{q}$-gluon 
configuration requires, in addition to the $q\bar{q}$-photon vertex, the 
knowledge of the $q\bar{q}$-gluon vertex (see Sect.~\ref{sect:hfs}). The 
corresponding spinor matrix elements are easily obtained from 
Eqs.~(\ref{ma1})$-$(\ref{ma4}) if the configuration is rotated in such a 
way that the gluon momentum is parallel to the $z$-axis. In the high-energy 
limit, such a rotation corresponds to the substitutions 
\be
k_\perp\,\to\,k_\perp-q_\perp(k_+/q_+)\quad,\quad k_\perp'\,\to\,
k_\perp'-q_\perp(k_+'/q_+)\,,
\ee
where $q$ is now interpreted as the gluon momentum. 

Note that similar matrix elements are commonly used in light-cone 
perturbation theory (see, e.g., Tables I$\!$I and I$\!$I$\!$I of~\cite{bl}). 

\newpage

\begin{figure}[ht]
\begin{center}
\vspace*{-.5cm}
\parbox[b]{14cm}{\psfig{width=13cm,file=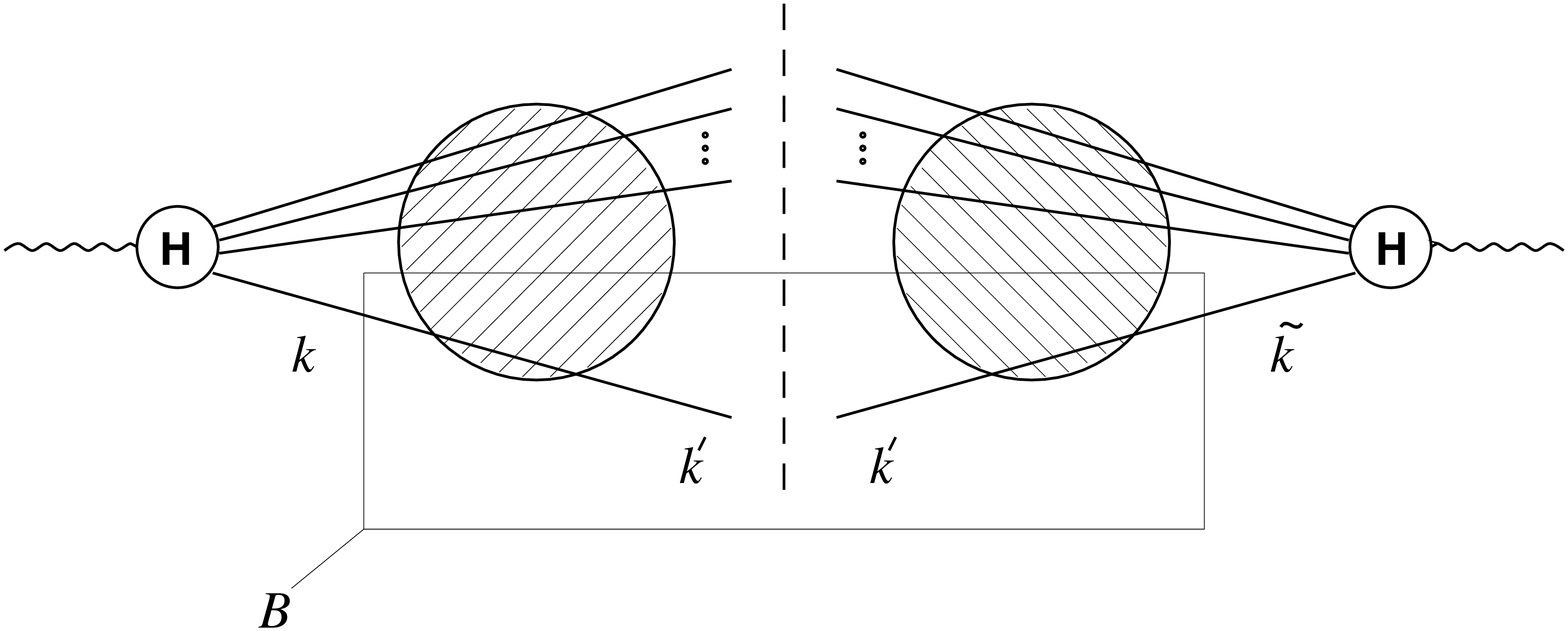}}\\
\end{center}
\refstepcounter{figure}
\label{fig:sp}
{\bf Figure \ref{fig:sp}:} Symbolic representation of the square of the 
amplitude for a hard diffractive process. The box separates the 
contributions associated with the soft parton and responsible for the 
differences between diffractive distributions for scalars, spinors and 
vector particles. 
\end{figure}

\mychapter{Derivation of Diffractive Quark and Gluon Distribution}
\label{sect:dqgd}
The explicit formulae for diffractive quark and gluon distributions, 
Eqs.~(\ref{fxsp}) and (\ref{fxg}), can be derived by appropriately 
adapting the scalar calculation of Sect.~\ref{sect:trf} to the case of 
spinor or vector particles. The most economic procedure is to first 
identify the piece of the old squared amplitude $|T|^2$ that depends 
explicitly on the spin of the soft parton. This piece, which is essentially 
just the squared scattering amplitude of the soft parton and the external 
field, is symbolically separated in Fig.~\ref{fig:sp}. The two independent 
integration variables for the intermediate momentum of the soft parton are 
denoted by $k$ and $\tilde{k}$ in the amplitude $T$ and its complex 
conjugate $T^*$ respectively. Note also that $k_+=\tilde{k}_+$ and 
$k_-=\tilde{k}_-$ at leading order. 

It is straightforward to write down the factor $B_s$ that corresponds to 
the box in Fig.~\ref{fig:sp} for the scalar case. Since the off-shell 
denominators $k^2$ and $\tilde{k}^2$ as well as the two eikonal factors 
and energy $\delta$-functions are present for all spins of the soft parton, 
they are not included into the definition of $B$. All that remains are the 
explicit factors $2k_0\simeq k_+$ from the effective vertex, 
Eq.~(\ref{hev}). Therefore, the result for the scalar case reads simply 
\be
B_s=k_+^2\, .\label{bsc}
\ee

The next step is the calculation of the corresponding expressions 
$B_q$ and $B_g$, given by the box in Fig.~\ref{fig:sp}, in the case where 
the produced soft parton is a quark or a gluon. Introducing factors $B_q 
/B_s$ and $B_g/B_s$ into Eq.~(\ref{fx}) will give the required diffractive 
parton distributions $(df^D_q/d\xi)$ and $(df^D_g/d\xi)$.

\subsubsection*{Diffractive quark distribution}\label{qua}
Observe first, that the analogue of Eq.~(\ref{hev}) for spinors is simply 
\be
V_q(p',p)=2\pi\delta(p_0'-p_0)\,\frac{\gamma_+}{2}\,\left[\tilde{U}(
p_\perp'-p_\perp)-(2\pi)^2\delta^2(p_\perp'-p_\perp)\right]\, .\label{hevs}
\ee
The Dirac structure of $V_q$ follows from the fact that, in the high-energy 
limit, only the light-cone component $A_-$ of the gluon field contributes. 
The normalization is consistent with Eqs.~(\ref{v1}) and (\ref{v2}). 

Consider now the Dirac propagator with momentum $k$. The hard part $T_H$ 
requires the interpretation of the quark line as an incoming parton, which 
collides head on with the photon. Therefore, it is convenient to switch to 
the Breit frame and to define a corresponding on-shell momentum $l$, given 
by $l_-=-k_-$, $l_\perp=-k_\perp$ and $l_+=l_\perp^2/l_-$. The propagator 
can now be written as 
\be
\frac{1}{\ks}=-\frac{\sum_su_s(l)\bar{u}_s(l)}{k^2}-\frac{\gamma_-}{2l_-}
\, .\label{prop}
\ee
This is technically similar to Sect.~\ref{sect:qq}, where an on-shell 
momentum was defined by adjusting the minus component of $k$. In the present 
treatment, however, the plus component of $l=-k$ is adjusted, so that a 
partonic interpretation in the Breit frame becomes possible. It has now 
to be shown that the second term on the r.h. side of Eq.~(\ref{prop}) 
can be neglected, since it is suppressed in the Breit frame by the hard 
momentum $l_-=yP_-$. This can be intuitively understood by observing that 
this term represents a correction for the small off-shellness of the quark, 
which is neither important for $T_H$, nor for the soft high-energy 
scattering off the external field. 

To see this more explicitly, write the relevant part of the amplitude in 
the form 
\be
\bar{u}(k')\gamma_+\frac{1}{\ks}T_H=-\bar{u}(k')\gamma_+\left(
\frac{\sum_su_s(l)\bar{u}_s(l)}{k^2}+\frac{\gamma_-}{2l_-}\right)T_H\equiv
-(C_1+C_2)\,.\label{c1c2}
\ee
It will now be shown that the $\gamma_-$ term $C_2$ is suppressed with 
respect to $C_1$. 

The spinor matrix element in $C_1$ can be estimated using the relation 
\be
\sum_s\bar{u}_s(p')\gamma_+u_s(p)=\frac{4(p_\perp'p_\perp)}{\sqrt{
p_-'p_-}}\, ,\label{blr} 
\ee
valid in the limit where the minus components of $p$ and $p'$ become large 
(cf. Table I$\!$I of \cite{bl}). In the soft region, $k^2\sim k_\perp^2\sim 
k_\perp'^2\sim \Lambda^2$, and $-k_-\sim k_-'\sim Q$ are the large 
components in the Breit frame. Thus, $[\bar{u}(k')\gamma_+u(l)]\sim 
\Lambda^2/Q$, and the first term on the r.h. side of Eq.~(\ref{c1c2}) is 
estimated to be 
\be
C_1\sim \frac{1}{Q}\,\Big[\bar{u}(l)T_H\Big]\,,
\ee
where factors ${\cal O}(1)$ have been suppressed. 

Introducing a vector $a=(1,0,0_\perp)$, so that $\gamma_-/2=\as$, the 
$\gamma_-$ term can be written as 
\be
C_2=\sum_s\,\Big[\bar{u}(k')\gamma_+u_s(a)\Big]\,\frac{1}{l_-}\,\Big[
\bar{u}_s(a)T_H\Big]\,.
\ee
Since $[\bar{u}(k')\gamma_+u(a)]$ vanishes for $k_+'=k_\perp'^2/k_-'\to 0$, 
in the Breit frame, where $k_-'$ is large, the estimate $[\bar{u}(k') 
\gamma_+u(a)]\sim|k_\perp'|/\sqrt{k_-'}\sim \Lambda/\sqrt{Q}$ can be made. 
With $l_-\sim Q$, one obtains 
\be
\frac{C_2}{C_1}\,\sim\,\frac{\Lambda}{\sqrt{Q}}\,\frac{[\bar{u}(a)T_H]}
{[\bar{u}(l)T_H]}\,\sim\,\frac{\Lambda}{Q}\,,\label{lq}
\ee
where it has been assumed that no specific cancellation makes 
$\bar{u}(l)T_H$ small, i.e., $[\bar{u}(l)T_H]/[\bar{u}(a)T_H]\sim\sqrt{Q}$. 
Equation~(\ref{lq}) establishes the required suppression of the 
$\gamma_-$ term in Eq.~(\ref{prop}). 

To proceed with the evaluation of $B_s$ note that, when the soft part is 
separated in Fig.~\ref{fig:sp}, the spinor $\bar{u}_s (l)\simeq\bar{v}_{-s} 
(l)$ from Eq.~(\ref{prop}) has to be considered a part of the hard amplitude 
$T_H$. Therefore the analogue of Eq.~(\ref{bsc}) reads 
\be
B_s=\sum_{s,s'}\,\bar{u}_{s'}(k')\,\frac{\gamma_+}{2}\,u_s(l)\,
\bar{u}_s(\tilde{l})\,\frac{\gamma_+}{2}\,u_{s'}(k')\, ,
\ee
where $\tilde{l}$ is defined analogously to $l$, but using the momentum 
$\tilde{k}$ instead of $k$. The spin summation decouples from the hard part 
if the measurement is sufficiently inclusive. 

The above expression can be evaluated further to give 
\be
B_s=\frac{1}{2}k_+\sum_s\bar{u}_s(\tilde{l})\gamma_+u_s(l)=k_+\,
\frac{2(l_\perp\tilde{l}_\perp)}{\sqrt{l_-\tilde{l}_-}}\, ,
\ee
where the last equality again uses Eq.~(\ref{blr}). Simple kinematics leads 
to the result 
\be
B_s=k_+^2\,\frac{2(k_\perp\tilde{k}_\perp)}{k_\perp'^2}\left(
\frac{\xi-y}{y}\right)\, .\label{bsp}
\ee
Comparing this with Eq.~(\ref{bsc}), the diffractive quark distribution, 
Eq.~(\ref{fxsp}), is straightforwardly obtained from the scalar case, 
Eq.~(\ref{fx}). 

Note that the virtual fermion line corresponds to a right-moving quark 
with momentum $k$ in the proton rest frame and to a left-moving antiquark 
with momentum $l$ in the Breit frame. Therefore, the above result has, in 
fact, to be interpreted as a diffractive antiquark distribution. The 
diffractive quark distribution is identical.

\subsubsection*{Diffractive gluon distribution}\label{glu}
To obtain the diffractive gluon distribution, the procedure of the last 
section has to be repeated for the case of an outgoing soft gluon with 
momentum $k'$ in Fig.~\ref{fig:sp}. Calculating the contribution separated 
by the box will give the required quantity $B_g$, in analogy to 
Eqs.~(\ref{bsc}) and (\ref{bsp}). 

It will prove convenient to introduce two light-like vectors $m$ and $n$, 
such that the only non-zero component of $m$ is $m_-=2$ in the proton rest 
frame, and the only non-zero component of $n$ is $n_+=2$ in the Breit 
frame. Since Breit frame and proton rest frame are connected by a boost 
along the $z$-axis with boost factor $\gamma=Q/(m_{\pro}x)$, the product of 
these vectors is $(mn)=2\gamma$.

Furthermore, two sets of physical polarization vectors, $e_{(i)}$ and 
$\epsilon_{(i)}$ (with $i=1,2$) are defined by the conditions $ek=\epsilon 
k=0$, $e^2=\epsilon^2=-1$, and $em=\epsilon n=0$. An explicit choice, 
written in light-cone co-ordinates, is
\be
e_{(i)}=\Big(\,0,\,\frac{2(k_\perp\epsilon_{(i)\perp})}{k_+},\,
\epsilon_{(i)\perp}\,\Big)\qquad\mbox{and}\qquad \epsilon_{(i)}=\Big(\,
\frac{2(k_\perp \epsilon_{(i)\perp})}{k_-},\,0,\,\epsilon_{(i)\perp}\,\Big)
\, ,\label{eepsdef}
\ee
where the transverse basis $\epsilon_{(1)\perp}=(1,0)$ and 
$\epsilon_{(2)\perp}=(0,1)$ has been used. Note that the above equations 
hold in the proton rest frame, in the Breit frame, and in any other frame 
derived by a boost along the $z$-axis.

These definitions give rise to the two following representations for the 
metric tensor: 
\begin{eqnarray}
g^{\mu\nu}&=&\left(\sum_ie_{(i)}^\mu e_{(i)}^\nu+\frac{m^\mu k^\nu}{(mk)}+
\frac{k^\mu m^\nu}{(mk)}-\frac{m^\mu m^\nu}{(mk)^2}k^2\right)\label{gm}
\\
&=&\left(\sum_i\epsilon_{(i)}^\mu\epsilon_{(i)}^\nu\,+\,\frac{n^\mu k^\nu}
{(nk)}\,+\,\frac{k^\mu n^\nu}{(nk)}\,-\,\frac{n^\mu n^\nu}{(nk)^2}k^2\right)
\, .\label{gn}
\end{eqnarray}

The amplitude for the process in Fig.~\ref{fig:sp}, with the lowest parton 
being a gluon in Feynman gauge, is proportional to
\be
A=\epsilon^\mu(k')\,V_g(k',k)_{\mu\nu}T_H^\nu=\epsilon^\mu(k')\,V_g(k',k
)_{\mu\nu}g^{\nu\rho}g_{\rho\sigma}T_H^\sigma\, ,\label{a1}
\ee
where $V_g^{\mu\nu}$ is the effective vertex for the scattering of the 
gluon off the external field. Next, the first and second metric tensor 
appearing in this expression for $A$ are rewritten according to 
Eq.~(\ref{gm}) and Eq.~(\ref{gn}) respectively. In this situation, only the 
first terms from Eqs.~(\ref{gm}) and (\ref{gn}) contribute at leading order 
in $x$ and $\Lambda/Q$. The intuitive reason for this is the relatively 
small virtuality of $k$, which ensures that for both the hard amplitude 
$T_H$ and the soft scattering vertex $V$ only the appropriately defined 
transverse polarizations are important. 

To see this explicitly, consider the expression 
\[
A=\epsilon V_g\left[\sum e\,e+\frac{m\,k}{(mk)}+\frac{k\,m}{(mk)}-
\frac{m\,m}{(mk)^2}(k^2)\right]\,\left[\sum \epsilon\,\epsilon+\frac{n\,k}
{(nk)}+\frac{k\,n}{(nk)}-\frac{n\,n}{(nk)^2}(k^2)\right]T_H\,,
\]
\be
\label{aina}
\ee
where the appropriate contractions of vector indices are understood. 
Several estimates involving products of $V_g$ and $T_H$ with specific 
polarization vectors will be required. 

Note that both $\epsilon$ and $n$ are ${\cal O}(1)$ in the Breit frame. 
For appropriate polarization, the amplitude $(\epsilon T_H)$ involves no 
particular cancellation, i.e., it has its leading (formal) power behaviour 
in the dominant scale $Q$. Therefore, $(nT_H)$ is not enhanced with 
respect to $(\epsilon T_H)$,
\be
\frac{(nT_H)}{(\epsilon T_H)}\sim{\cal O}(1)\,.\label{r1}
\ee
By analogy, it can be argued that $(\epsilon V_gm)$ is not enhanced with 
respect to $(\epsilon V_ge)$: since both $e$ and $m$ are ${\cal O}(1)$ in 
the proton rest frame and $(\epsilon V_ge)$ has the leading power behaviour 
for appropriate polarizations $\epsilon$ and $e$, the following estimate 
holds,
\be
\frac{(\epsilon V_gm)}{(\epsilon V_ge)}\sim {\cal O}(1)\,.\label{r2}
\ee

Gauge invariance requires $(kT_H)$ to vanish if $k^2=0$. Since the 
amplitude $T_H$ is dominated by hard momenta ${\cal O}(Q)$, and 
$k^2\sim\Lambda^2\ll Q^2$, this leads to the estimate 
\be
\frac{(kT_H)}{(\epsilon T_H)}\sim\frac{k^2}{Q}\,.\label{r3}
\ee
Analogously, from $(\epsilon V_gk)=0$ at $k^2=0$, the suppression of this 
quantity at small virtualities $k^2$ can be derived,
\be
\frac{(\epsilon V_gk)}{(\epsilon V_ge)}\sim\frac{k^2}{k_+}\,.\label{r4}
\ee
For this estimate it is also important that none of the soft scales 
involved in $V_g$, like $k_\perp^2$ or the gauge field $A_\mu$, can appear 
in the denominator to compensate for the dimension of $k^2$. 

All the vector products $nk$, $mk$, $ne$, $m\epsilon$, $mn$, and 
$e\epsilon$ can be calculated explicitly. Using the relations in 
Eqs.~(\ref{r1}) -- (\ref{r4}), it is now straightforward to show that 
$(\epsilon V_ge)(e\epsilon)(\epsilon T_H)$ is indeed the leading term in 
Eq.~(\ref{aina}). The other terms are suppressed by powers of $Q$ or $k_+$. 

Thus, the leading contribution to $|A|^2$, with appropriate polarization 
summation understood, reads 
\be
|A|^2=\!\!\!\!\!\sum_{i,j,i',j',l}\!\left[\Big(\epsilon_{(l)}(k')V_g
e_{(i)}\Big)\Big(e_{(i)}\epsilon_{(j)}\Big)\Big(\epsilon_{(j)}T_H\Big)
\right]\,\left[\Big(\epsilon_{(l)}(k')V_ge_{(i')}\Big)\Big(e_{(i')}
\epsilon_{(j')}\Big)\Big(\epsilon_{(j')}T_H\Big)\right]^*\!\!,
\ee
where the arguments $k$ and $\tilde{k}$ of the polarization vectors in the 
first and second square bracket respectively have been suppressed. 

In the high-energy limit, the scattering of a transverse gluon off an 
external field is completely analogous to the scattering of a scalar or a 
spinor,
\be
\epsilon_{l'}(p')\,V_g(p',p)\,\epsilon_l(p)=2\pi\delta(p_0'-p_0)\,2p_0\,
\delta_{l'l}\,\tilde{U}^{\cal A}(p_\perp'-p_\perp)\,,\label{eve}
\ee
the only difference being the non-Abelian eikonal factor, which is now 
in the adjoint representation. 

In analogy to the spinor case, the polarization sum decouples from the hard 
part for sufficiently inclusive measurements, so that the squared amplitude 
is proportional to
\be
|A|^2=|T_H|^2\,\Big(\epsilon_{(l)}(k')V_ge_{(l)}(k)\Big)\Big(\epsilon_{(l)}
(k')V_ge_{(l)}(\tilde{k})\Big)^*\sum_{i,j}\Big(e_{(i)}(k)\epsilon_{(j)}(k)
\Big)\Big(e_{(i)}(\tilde{k})\epsilon_{(j)}(\tilde{k})\Big)\, .\label{aaf}
\ee
Note that there is no summation over the index $l$. Recall the definition 
of $B$, the soft part of the amplitude square, given at the beginning of 
the last section and illustrated in Fig.~\ref{fig:sp}. The corresponding 
expression in the case of a soft gluon can now be read off from 
Eqs.~(\ref{eve}) and (\ref{aaf}): 
\be
B_g=k_+^2\sum_{i,j}\Big(e_{(i)}(k)\epsilon_{(j)}(k)\Big)\Big(
e_{(i)}(\tilde{k})\epsilon_{(j)}(\tilde{k})\Big)\, .
\ee
This is further evaluated using the explicit formulae in 
Eq.~(\ref{eepsdef}) and the identity 
\be
\sum_{i}\epsilon_{(i)\perp}^a\epsilon_{(i)\perp}^b=\delta^{ab}\qquad
(a,b\in\{1,2\})\, .
\ee
Comparing the resulting expression, 
\be
B_g=k_+^2\left(\delta^{ij}+\frac{2k_\perp^ik_\perp^j}{k_\perp'^2}\left( 
\frac{1-\beta}{\beta}\right)\right)\left(\delta^{ij}+\frac{2\tilde{k}_\perp^i
\tilde{k}_\perp^j}{k_\perp'^2}\left(\frac{1-\beta}{\beta}\right)\right)\,,
\ee
to Eq.~(\ref{bsc}), the diffractive gluon distribution of Eq.~(\ref{fxg}) 
is obtained. 

Note that the factor $N_c$ appearing in the denominator of Eq.~(\ref{fx}) 
has been replaced by the dimension of the adjoint representation, 
$N_c^2-1$. 

\newpage

\mychapter{Inclusive Parton Distributions}\label{sect:isf}
This appendix is devoted to the calculation of inclusive parton 
distributions and inclusive structure functions in the semiclassical 
framework. Inclusive DIS was discussed in Sect.~\ref{sect:qq}, where the 
cross section for $q\bar{q}$ production off a colour field was derived, at 
the end of Sect.~\ref{sect:trf}, where the parton model interpretation of 
this cross section was outlined, and in Sect.~\ref{sect:edsf}, where it was 
part of the semiclassical analysis of diffractive and inclusive structure 
function data. Here, the above scattered information is collected, and a 
more coherent account, including further technical details, is given 
(cf.~\cite{bgh}). 

Recall the leading twist cross section $\sigma_T$ for $q\bar{q}$ pair 
production by a transversely polarized photon obtained in 
Sect.~\ref{sect:qq} (cf.~Eq.(\ref{stt})). The corresponding transverse 
structure function reads 
\bea
F_T(x,Q^2)&=&{4\over 3(2\pi)^3 }\left(\ln\frac{Q^2}{\mu^2}-1\right)\ 
\int_{x_{\perp}}\mbox{tr}\left(\partial_{\y}W_{\x}(0)
              \partial_{\y}W_{\x}^\dagger(0)\right)
\nonumber \\ \nonumber \\
&&+{2\over (2\pi)^7 }\int_0^{\mu^2}dN^2\int
d k_\perp'^2 \int_{x_{\perp}} \left|\int d^2 k_\perp \frac{k_\perp 
\tilde{W}_{x_\perp}(k_\perp'-k_\perp)}{N^2 + k_\perp^2}\right|^2\!\!.
\label{ftsc}
\eea
To map this calculation onto the conventional parton model framework, 
identify the result as $F_T(x,Q^2)=2xq(x,Q^2)$. The corresponding quark 
distribution reads 
\bea
xq(x,Q^2) &=& {2\over 3(2\pi)^3}\left(\ln\frac{Q^2}{\mu^2}-1\right)
              \int_{\x} \mbox{tr}\left(\partial_{\y}W_{\x}(0)
              \partial_{\y}W_{\x}^\dagger(0)\right)
\nonumber\\
\nonumber\\
            &&+{2\over (2\pi)^4}\int_0^{\mu^2} N^2\,dN^2 \int_{\y} 
              K_1(y N)^2\int_{\x} \mbox{tr}\left(W_{\x}(\y)
              W_{\x}^\dagger(\y)\right)\;.\label{qi}
\eea
Here the modified Bessel function $K_1$ has been introduced so that, in 
both terms, the functions $W$ appear in co-ordinate space. This makes it 
particularly clear that the first term is only sensitive to the short 
distance behaviour, while the second term depends on the non-perturbative 
long-distance structure of the colour field. Note that the sum of both 
terms is independent of $\mu^2$. 

The corresponding gluon distribution at small $x$ is most easily calculated 
as 
\be
xg(x,Q^2) = \frac{3\pi}{\alpha_s }\,\cdot\,\frac{\partial F_T(x,Q^2)}
{\partial\ln Q^2} = {1\over 2\pi^2 \alpha_s}\int_{\x} \mbox{tr}\left(
\partial_{\y}W_{\x}(0)\partial_{\y}W_{\x}^\dagger(0)\right)
\,.\label{gi}
\ee
Equations (\ref{qi}) and (\ref{gi}) can serve as the starting point for a 
conventional partonic analysis of inclusive DIS. 

To gain more physical insight into the correspondence of the semiclassical 
and the parton model approach, return to the starting point, 
Eq.~(\ref{ftsc}). It is instructive to view $F_T$ as a sum of two terms: 
$F_T^{\,\mbox{\scriptsize asym}}$, the contribution of asymmetric 
configurations where quark or antiquark are slow, $\alpha < \mu^2/Q^2$ or 
$1\!-\!\alpha < \mu^2/Q^2$ (Fig.~\ref{fig:f2}a), and $F_T^{\,
\mbox{\scriptsize sym}}$, the contribution of symmetric configurations 
where both quark and antiquark are fast, $\alpha, 1\!-\!\alpha > \mu^2/Q^2$ 
(Fig.~\ref{fig:f2}b). In a frame where the proton is fast, say, the Breit 
frame, the asymmetric and symmetric contribution to $F_T$ correspond to 
photon-quark scattering and photon-gluon fusion respectively. 

The symmetric part is dominated by small $q\bar{q}$ pairs, i.e., by the 
short distance contribution to the Wilson-loop trace, 
\be 
\int_{\x} \mbox{tr}\left(W_{\x}(\y)
              W_{\x}^\dagger(\y)\right)
= {1\over 2} y^2 \int_{\x} \mbox{tr}\left(\partial_{\y}W_{\x}(0)
              \partial_{\y}W_{\x}^\dagger(0)\right)
  +{\cal O}(y^4)\; .
\ee
The corresponding contribution to the structure function is related to the 
first term on the r.h. side of Eq.~(\ref{qi}), which generates the gluon 
distribution, Eq.~(\ref{gi}). It can also be written as 
\be
F_T^{\,\mbox{\scriptsize sym}}(0,Q^2) = {e_q^2\over 2\pi^3} \int_0^1 dz 
P_{qg}(z)\left(\ln{Q^2\over \mu^2} - 1\right)
\int_{\x} \mbox{tr}\left(\partial_{\y}W^{\cal F}_{\x}(0)
\partial_{\y}W_{\x}^{{\cal F} \dagger}(0)\right)\,,
\ee
where $P_{qg}(z)$ is the conventional gluon-quark splitting function. 

The other splitting functions appear if $\alpha_s$ corrections to $F_T$ and 
$F_L$, associated with higher Fock states of the virtual photon, are 
considered in the semiclassical approach. For example, the $q\bar{q}g$ 
parton configuration involves, in the case where one of the quarks carries 
a small fraction of the photon momentum, a $\ln Q^2$ term associated with 
$P_{qq}(z)$. 

The splitting function $P_{gg}(z)$ is most easily derived by considering 
an incoming virtual scalar which couples directly to the gluonic action 
term $F_{\mu\nu}F^{\mu\nu}$. Such a current was used previously 
in~\cite{mues} to study small-$x$ saturation effects. The relevant lowest 
order Fock state consists of two gluons. As reported in~\cite{bgh}, it can 
be checked explicitly that the semiclassical calculation of the 
corresponding high-energy scattering process yields the usual gluon-gluon 
splitting function. 

Since the semiclassical approach exactly reproduces the well-known DGLAP 
splitting functions, the large logarithms $\ln\left(Q^2/\mu^2\right)$ can 
be resummed in the conventional way, by means of the renormalization 
group. To this end, the parton distributions $q(x,Q^2)$ and $g(x,Q^2)$ are 
evaluated using DGLAP evolution equations, with the input distributions 
$q(x,Q_0^2)$ and $g(x,Q_0^2)$ given by Eqs.~(\ref{qi}) and (\ref{gi}). 
Here $Q_0^2$ is some small scale where logarithmic corrections are not yet 
important. The parton model description of the structure function at 
leading order includes only photon-quark scattering. The leading 
logarithmic term from the photon-gluon fusion process appears now as part 
of the resummed quark distribution. 

The large hadron model of Sect.~\ref{sect:lh} provides an expression for 
the basic function $\int_{x_\perp}\mbox{tr}\left(W_{\x}(\y)W_{\x}^\dagger 
(\y)\right)$. Inserting the explicit formula of Eq.~(\ref{ww0}) into 
Eqs.~(\ref{qi}) and (\ref{gi}), the following compact expressions for the 
inclusive parton distributions at a  low scale $Q_0^2$ are obtained 
\pagebreak[0]
\begin{eqnarray}
xq(x,Q_0^2) & = & \frac{a \Omega N_c}
{3 \pi^3} \left(\ln\frac{Q_0^2}{a} - 0.6424\right)\; , \label{qiinp}\\
xg(x,Q_0^2) & = & \frac{ 2 a \Omega N_c}
{\pi^2 \alpha_s(Q_0^2)}\; \label{giinp}. 
\end{eqnarray}

\begin{figure}[t]
\begin{center}
\parbox[b]{15.5cm}{\psfig{file=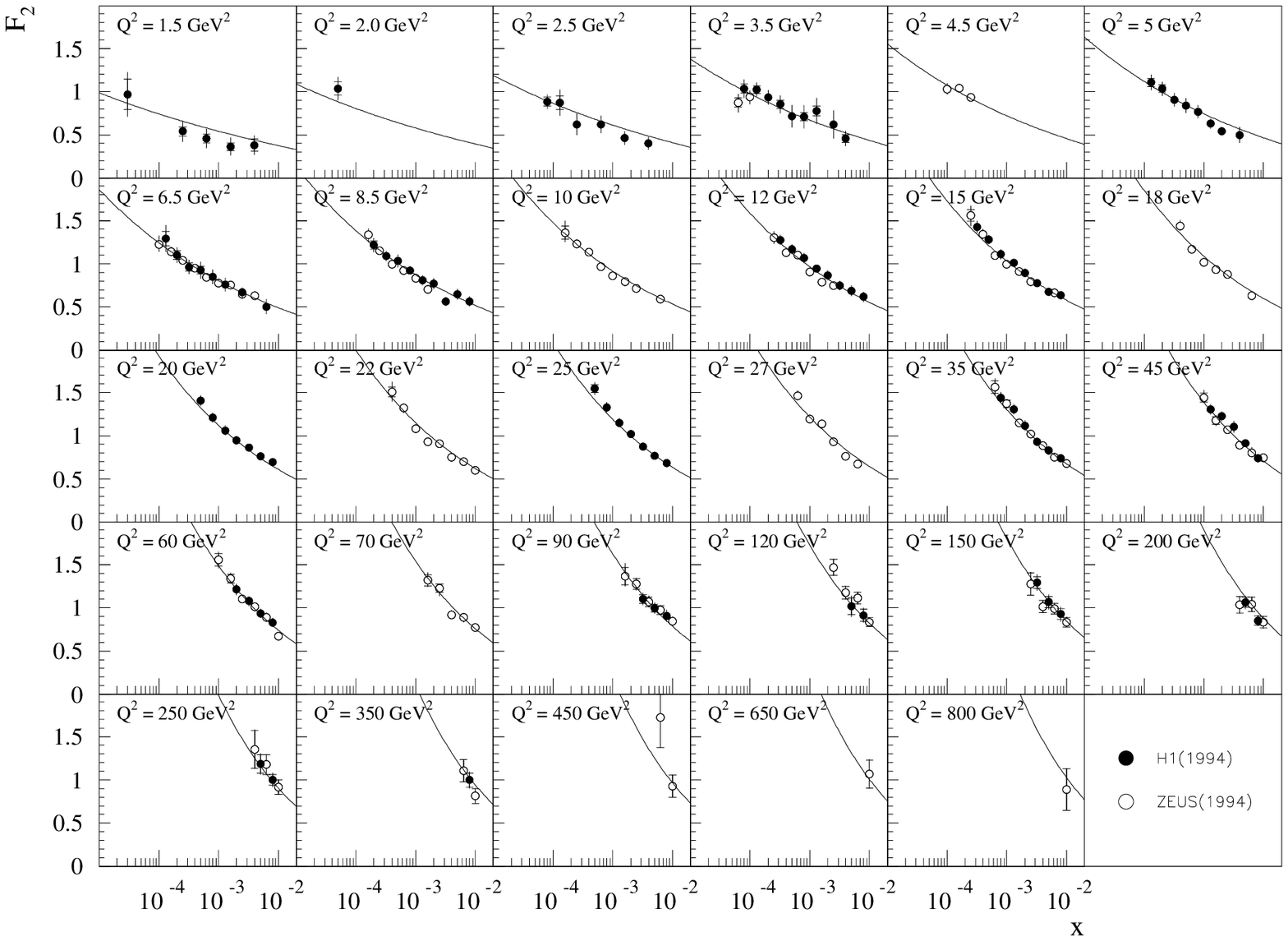,width=15.5cm}}\\
\end{center}
\refstepcounter{figure}
\label{fig:f2i}
{\bf Figure \ref{fig:f2i}:}
The inclusive structure function $F_2(x,Q^2)$ at small $x$ computed in the 
semiclassical approach with data from~\cite{incl}. The data with $Q^2 = 
1.5$~GeV$^2$ are not included in the fit.
\end{figure}

As discussed in Sect.~\ref{sect:edsf}, the analysis of \cite{bgh} introduces 
a soft energy dependence into these input distributions by ascribing a 
logarithmic growth to the total normalization, 
\begin{equation}
\Omega \to \Omega \left(L - \ln x \right)^2.
\end{equation}
Note also the observation of \cite{bha} that the small-$x$ structure 
function is well described by a simple $\ln(1/x)$ with an additional 
$\ln Q^2$ enhancement, similar to the effects of Altarelli-Parisi 
evolution. 

The DGLAP evolution of the above inclusive parton distributions and 
corresponding diffractive distributions given in Sect.~\ref{sect:edsf} 
provides predictions for both the inclusive and diffractive structure 
functions. In the numerical analysis, the inclusive distributions are 
multiplied with $(1-x)$ to ensure vanishing of the distributions in the 
limit $x\to 1$, which is required for the numerical stability of the DGLAP 
evolution. 

A combined fit to small-$x$ HERA data gives a good description of the 
experimental results. While plots of $F_2^{D(3)}$ are given 
Figs.~\ref{fig:f2dh1} and \ref{fig:f2dzeus}, here, corresponding results for 
the inclusive structure function are presented (see Fig.~\ref{fig:f2i}). 
Since the underlying model is only valid in the small-$x$ region, data 
points above $x=0.01$ are not considered. To appreciate the quality of the 
fits, recall that, within the large hadron model, the diffractive 
(Eqs.~(\ref{qdinp}) and (\ref{gdinp})) and inclusive (Eqs.~(\ref{qiinp}) 
and (\ref{giinp})) parton distributions are highly constrained with respect 
to each other. 

\newpage
\end{appendix}

\end{document}